# Long-term Variations of Cyclone-Influenced Exchange Flows in the Coastal Arctic in 2000-2025

*Chunyan Li[1,2]

[1]Department of Oceanography and Coastal Sciences, Louisiana State University, Baton Rouge, LA, U.S.A.

[2]Coastal Studies Institute, Louisiana State University, Baton Rouge, LA, U.S.A.

*Corresponding: cli@lsu.edu

## Abstract

Summertime synoptic weather regulates Arctic land-ocean exchange through meteorological tides during the ice-free period, when the predominantly easterly regime of the Beaufort High is punctuated by episodic Arctic cyclones, changing coastal winds to westerlies. We analyze wind-driven exchange flows in a northern Alaska lagoon using a 26-year reconstruction (2000–2025) validated by new *in situ* observations and FVCOM simulations. Phase-space EOF analysis of ERA5 data links deep cyclonic anomalies to 89% of Extreme Eastward Throughflow (EET) events (24 of 27). Coastal easterly occurrence frequencies show a consistent decrease (up to 38.3% at > 10 m/s) while low-to-moderate westerly occurrence frequencies increase by up to 30.9% but decrease by 40.6% at extreme westerly wind speeds (>10 m/s). Offshore ERA5 wind occurrence frequencies increased, with extreme offshore events (> 10 m/s) increasing by 82.4%. Local ASOS observations reveal a near-monotonic ~88% reduction in atmospheric forcing variability (East-West Wind Excursion variability drops from $2.98\times10^4$ to $0.35\times10^4$ km). The In-Out Flow Excursion variability declines steadily by 70% from $1.40\times10^3$ to $<0.40\times10^3$ km, while the positive (outward) flow ratio variability contracts steadily from 7.4% to below 2.2%. Extreme Westward Throughflow (EWT) events were infrequent, but were concentrated in the most recent period, with five of six events occurring in 2022-2024 and one in 2013. These results indicate that the summertime coastal regime remains predominantly easterly, but exhibits progressively lower variability, while the wind distribution shifts toward more frequent low-to-moderate westerlies. This evolution of weather-forcing state may impact flushing and biogeochemical connectivity of Arctic lagoon ecosystems.

## Introduction

Land-ocean exchange governs the global transport of water, sediments, and nutrients through terrestrial and aquatic boundaries[1], sustaining fundamental Earth system cycles. Governed primarily by estuarine and lagoonal hydrodynamic processes, this cross-shelf transport controls bay flushing rates, and associated regional biogeochemical dynamics[2]. The main drivers include river discharge, tides, and weather.

Among global coastal systems, land-ocean exchange plays a critical role in the Arctic[3], where it influences freshwater content[4-9], coastal geomorphology, and nutrient cycling[10,11]. Unlike temperate regions, Arctic coastal exchange remains understudied due to harsh environmental conditions that severely hinder field observations and data collection. However, the Arctic amplification - the warming of the region at a rate much faster than the global average[12-15] - is altering the Arctic hydrodynamics. Key drivers include lengthening ice-free season[16,17], intensifying storm activity[18,19] associated with the Arctic cyclones[20] and synoptic weather variations[21], and shifting heat budget[22], all of which impact coastal circulation and cross-shelf exchange processes.

The scarcity of long-term in situ measurements negatively impacts the ability to quantify these changes and predict the consequences. Understanding the Arctic land-ocean exchange is essential for assessing ecosystem responses, improving climate models[19,23,24], and addressing the broader implications of a rapidly warming Arctic[15].

Tidal forcing across much of the Arctic is micro-tidal. For example, the Beaufort Sea along the Alaskan and Canadian coastlines has a tidal range of only 0.2 to 0.5 m[25,26], and the Chukchi Sea is also mostly micro-tidal. Similarly, the Laptev, Kara, and East Siberian Seas experience predominantly micro-tidal conditions[26,27]. This minimal tidal influence leaves weather-driven atmospheric forcing as the primary mechanism governing land-ocean exchange, particularly during ice-free period. As Arctic amplification of climate change continues[15,19,23,24,28], atmospheric forcing is becoming an even more significant driver of evolving Arctic coastal exchange dynamics.

While numerical modeling and observational research on the Arctic ocean hydrodynamics have increased in recent decades[29,30], most studies focus on deep-water processes. Coastal studies have primarily examined short-term exchange events, such as the correlation between exchange flows and passing Arctic cyclones[31,32], which transiently replace the more persistent Beaufort High pressure system, which is associated with the Beaufort Gyre in the ocean, and is one of the major components of the Arctic climate variability across seasonal to decadal scales[33]. These lagoons serve as essential biogeochemical filters[34] and critical habitats for species like Arctic cod[35,36], yet their long-term response to changing weather patterns remains a significant knowledge gap.

The strong correlation between exchange flows and wind velocity[31] provides a theoretical basis for examining long-term variations. Specifically, the exchange flow can be expressed as a functional mapping[37] $Q$:

$$Q = \mathcal{F}(w) \tag{1}$$

where $w$ represents weather information, and $\mathcal{F}$ denotes a mathematical operator representing the ocean response characterized by exchange flow dynamics. This relationship (Equation (1)) emerges from coastal hydrodynamics under atmospheric forcing, which can be modeled with weather as the input and exchange flow as the output (more details in the Method section). A validated form of this mapping allows for the retrospective estimation of past exchange flows, providing an effective tool for assessing long-term climate-driven oceanic responses.

The goal of the present study is to use observational data and an empirical regression model to shed light on the long-term variability of exchange flows in an Arctic lagoon over the past 26 years (2000-2025). Water velocity time series were measured using a bottom-mounted acoustic Doppler current profiler (ADCP) deployed at the deep channel of Eluitkak Pass, located approximately 17 km northeast of the Automated Surface Observing System (ASOS) weather station at the Wiley Post-Will Rogers Memorial Airport in Utqiaġvik (formerly Barrow), Alaska. The regression model used here is derived from a Taylor series expansion of the solution from the 2D depth-averaged momentum equations, with coefficients determined by empirical regression against observations and validated by additional observations and the 3D FVCOM hydrodynamic simulations. This approach offers a computationally efficient alternative to full multi-decadal numerical simulations. This formulation approach has been previously validated across similar multi-inlet lagoonal systems using extensive multi-year field arrays and high-resolution numerical model simulations[38]. Our primary focus is to evaluate how long-term shifts in the Arctic Beaufort High and passing Arctic cyclones have altered lagoon-ocean exchange flows over the 26-year record.

Our study site is a coastal lagoon (bounded roughly by 156˚36’W, 155˚W, 70˚48’N, and 71˚23’N) at the southern boundary of the Beaufort Sea in the northernmost Alaska. The bay has a flipped “L” shape, with each leg spanning ~55 km (Fig. 1a). The outer bay is enclosed by a series of barrier islands, separated by tidal inlets of varying widths and depth (Fig. 1b). While much of the bay and its tidal inlets are very shallow (1–2 m), the northwestern Eluitkak Pass is approximately 400 m wide and 16 m deep (Fig. 1c) which is the deepest of the entire lagoon. This dramatic bathymetric contrast makes Eluitkak Pass the primary conduit for water exchange driven by transient weather systems, especially given the region's minimal 0.2 m tidal range[31].

## Results

### *Wind Regimes at the Local ASOS Station at Utqiaġvik*

The dominant local wind at the nearby airport of Utqiaġvik, 17 km southwest of the Eluitkak Pass (Fig. 1) is easterly, as a result of the quasi-permanent presence of the Beaufort High atmospheric system at sea-level. This is the case year-round, with more variations in the summer (July-August). It is shown in Figure 2a that the averaged December wind-rose is dominated by a single mode: the northeast wind, indicative of the dominance of the Beaufort High over the Beaufort Sea area, north of the northern coast of Alaska. In contrast, the summertime wind-rose is bi-modal (Fig. 2b): easterly wind (still dominant) due to the Beaufort High and westerly wind due to transient Arctic summertime cyclones which produce wind from the western quadrants if the cyclone is located north of the coastline of the northern Alaska (replacing or displacing the Beaufort High). It has been shown in earlier work[31,32] that easterly wind under Beaufort High generates westward throughflow (Fig. 2c) in 60-80% of the time, punctuated by sporadic Arctic cyclone induced eastward throughflow (Fig. 2d).

The overall summertime East-West Wind Excursion (EWWE, see Method section) in the atmosphere is negative (more influence of easterly winds, consistent with high pressure dominated situation) as can be seen from Table 1 and Supplemental Materials (Fig. S2-S56 and Section S6). It is obvious that the EWWE computed from the ASOS weather station data fluctuates more in the early years of the 26-year study period in this century and has been continuously approaching a more stable state over the years (Fig. 3 and Supplemental Materials Section S6). To compare the variations during the first half and latter half of this period, the EWWE in the first 13 years has an average of $-7.3 \times 10^4$ km with a standard deviation of $2.5 \times 10^4$, while the last 13 years has a smaller magnitude average of $-6.2 \times 10^4$ with much reduced standard deviation of only $1.1 \times 10^4$ (Table 1, Supplemental Materials Section S6).

*Wind-driven Velocity at Eluitkak Pass*

The summertime (July–August) along-channel subtidal velocity at the Eluitkak Pass was reconstructed for the 26-year period from 2000 to 2025 (Figs 4–5) by using the wind velocity time series from Utqiaġvik and the regression model (see the Method section for details). A Fourier low-pass filter[39] with a 30-hour cutoff (0.8 CPD) was applied to the wind velocity time series from Utqiaġvik and this low-pass filtered wind data is provided as the input to compute the wind-driven along-channel velocity using the regression model[31]. This subtidal along-channel velocity at the Eluitkak Pass typically ranges between $-0.5$ and 0.5 m/s, where positive values denote outward flow from Elson Lagoon to the Beaufort Sea and negative values denote inward flow into the coastal lagoon at the Pass.

An extreme event is defined as occurring when the low-pass filtered along-channel velocity magnitude at the Eluitkak Pass exceeds a threshold of 0.5 m/s. Over the 26-year study period, 33 such events are identified (Table 2 and Figs 4-5), averaging ~ 1.3 events per year. These extremes are strongly coupled with the positioning of the sea-level atmospheric pressure centers:

For Extreme Inward Flows at the Eluitkak Pass (27 out of the 33 events): The strongest such event was in 2000 when the low-pass filtered flow velocity reached close to 2 m/s (Fig. 4). Most summertime extreme inward flow events at the Eluitkak Pass occurred when an Arctic cyclone was centered north of the Alaskan coastline or near the Canadian Archipelago, generating winds from the western quadrants (i.e. westerly, northwesterly, or southwesterly winds) at the lagoon (Fig. 6a and Supplemental Materials, Fig. S2-S56).

For Extreme Outward Flows at the Eluitkak Pass (6 of the 33 events)**:** The summertime outward extremes at the Eluitkak Pass occurred when a high-pressure center was positioned north of the Alaskan coastline over the Beaufort Sea, resulting in wind at the coast from the eastern quadrants (i.e., northeasterly, easterly, or southeasterly winds, Fig. 6b and Supplemental Materials, Fig. S2-S56). Note that the extreme outward flow at the Eluitkak Pass is indicative of an Extreme Westward Throughflow (EWT) of the Elson Lagoon[31] (Fig. 2c), and the extreme inward flow at the Eluitkak Pass is associated with an Extreme Eastward Throughflow (EET) of the Elson Lagoon[31,32] (Fig. 2d).

*Variability of Extreme Events*

Over the entire study period (2000-2025), the annual average number of extreme events is 1.3 with a standard deviation of 1.3 (Table 1). During the first half of the 26-year study period, there were no Beaufort High related extreme events and the average is 1.1 events per year with a standard deviation of 1.5. In contrast, the average number of extreme events in the second half of the 26-year study period is 1.5 events per year with a standard deviation of 1.1 (Table 1). There has been an increase in average events but a decrease in standard deviation of extreme events, consistent with the variations in the standard deviation of the indices (see discussion below). It should be noted that excluding all positive extreme flow events at the Eluitkak Pass under Beaufort High's influence in the last half of the 26-year period, the average extreme inward flow events at the Eluitkak Pass is 1 per year with an even more reduced standard deviation of 0.7 per year, which is ~ half of that of the first half of the 26-year period. All of this points to a continuous regime evolution over the 26-year period.

A notable evolution is seen between the westward and eastward throughflows in the lagoon or, equivalently, the inward and outward extreme flows at the pass. Historically, extreme inward flows (and EET) dominate. In fact, the first extreme outward flow (and EWT) event in the record did not occur until July 9, 2013. While these events appear to occur randomly, the later years of the study period show a clustering of occurrences: five of the six extreme outward events at the Pass recorded since 2000 took place in 2022 (1 event), 2023 (1 event), and 2024 (3 events), making all six such extreme events occur in the second half of the study period (Table 2).

This evolution is also captured by the Extreme Flow Index (EFI, Table 1), an annually integrated measure of extreme event magnitude (see Method section). The mean EFI is $-32.1 \pm 54$ km, indicating both high interannual variability and a bias toward inward flux at the Eluitkak Pass and associated EET in the lagoon driven by westerly winds from Arctic cyclones. The years 2022 and 2024 were notable exceptions where the EFI turned positive, as most extreme events in those years were driven by high-pressure systems. The year 2023 has an EFI of $-8.8$ km because two opposing extreme events nearly neutralized each other, and 2025 recorded an EFI of 0 as no extreme events occurred.

The first half of the 26-year period has an average EFI of $-43.9$ km and a standard deviation of $65.3$ km (Table 1), both having much greater magnitude than the corresponding overall 26-year statistics. For the second half, the average EFI of $-20.4$ km has a greatly reduced magnitude with a standard deviation of 38.9, only ~60% of the first 13 years. This again shows the reduction in standard deviation over time.

*Integrated Exchange: IOFE and PR*

Beyond extreme events, the overall assessment of the summertime exchange flow in the lagoon water is characterized by the In-Out Flow Excursion (IOFE, see Method section) at the pass. The IOFE remained positive (outward excursion) for every summer in the study period, with an average value of $4.1 \pm 0.8$ ($10^3$ km). The relatively small standard deviation highlights the robustness of this result and the overall dominance of the Beaufort High system. While sporadic Arctic cyclones can reverse the flow, the persistent influence of the Beaufort High ensures that the net water particle excursion is outward-dominated at the Eluitkak Pass and thus westward throughflow dominated in the Elson Lagoon (Fig. 2c,d). For the first half of the study period, the IOFE averaged to 4.1 ($10^3$ km) with a standard deviation of 1.0 ($10^3$ km), while the second half averaged to a similar value of 4.2 ($10^3$ km) but with a half standard deviation of only 0.5 ($10^3$ km) (Table 2). This is consistent with the EWWE discussed earlier.

Correspondingly, the Positive Ratio (PR)—the ratio between the time of positive (outward) and negative (inward) low-pass filtered flows in the water at the Eluitkak Pass —was consistently greater than 70% for all years. The mean PR was approximately $78\% \pm 5\%$, showing remarkably low variability. The minimum PR occurred in 2002 (70%) and the maximum in 2007 (90%). While the overall condition remains stochastic without a clear linear trend, the recent shift toward positive EFI values in 2022 and 2024 shows that more extreme events are contributing to the westward throughflow phase of the Elson Lagoon. Although this parameter had small variability, the first half and second half of the period still show differences: for the first half, the PR has an average of $\sim80\%$ and a standard deviation of $\sim6\%$. In contrast, the second half has an average of $\sim77\%$ and a standard deviation of only $\sim2\%$ (Table 2). Again, the variability in the second half is reduced, representing a mere 1/3 of the first half.

The 13-year partition is for illustration purposes. The long-term change is rather continuous and almost monotonic. More specifically, a five-year moving (or rolling) standard deviation is computed. It reveals a striking, continuous decline in the variability across all three indices (EWWE, IOFE, and PR) throughout the 26-year record (Fig. 7 and Supplemental Materials Section S6). For the East-West Wind Excursion, the standard deviation progressively drops from an early peak of $2.98 \times 10^4$ km down to $0.35 \times 10^4$ km in the final five-year period, marking an approximate 88% reduction in atmospheric forcing variability. The standard deviation of the In-Out Flow Excursion experiences a parallel continuous decline from $1.40 \times 10^3$ km to under $0.40 \times 10^3$ km (over a 70% reduction), while the variability of the positive flow ratio contracts steadily from 7.4% down to below 2.2%. This result indicates that the local wind on the coast has reduced its variability substantially almost monotonically this century (Fig. 7).

*Major Modes and Patterns from EOF Analysis*

The first EOF mode (EOF 1[40,41]) of the sea level air pressure anomaly from the ERA5 reanalysis data, accounting for 65.9% of the variance (Fig. 8a), represents the spatial pattern that accounts for the maximum possible variance in the Arctic SLP anomaly field. This mode shows a center of sea-level air pressure anomaly over the Beaufort Sea region, quantifying the departure of the Beaufort Sea High from its mean with alternating positive and negative deviations. The associated time series (PC1) is reconstructed as:

$$PC_1(t) = \mathbf{u}_1 \sigma_1 \qquad (2)$$

where $\sigma_1$ is the first singular value of the data matrix $\mathbf{X}$ (see Method section for details) - it is a scalar representing the amplitude or strength of the mode. Its square is proportional to the variance explained; $\mathbf{u}_1$ in Equation (2) is a dimensionless, unit vector representing the temporal

evolution – it is the first left singular vector derived from the Singular Value Decomposition of the data matrix in the EOF analysis[40].

In this study, the PC1 (Fig. 9) amplitude is expressed in physical units (hPa) for convenience. Positive values of PC1 correspond to a strengthening of the spatial pattern of EOF1, while negative values represent the reversed pattern of EOF1. EOF1 exhibits a uniform negative structure across the Beaufort Sea (Fig. 8a). Consequently, a positive PC1 amplitude corresponds to a negative pressure anomaly (cyclonic biased conditions), while a negative PC1 corresponds to a positive anomaly (anticyclonic biased conditions). This relationship is consistent with our observations, where high positive values of PC1 (Fig. 9) coincide with the 27 identified extreme inward exchange events at the Eluitkak Pass or Extreme Eastward Throughflow inside the lagoon driven by cyclonic forcing.

EOF2 (explaining 16.7%) exhibits a zonal structure with contours roughly oriented parallel to latitude (Fig. 8b), representing fluctuations in the regional geostrophic zonal wind. In contrast, EOF3 (explaining 8.3%) captures the meridional variability (Fig. 8c) and the longitudinal migration of pressure centers across the Beaufort Sea shelf. These modes express the change in the Beaufort High position and occurrence of summertime Arctic cyclone which move into the Beaufort Sea, replacing or displacing the more persistent Beaufort High. The fact that EOF1 explains ~ 66% of the variability indicates that the dominant factor is whether the Beaufort High or an Arctic cyclone is occupying the Beaufort Sea. EOF2 and EOF3 are mainly modifying the positions of the center of high or low.

*Phase-Space Analysis*

The PC1-PC2 phase-space analysis (Fig. 10) reveals a high degree of dynamical separation between the two exchange regimes, demonstrating that hydrodynamic direction is a deterministic response to the synoptic state. A natural way to characterize these drivers is the examination of the four quadrants of the PC1-PC2 phase-space based on the reinforcement of the leading EOF modes:

Quadrant 1 (Q1; PC1 > 0, PC2 > 0): represents the deep cyclonically biased (ocean-low/land-high) regime. Here, the positive polarities of both PCs reinforce a low-pressure center over the Beaufort/Chukchi shelf, creating the steepest ocean-low/land-high gradient and driving coastal westerlies.

Quadrant 2 (Q2; PC1 < 0, PC2 > 0): a weaker, transitional regime where opposing PC signs result in a hybrid or shifted pressure center. In this case, EOF1 and EOF2 are in their negative and positive phases, respectively, creating a partial offset between the two modes, especially over the central Beaufort Sea region away from the coast (Fig. 8a,b).

Quadrant 3 (Q3; PC1 < 0, PC2 < 0): represents the deep anticyclonically biased (ocean-high/land-low) regime. The negative reinforcement of both modes strengthens the Beaufort High over the ocean, driving the easterly anomalies responsible for extreme outward flow at the pass.

Quadrant 4 (Q4; PC1 > 0, PC2 < 0): another weaker regime characterized by a pressure gradient where the cyclonic tendency of PC1 and the anticyclonic tendency of PC2 are partially neutralized between the two, as EOF1 and EOF2 are in their positive and negative phases, respectively.

Mapping the exchange events onto this space confirms a binary forcing structure. Of the 27 events with inward flow at the Eluitkak Pass and Extreme Eastward Throughflow inside the lagoon, 24 (89%) are clustered within the land-high westerly domain of Q1, with only three events falling into the transitional regimes of Q2 and Q4. Notably, the EET events exhibit tight clustering within the positive PC2 domain, underscoring the necessity of the latitudinal pressure squeeze to drive the EET.

Conversely, the Eluitkak Pass' outward (or lagoon's EWT) events form a distinct cluster primarily within the negative PC2 domain (Quadrants 3 and 4). While Q3 hosts two of the six extreme ocean-high outward pulses at the Eluitkak Pass (corresponding to the EWT), the distribution of four events in Q4 reflects a gradient that continues to drive significant seaward hydrodynamic responses. This spatial partitioning suggests that while the system is transitioning toward a Q1 baseline, it remains highly responsive to the high-amplitude anticyclonic forcing represented by Q3 and Q4.

It is worth noting that while large-scale atmospheric teleconnections—such as the Arctic Oscillation (AO), Pacific Decadal Oscillation (PDO), or North Atlantic Oscillation (NAO)—modulate the broad-scale background positioning of Arctic storm tracks and the Beaufort High, their local hydrodynamic influence on coastal lagoons is expressed directly through these regional sea-level pressure phase-space trajectories. By mapping synoptic weather states into the PC1–PC2 phase space, this diagnostic implicitly integrates the localized atmospheric pressure gradients resulting from broader climate modes, providing a direct deterministic mapping from large-scale atmospheric configurations to local coastal momentum balances.

*Mean occurrence frequency variation*

The comparison of mean annual wind occurrence frequency between the first (2000–2012) and second (2013–2025) half of the 26-year study period in the region reveals a significant change over time of the wind field (Table 3). Note that the comparison between the first and second 13-year periods in Table 3 is for convenience of discussion. The year-to-year variations of the occurrence frequencies from the ERA5 reanalysis data are presented in Fig. 11 for both westerlies and easterlies in the coastal region (Fig. 11a) and the wind magnitude for the offshore Beaufort Sea region (Fig. 11b) at all wind threshold values. In the offshore wind occurrence frequency analysis, only wind magnitude is used because the Beaufort Sea region is large enough to encompass frequent opposing wind directions, such as those from the rotating Beaufort High pressure system or a passing Arctic cyclone. The upward swing of the coastal westerlies and downward swing of coastal easterlies in recent years are remarkable at most wind threshold values (Fig. 11a,b). These changes suggest an evolution in the atmospheric forcing that drives hydrodynamic exchange at Eluitkak Pass and coastal lagoon:

The most striking trend is the consistent decline in the occurrence frequency of coastal easterly winds across all intensity thresholds (the "Change of the average" column of the top 10 lines in Table 3). While low-magnitude easterlies (> 1 m/s) decreased by 12.1% (Table 3), the reduction becomes more pronounced at higher intensities. Notably, extreme easterly winds (> 10 m/s) have seen a dramatic 38.3% reduction in annual frequency. This suggests that the ocean-high (anticyclonic) regime, which dominates the Beaufort Sea shelf and drives outward flow at the Eluitkak Pass and westward throughflow in Elson Lagoon, now has less influence along the coastal northern Alaska in the recent decade.

In remarkable contrast to the easterlies, coastal westerly winds have seen an increase in occurrence frequency for most thresholds. Within the moderate intensity range (1 to 9 m/s) in Table 3, the frequency of westerlies has increased by up to 31%. However, a distinct tipping point is observed at the extreme end of the spectrum: the most extreme westerly winds (> 9 m/s and > 10 m/s) have actually decreased by 31.0% and 40.6% in occurrence frequency, respectively. This indicates that while the *average* forcing for inward flow at the Eluitkak Pass and eastward throughflow in the Elson Lagoon are more frequent, the absolute peak-velocity events are moderating at the coast. The reduction of EFI magnitude is consistent with the reduction in occurrence frequency of strong westerly winds at > 9 m/s threshold (Table 3). This, however, is not in contradiction with the increased occurrence frequency of low-to-moderate westerly winds (1-9 m/s, Table 3), which is consistent with the reduced EWWE.

The fact that extreme coastal easterly occurrence frequencies declined substantially across the epochal comparison (Table 3) indicates that the recent cluster of outward-flow (EWT) events does not reflect a sustained upward trend, but rather episodic interannual fluctuations operating within an overall declining easterly climatology in the coastal region.

*Variability Analysis*

Consistent with the analysis of the local ASOS observations (Figs 3,7 and Table 1), the analysis of annual standard deviation (StdDev) from the ERA5 data reveals a striking stabilization of coastal wind patterns. But this is contrasted by increasing volatility at high intensity end in the open ocean (Table 4).

The most significant finding is the widespread reduction in variability across nearly all coastal wind thresholds: (a) Easterly Winds **-** The variability of moderate easterlies (2–6 m/s) has collapsed by 34% to 46% (Table 4). This indicates that the decline in easterly frequency observed in the mean analysis is not just a fluctuating trend, but a consistent, year-over-year reduction. (b) Westerly Winds – The variability of westerlies has also decreased, particularly at higher intensities. The standard deviation for westerlies > 9 m/s and > 10 m/s dropped by 50.7% and 47.9%, respectively, much greater than the decrease of the variability for the easterlies at these thresholds (9.2% and 31.1% decrease, respectively, Table 4). This suggests that the occurrence of these winds has become more locked-in compared to the early 2000s.

In contrast to the coastal stabilization, the open ocean wind magnitude (OceanMag) is experiencing a surge in variability at both low and high ends of the spectrum (Table 4): (a) the standard deviation (SD) for extreme offshore winds (> 10 m/s) increased by 39.4%; (b) combined with the previous finding that the mean frequency of these winds increased by over 80%, this indicates that the offshore environment is experiencing more frequent and more erratic extreme wind events.

**Discussion**

To facilitate the following discussion, key terms and index definitions are summarized below:

EFI (km): Extreme Flow Index at the inlet (Eluitkak Pass), computed from the regression-derived velocity time series. It represents the net integrated excursion during each summer ice-free period (July–August) by low-pass filtered (non-tidal) flow velocities exceeding $|u| > 0.5$ m/s. Velocities below 0.5 m/s do not contribute to EFI, providing a direct metric of annual extreme event magnitude (positive is out of the inlet).

IOFE ($10^3$ km) – In-Out Flow Excursion index at the inlet, computed from the regression-derived velocity time series without a velocity threshold. It represents the net integrated excursion of all low-pass filtered (non-tidal) flows during each summer ice-free period (July–August), providing a measure of total seasonal flow intensity (positive is out of the inlet).

EWWE ($10^4$ km) – East-West Wind Excursion index, derived from surface wind observations at the Utqiaġvik airport ASOS station. It measures the net integrated spatial displacement of sea-level air driven by low-pass filtered winds during each summer ice-free period (July–August), serving as a proxy for seasonal directional wind work in the east-west axis (positive is toward the east).

PR – Positive flow ratio, defined as the fraction of time during each summer ice-free period (July–August) that the low-pass filtered (non-tidal) inlet velocity is positive (outward flow from the lagoon into the Beaufort Sea).

EET – Extreme Eastward Throughflow, the low-pass filtered velocity inside the lagoon is toward the east when the inward flow at the Eluitkak Pass satisfies $u < -0.5$ m/s (inward into the lagoon system).

EWT – Extreme Westward Throughflow, the low-pass filtered velocity inside the lagoon is toward the west when the inward flow at the Eluitkak Pass satisfies $u > 0.5$ m/s (outward from the lagoon to coastal ocean).

The simultaneous stabilization of coastal wind regimes and the shift in offshore extremes point toward a reorganization of the regional atmospheric engine. At the coast, the increased occurrence frequency of low-to-moderate westerlies, combined with the collapse of high-magnitude westerlies (> 9 and > 10 m/s thresholds) and a ~50% decrease in variance, signals a transition toward a mellowed, low-variance baseline (Table 5). The land-high or ocean-low westerly regime has evolved into a more persistent but energetically depleted feature of the summertime climate, explaining why increased westerly occurrence frequency at smaller speed thresholds has paradoxically failed to drive more extreme eastward throughflow (EET). Conversely, the land-low or ocean-high easterly regime shows an almost across-the-board decrease in both occurrence frequency and variance, though with less relative decline at high end compared to the westerlies. In contrast, the offshore environment is characterized by increased occurrence frequency and elevated volatility at extreme thresholds (Table 5).

The EOF-based phase-space analysis provides a novel diagnostic for the hydrodynamic sensitivity of the lagoon-ocean exchange flows across the first 26-year period of the 21$^{st}$ century. The continuous 5-year rolling standard deviation analysis reveals a long-term, secular decline in environmental variability across atmospheric and hydrodynamic indices. While the early portion of the time series exhibited higher overall variance and more frequent extreme inward-exchange (EET) events, the long-term trend reflects a progressive dampening of this baseline variability. Concurrently, extreme outward-flow (EWT) events have emerged as the primary source of episodic variability. With 83% of all recorded EWT pulses clustering between 2022 and 2024, the data demonstrates that while the ocean-high or land-low state has become less frequent on a mean seasonal basis, individual anticyclonic events remain capable of driving high-amplitude hydrodynamic exchange in the Arctic summer.

The subtidal circulation at Eluitkak Pass reveals a compelling paradox between mean-state stability and event-scale behavior over the 2000–2025 record. On a seasonal scale, the time-integrated fraction of outward flow at the Eluitkak Pass has been stable (78% ± 5%), consistent with the Beaufort High remaining a dominant and persistent feature of the summertime Arctic atmosphere. However, this mean seasonal stability masks distinct, threshold-dependent shifts at the event scale. Easterly wind occurrence frequencies declined across all thresholds, accompanied by a reduction in their interannual variability, with the largest reductions in mean frequency occurring at speeds exceeding 9–10 m/s (Tables 3 and 4).

Coastal hydrodynamics therefore reflect a redistribution of wind occurrence within a predominantly easterly summertime regime. The increasing frequency of low-to-moderate westerlies occurs concurrently with a broad decline in easterly occurrence and a substantial reduction in coastal wind and exchange-flow variability. Continuous 5-year rolling metrics reinforce this transition across the 26-year record. Within this overall stabilized background, phase-space analysis shows results consistent with the finding that low-to-moderate westerly winds have become ~30% more frequent (except at high speeds > 9 m/s) while experiencing a ~50% variance reduction, providing a persistent but energetically mellowed driver for eastward throughflow. Nevertheless, episodic Beaufort High anticyclonic forcing can still trigger intense outward-flow (EWT) pulses—as observed in the 2022–2024 cluster, demonstrating that individual high-impact events can occur even within a long-term declining easterly climatology.

The present study is focused on the impact of weather system on the exchange flows through an Arctic lagoon. This weather induced hydrodynamic response in the coastal ocean and estuaries is sometimes referred to as meteorological tides[42,43], different from the astronomical tides which have higher frequencies (shorter time scales). The observed changes should not be interpreted as evidence for a monotonic increase in Arctic cyclone activity or a complete transition from Beaufort High to cyclone-dominated conditions. Rather, the results indicate a redistribution of coastal wind occurrence and variability, with increased low-to-moderate westerly occurrence frequency but reduced occurrence of the easterlies and strongest westerlies, while Beaufort Sea wind magnitude increases. The relatively small number of extreme outward-flow events in the lagoon also limits inference about their long-term trend.

While this study reveals long-term variations of meteorological conditions and associated meteorological tides over the 26-year record, several limitations warrant mention. First, the analysis is restricted to the ice-free summer months (July–August) to isolate weather-driven hydrodynamics without the confounding mechanics of landfast or moving sea ice. As a result, the findings cannot be directly extrapolated to shoulder freeze-up/thaw seasons or ice-covered conditions. Second, our empirical model utilizes a 2D depth-averaged barotropic framework. Although validated against bottom-mounted ADCP profiles and 3D FVCOM hydrodynamic simulations, this formulation does not include the details of vertical stratification within the tidal inlets.

This continuous evolution in atmospheric forcing and subtidal flow dynamics potentially alters sedimentation patterns and lagoon flushing rates, serving as a critical indicator of the broader Arctic hydrological cycle[44] and its shifting connectivity to the Beaufort Sea. Resultant changes in circulation and residence times affect the capacity of lagoons to function as biogeochemical filters and critical nurseries for Arctic cod[35,36]. As these forage fish provide the primary trophic link to Arctic marine megafauna[45], the observed shift toward a reconfigured forcing state signals a bottom-up threat to the stability of the broader high-latitude marine ecosystems[46,47]. These findings underscore the necessity of accounting for multi-scale atmospheric reorganization and continuous variance shift when predicting the future of the high-latitude coastal domain; focusing solely on open-ocean mean climatology is insufficient and risks overlooking the complex, nonlinear, multifaceted dynamics inherent to coastal exchange zones.

## Method

*Regional Domains*

To differentiate between localized coastal forcing and broader shelf-scale dynamics, we defined two primary analytical polygons based on the ERA5 high-resolution atmospheric grid (Fig. 11d). These regions were used to extract the wind and sea-level pressure (SLP) data for the time series analysis to examine temporal evolution of the weather systems which influence the coastal exchange flows in Arctic lagoons.

ERA5 data was selected within a geographic bounding box: 65°N–80°N, 120°W–170°W, covering the coastal northern Alaska and the southern Beaufort Sea. The region is defined based on the regions of weather influence (ROWI) as shown in the Supplemental Materials. This ERA5 reanalysis dataset is obtained for the atmospheric forcing for the summer months of July and August, spanning the 26-year period from 2000 to 2025.

The ocean region encompasses the high-latitude Beaufort Sea and central Arctic basin to monitor the development and variation of the Beaufort High. The latitude range is 71.5 N to 80.0 N while the longitude range is 170.0 W to 126.0 W (Fig. 11d), encompassing 6,195 grid points at 0.25° resolution for the ERA5 reanalysis data.

The coastal polygon is strategically shaped to follow the northern Alaskan coastline, capturing the region directly influencing the lagoon-ocean exchange at Eluitkak Pass. The defining vertices (Lat, Lon) are (71.5 N, 157.0 W), (70.0 N, 143.0 W), (69.5 N, 143.8 W), and (71.0 N, 157.0 W) (Fig. 11d), covering 132 grid points at 0.25° resolution. Inland areas are deliberately excluded because terrestrial topography introduces localized drag and thermal effects that decouple inland meteorology from the coastal weather systems driving the exchange.

The time series hourly wind data from the Utqiaġvik's airport, Alaska at 156.7922° W, 71.2826° N was used to generate the wind-rose plots, analyze the wind regimes and their variabilities, compute the surface air (wind) excursion and related indices, and run the regression model (see below) to generate the exchange flows through the western inlet of the Elson Lagoon. This ASOS station is about 17 km southwest of the station of inlet water velocity measurements using a bottom-mounted RDI 1200 kHz ADCP.

*ERA5 Data Extraction and Analysis*

We extracted SLP, 10m wind velocity at a horizontal grid resolution of 0.25° × 0.25° (approximately 31 km) and a temporal resolution of 1 hour. The ERA5 dataset was selected for its high-fidelity representation of Arctic mesoscale cyclones and its ability to resolve the sharp pressure gradients often found along the Northern Alaskan coastline. The study area focuses on the Beaufort Sea and the adjacent Northern Alaskan continental shelf.

The SLP data, denoted as $F(x,t)$, represents pressure at $M$ spatial locations and $N$ time steps. To focus on variability rather than the mean state, we first calculate the anomalies $A(x,t)$ by subtracting the long-term temporal mean $\bar{F}(x)$ from the raw data:

$$A(x,t) = F(x,t) - \bar{F}(x) \qquad (3)$$

The anomaly data from Equation (3) is then organized into a matrix **X** of size $N \times M$, where each row represents a time step and each column represents a spatial grid point. While EOFs can be calculated via a covariance matrix, we employ Singular Value Decomposition (SVD) for numerical efficiency and stability. The matrix **X** is decomposed as:

$$\mathbf{X} = \mathbf{U\Sigma V}^{\boldsymbol{T}} \qquad (4)$$

where $\mathbf{V}$ (size $M \times M$) contains the EOFs, which are the orthogonal spatial patterns; $\mathbf{U\Sigma}$ (size $N \times M$) represents the Principal Components (PCs), which are the time-varying amplitudes of each mode; $\mathbf{\Sigma}$ is a diagonal matrix of singular values ($\sigma_i$). The variance explained by the $i$-th mode derived from Equation (4) is proportional to $\sigma_i^2$.

It should be noted that the two key observations at the end of Section 2.2 guided the spatial selection of the ERA5 dataset, defining the region of weather influence (ROWI). The domain spans $15°$ in latitude ($65°$N to $80°$N) and $50°$ in longitude ($170°$W to $120°$W), covering an expansive area of $2.78$ million km$^2$—which represents over half of a full Arctic quadrant and more than one-eighth of the total Arctic domain. This domain captures the synoptic high- and low-pressure systems that directly drive exchange flows in the coastal lagoons of northern Alaska.

To test the sensitivity of this spatial selection (as detailed in the Supplemental Materials), the ERA5 domain is expanded by over 15% in area to over $3.21$ million km$^2$ ($65°$N to $82°$N and $172°$W to $118°$W). The results showed only negligible shifts in EOF variance percentages (e.g., EOF1 shifts slightly from 65.9% to 64.4%), keeping the essential spatial structures intact and leaving all primary time-series indices virtually unchanged. Expanding the domain further to include regions such as Russia or Europe would introduce distant teleconnections that obscure the direct synoptic linkage between the Beaufort Sea and the northern Alaska coast. Furthermore, incorporating those expansive landmasses introduces complex terrestrial topographic effects that unnecessarily convolute and obscure the local dynamics that directly influence the exchange flows at the coastal lagoons. Rather than helping disentangle dynamics, an overly broad domain complicates them.

*Flow velocity measurements*

A bottom-mounted RDI 1200 kHz ADCP was deployed inside the Eluitkak Pass at (~71.3593° N, 156.3561° W) on 29 July 2014. The ADCP was moored on the bottom with upward facing transducers in approximately 13 m of water. It was retrieved on August 3 and redeployed a day later, on 4 August with valid data until 9 August. The first deployment was set up to sample every 80 s, with 45 samples every hour. The hourly-averaged results with a vertical bin size of 1.0 m were recorded. The first bin with velocity data was 1.53 m above the bottom. During the second deployment, the samplings were done at 6 s intervals. An average was calculated every 50 samples to produce the ensemble data every 5 min.

*Phase-Space Characterization and Regime Classification*

The leading principal components (PC1 and PC2) were used to construct a two-dimensional phase-space, representing the instantaneous state and evolution of the regional SLP field. This phase-space was partitioned into four quadrants based on the polarities of the PCs, with each quadrant corresponding to a distinct synoptic regime (e.g., the first quadrant representing a deep cyclonic-biased state). To link atmospheric forcing with the lagoon's hydrodynamic response, we mapped Extreme Westward Throughflow (EWT) and Extreme Eastward Throughflow (EET) events—defined by a 0.5 m/s velocity threshold—onto this phase-space. This approach allowed us to quantify the chronological clustering of extreme states.

*FVCOM Numerical Modeling*

The unstructured grid FVCOM[29,48] was used for a numerical simulation of wind-driven flows in Elson Lagoon. The bathymetric data used in the model were a combination of the dataset from the National Centers for Environmental Information (formerly National Geophysical Data Center) and high-resolution measurements in the Eluitkak Pass. The source of bathymetry data elsewhere in the model (basically the Beaufort Sea and Chukchi Sea) was from the HYbrid Coordinate Ocean Model (HYCOM). The HYCOM datasets with water level and flow were interpolated onto our model grid for the initial and boundary conditions. The wind velocity time series and atmospheric pressure were specified using the weather data at the airport in Utqiaġvik and assumed spatially uniform in the computational domain. The simulation[32] was run for 15 July to 31 August 2014, encompassing the observational period. The mesh had 15,384 nodes and 29,015 cells with the highest resolution of ~20 m in the horizontal and 40-sigma layers in the vertical. The model used a 0.2 s external time step and a 1.0 s internal time step.

*Regression model*

In earlier studies[31,38], a regression model between wind velocity components and the velocity at the tidal pass was proposed based on the consideration of the functional relationship of the flow velocity and the wind stress for wind-driven circulations in a shallow water lagoon. Each of the variables in the 2-D depth-averaged momentum equations has a high frequency component (tidal or higher frequency oscillations) and the low frequency subtidal component as expressed by

$$u = u_H + u_L, v = v_H + v_L, \zeta = \zeta_H + \zeta_L \tag{5}$$

The subscripts $H$ and $L$ in Equation (5) denote the high frequency and low frequency components, respectively. Here $u$, $v$, and $\zeta$ are the velocity components and water level, respectively. Averaging over tidal cycles, the high frequency components will be eliminated, and the remaining low frequency components satisfy the tidally averaged or low-pass filtered momentum equations. The solution for these low frequency components $(u_L, v_L, \zeta_L)$ can be expressed by the general format:

$$u_L = u_L(\tau_{axL}, \tau_{ayL}), \; v_L = v_L(\tau_{axL}, \tau_{ayL}) \tag{6}$$

$$\zeta_L = \zeta_L(\tau_{axL}, \tau_{ayL}) \quad (7)$$

In our case, in Equations (6) and (7) we are interested mainly in the along-channel velocity $u_L$. Using the Taylor series expansion[38] only for the along-channel velocity $u_L$ in terms of the variables $\tau_{ax}$ and $\tau_{ay}$ to the first order:

$$u_L = u_L(0,0) + \frac{\partial u_L}{\partial \tau_{axL}} \tau_{axL} + \frac{\partial u_L}{\partial \tau_{ayL}} \tau_{ayL} \quad (8)$$

All the derivatives in Equation (8) are evaluated at (0,0) for $(\tau_{axL}, \tau_{ayL})$. In the above equations, the wind stress terms can be further expressed by wind velocity components using the quadratic bottom friction. A Taylor series expansion[38] can be used to express the stress in terms of a linear term and a higher order nonlinear term, i.e.

$$\tau_{axL} = \alpha w_x + \beta w_x^2 + \text{higher order terms} \quad (9)$$

$$\tau_{ayL} = \gamma w_y + \delta w_y^2 + \text{higher order terms} \quad (10)$$

where $\alpha, \beta, \gamma, \delta$ are all constants of the Taylor series expansions.  Substitute Equations (9) and (10) into (8) and neglect the higher order terms we obtain:

$$u_L = u_L(0,0) + \frac{\partial u_L}{\partial \tau_{axL}} (\alpha w_x + \beta w_x^2) + \frac{\partial u_L}{\partial \tau_{ayL}} (\gamma w_y + \delta w_y^2) \quad (11)$$

These can be rewritten as:

$$u_L = \mathrm{A} w_x + B w_x^2 + \mathrm{C} w_y + D w_y^2 + E \quad (12)$$

in which, $A = \alpha \frac{\partial u_L}{\partial \tau_{axL}}$, $B = \beta \frac{\partial u_L}{\partial \tau_{axL}}$, $C = \gamma \frac{\partial u_L}{\partial \tau_{ayL}}$, $D = \delta \frac{\partial u_L}{\partial \tau_{ayL}}$, and $E = u_L(0,0)$.

The wind-driven subtidal flow velocity at the Eluitkak Pass is reconstructed using this empirical regression model and correlation coefficients in Equations (11) and (12) established[31] with $A$, $B$, $C$, $D$, and $E$ equal -0.045, 0.0004, -0.0125, -0.0038, and 0.1702, respectively. The model demonstrates high fidelity with an $R^2$ of 0.96 between the observed and predicted velocities. As established earlier[31], winds from the eastern quadrants—typical of the Beaufort High—drive outward subtidal flows, whereas the intrusion of Arctic cyclones into the Beaufort Sea triggers flow reversals. This robust relationship allows for the conversion of long-term wind time series into flow conditions at the Eluitkak Pass, enabling a long-term characterization of synoptic weather impacts on coastal exchange. This regression model has been successfully applied to a similar shallow lagoon at multiple inlets, validated by numerical model results and in situ observations over several years[38].

*Definition of the integrated flow index*

To quantify the intensity of exchange flows through a tidal pass, we define the In-out Flow Excursion (IOFE), which is an integration over the summertime (July-August) of the outward flow velocity at the Eluitkak Pass, which yields a net distance or excursion of water:

$$IOFE(y) = \int_0^T u_L dt \qquad (13)$$

Here $t$ is time, $T$ is the length of time from July 1 to August 31, $y$ is the year (from 2000 to 2025). The Extreme Flow Index (EFI) is defined by the net distance or excursion of water for along channel flow velocity being higher than 0.5 m/s:

$$EFI(y) = \int_0^T u_L|_{|u|\geq 0.5}\, dt \qquad (14)$$

Likewise, for discussion purposes and for better insight of wind effect, we define the East-West Wind Excursion (EWWE) index as:

$$EWWE(y) = \int_{0}^{T} w_x dt \qquad (15)$$

Since the Eluitkak Pass is located on the coast of northern Alaska, if the dominant wind north of the coast is determined by a high pressure over the Beaufort Sea, wind at the tidal pass would be from the east quadrants, whereas if a low pressure (cyclone) is over the Beaufort Sea (north of the coast), the dominant wind at the study site would be from the western quadrants. In the present study, the most convenient unit for $EFI$, $IOFE$, and $EWWE$ in Equations (13)—(15) are km, $10^3$ km, and $10^4$ km, respectively (Table 2).

*Wind Frequency Analysis*

To quantify the long-term evolution of atmospheric forcing, an analysis was performed for the occurrence frequency of the ERA5 wind fields across the 26-year study period (2000–2025). Continuous changes were evaluated using 5-year rolling calculations of standard deviation to capture the evolution. For concise discussion purposes, the record was partitioned into two equal 13-year benchmark periods (2000–2012 and 2013–2025).

We define a series of wind thresholds ranging from 1 to 10 m/s at 1 m/s increments. For each year, the total number of hours exceeding these thresholds was calculated for three variables: (1) Coastal Westerlies ($u > 0$) - Representing the cyclonically-biased forcing; (2) Coastal Easterlies ($u < 0$) - Representing the anticyclonically-biased forcing; (3) Ocean Wind Magnitude ($|U|$) - Representing the total kinetic energy of the offshore atmospheric environment.

To evaluate evolution in the wind field, we computed both the long-term annual mean frequencies across the comparative 13-year halves and a 5-year moving standard deviation across the full 26-year time series. The percentage change between the two benchmark halves highlights baseline changes, while the 5-year rolling standard deviation provides a continuous, unconstrained metric for the progressive stabilization or volatilization of specific wind regimes over time. This dual-metric approach allows for a rigorous differentiation between changes in the average atmospheric state and changes in the frequency of high-amplitude extreme events.

*Model Formulation and Hydrodynamic Validation*

The empirical regression model used here is grounded in the 2D depth-averaged momentum equations and was previously established and validated across a shallow, multi-inlet bay system using 5–6 horizontal ADCPs over a multi-year period (2013–2015) alongside 3D numerical models, consistently yielding $R^2$ values between 0.50 and 0.80 (Li et al., 2019).

For the present study, our 2014 bottom-mounted ADCP deployment at Eluitkak Pass captured over 10 days of continuous, high-frequency velocity profiles consisting of thousands of individual ensemble measurements (sampled every 80 s during July 29–August 3, and every 5 min during August 4–9), yielding ~240 hourly averaged profile observations. Crucially, this deployment spanned both typical ambient flows and a rare, major wind-driven extreme exchange event (subtidal pass velocity reaching ~1.0 m/s). This rare field capture of an extreme event provides a robust empirical foundation to calibrate and validate the momentum balance underlying the regression model.

The regression coefficients are from ADCP observations done in 2013[31]. The overall fidelity of the regression was evaluated by comparing the new in situ ADCP observations from 2014, 3D

FVCOM numerical model simulations, and the empirical regression model output (Fig. S1 in Supplemental Materials). The three datasets demonstrate strong statistical agreement: 1) ADCP Field Data from 2014 vs. 3D FVCOM Simulation -- $r = 0.95$ ($R^2 = 0.91$); 2) ADCP Field Data from 2014 vs. Regression Model: $r = 0.77$ ($R^2 = 0.59$); 3) 3D FVCOM Simulation vs. Regression Model: $r = 0.77$ ($R^2 = 0.60$).

This multi-tiered validation confirms that the regression model effectively captures the subtidal exchange dynamics driven by synoptic-scale forcing, supporting its application across the 26-year study period.

**Table 1.** Exchange and Excursion Parameters. Here Yr = Year; N = total number of extreme summer events in a year; EFI = Extreme Flow Index in km in the ocean; IOFE = in-out flow excursion in $10^3$ km in the ocean (outward is positive); EWWE = east-west wind excursion in $10^4$ km in the atmosphere; PR = positive-ratio (fraction of positive flow, or flow out of the lagoon onto Beaufort Sea). The left and right columns are for the first 13 years and last 13 years, respectively. The means and standard deviations for the first and last 13 years are shown at near bottom. The overall means and standard deviations are shown at the bottom. Wind data is from Utqiaġvik, about 17 km southwest of the station for water velocity measurements using an ADCP.

| **Average of annual means for ice-free period of the first 13 years (2000-2012)** | | | | | | **Average of annual means for ice-free period of the last 13 years (2013-2025)** | | | | | |
|---|---|---|---|---|---|---|---|---|---|---|---|
| **Yr** | **N/ Yr** | **EFI** (km) | **IOFE** ($10^3$ km) | **EWWE** ($10^4$ km) | **PR** | **Yr** | **N/ Yr** | **EFI** (km) | **IOFE** ($10^3$ km) | **EWWE** ($10^4$ km) | **PR** |
| 2000 | 2 | -125.1 | 4.2 | -7.7 | 0.78 | 2013 | 3 | -12.9 | 4.5 | -6.2 | 0.79 |
| 2001 | 0 | 0 | 4.0 | -8.0 | 0.79 | 2014 | 1 | -15.7 | 4.7 | -7.3 | 0.81 |
| 2002 | 2 | -60.1 | 2.0 | -4.7 | 0.70 | 2015 | 1 | -101.4 | 4.8 | -7.1 | 0.78 |
| 2003 | 5 | -208.9 | 2.5 | -5.6 | 0.71 | 2016 | 0 | 0 | 4.6 | -8.3 | 0.72 |
| 2004 | 2 | -51.5 | 5.2 | -10.0 | 0.85 | 2017 | 1 | -25.9 | 3.5 | -5.8 | 0.78 |
| 2005 | 0 | 0 | 4.8 | -9.5 | 0.84 | 2018 | 2 | -16.8 | 4.2 | -7.7 | 0.76 |
| 2006 | 0 | 0 | 3.6 | -5.0 | 0.77 | 2019 | 2 | -86.9 | 3.6 | -6.3 | 0.75 |
| 2007 | 0 | 0 | 5.3 | -11.9 | 0.90 | 2020 | 1 | -44.7 | 3.8 | -5.0 | 0.77 |
| 2008 | 1 | -99.8 | 4.3 | -5.6 | 0.83 | 2021 | 1 | -15.7 | 4.0 | -4.8 | 0.73 |
| 2009 | 0 | 0 | 4.1 | -7.3 | 0.80 | 2022 | 1 | **28.6** | 3.5 | -5.7 | 0.76 |
| 2010 | 2 | -25.0 | 4.9 | -8.5 | 0.82 | 2023 | 2 | -8.8 | 4.3 | -5.2 | 0.76 |
| 2011 | 0 | 0 | 4.9 | -8.3 | 0.83 | 2024 | 4 | **35.2** | 4.2 | -5.4 | 0.77 |
| 2012 | 0 | 0 | 3.7 | -2.7 | 0.75 | 2025 | 0 | 0 | 4.5 | -5.5 | 0.79 |
| **mean** | **1.1** | **-43.9** | **4.1** | **-7.3** | **0.8** | **mean** | **1.5** | **-20.4** | **4.2** | **-6.2** | **0.77** |
| **std** | **1.5** | **65.3** | **1.0** | **2.5** | **0.06** | **std** | **1.1** | **38.9** | **0.5** | **1.1** | **0.02** |
| **Overall mean** | | | | | | | **1.3** | **-32.1** | **4.1** | **-6.7** | **0.78** |
| **Overall standard deviation** | | | | | | | **1.3** | **54.0** | **0.8** | **2.0** | **0.05** |

**Table 2**. Extreme Flow Events. Day of Year (DOY) is defined as the day in the year starting from January 1 as Day 1.

| Year | DOY | Date | u (m/s) | DOY | Date | u (m/s) | DOY | Date | u (m/s) | No. |
|---|---|---|---|---|---|---|---|---|---|---|
| **2000** | 187.4 | 5-Jul | -0.68 | 224.3 | 11-Aug | -1.76 | - | - | - | 2 |
| **2001** | - | - | - | - | - | - | - | - | - | 0 |
| **2002** | 227.5 | 15-Aug | -0.80 | 229.0 | 17-Aug | -0.54 | - | - | - | 2 |
| **2003** | 199.8 | 18-Jul | -0.66 | 209.2 | 28-Jul | -0.77 | 210.9 | 29-Jul | -1.08 | |
| | 217.5 | 5-Aug | -0.69 | 218.9 | 6-Aug | -0.56 | - | - | - | 5 |
| **2004** | 194.4 | 12-Jul | -0.58 | 213.6 | 31-Jul | -0.78 | - | - | - | 2 |
| **2005** | - | - | - | - | - | - | - | - | - | 0 |
| **2006** | - | - | - | - | - | - | - | - | - | 0 |
| **2007** | - | - | - | - | - | - | - | - | - | 0 |
| **2008** | 213.0 | 31-Jul | -0.81 | - | - | - | - | - | - | 1 |
| **2009** | - | - | - | - | - | - | - | - | - | 0 |
| **2010** | 193.5 | 12-Jul | -0.51 | 214.8 | 2-Aug | -0.57 | - | - | - | 2 |
| **2011** | - | - | - | - | - | - | - | - | - | 0 |
| **2012** | - | - | - | - | - | - | - | - | - | 0 |
| **2013** | 190.6 | 9-Jul | 0.52 | 206.5 | 25-Jul | -0.53 | 210.6 | 29-Jul | -0.61 | 3 |
| **2014** | 214.2 | 2-Aug | -0.56 | - | - | - | - | - | - | 1 |
| **2015** | 239.8 | 28-Aug | -1.00 | - | - | - | - | - | - | 1 |
| **2016** | - | - | - | - | - | - | - | - | - | 0 |
| **2017** | 203.7 | 22-Jul | -0.81 | - | - | - | - | - | - | 1 |
| **2018** | 201.2 | 20-Jul | -0.53 | 243.8 | 31-Aug | -0.52 | - | - | - | 2 |
| **2019** | 213.9 | 1-Aug | -0.90 | 236.6 | 24-Aug | -0.68 | - | - | - | 2 |
| **2020** | 210.0 | 28-Jul | -0.52 | - | - | - | - | - | - | 1 |
| **2021** | 234.3 | 22-Aug | -0.56 | - | - | - | - | - | - | 1 |
| **2022** | 230.3 | 18-Aug | 0.54 | - | - | - | - | - | - | 1 |
| **2023** | 235.5 | 23-Aug | -0.83 | 239.3 | 27-Aug | 0.60 | - | - | - | 2 |
| **2024** | 192.5 | 10-Jul | 0.51 | 197.1 | 15-Jul | 0.54 | 208.6 | 26-Jul | 0.54 | |
| | 214.5 | 1-Aug | -0.58 | - | - | - | - | - | - | 4 |
| **2025** | - | - | - | - | - | - | - | - | - | 0 |
| | | | | | | | | | **Total** | 33 |

**Table 3.** Mean frequency analysis (average hours/year) of ERA5 data: the first half (2000-2012) vs the second half (2013-2025) of the 26-year study period. Here the East and West in the first column are the threshold for the easterly and westerly winds, respectively, averaged in the coastal polygon. The OceanMag is the threshold of averaged wind speed in the Beaufort Sea polygon. The second and third columns are the mean cumulative hours exceeding the wind threshold (in hours) for the first and second 13-year periods, respectively. The last column shows the percentage changes between the two 13-year periods.

| Threshold (m/s) | Average Annual Frequency in 2000-2012 | Average Annual Frequency in 2013-2025 | Change of the average | Note |
|---|---|---|---|---|
| East >1 | 890.85 | 782.85 | **-12.1%** | Cross the board decrease of coastal easterly wind frequencies |
| East >2 | 771.77 | 654.85 | **-15.2%** | |
| East >3 | 635.85 | 530.54 | **-16.6%** | |
| East >4 | 498.54 | 400.92 | **-19.6%** | |
| East >5 | 349.77 | 280.77 | **-19.7%** | |
| East >6 | 218.92 | 180.38 | **-17.6%** | |
| East >7 | 112.38 | 99.46 | **-11.5%** | |
| East >8 | 55.62 | 45.46 | **-18.3%** | |
| East >9 | 22.38 | 16.00 | **-28.5%** | |
| East >10 | 8.23 | 5.08 | **-38.3%** | |
| West >1 | 388.69 | 473.54 | **+21.8%** | Increase of coastal westerly wind frequencies |
| West >2 | 299.85 | 380.62 | **+26.9%** | |
| West >3 | 223.15 | 292.23 | **+30.9%** | |
| West >4 | 161.62 | 211.23 | **+30.7%** | |
| West >5 | 116.23 | 143.00 | **+23.0%** | |
| West >6 | 78.31 | 91.85 | **+17.3%** | |
| West >7 | 47.54 | 56.15 | **+18.1%** | |
| West >8 | 26.69 | 29.23 | **+9.5%** | |
| West >9 | 16.38 | 11.31 | **-31. 0%** | Decrease in high westerly wind |
| West >10 | 8.15 | 4.85 | **-40.6%** | |
| OceanMag >1 | 1488.00 | 1488.00 | 0.00% | Little change for small wind speed offshore |
| OceanMag >2 | 1488.00 | 1488.00 | 0.00% | |
| OceanMag >3 | 1465.54 | 1461.69 | **-0.3%** | |
| OceanMag >4 | 1250.62 | 1266.00 | **+1.2%** | |
| OceanMag >5 | 823.00 | 820.46 | **-0.3%** | |
| OceanMag >6 | 415.00 | 425.38 | **+2.5%** | Increase of frequencies for large wind speed offshore |
| OceanMag >7 | 186.38 | 197.69 | **+6.1%** | |
| OceanMag >8 | 67.15 | 94.38 | **+40.6%** | |
| OceanMag >9 | 21.77 | 42.23 | **+94.0%** | |
| OceanMag >10 | 7.85 | 14.31 | **+82.4%** | |

**Table 4.** Variability analysis: the first half (2000-2012) vs the second half (2013-2025) of the 26-year study period. Here the East and West in the first column are the threshold for the easterly and westerly winds, respectively, averaged in the coastal polygon. The OceanMag is the threshold of averaged wind speed in the Beaufort Sea polygon from the ice-free period (July-August). The second and third columns are the standard deviations (in hours) for the first and second 13-year periods, respectively, computed from annual ice-free period means. The last column shows the percentage changes of the standard deviation between the two 13-year periods.

| Threshold (m/s) | SD (hr) for 2000-2012 | SD (hr) for 2013-2025 | Change | Note |
|---|---|---|---|---|
| East >1 | 197.89 | 130.21 | **-34.2%** | Almost cross board decrease in variability for easterly wind in coastal zone |
| East >2 | 212.04 | 119.27 | **-43.8%** | |
| East >3 | 208.87 | 112.84 | **-46.0%** | |
| East >4 | 188.12 | 115.86 | **-38.4%** | |
| East >5 | 156.09 | 94.18 | **-39.7%** | |
| East >6 | 114.77 | 74.72 | **-34.9%** | |
| East >7 | 64.78 | 61.90 | **-4.4%** | |
| East >8 | 39.23 | 41.89 | **+6.8%** | |
| East >9 | 23.68 | 21.49 | **-9.2%** | |
| East >10 | 14.32 | 10.01 | **-30.1%** | |
| West >1 | 168.50 | 131.16 | **-22.2%** | Cross board decrease in variability for westerly wind in coastal zone |
| West >2 | 146.37 | 122.95 | **-16.0%** | |
| West >3 | 122.62 | 106.14 | **-13.4%** | |
| West >4 | 99.63 | 89.75 | **-9.9%** | |
| West >5 | 77.00 | 72.92 | **-5.3%** | |
| West >6 | 58.05 | 50.26 | **-13.4%** | |
| West >7 | 40.96 | 32.57 | **-20.5%** | |
| West >8 | 30.31 | 20.84 | **-31.2%** | |
| West >9 | 21.29 | 10.49 | **-50.7%** | |
| West >10 | 11.91 | 6.20 | **-47.9%** | |
| OceanMag >1 | 0.00 | 0.00 | - | Offshore wind speed frequency variability is complicated but now more variable for large wind |
| OceanMag >2 | 0.00 | 0.00 | - | |
| OceanMag >3 | 18.22 | 27.32 | **+50.0%** | |
| OceanMag >4 | 125.81 | 116.95 | **-7.0%** | |
| OceanMag >5 | 198.80 | 175.82 | **-11.6%** | |
| OceanMag >6 | 181.65 | 161.16 | **-11.3%** | |
| OceanMag >7 | 118.52 | 119.02 | **+0.4%** | |
| OceanMag >8 | 61.55 | 69.22 | **+12.5%** | |
| OceanMag >9 | 30.59 | 34.16 | **+11.7%** | |
| OceanMag >10 | 14.11 | 19.67 | **+39.4%** | |

**Table 5.** Summary of Changes in Wind Frequency and Variability between the first half and second half of the 26-year study period.

| | Coastal zone | | Offshore |
|---|---|---|---|
| | Easterly | Westerly | Magnitude |
| Frequency | All decrease | Increase in low-moderate wind, decrease at high-end. | Small change at low-moderate wind end, much greater increase at high-end |
| Variability | Almost all decrease, less decrease at high-end | All decrease, less decrease at low-moderate end, much more at high end | Mixed change, increase at low and high ends |
| Summary of changes | Decreased frequency & low variance baseline (ocean-high or land-low) | Mellowed & low variance baseline (land-high or ocean-low) | Higher energy & more erratic fluctuations |

**Data Availability**

The NCEP/NCAR Reanalysis 1 data used in this study is accessed via the website of the NOAA Physical Sciences Laboratory (PSL) at https://psl.noaa.gov/ (last accessed March 2026). The ERA5 atmospheric reanalysis data used in this study were provided by the European Centre for Medium-Range Weather Forecasts (ECMWF) and were retrieved from the Copernicus Climate Data Store (CDS) at https://cds.climate.copernicus.eu/ (last accessed March 2026). The data processed for the North Alaska and Beaufort Sea regions include hourly estimates of 10m u/v wind components and mean sea level pressure. The weather data for Utqiaġvik (formerly Barrow, Alaska) is from the Iowa Environmental Mesonet (https://mesonet.agron.iastate.edu/ASOS/). The acoustic Doppler current profiler data for flow velocity is accessible from https://zenodo.org/records/18602550.

**Code Availability**

The MATLAB codes generated during the current study are not publicly available due to ongoing research and student projects, but are available from the corresponding author on reasonable request.

**Author Contributions**

C.L. conceived the study, obtained the funding, directed and conducted the observations and numerical modeling, performed the statistical analyses, drew the graphics, and wrote the manuscript. The author read and approved the final manuscript.

**Acknowledgements**

Logistic support for the fieldwork to collect the data was provided by North Slope Borough and UMIAQ. The author is indebted to Kevin Boswell for introducing the study site, as well as for his collaboration and assistance with Arctic field logistics.

**Funding**

The study was funded by the North Pacific Research Board.

**Competing Interests**

The author declares no competing financial or non-financial interests.

**Figure captions.**

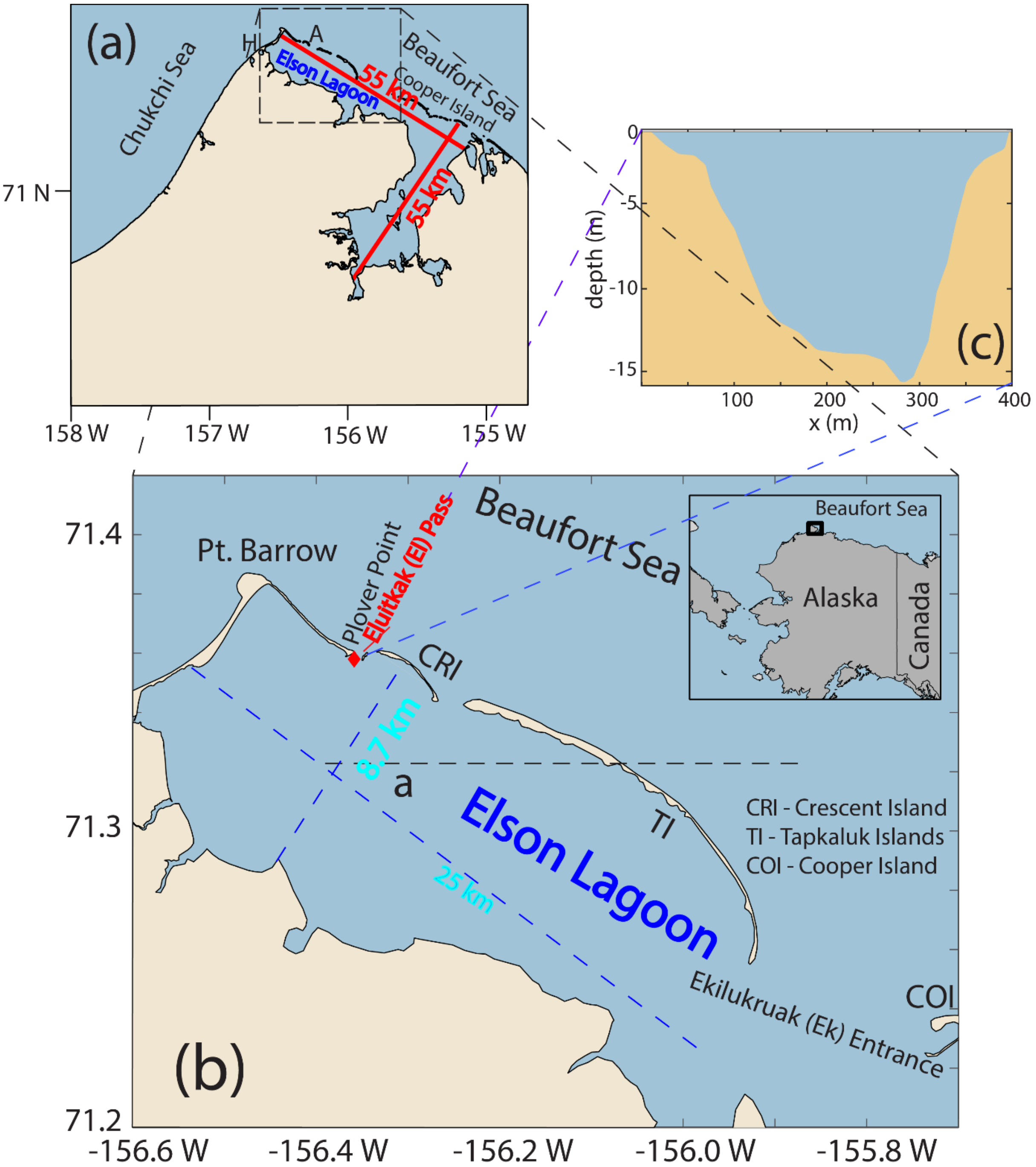

**Figure 1**. **Sampling site in the Arctic Alaska.** (a) The northwestern Alaska adjacent to Beaufort and Chukchi Seas. (b) Zoom in view of the study site, showing the location of major inlet (Eluitkak Pass) and sampling location (Plover Point) inside the pass which has the deepest depth (16 m) of the lagoon and the shallow Elson Lagoon (depth ~ 1-2 m). (c) Water depth profile across the narrowest and deepest inlet at the Eluitkak Pass. The other inlets are all very shallow (~ 1 m).

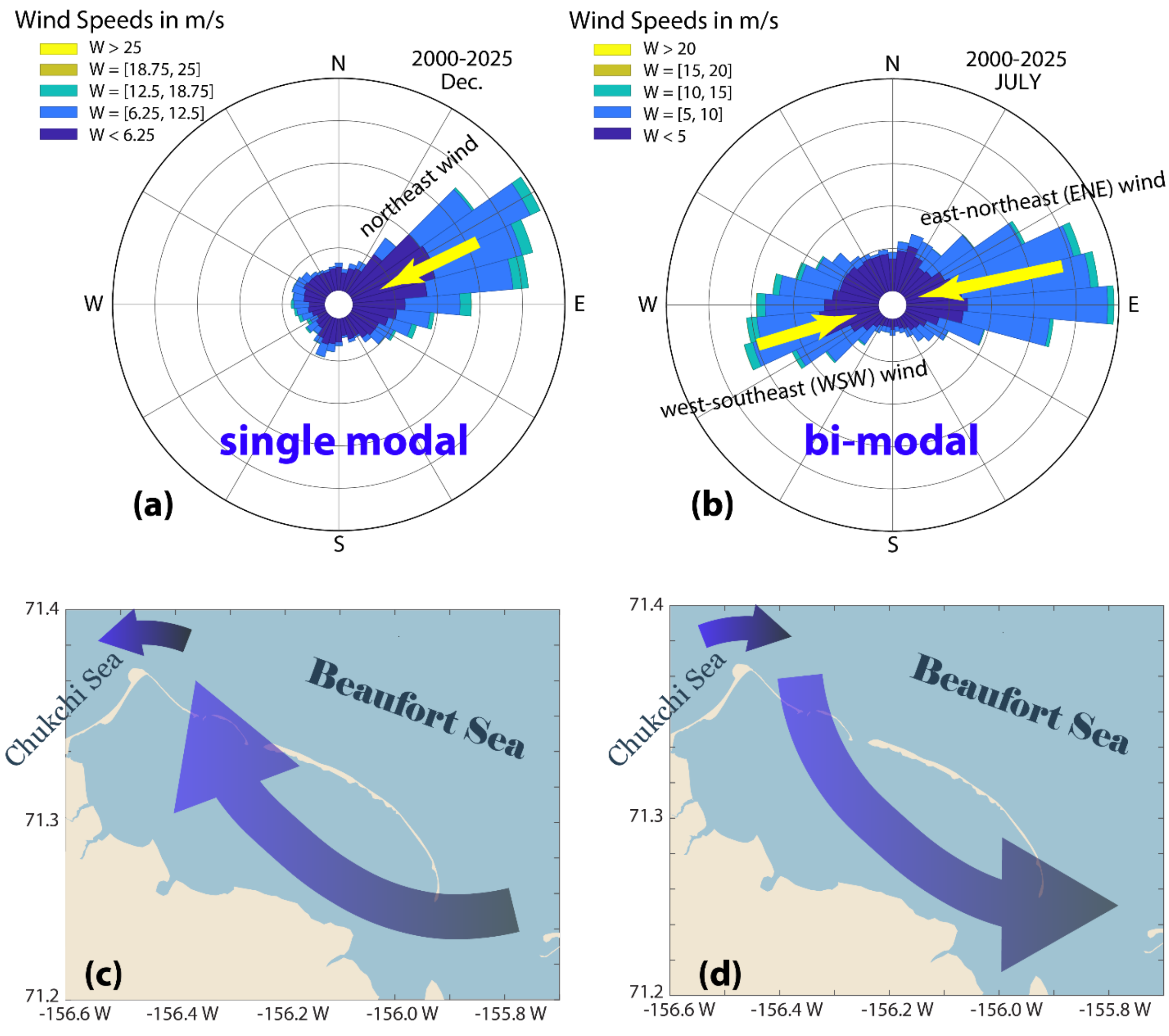

**Figure 2. Wind and flow regimes.** (a) Wind rose of the northern Alaska for the summer (July and August) and (b) for the whole year. (c) Westward throughflow under Beaufort High. (d) Eastward throughflow under Arctic cyclones.

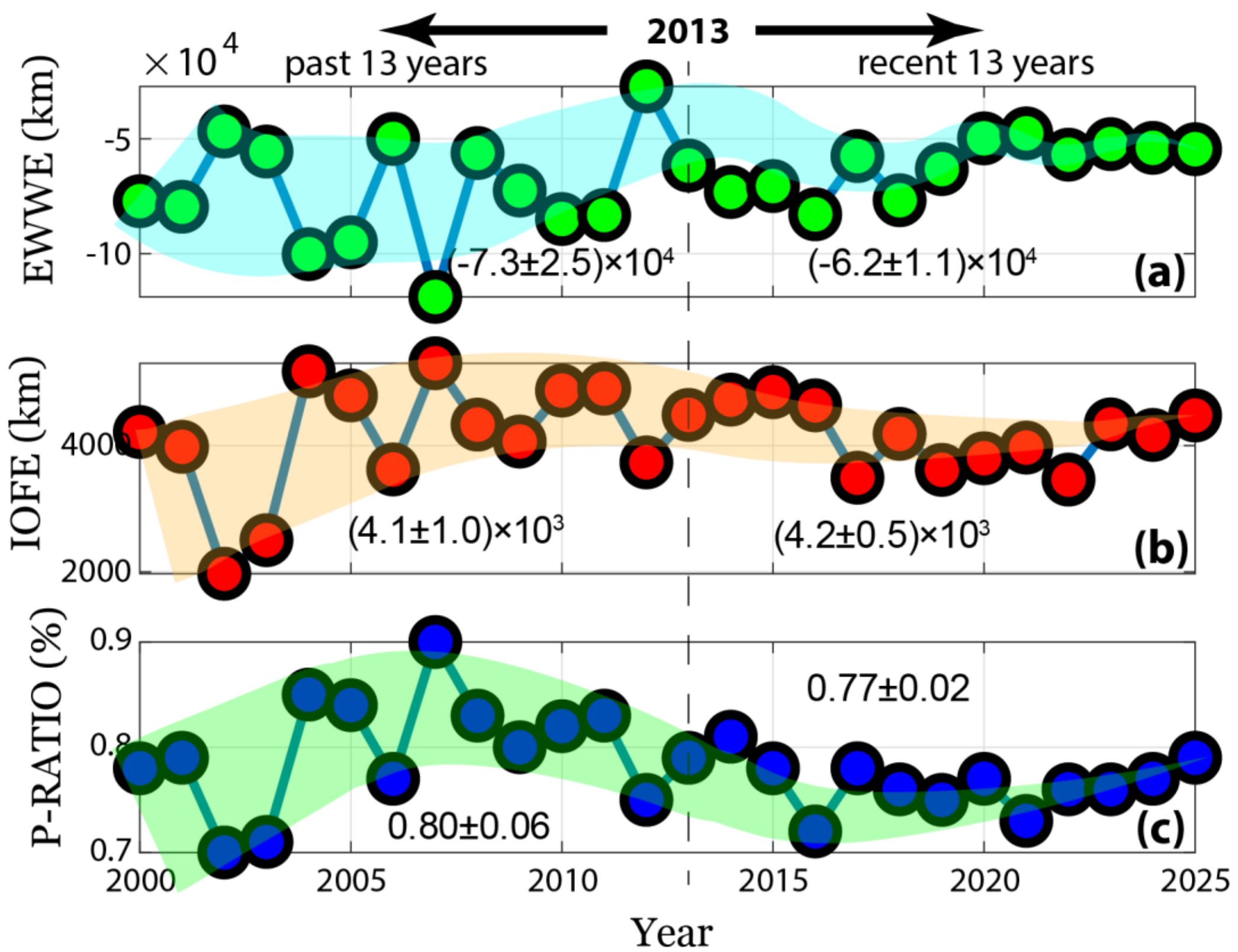

**Figure 3**. **Time series of indices.** These are (a) the East-West Wind Excursion (EWWE) in km, (b) In-Out Flow Excursion (IOFE) in km, and (c) the percentage of positive flow P-RATIO in % at the Eluitkak Pass. The time is between the year 2000 and 2025 for each summer (1 July to 31 August). The background belts qualitatively demonstrate the regime shift. For the quantified results, consult Table 1.

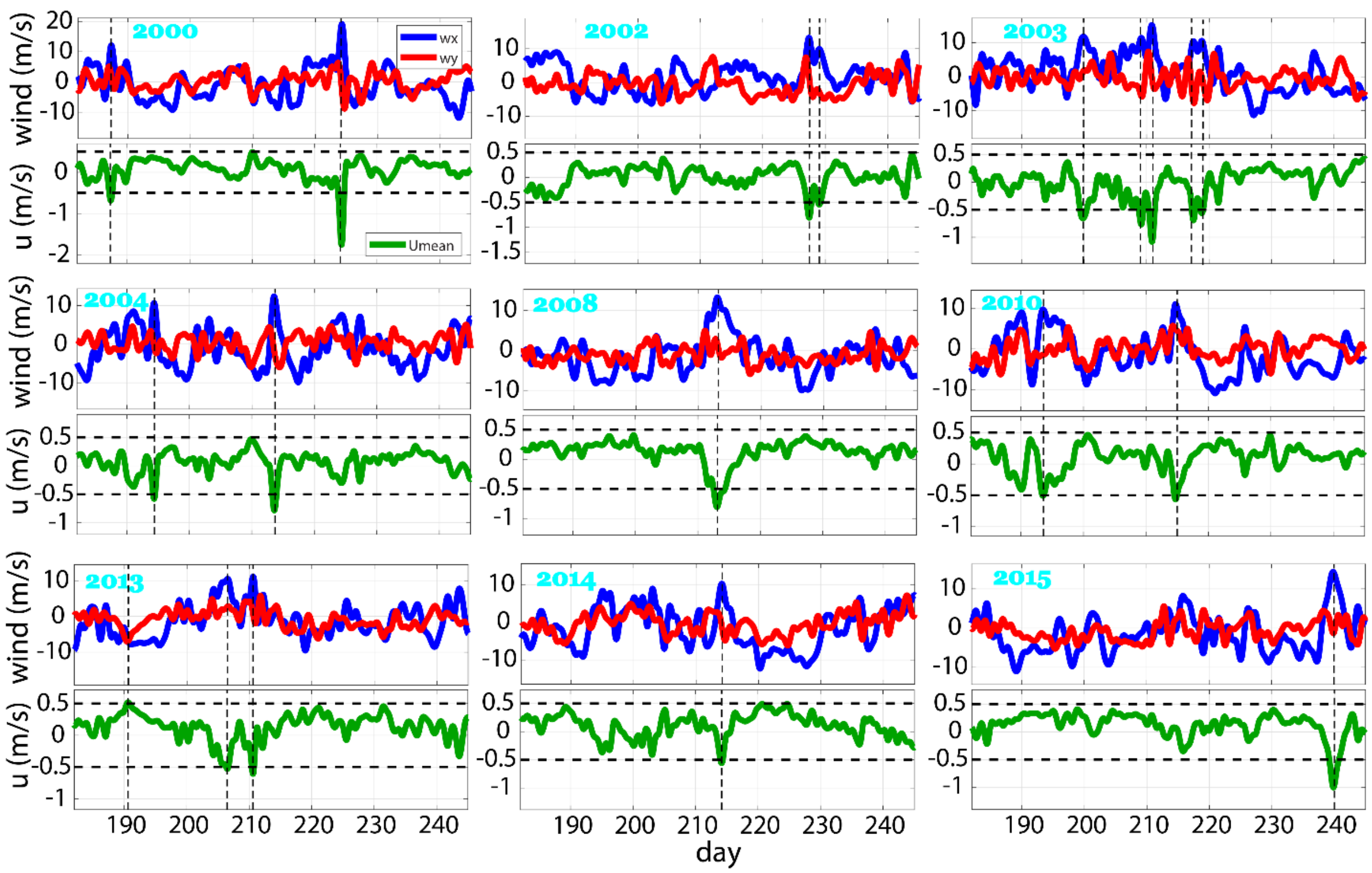

**Figure 4**. **Regression results of the exchange flow at the sampling location in the Eluitkak Pass (1 of 2) - for the years 2000 to 2011**. Observed wind velocity components in m/s are shown in the top panels with blue (east) and red (north) curves. The curves of the lower panels are corresponding subtidal flow velocity u in m/s. The horizontal dashed black lines in the lower panels show the thresholds (|u|>0.5 m/s) of extreme events. The vertical dashed lines show the extreme events with |u|>0.5 m/s. For the years without extreme events, results are shown in the Supplemental Materials (Fig. S57-S82).

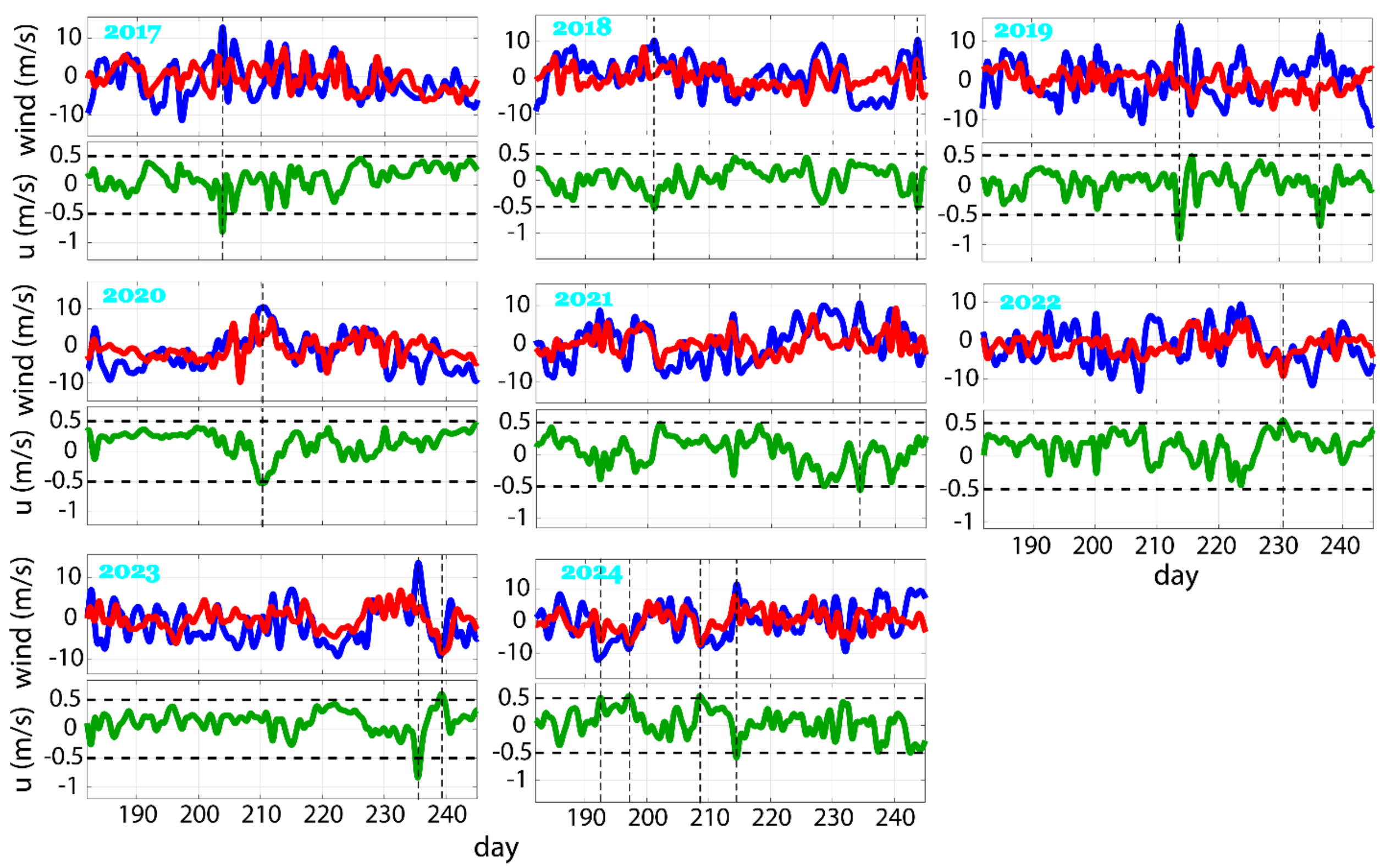

**Figure 5**. **Regression results of the exchange flow at the sampling location in the Eluitkak Pass (2 of 2) - for the years 20*12* to 20*25***. The figure style is the same as Figure 4. For the years without extreme events, results are shown in the Supplemental Materials (Fig. S57-S82).

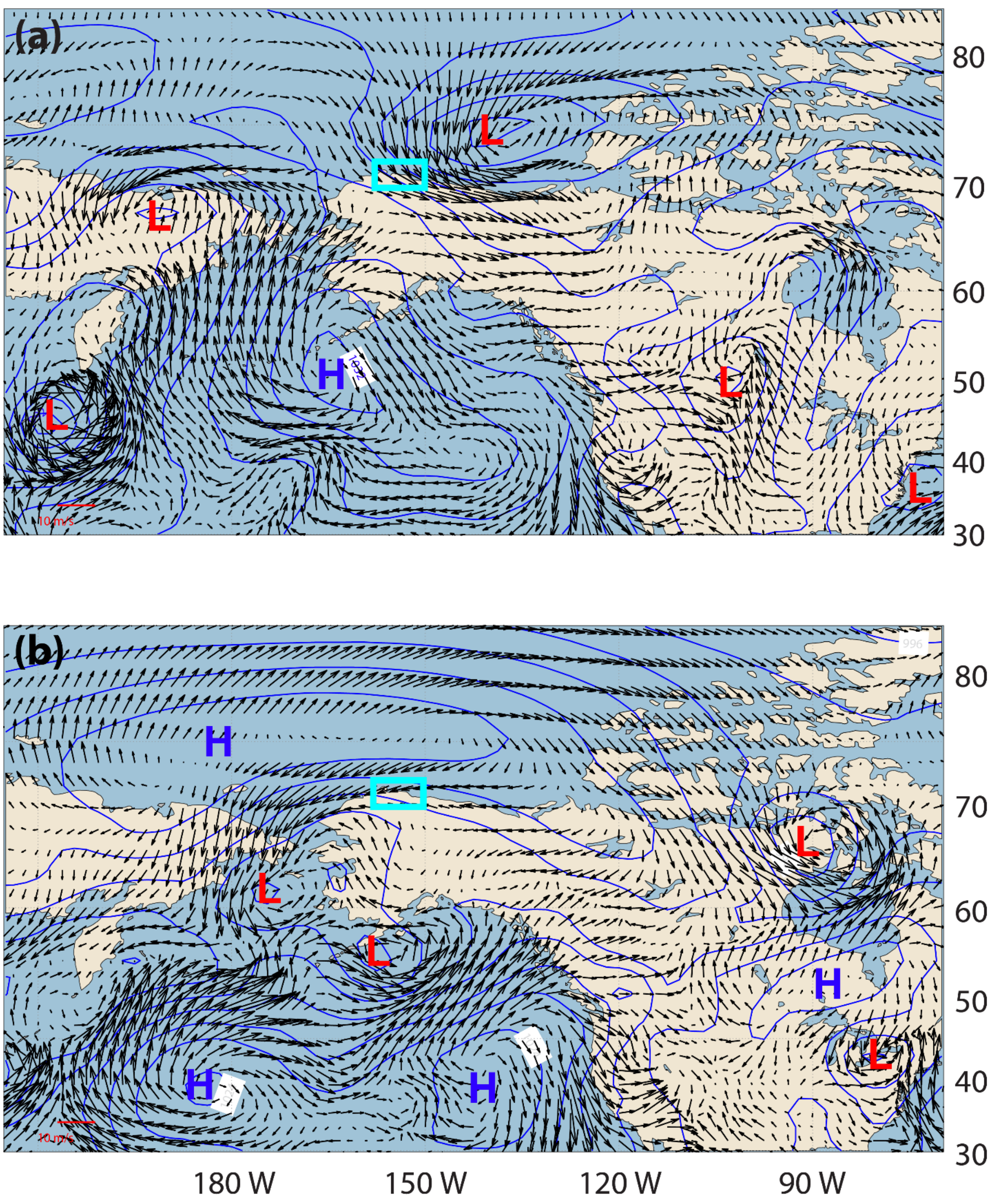

**Figure 6. Weather maps from the NCEP/NCAR reanalysis results**. Shown here are (a) one example for the weather map related to inward flow at the Eluitkak Pass on August 11, 2000; and (b) another example for the weather map related to outward flow at the Eluitkak Pass on July 10, 2024. Maps of other extreme events are included in the Supplemental Materials (Fig. S2-S34).

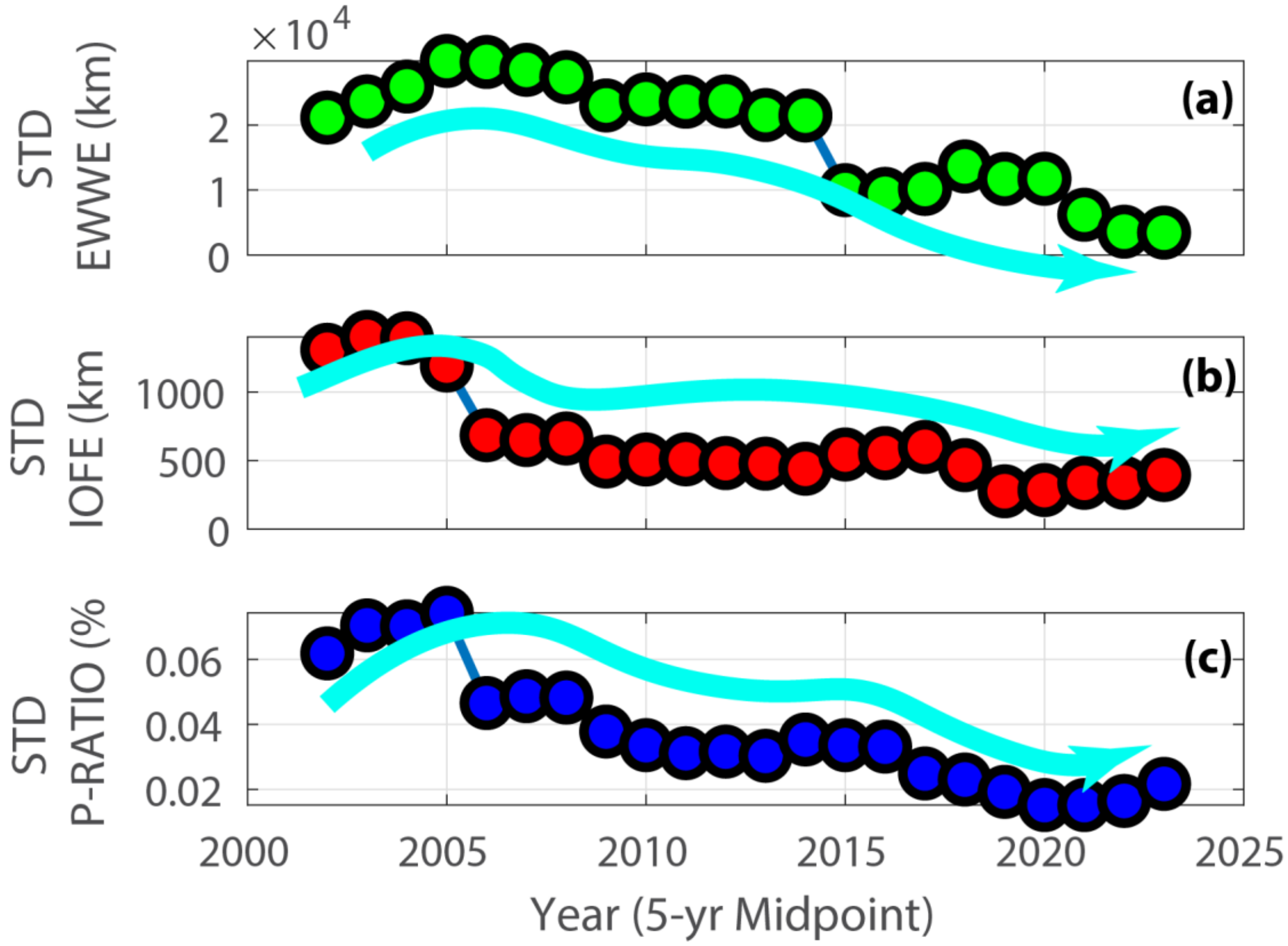

**Figure 7**. **Rolling standard deviation of the time series of indices**. The rolling standard deviation is computed every 5 years centered at every year for (a) the EWWE in km, (b) IOFE in km, and (c) P-RATIO (%) at the Eluitkak Pass. The time is between the year 2000 and 2025 for each summer (1 July to 31 August).

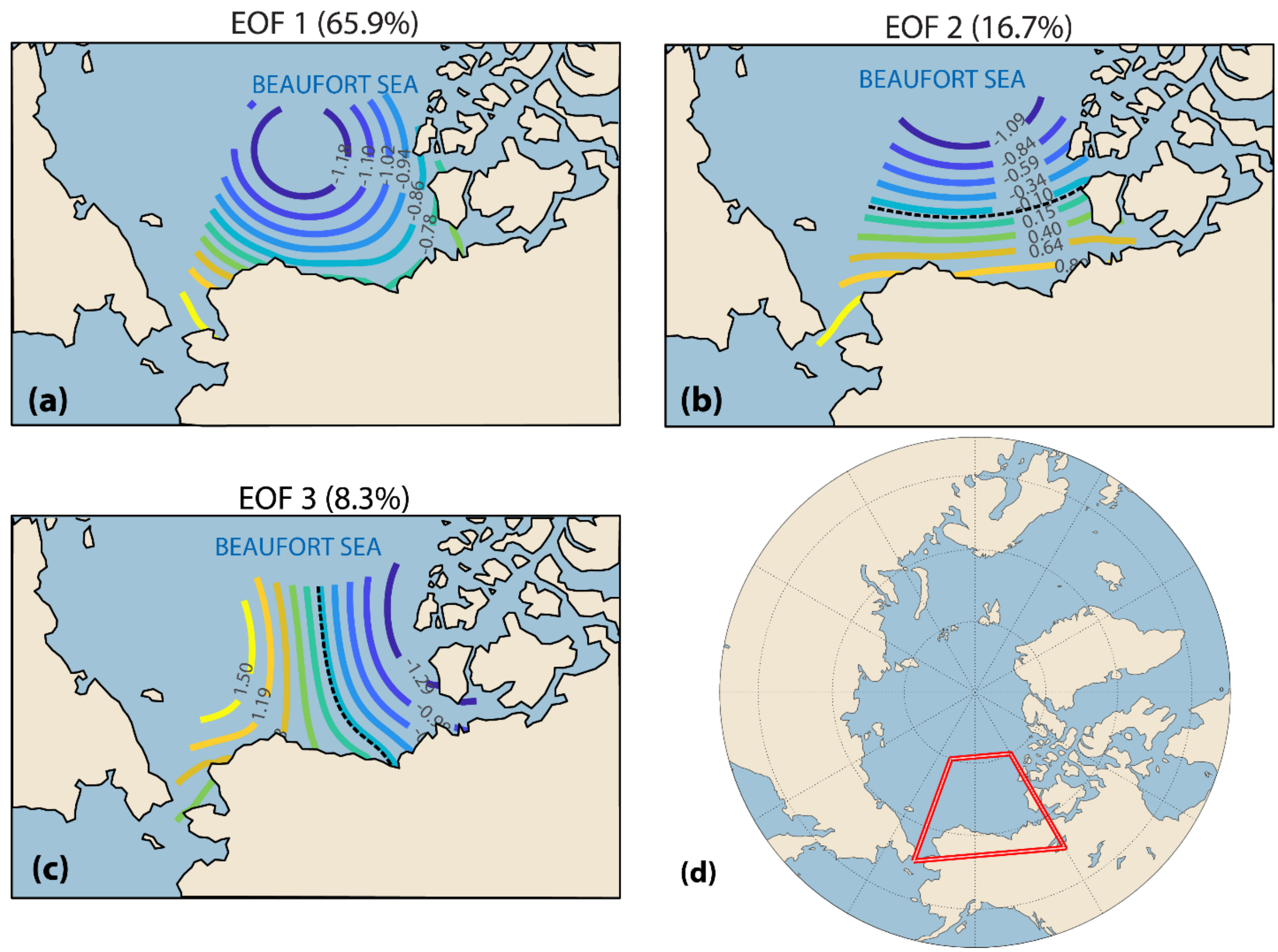

**Figure 8. Spatial patterns of the three leading EOF modes for Arctic Sea Level Pressure.** (a) EOF 1, representing 65.9% of the total variability, characterized by a uniform negative loading across the domain; (b) EOF 2, representing 16.7% of the variability, showing a cross-shelf gradient with negative loadings over the deep ocean and positive loadings toward the coast; (c) EOF 3, representing 8.3% of the variability, displaying a zonal dipole with positive loadings in the west and negative in the east; (d) Map of the study region, where the red trapezoid denotes the ERA5 data coverage area (65–80°N, 170–120°W; July-August, 2000-2025).

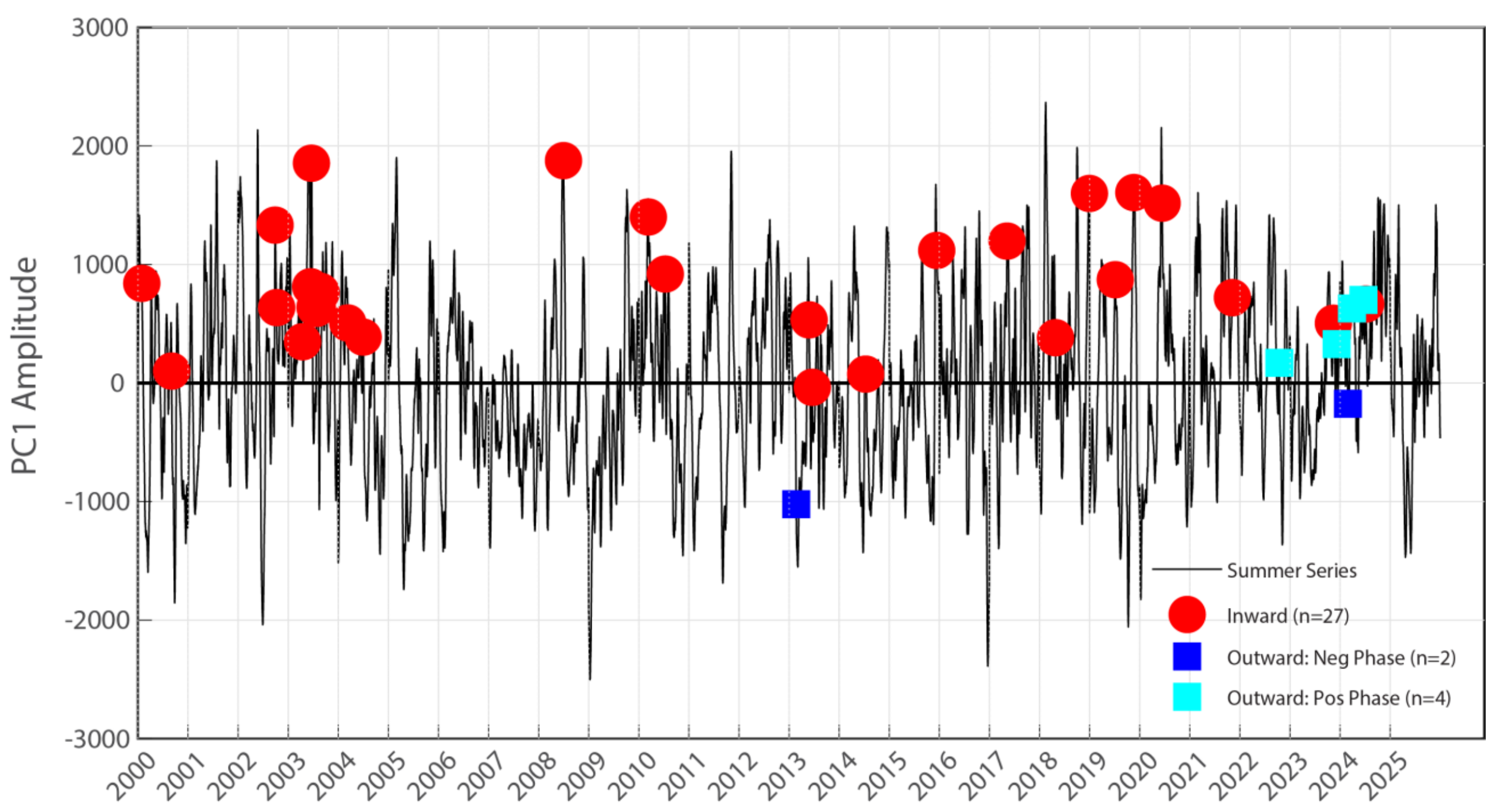

**Figure 9. Principal Component (PC) time coefficients for Mode 1 during summer periods (2000–2025)**. Filled red circles indicate the timing of 27 extreme inward flow events at Eluitkak Pass. Filled blue squares denote extreme outward flow events coinciding with the negative phase of PC1, while filled cyan squares represent outward flow events occurring during the positive phase of PC1.

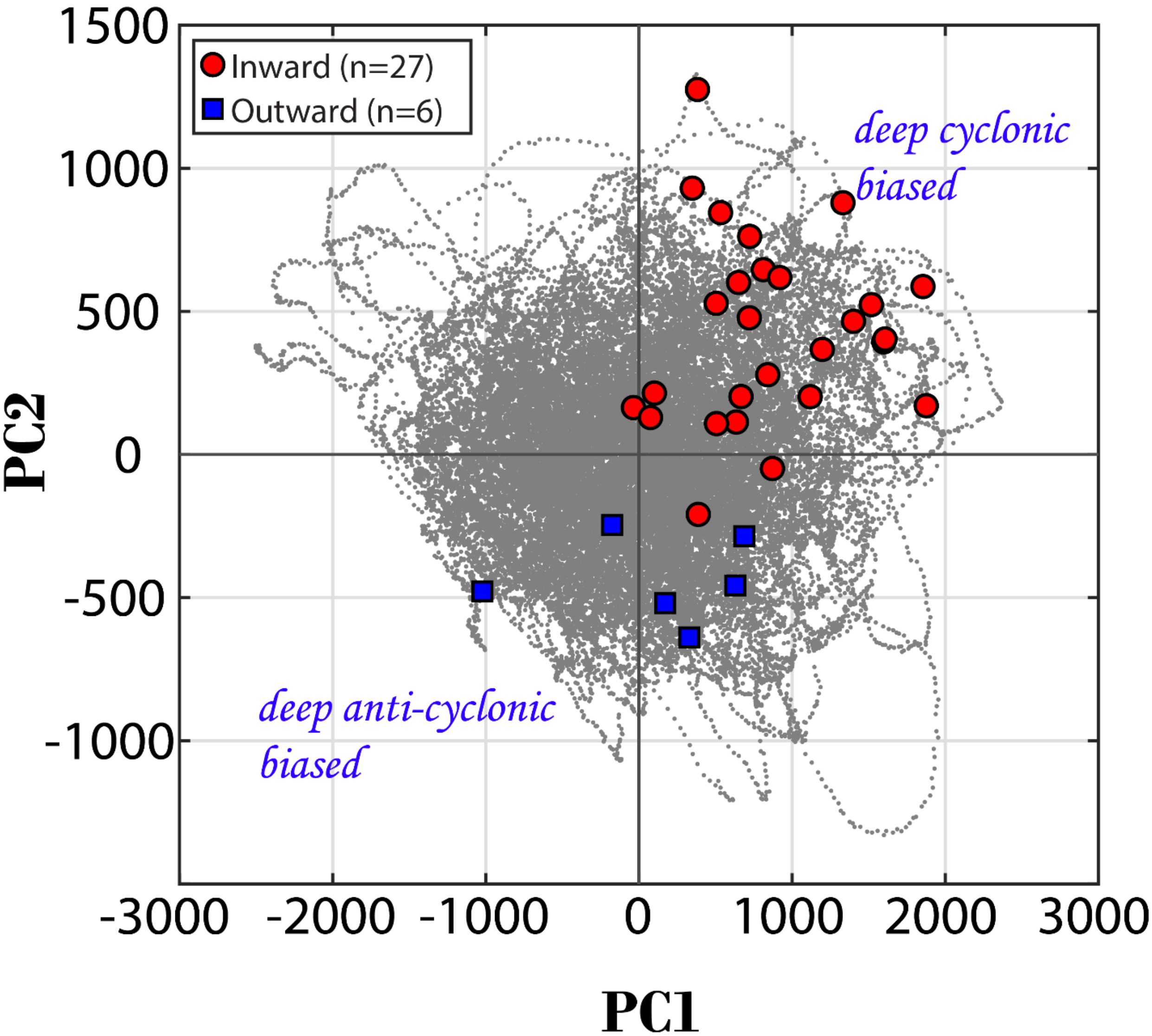

**Figure 10. Phase-space plots of PC1 vs. PC2 from the EOF analysis of sea-level pressure (SLP) anomalies.** Gray dots represent the total summertime phase-space trajectories. Filled red circles and blue squares indicate the positions corresponding to the hydrodynamic extreme inward and extreme outward flow events at Eluitkak Pass, respectively. The first quadrant represents a cyclonic anomaly bias, while the third quadrant represents an anticyclonic anomaly bias.

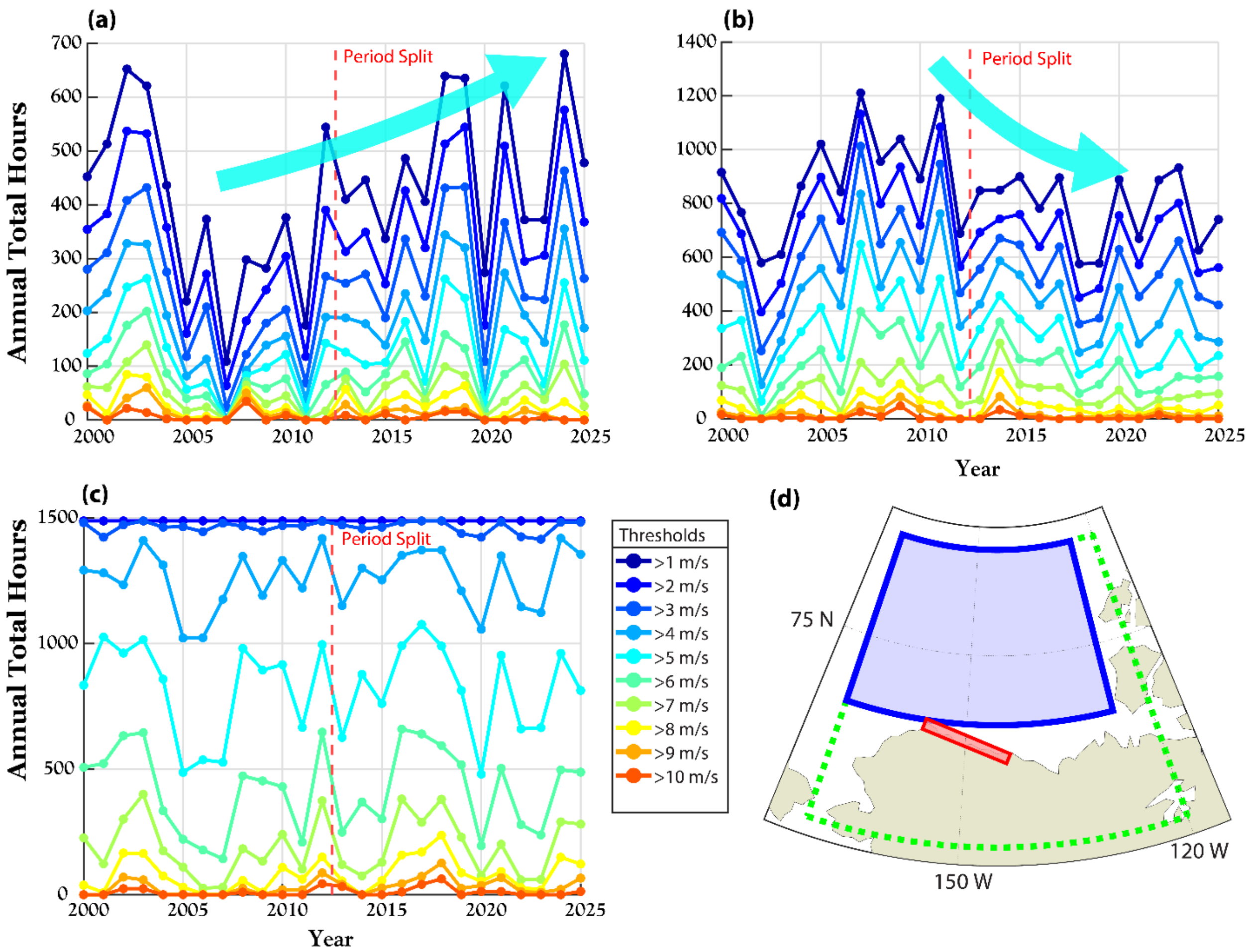

**Figure 11: Trends in wind frequency between 2000-2025**. (a) Time series of the mean annual frequency (hours/year) for coastal westerly wind thresholds ranging from 1 to 10 m/s. Results indicate a ~30% intensification in the frequency of moderate-to-strong westerlies, signaling a shift toward a more stabilized land-high baseline. (b) Time series of the mean annual frequency for coastal easterly wind thresholds (1–10 m/s). In contrast to the westerlies, a consistent decline in frequency is observed across all thresholds, particularly in extreme events (> 10 m/s), which have decreased by 38.32%. (c) Time series of open-ocean wind magnitude frequency. While low-magnitude winds remain stable, the offshore domain exhibits a surge in extreme events (> 9 m/s), increasing by over 80% in the recent 13-year epoch. (d) Map of the study area illustrating the three distinct analytical polygons used for the regional regime evolution analysis: the broader ERA5 atmospheric region, the offshore Ocean region, and the Coastal region adjacent to Eluitkak Pass.

# Regime Shift of Cyclone Influenced Exchange Flows in Coastal Arctic in 2000-2025
(Supplemental Materials)

## Table of Contents

## S1. Validation of model: results from the regression model, FVCOM simulation, and observations from a bottom mounted ADCP.

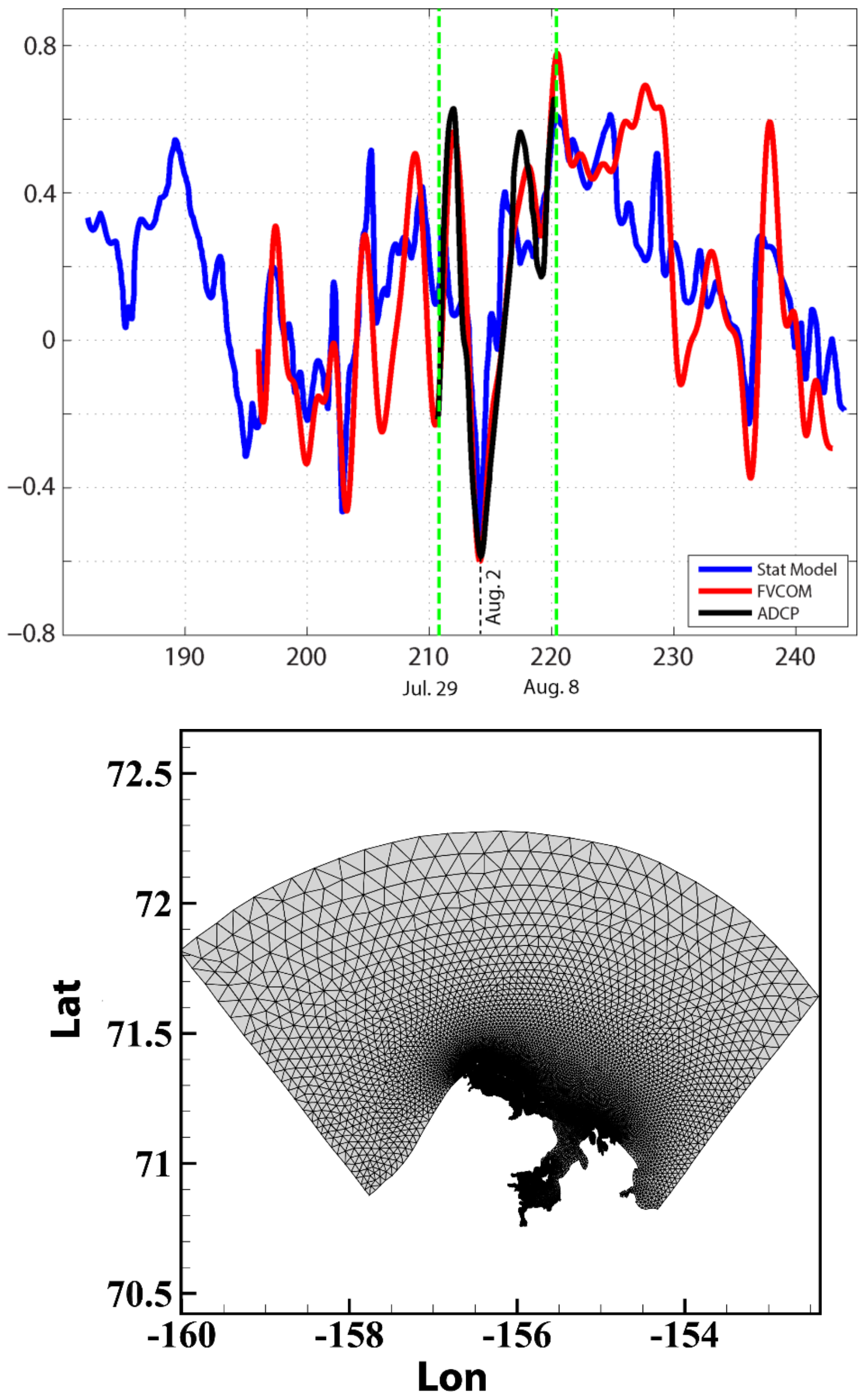

**Figure S1.** Comparison between the regression model, FVCOM output, and observations (top panel). The FVCOM model grid is shown in the lower panel.

## S2. Surface weather maps from reanalysis data for all the 33 events.

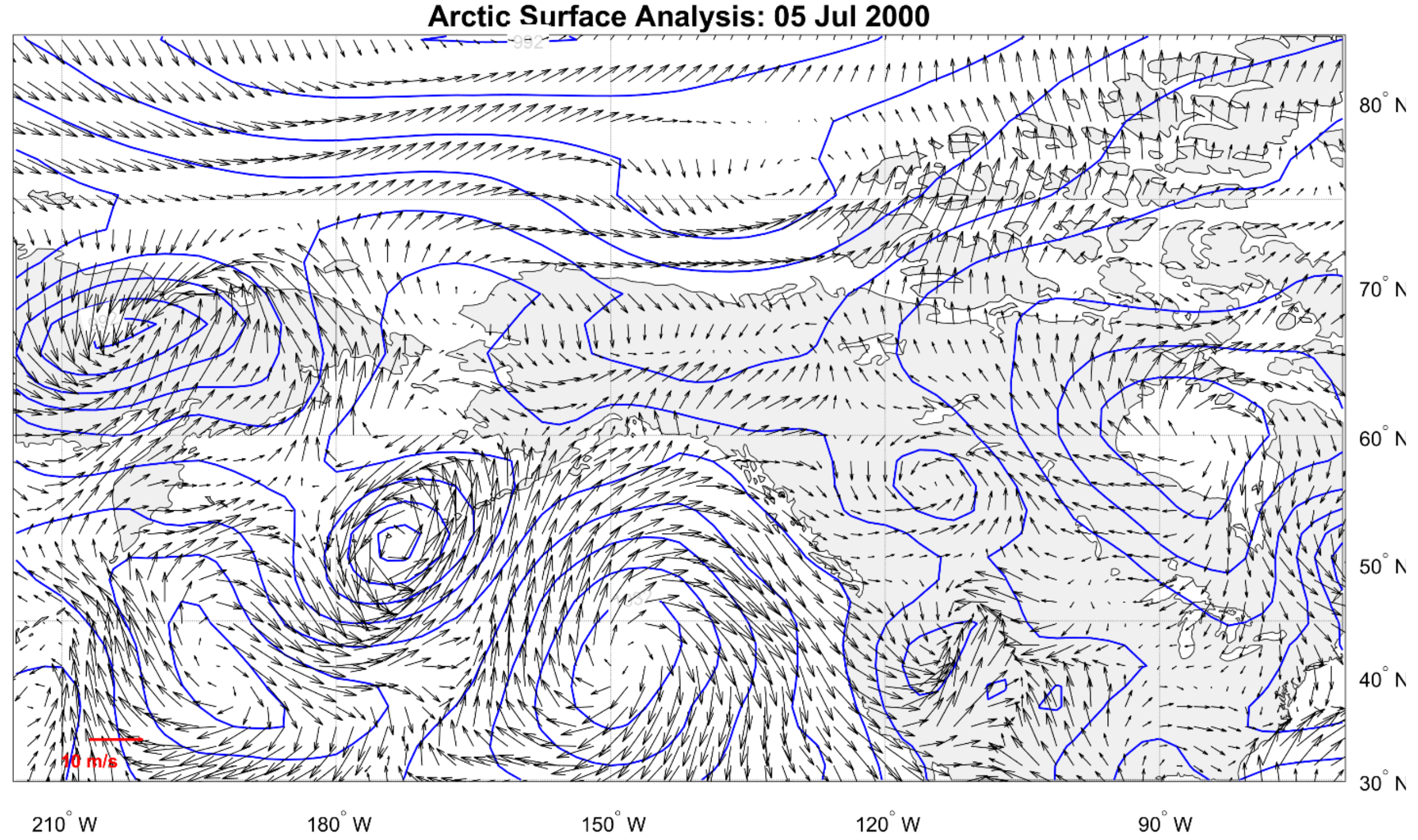

**Figure S2.** Surface analysis for 5 July 2000.

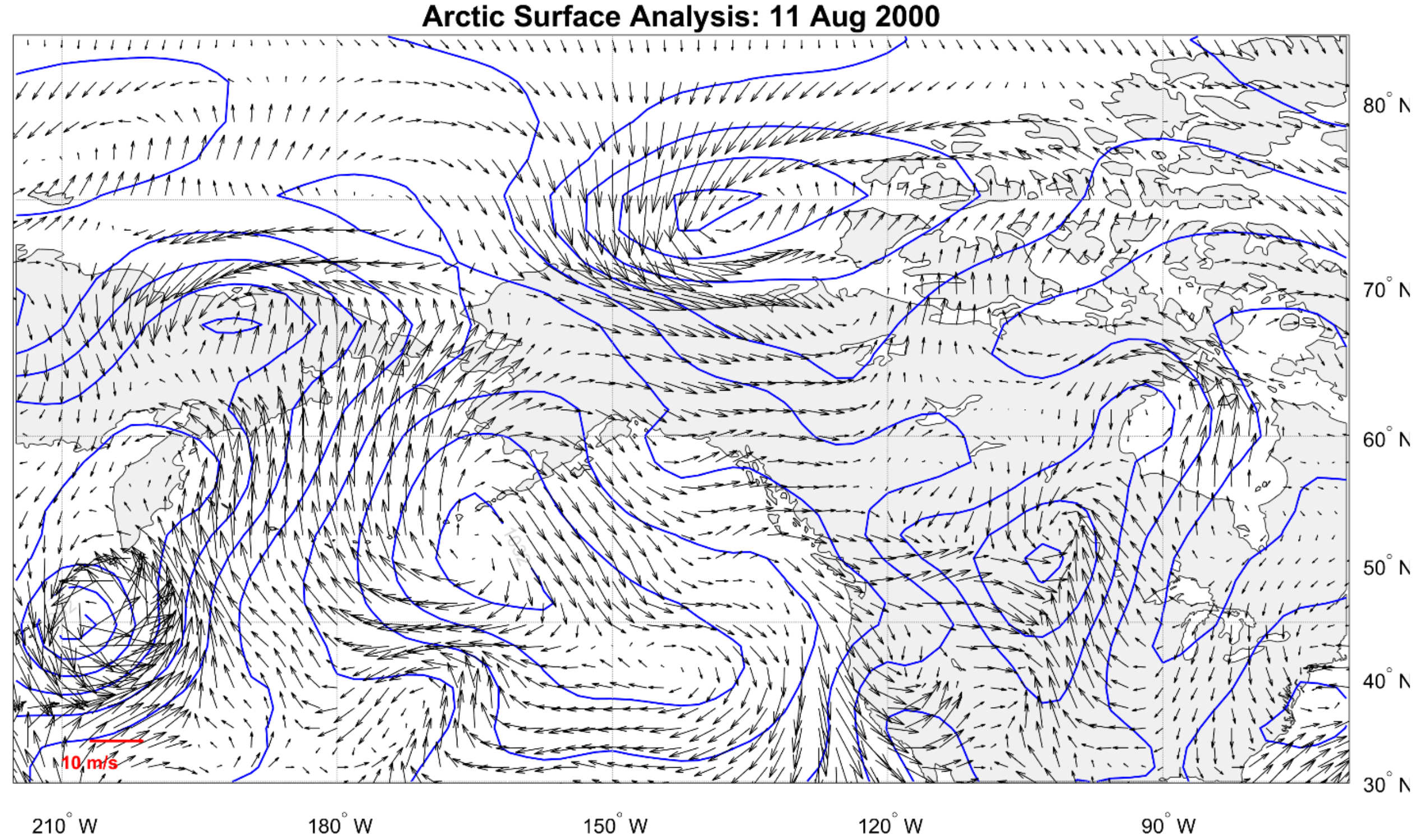

**Figure S3.** Surface analysis for 11 August 2000.

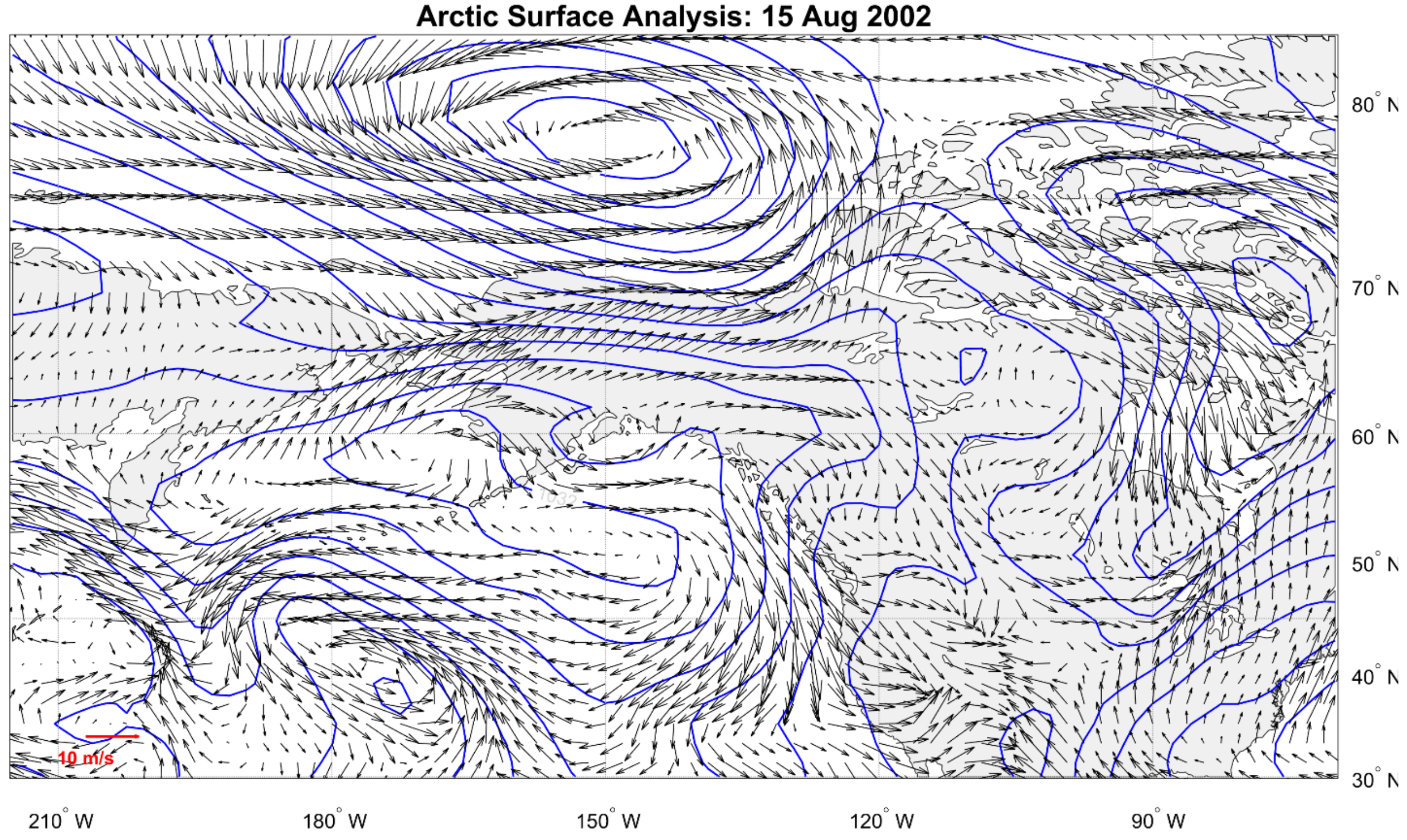

**Figure S4.** Surface analysis for 15 August 2002.

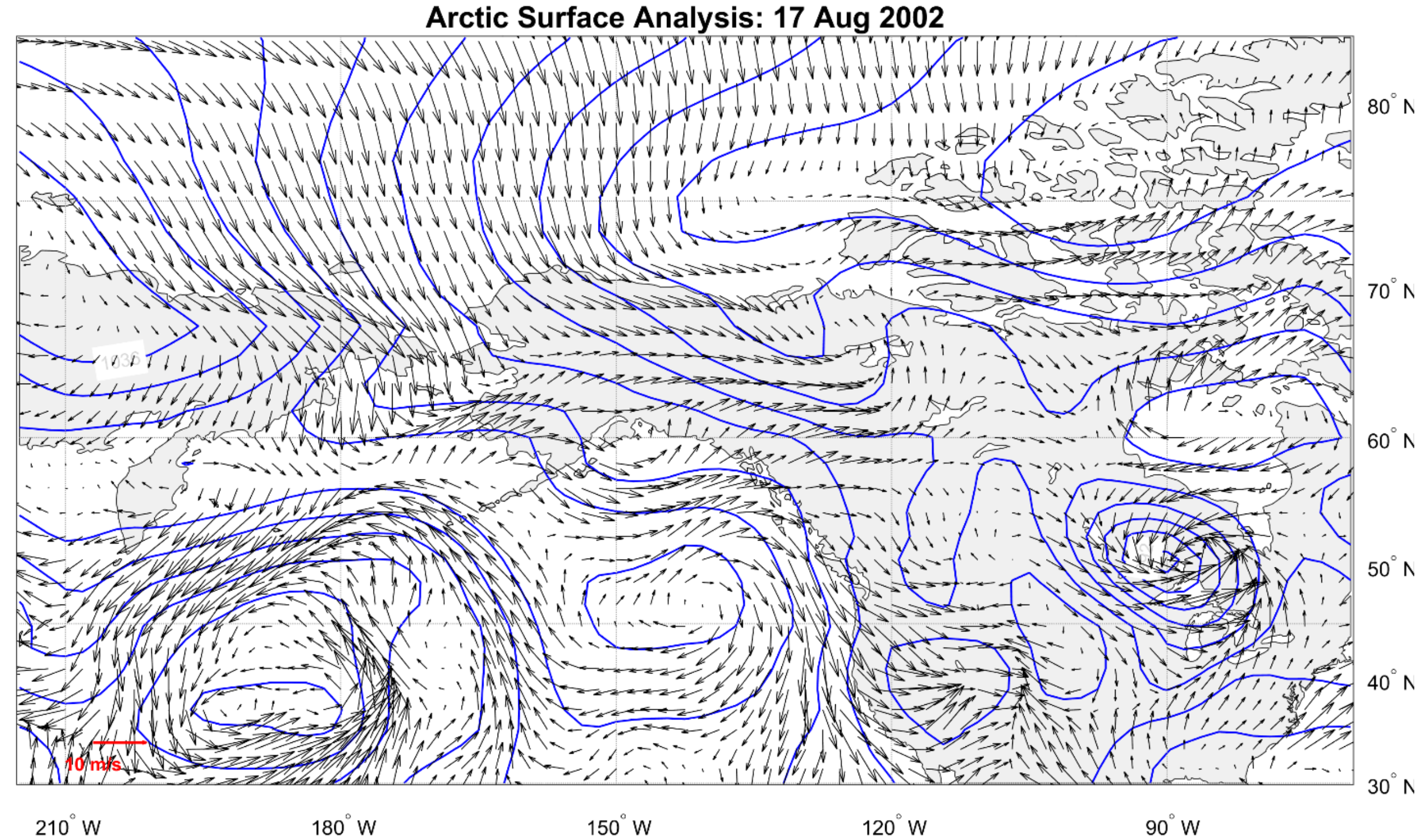

**Figure S5.** Surface analysis for 17 August 2002.

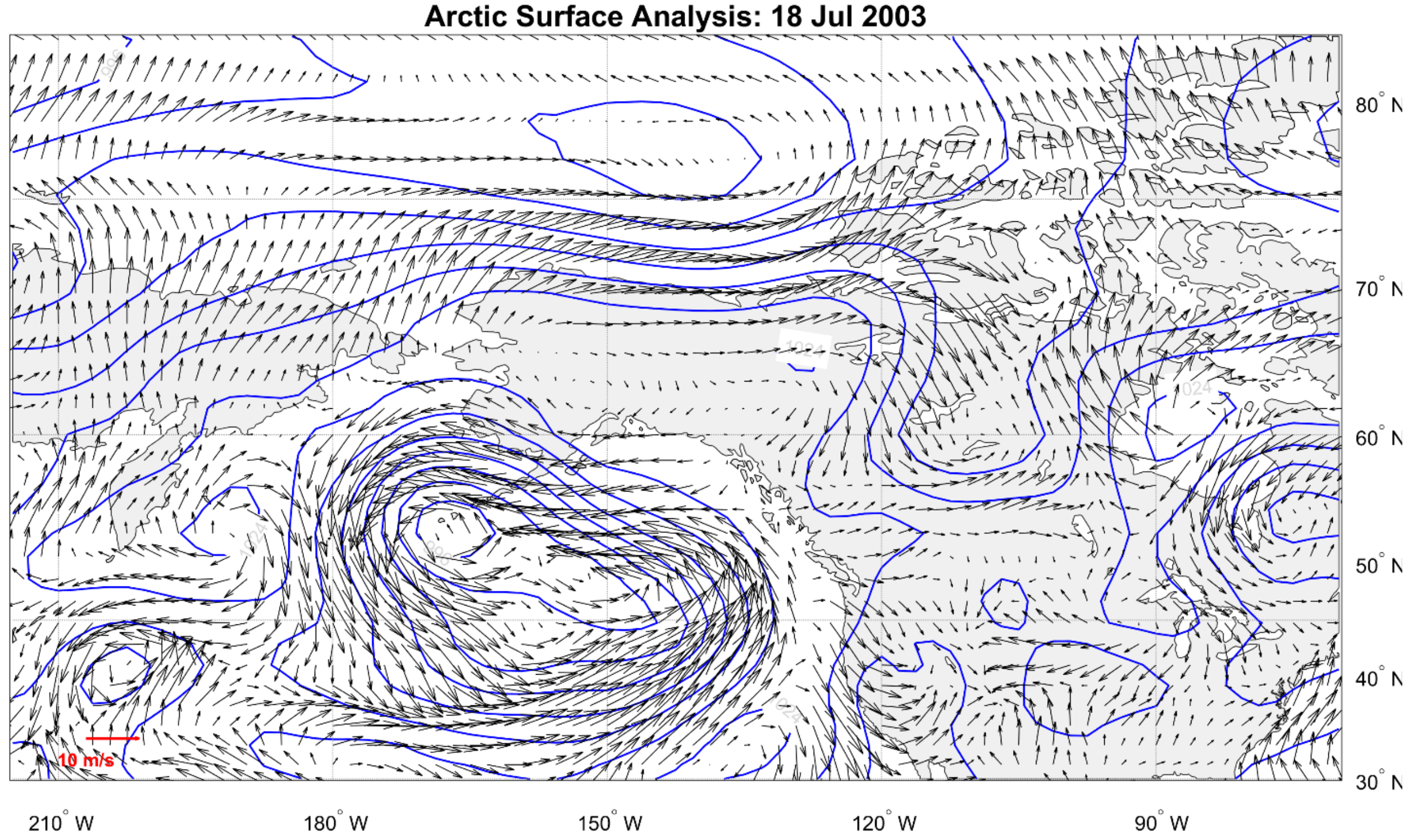

**Figure S6.** Surface analysis for 18 July 2003.

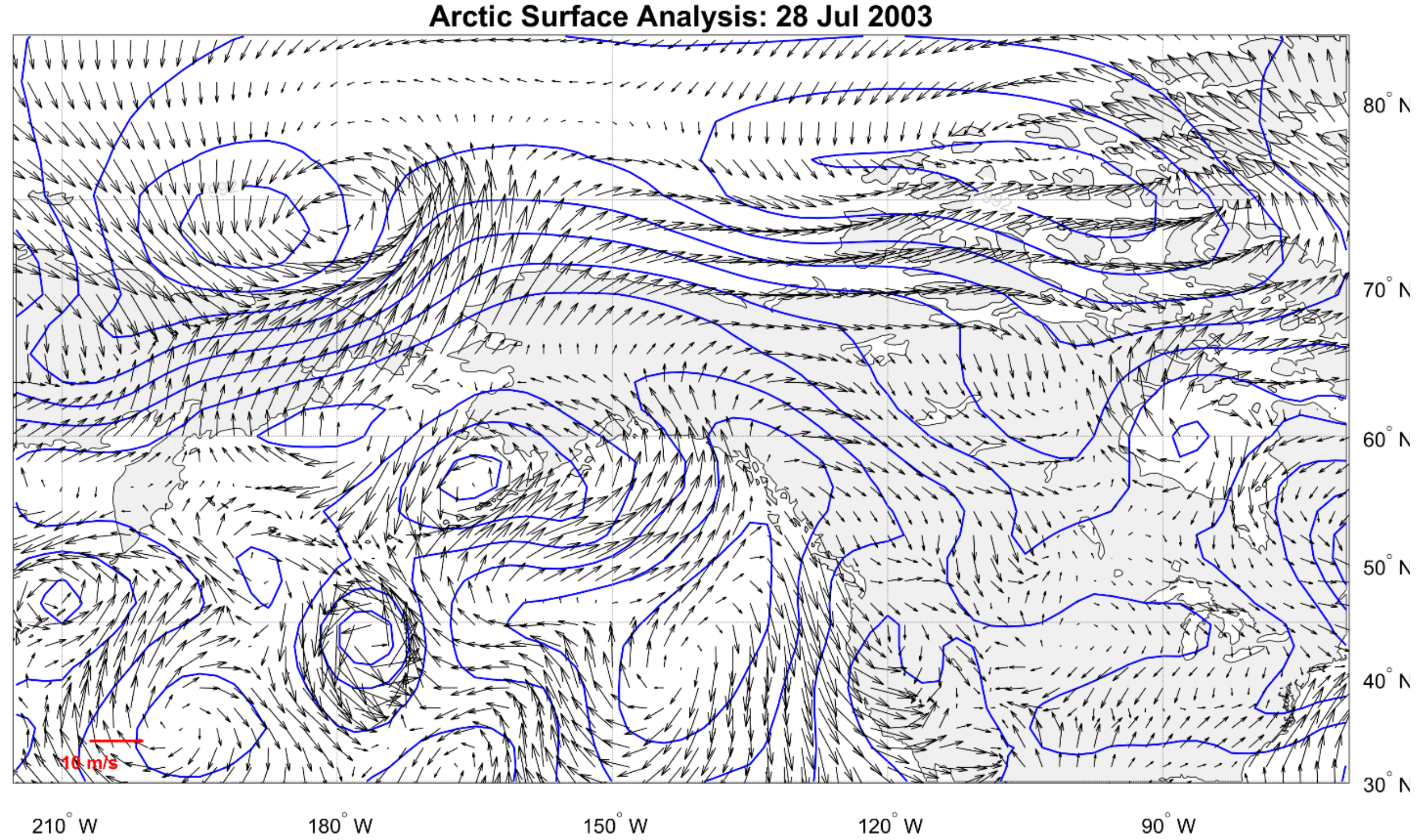

**Figure S7.** Surface analysis for 28 July 2003.

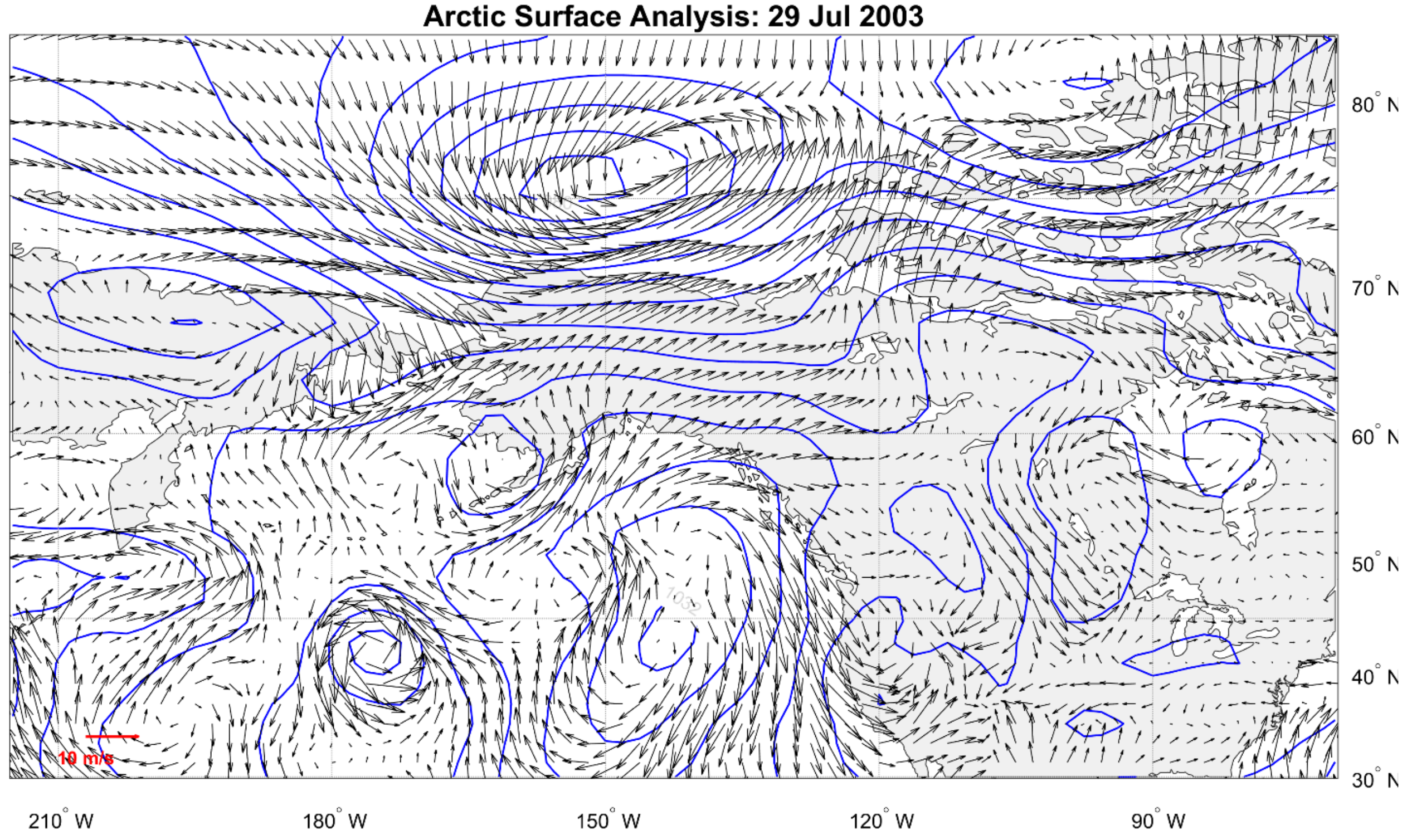

**Figure S8.** Surface analysis for 15 August 2002.

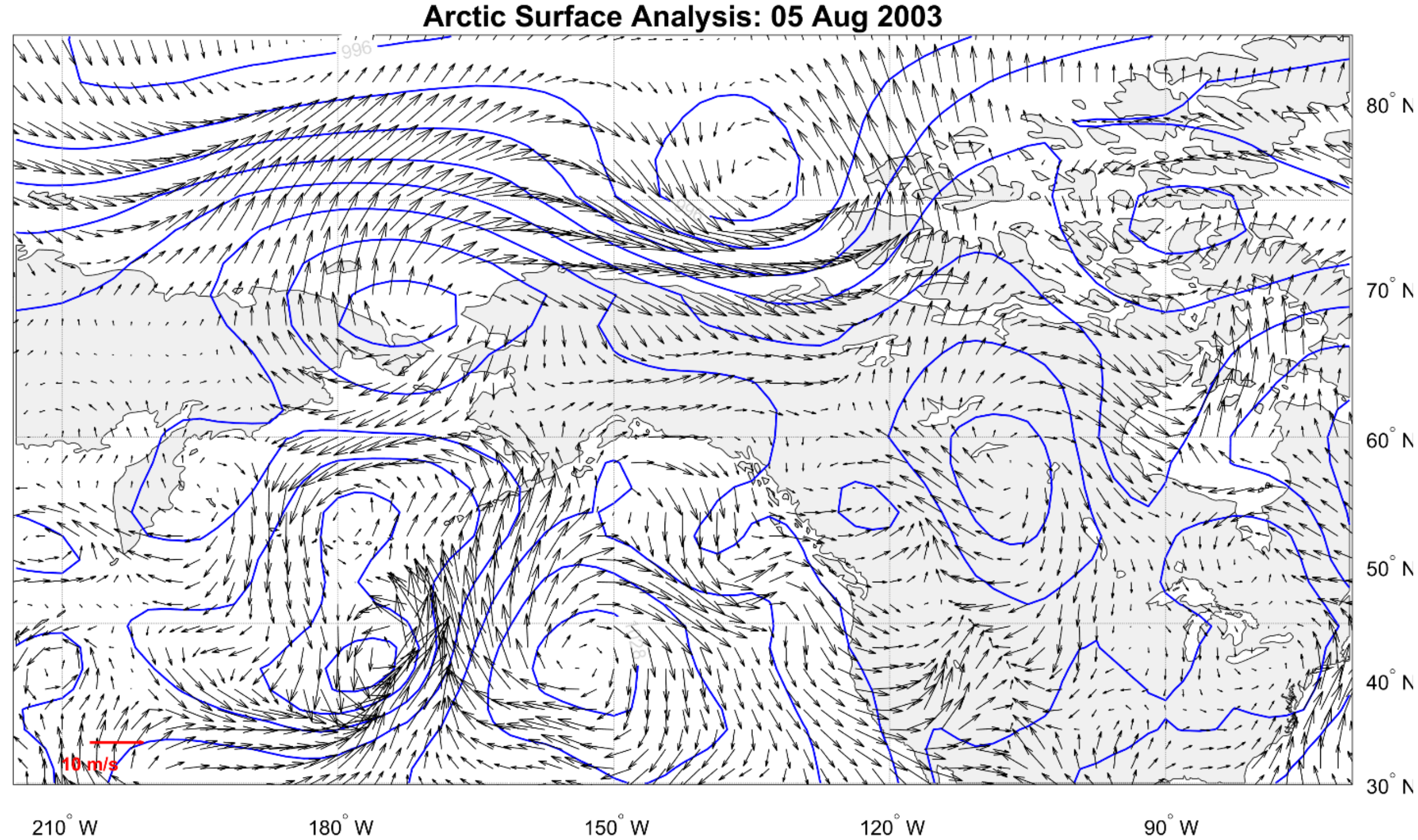

**Figure S9.** Surface analysis for 5 August 2003.

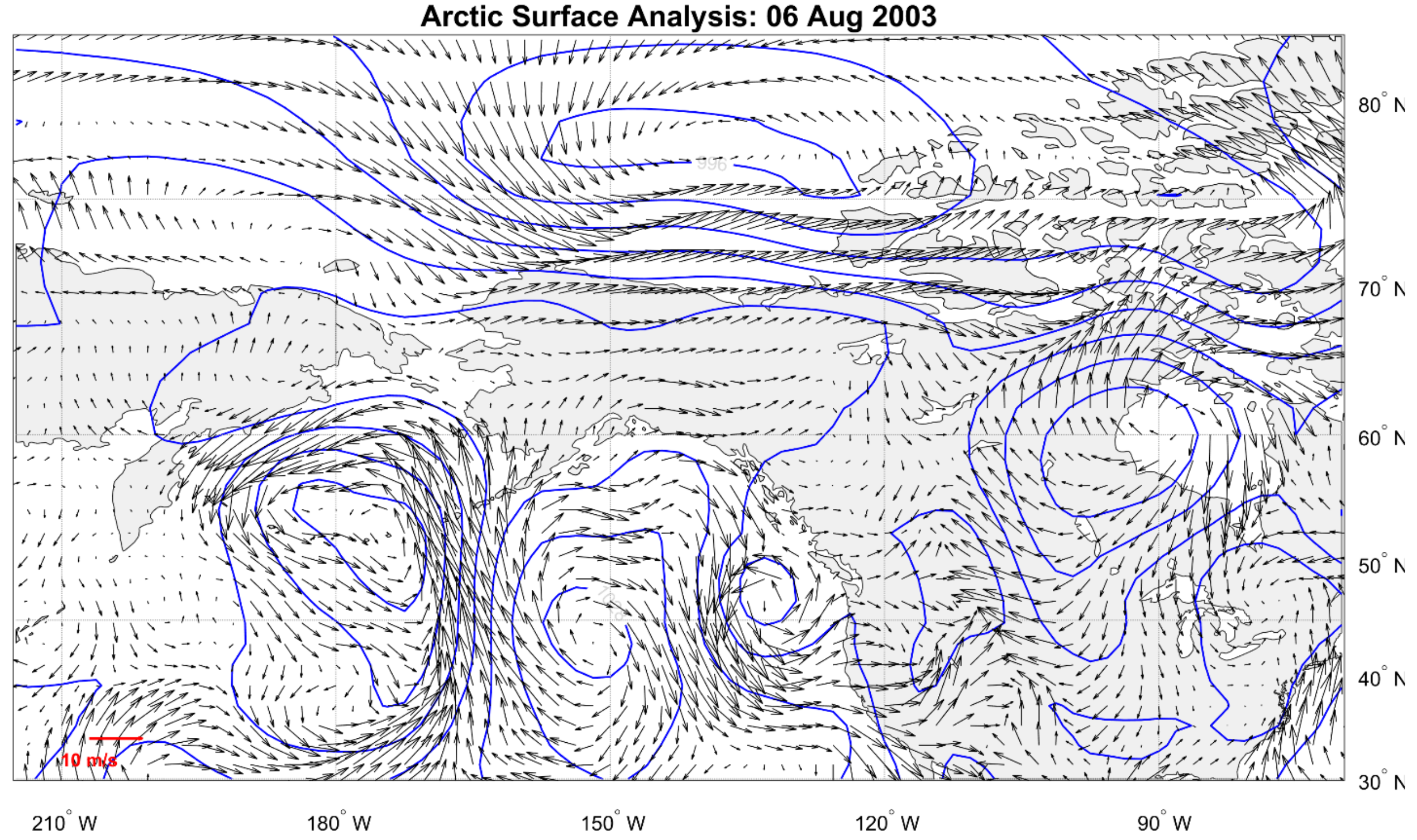

**Figure S10.** Surface analysis for 6 August 2003.

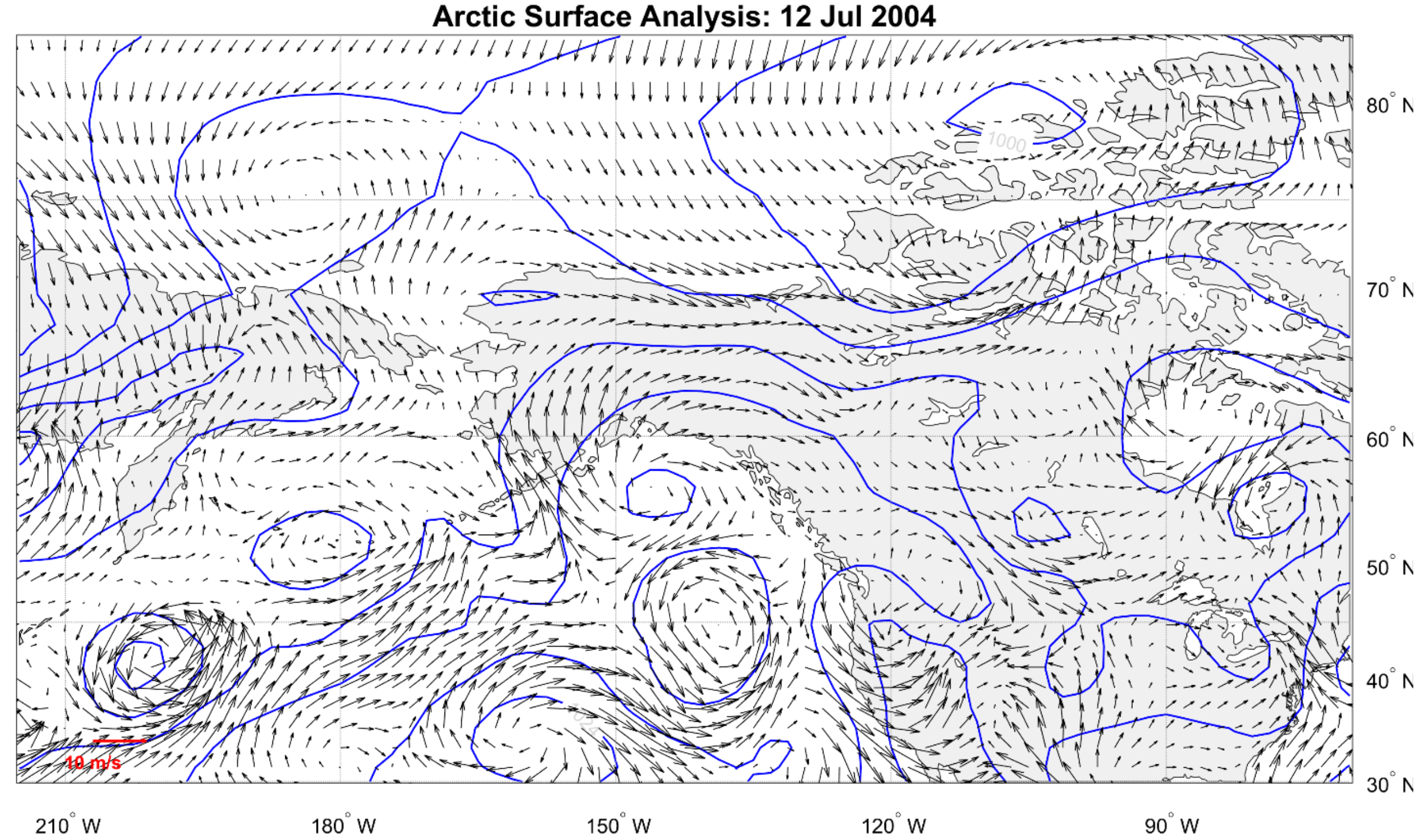

**Figure S11.** Surface analysis for 12 July 2004.

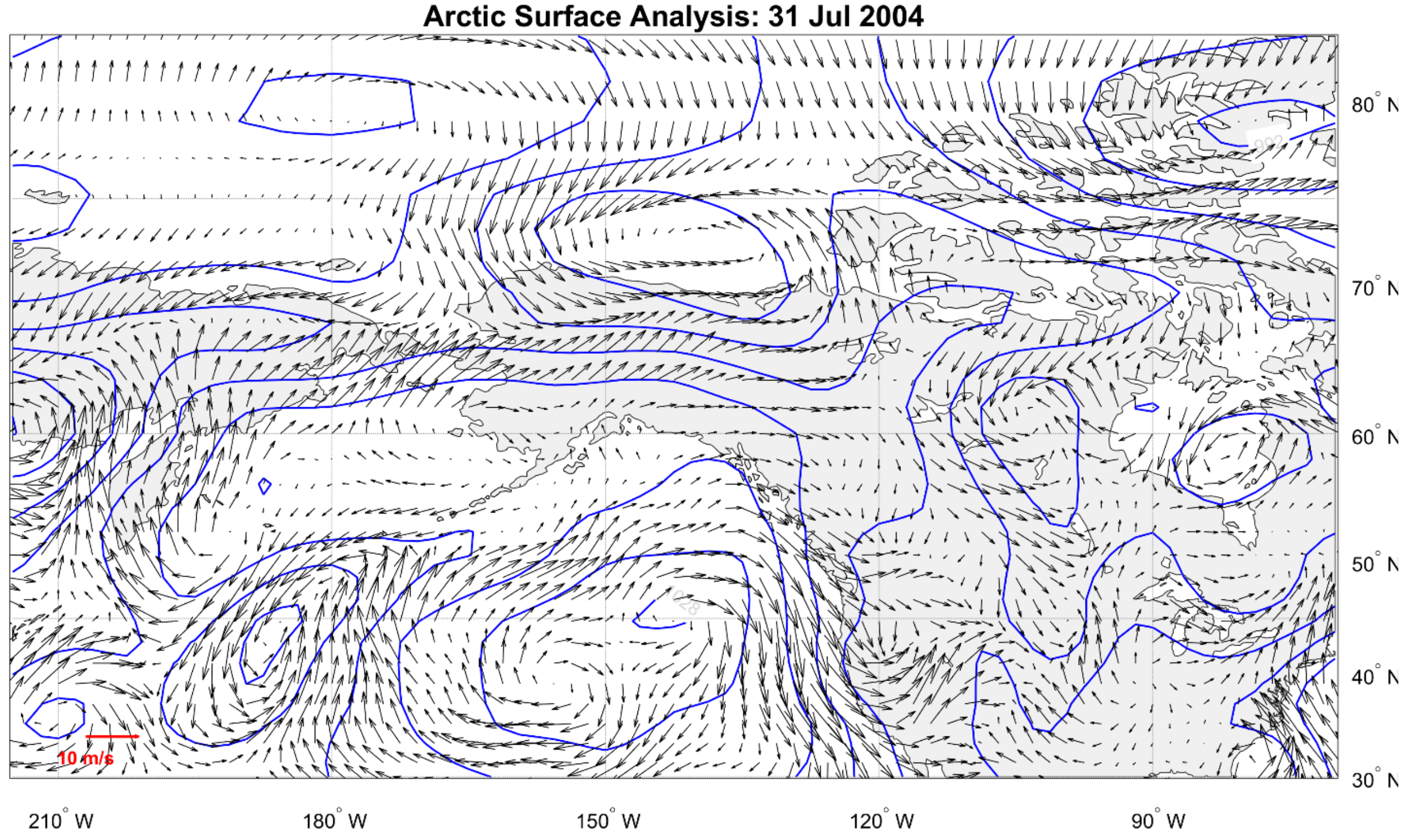

**Figure S12.** Surface analysis for 31 July 2004.

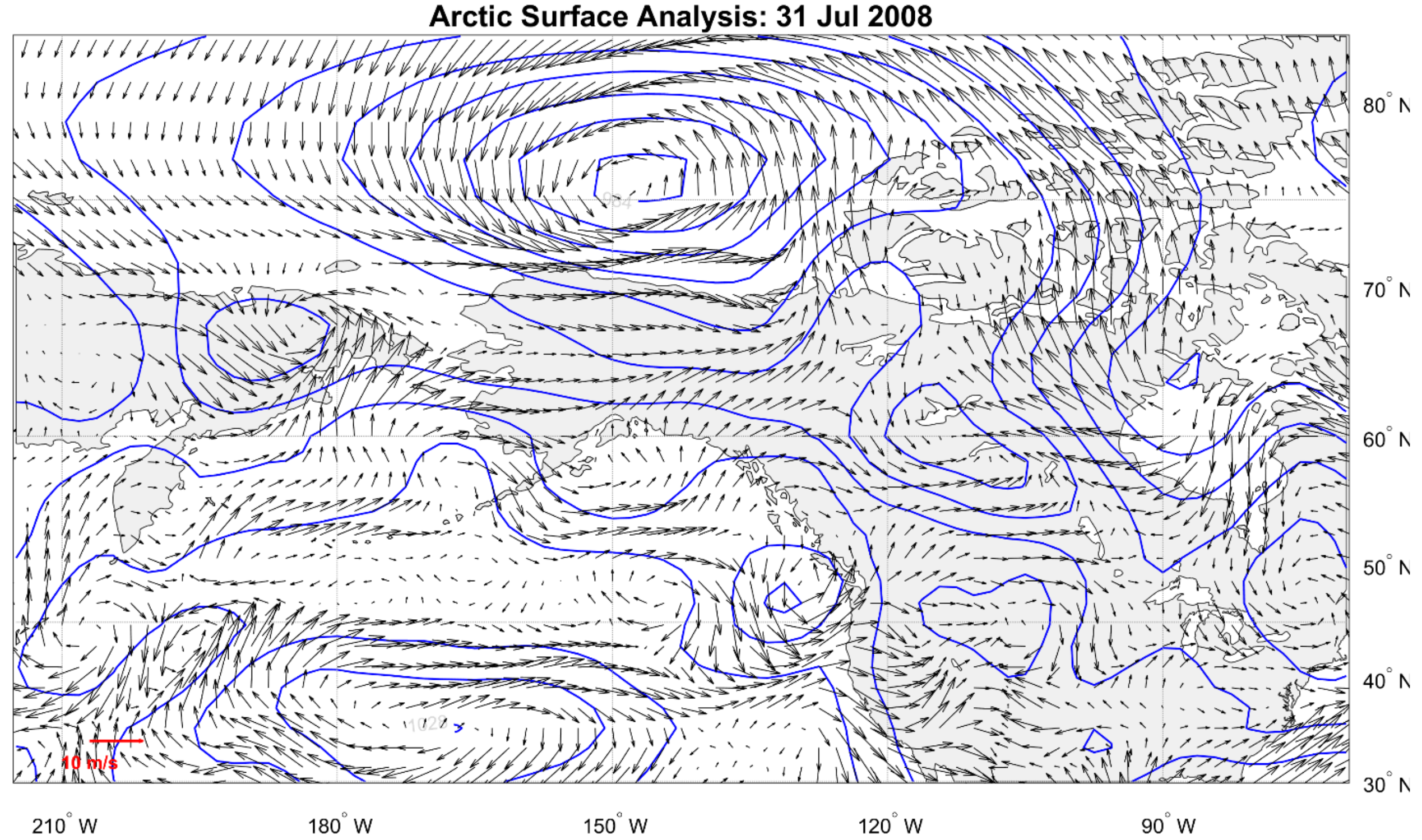

**Figure S13.** Surface analysis for 31 July 2008.

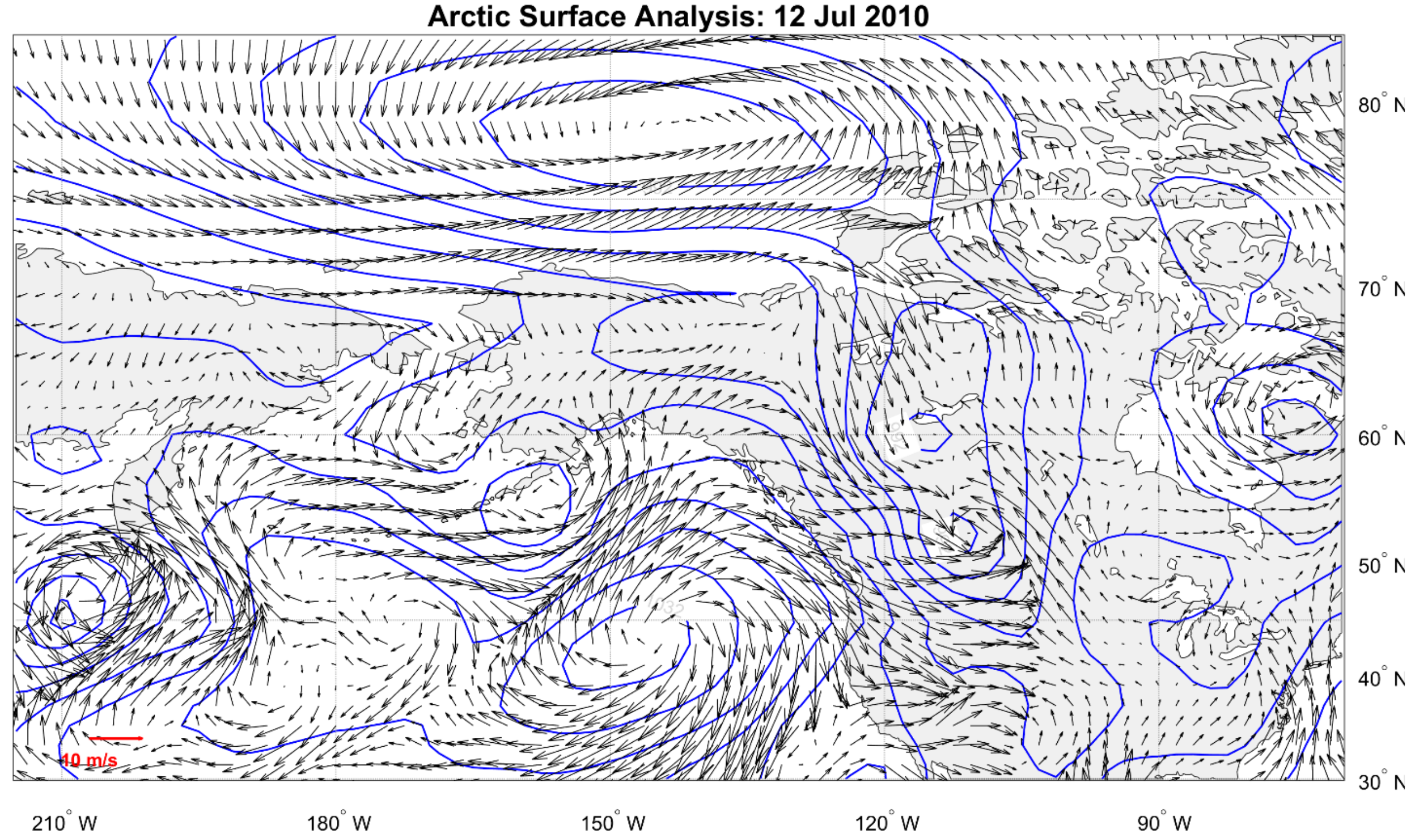

**Figure S14.** Surface analysis for 12 July 2010.

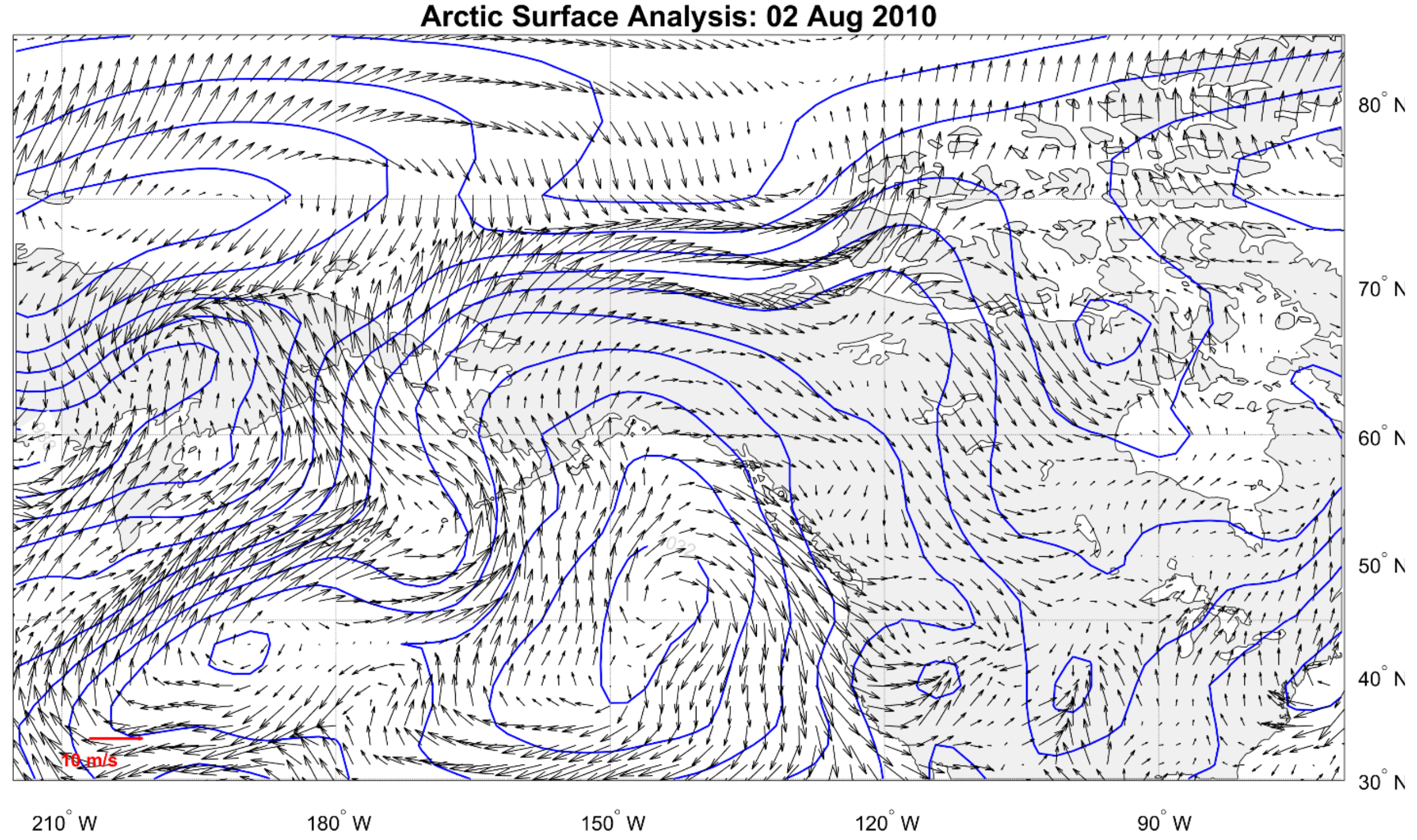

**Figure S15.** Surface analysis for 2 August 2010.

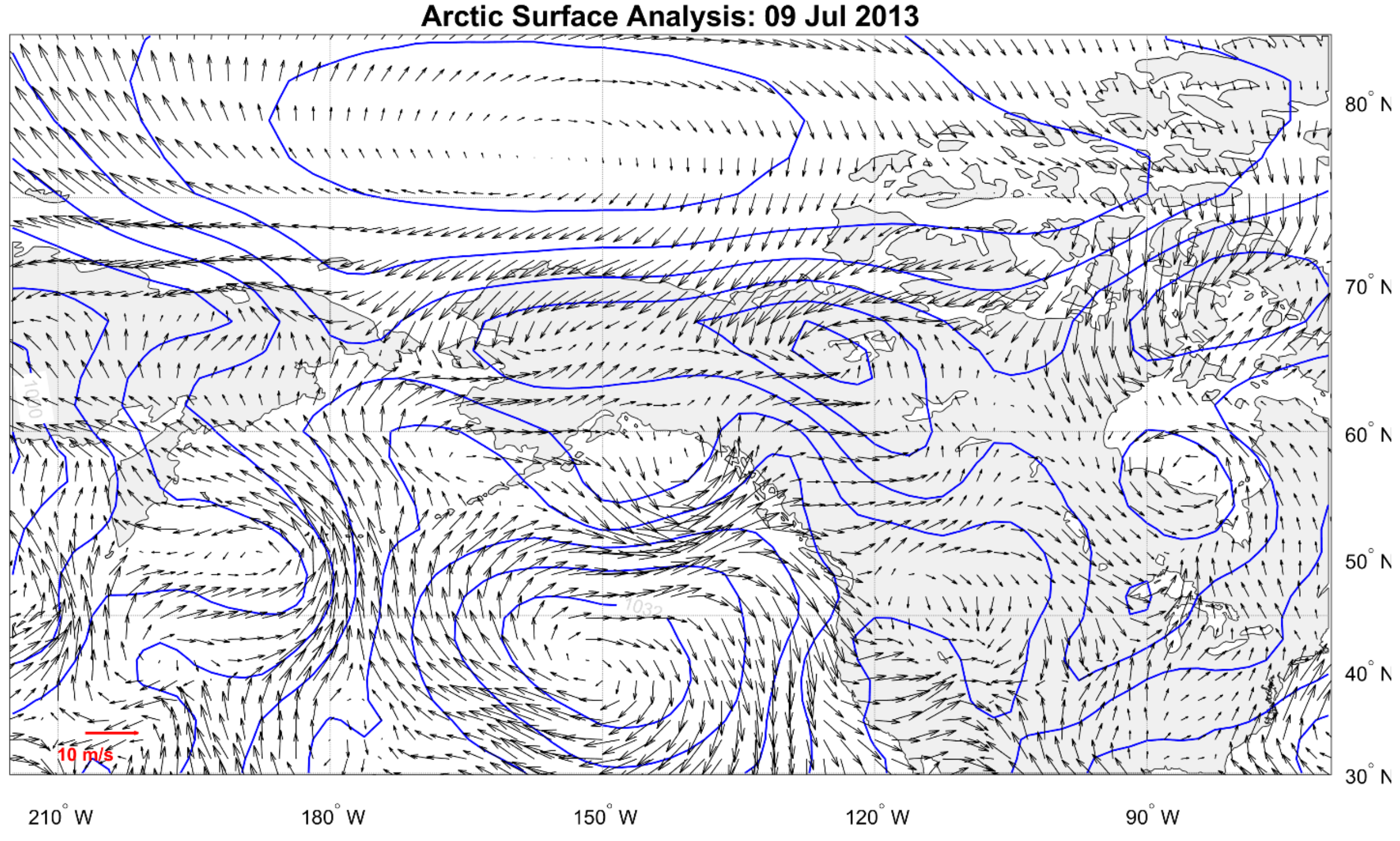

**Figure S16.** Surface analysis for 9 July 2013.

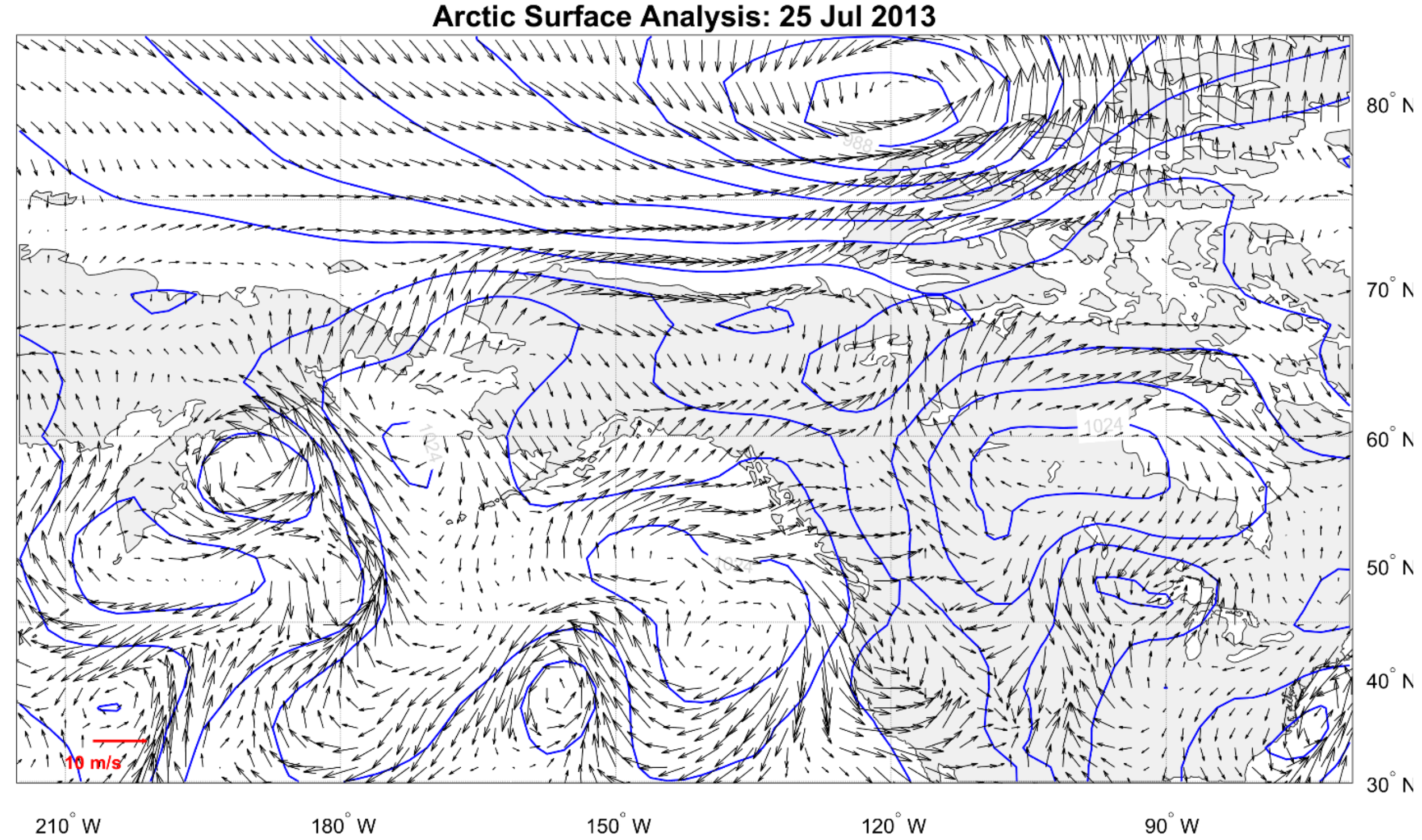

**Figure S17.** Surface analysis for 25 July 2013.

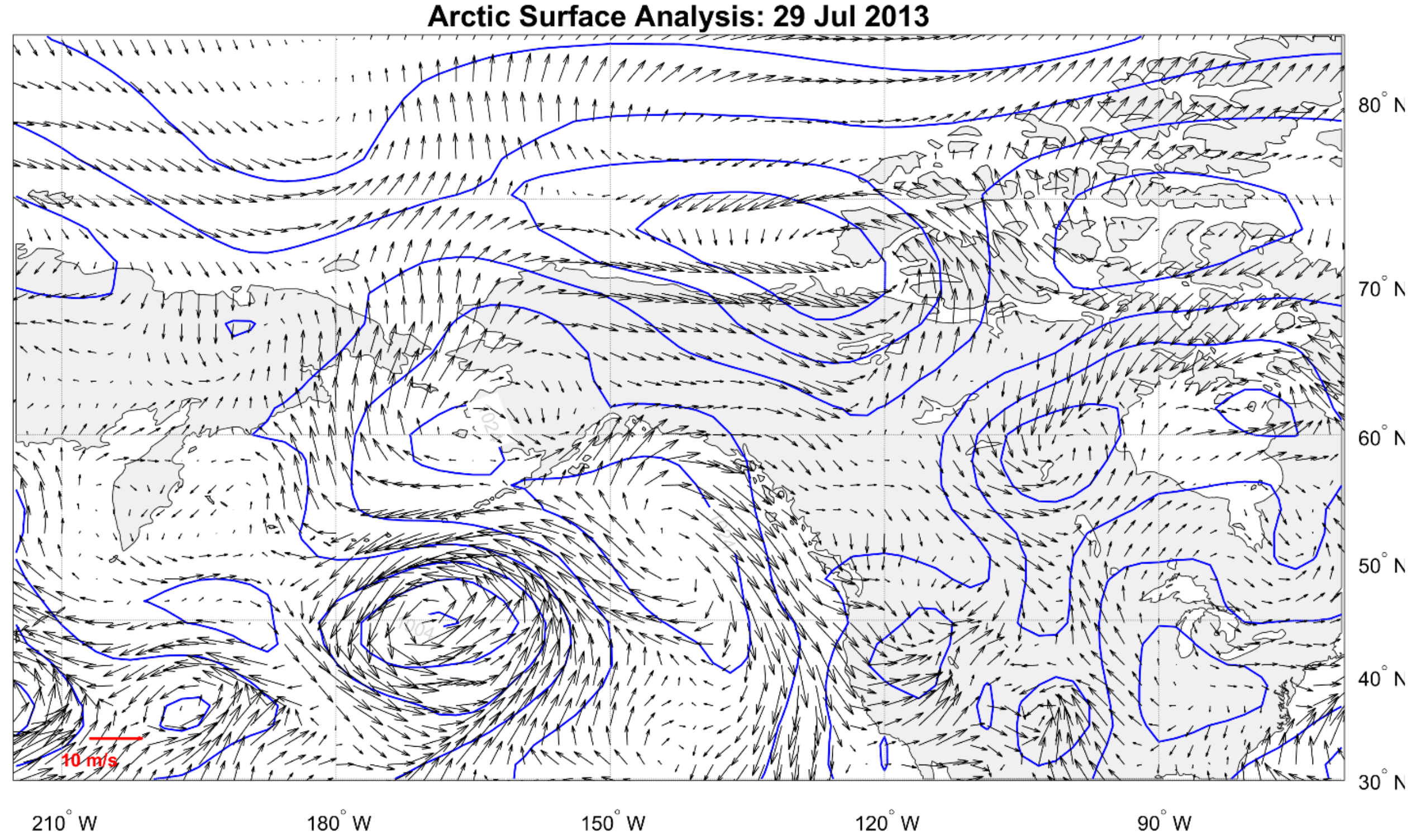

**Figure S18.** Surface analysis for 29 July 2013.

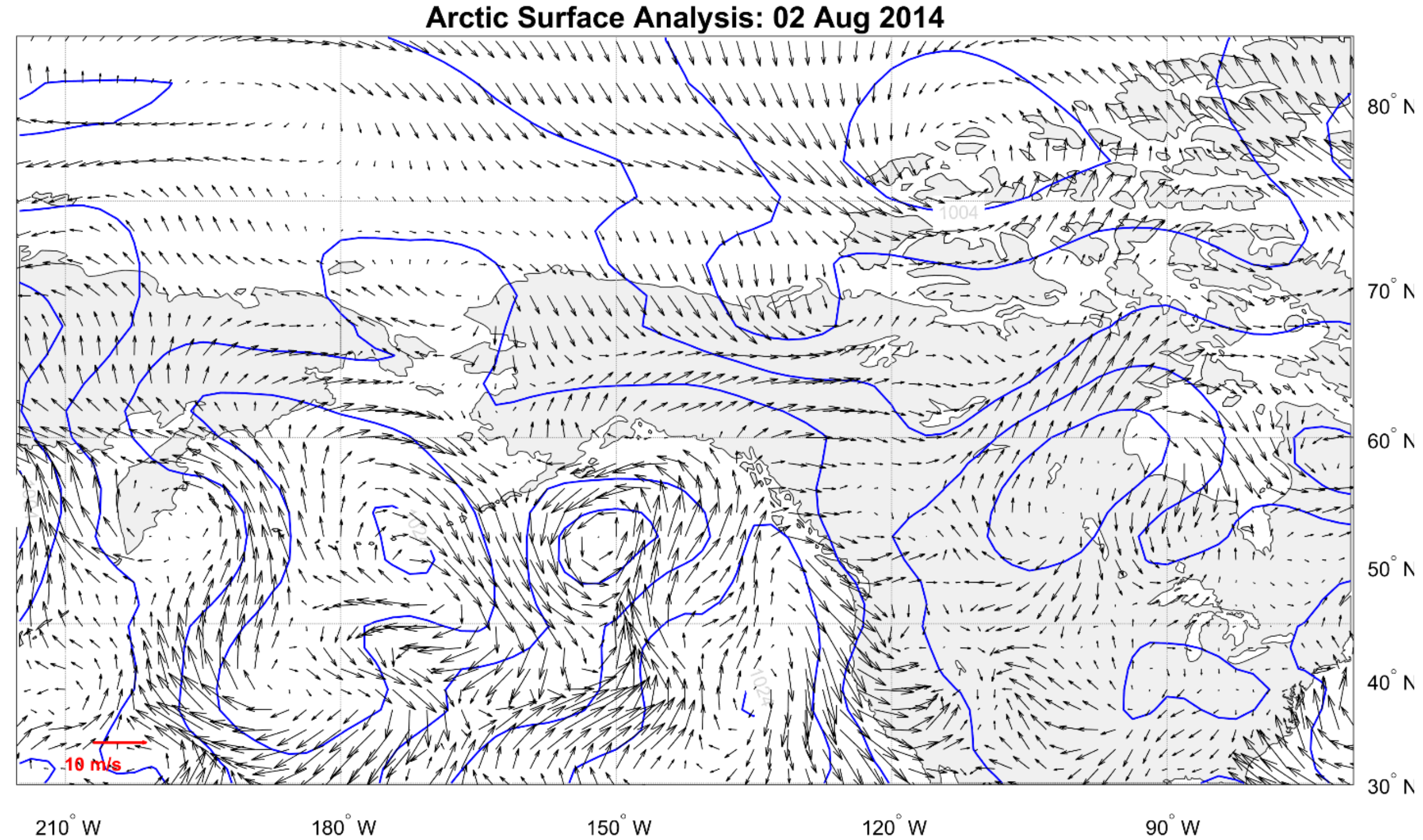

**Figure S19.** Surface analysis for 2 August 2014.

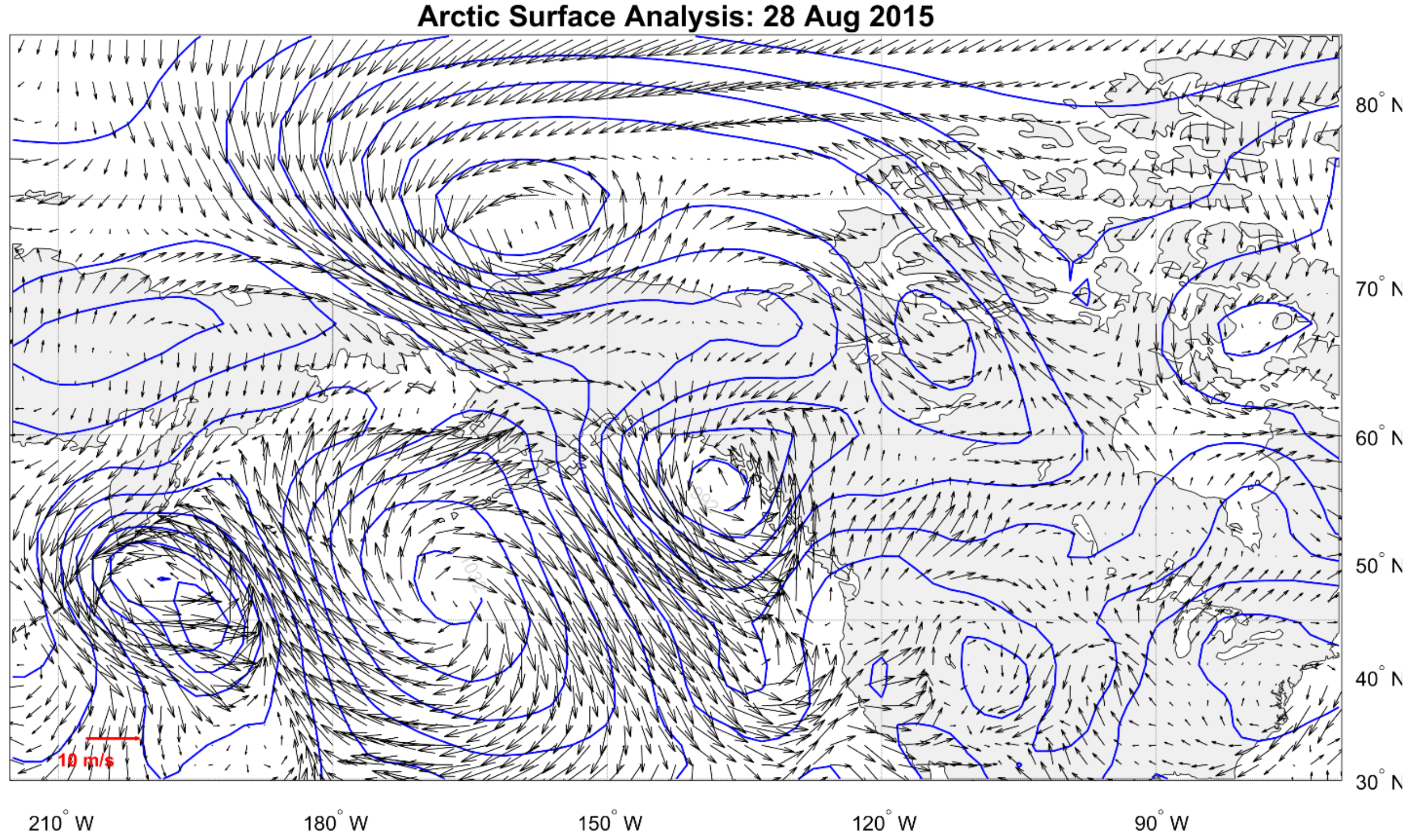

**Figure S20.** Surface analysis for 28 August 2015.

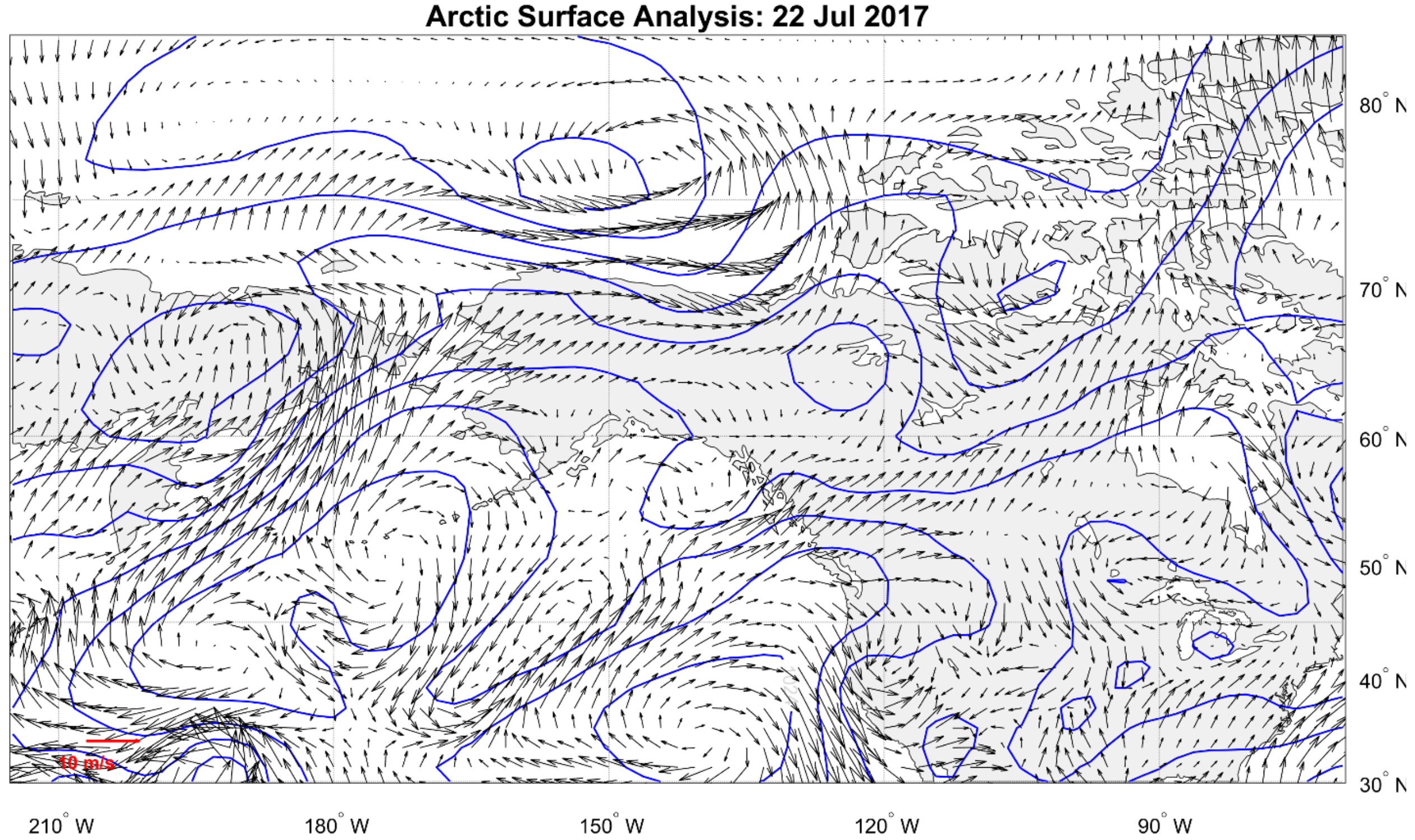

**Figure S21.** Surface analysis for 22 July 2017.

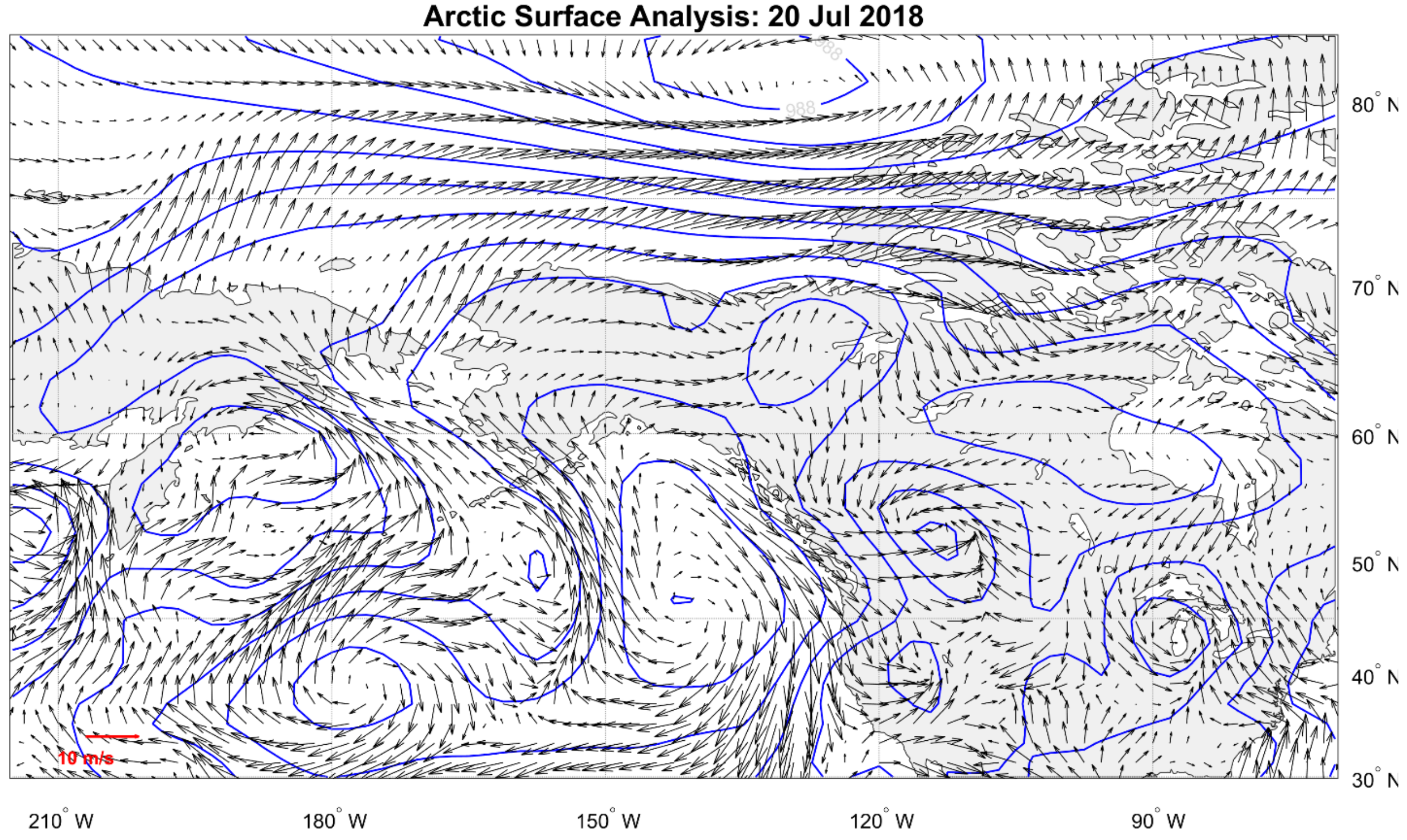

**Figure S22.** Surface analysis for 20 July 2018.

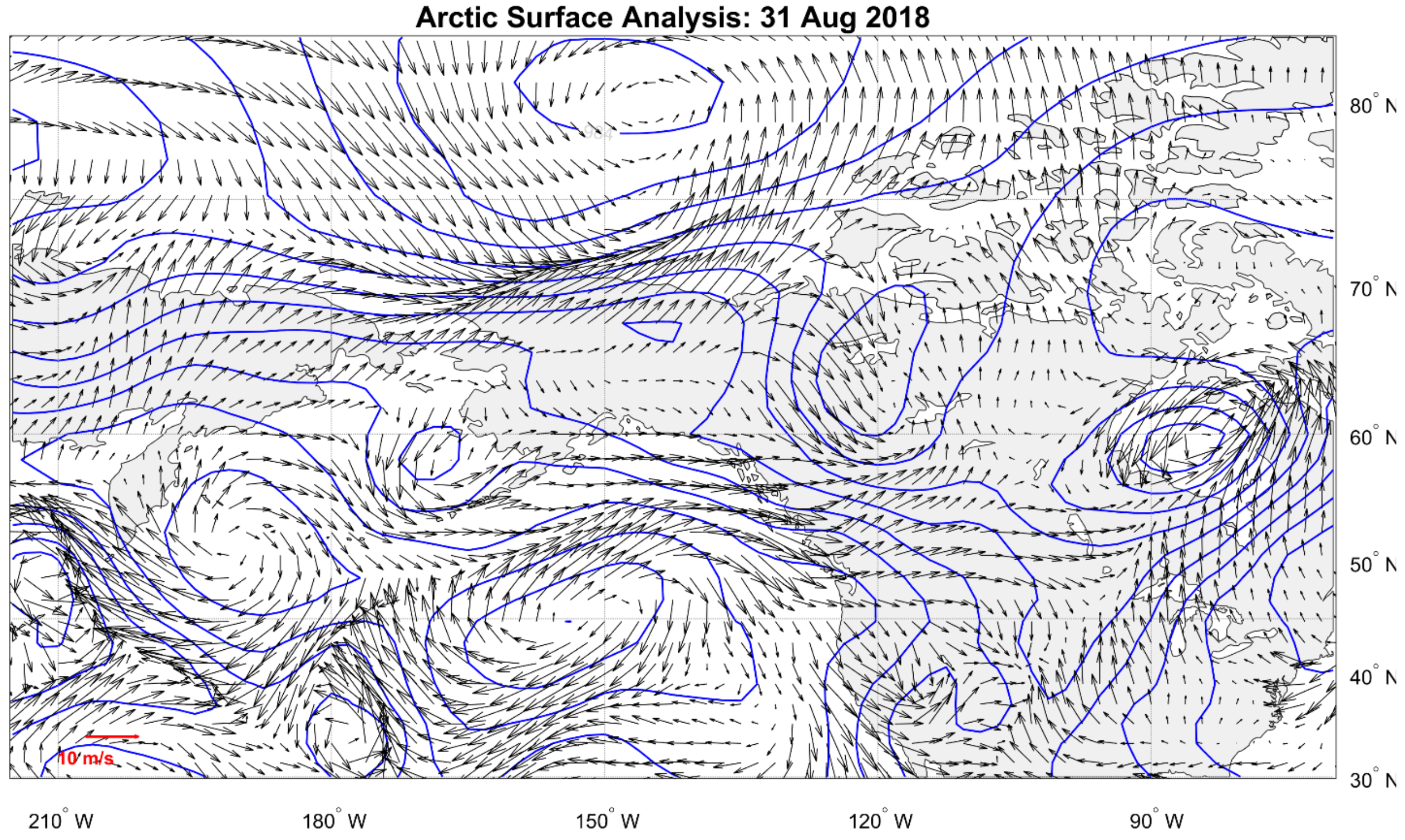

**Figure S23.** Surface analysis for 31 August 2018.

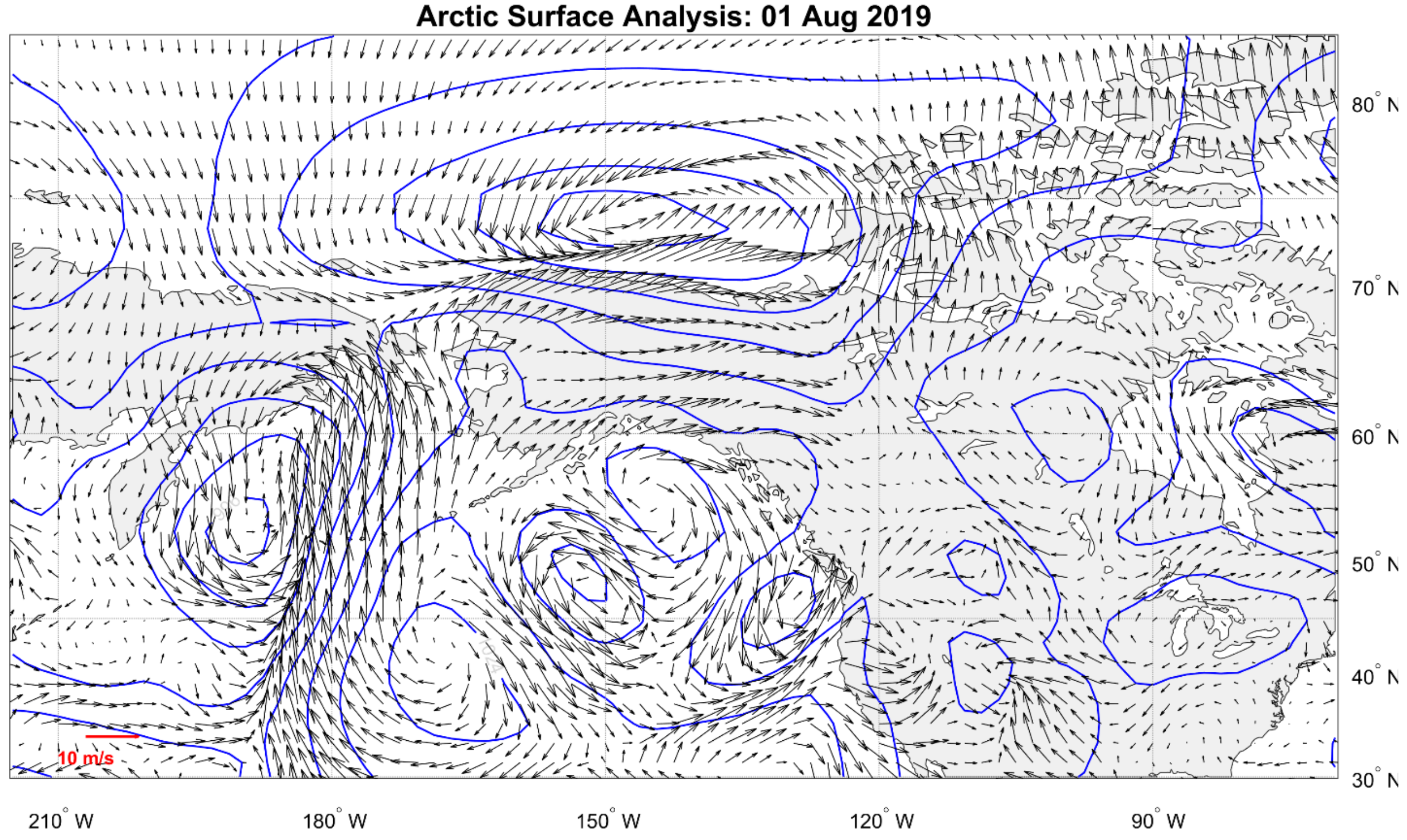

**Figure S24.** Surface analysis for 1 August 2019.

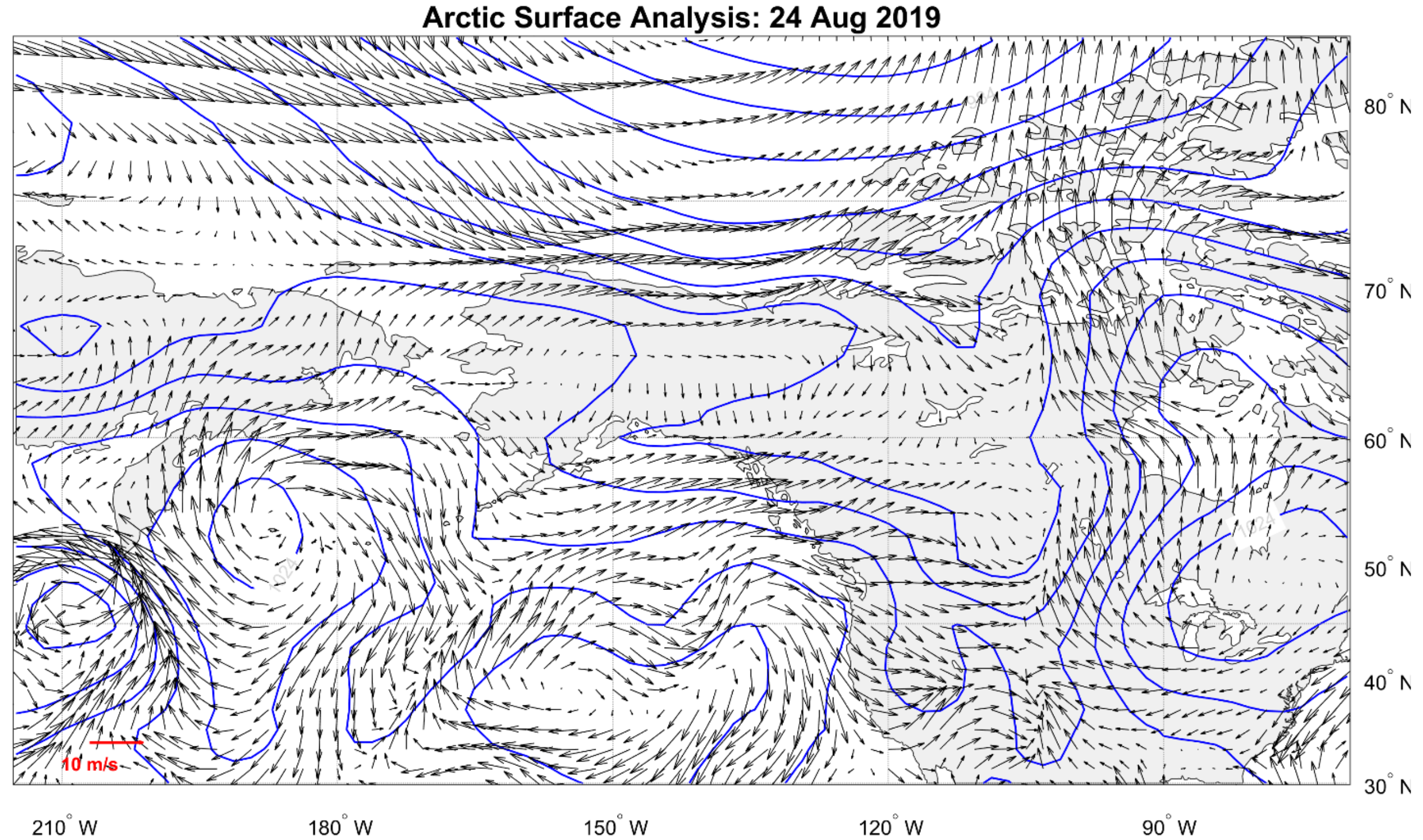

**Figure S25.** Surface analysis for 24 August 2019.

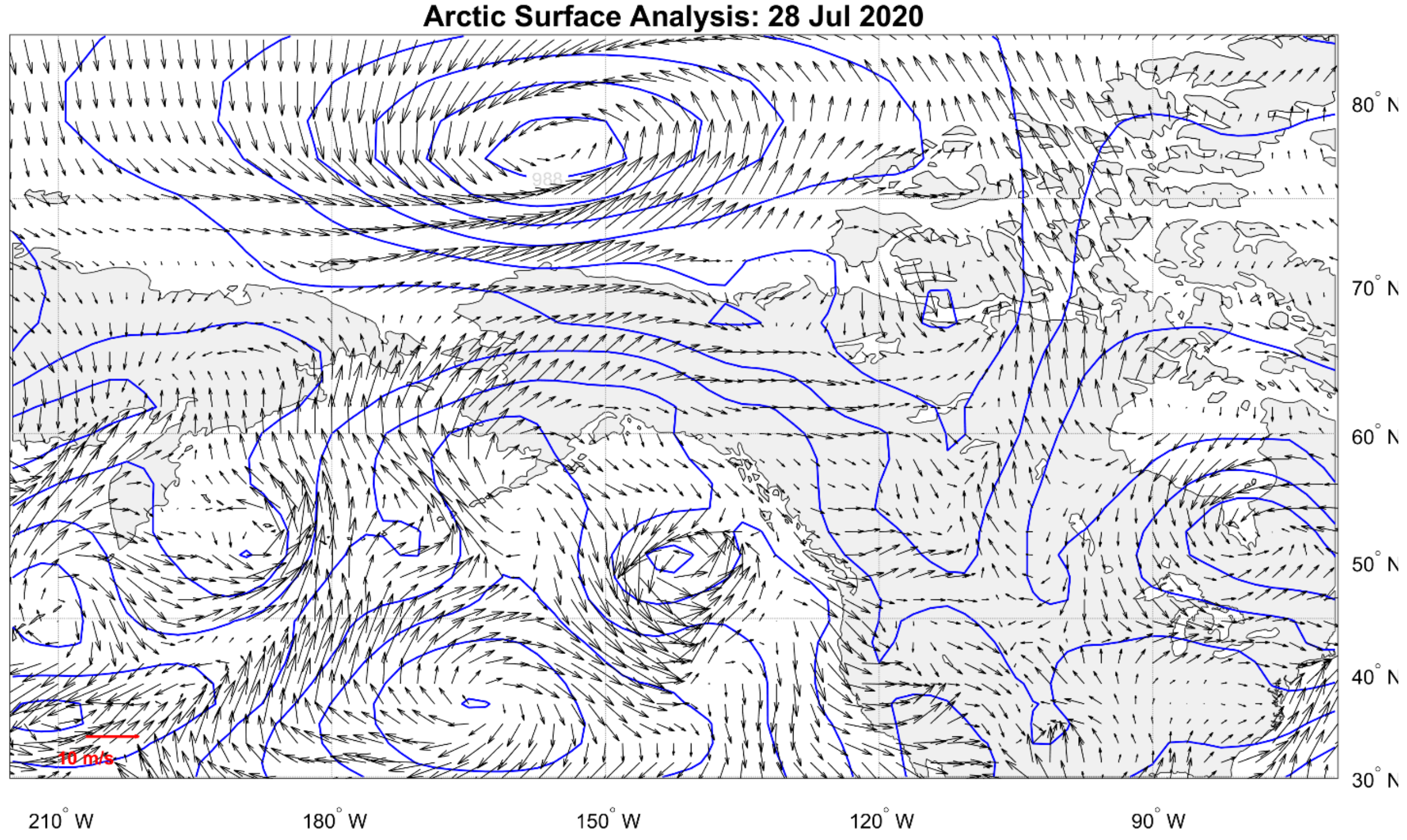

**Figure S26.** Surface analysis for 28 July 2020.

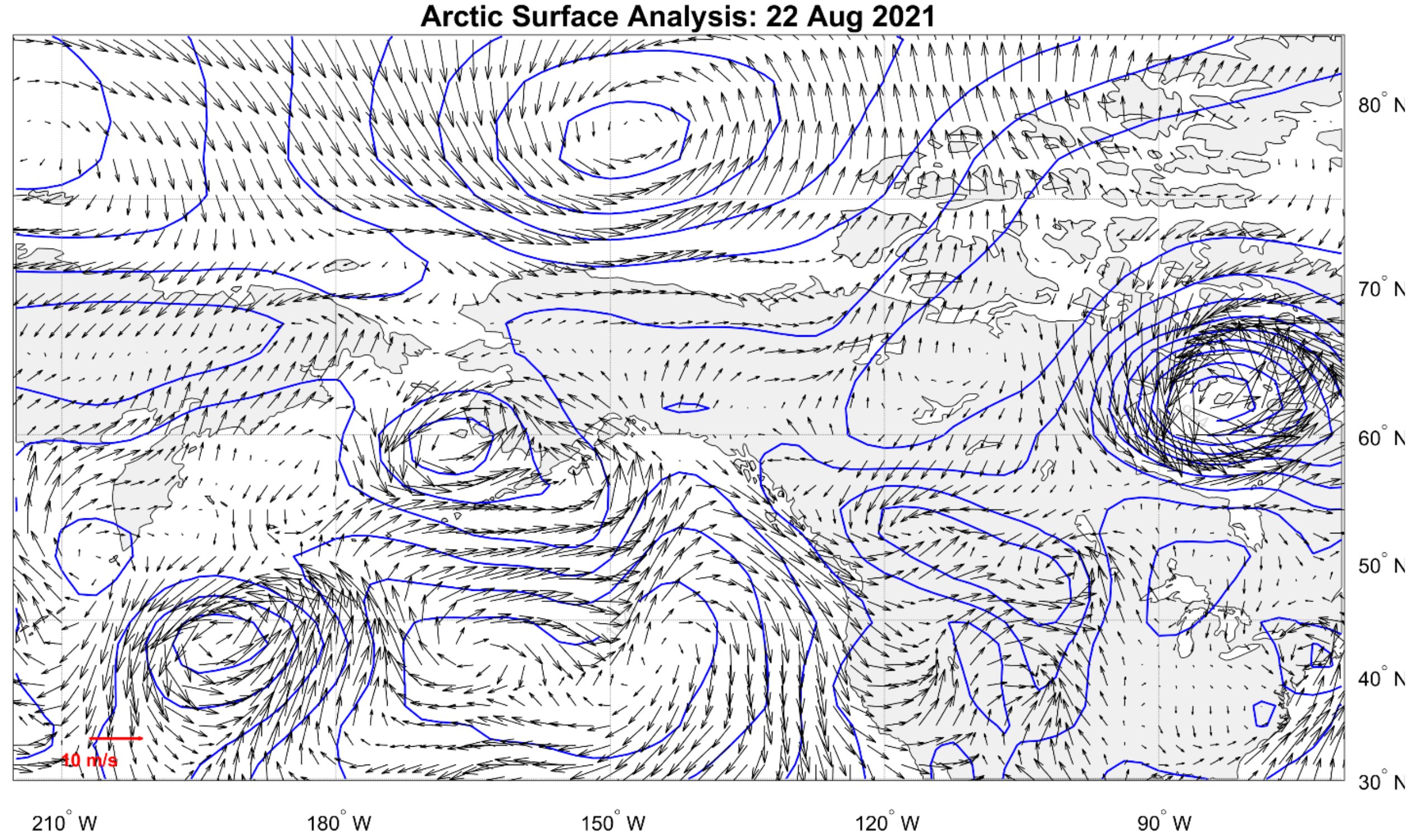

**Figure S27.** Surface analysis for 22 August 2021.

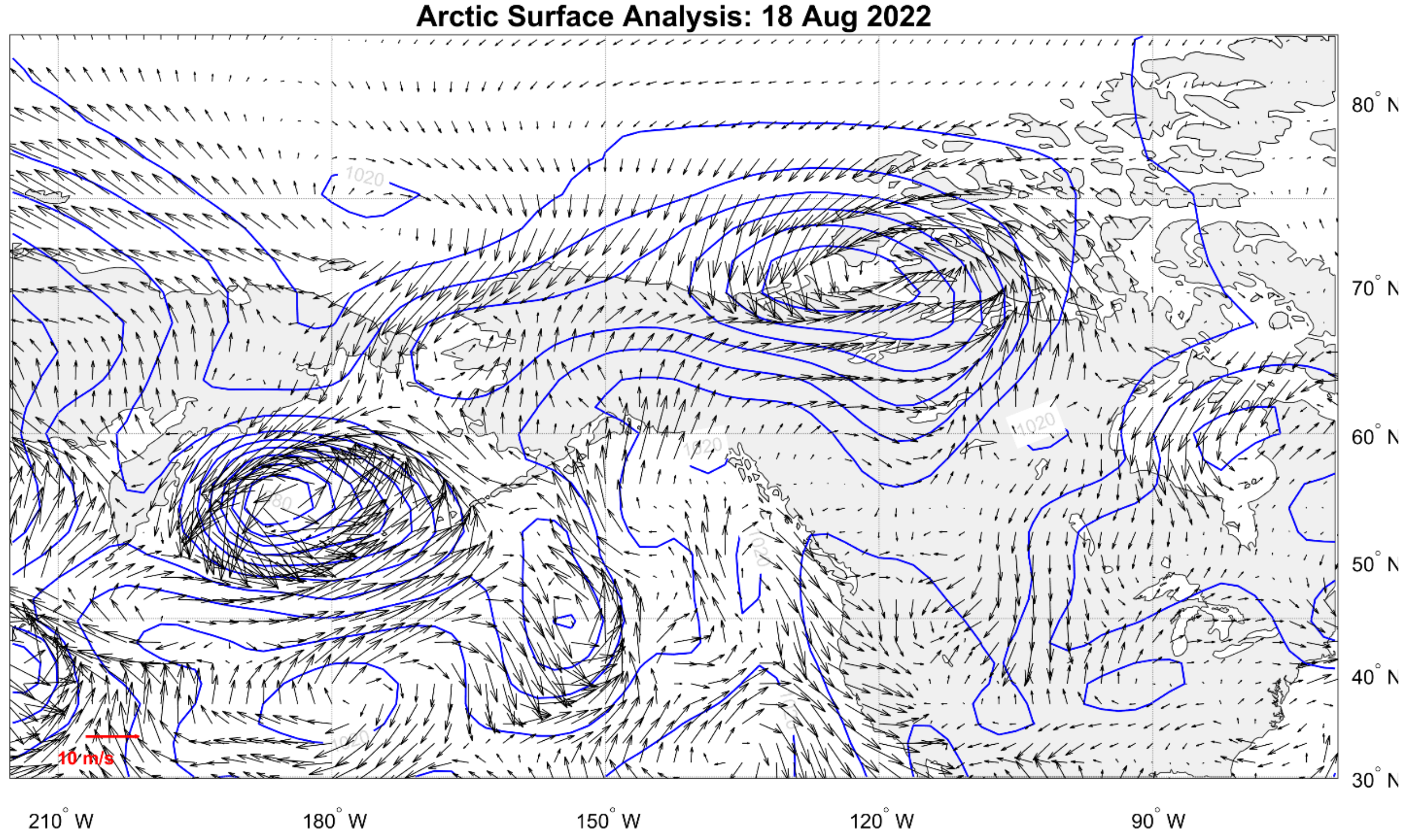

**Figure S28.** Surface analysis for 18 August 2022.

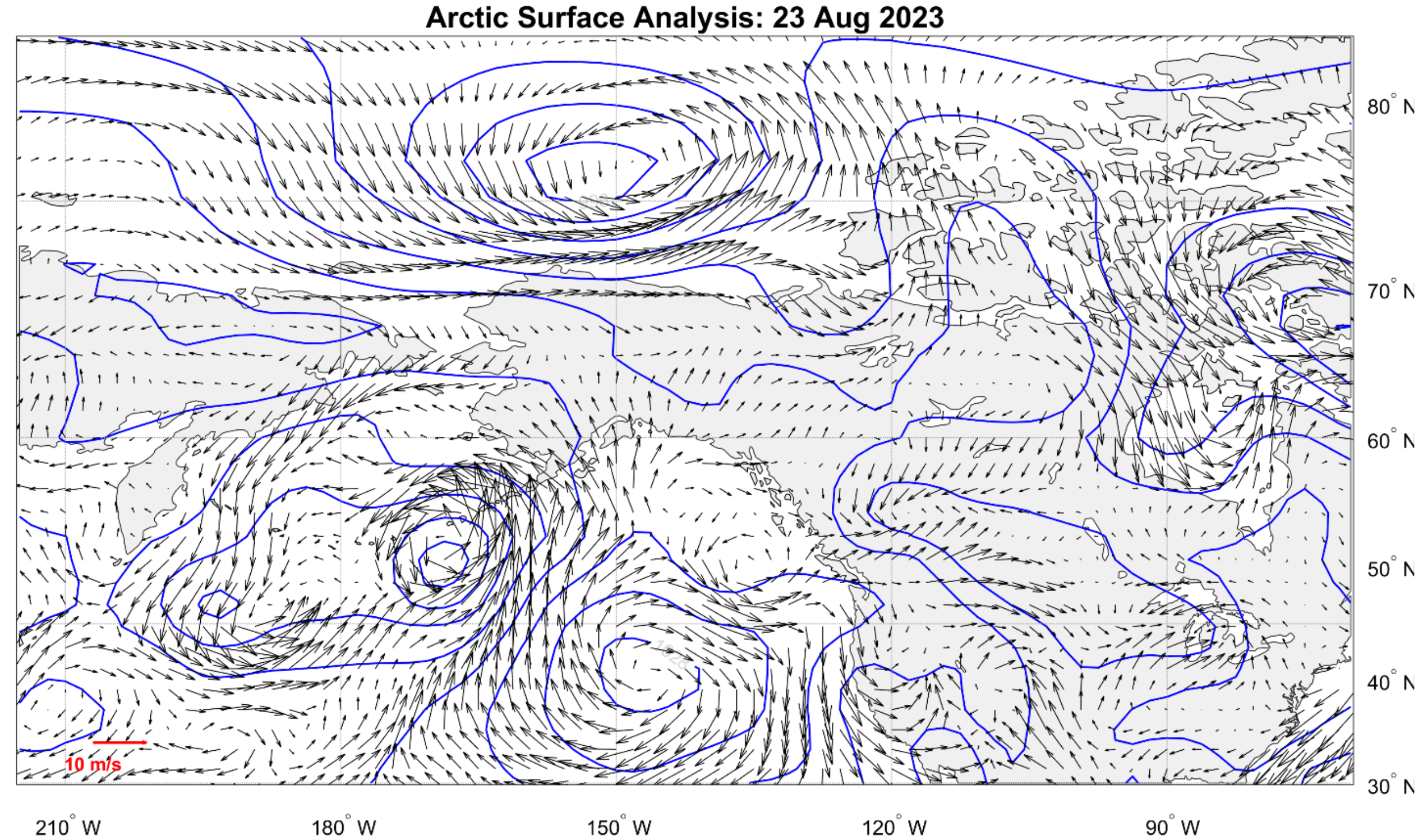

**Figure S29.** Surface analysis for 23 August 2023.

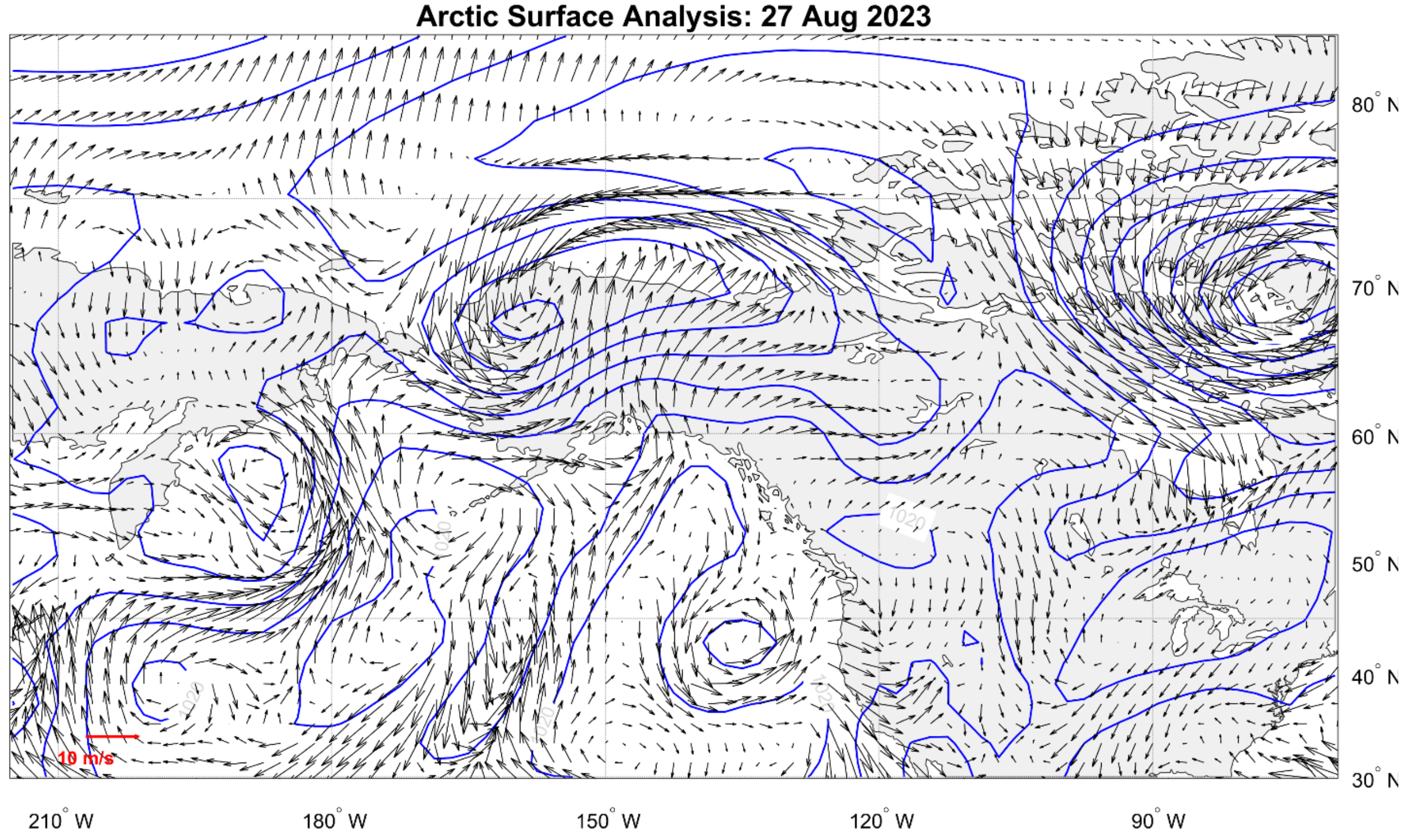

**Figure S30.** Surface analysis for 27 August 2023.

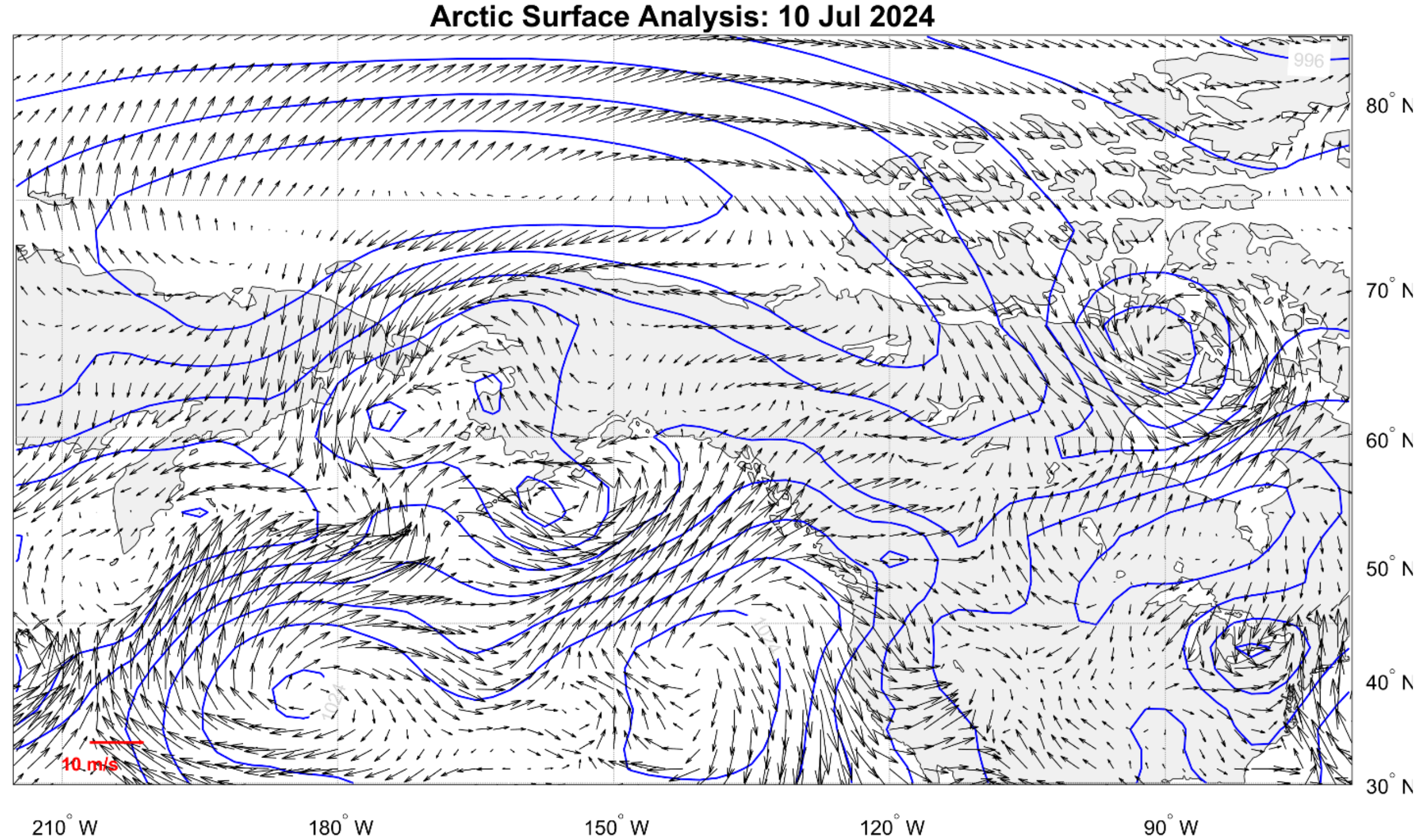

**Figure S31.** Surface analysis for 10 July 2024.

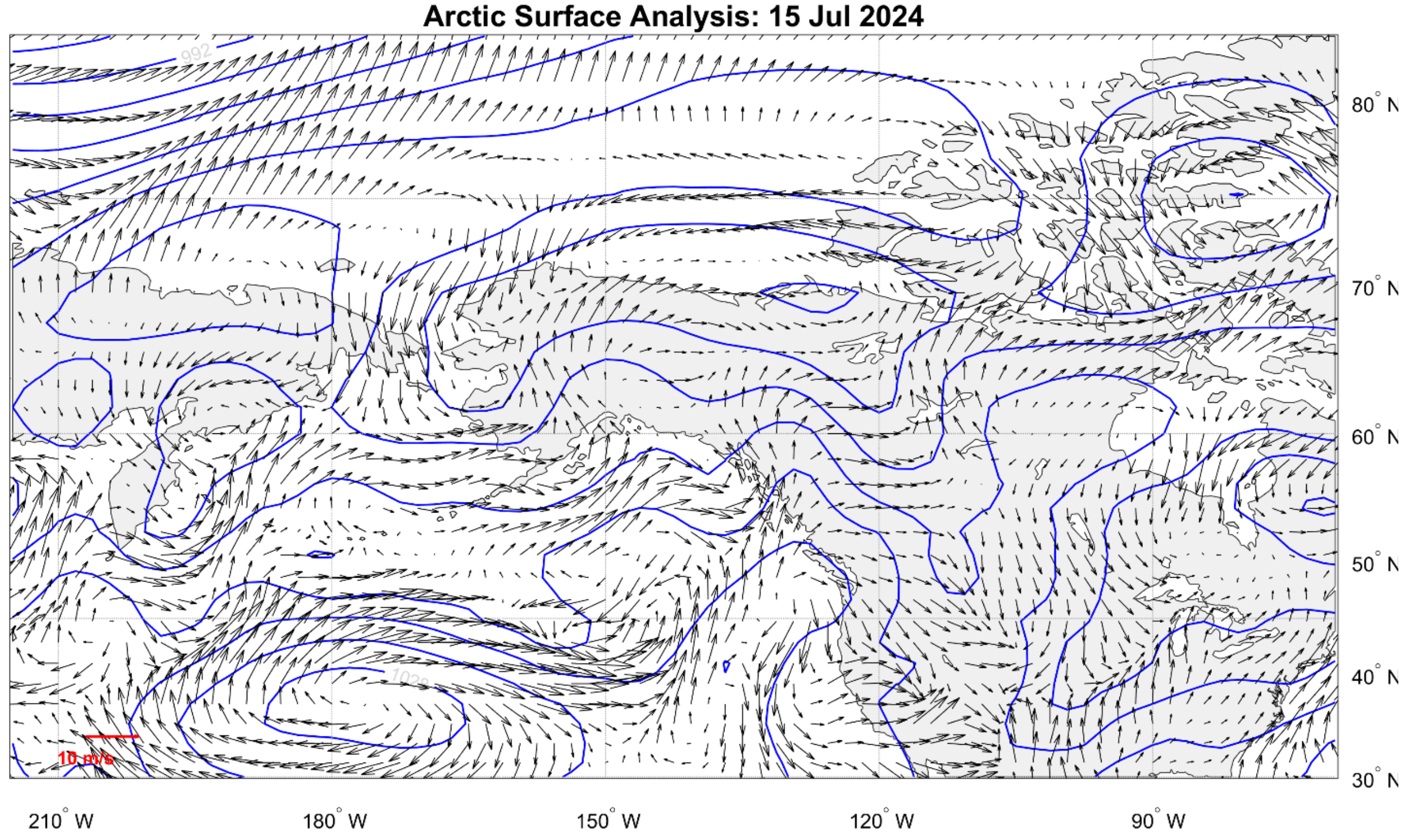

**Figure S32.** Surface analysis for 15 July 2024.

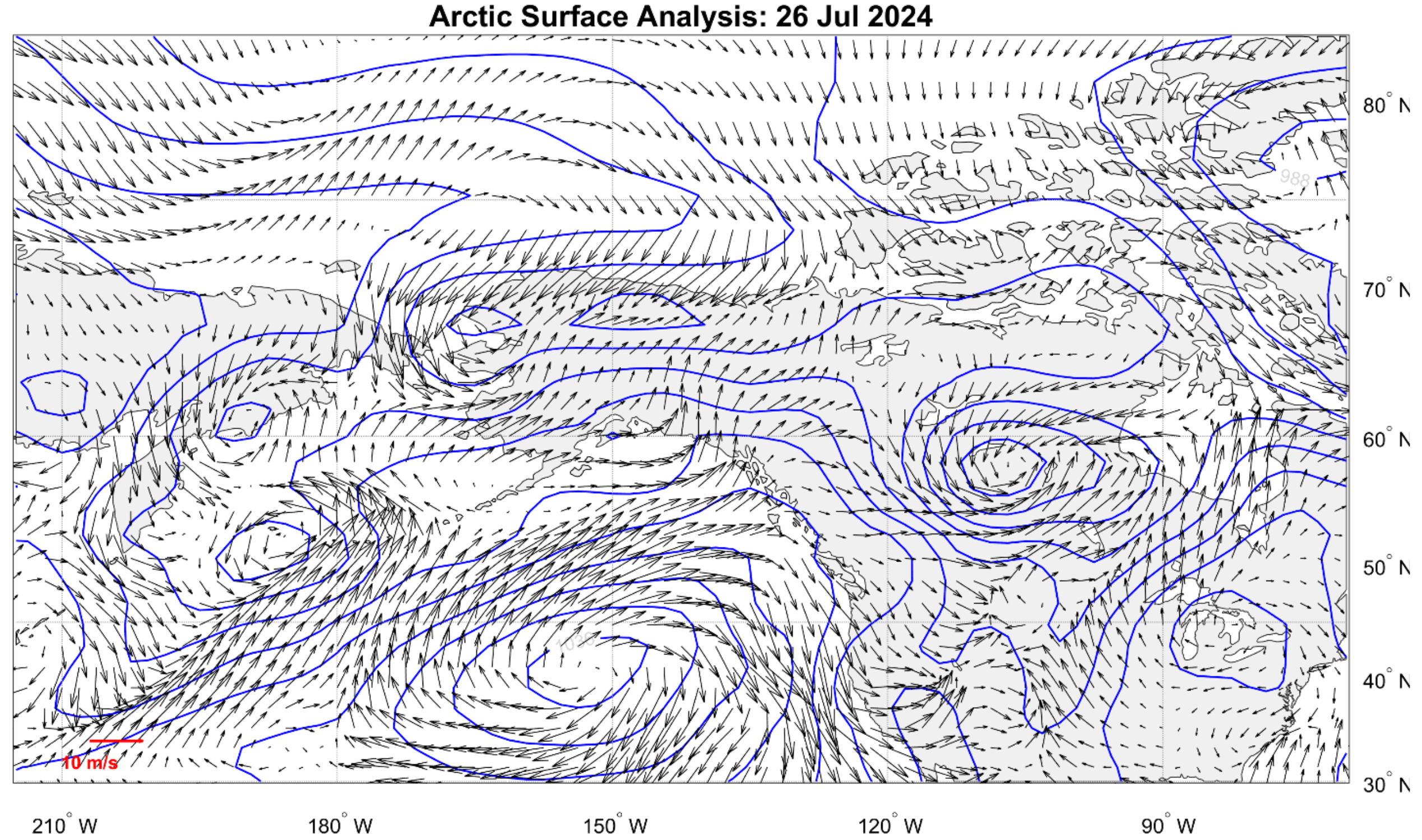

**Figure S33.** Surface analysis for 26 July 2024.

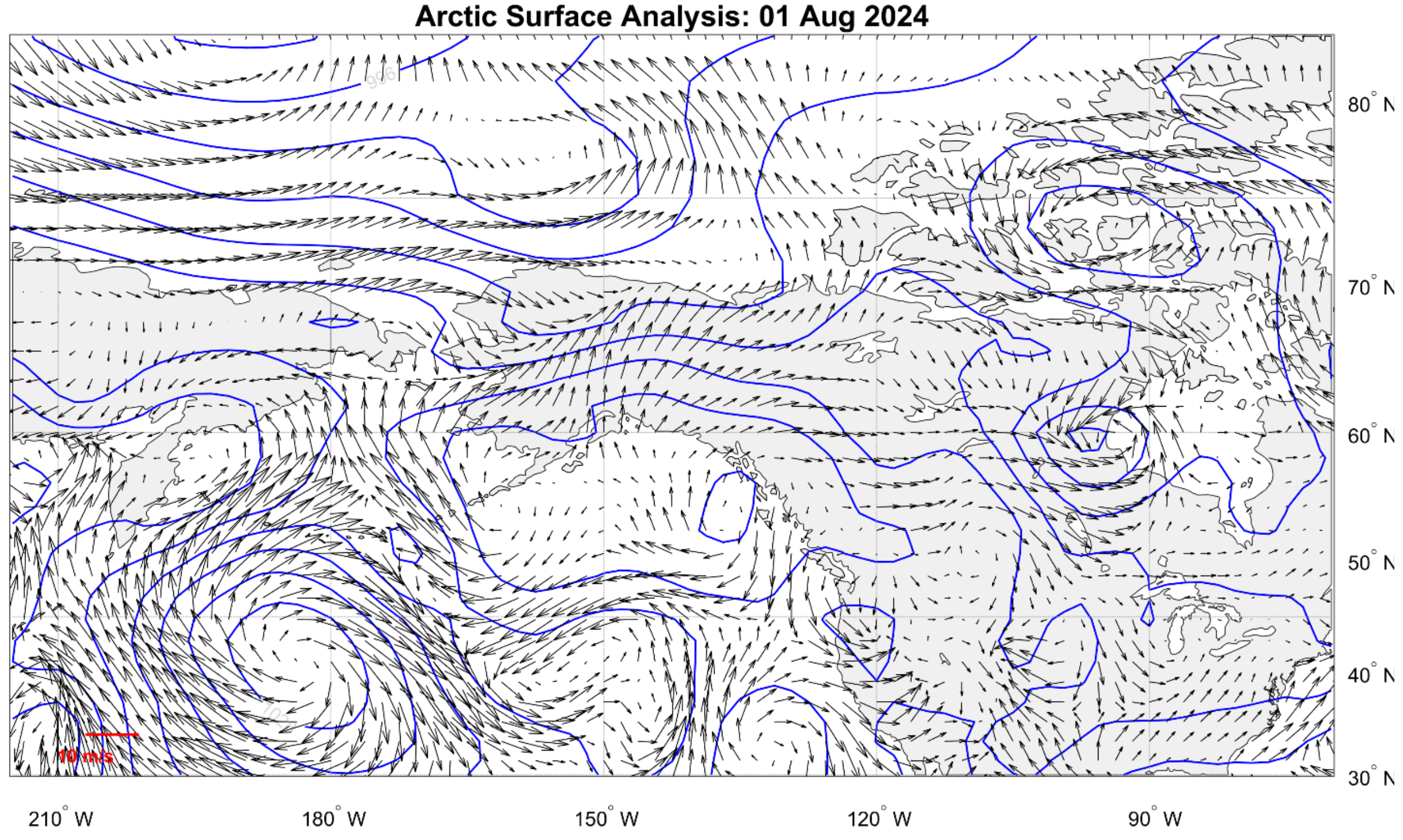

**Figure S34.** Surface analysis for 1 August 2024.

## S3. Surface weather maps from observations for the latter 22 events (for earlier events, such weather maps are not readily available).

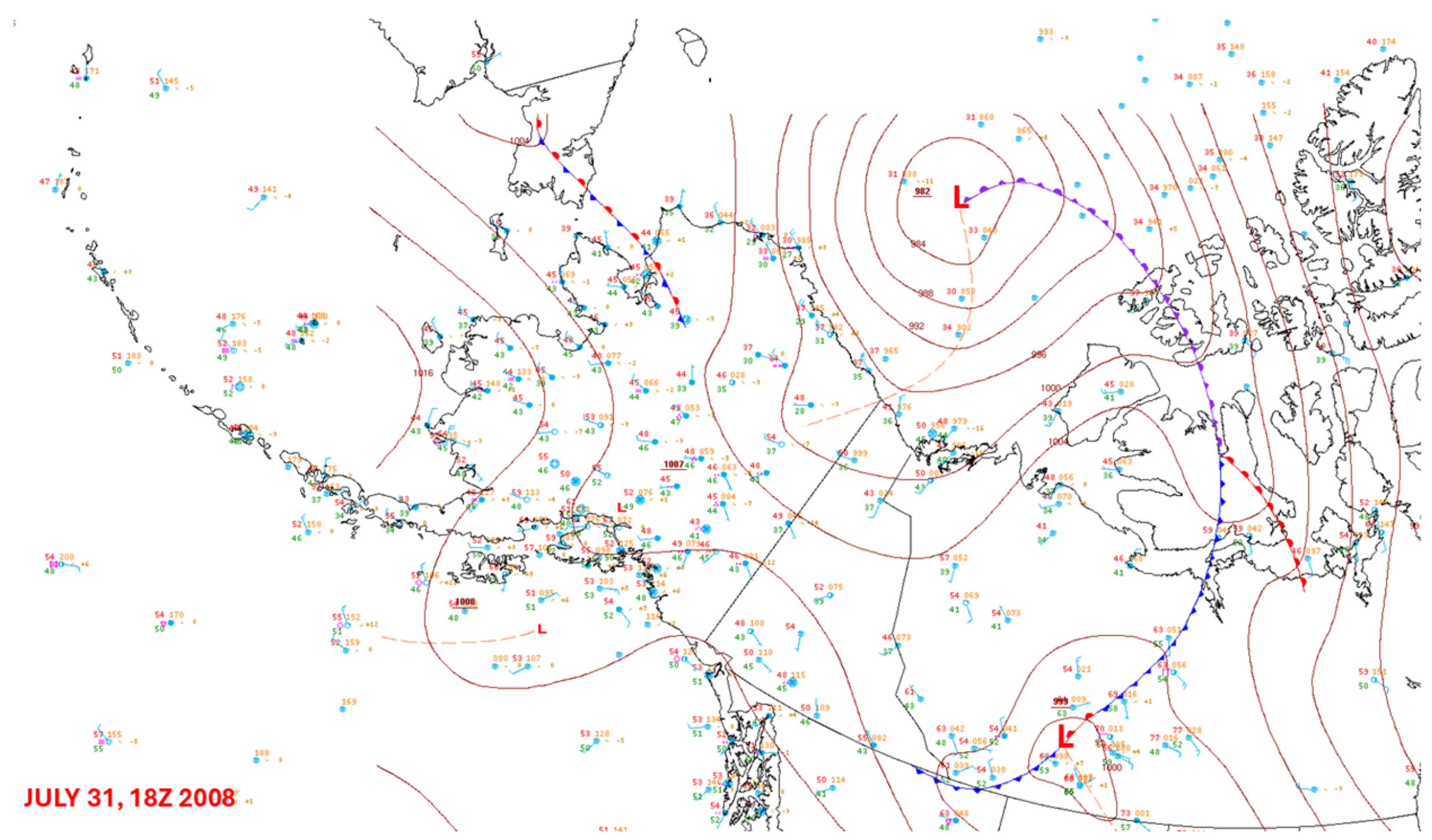

**Figure S35.** Surface weather map for 31 July 2008.

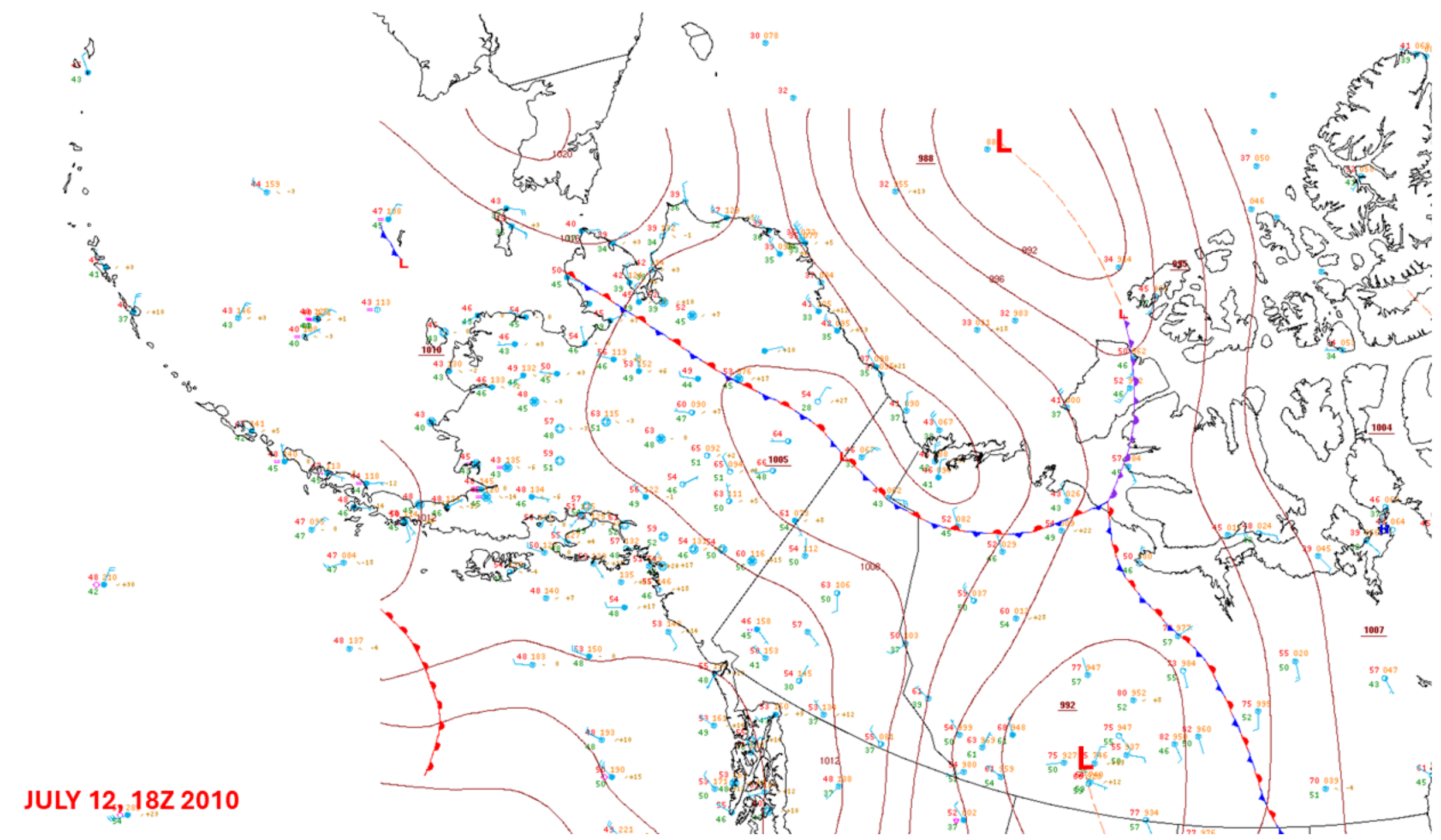

**Figure S36.** Surface weather map for 12 July 2010.

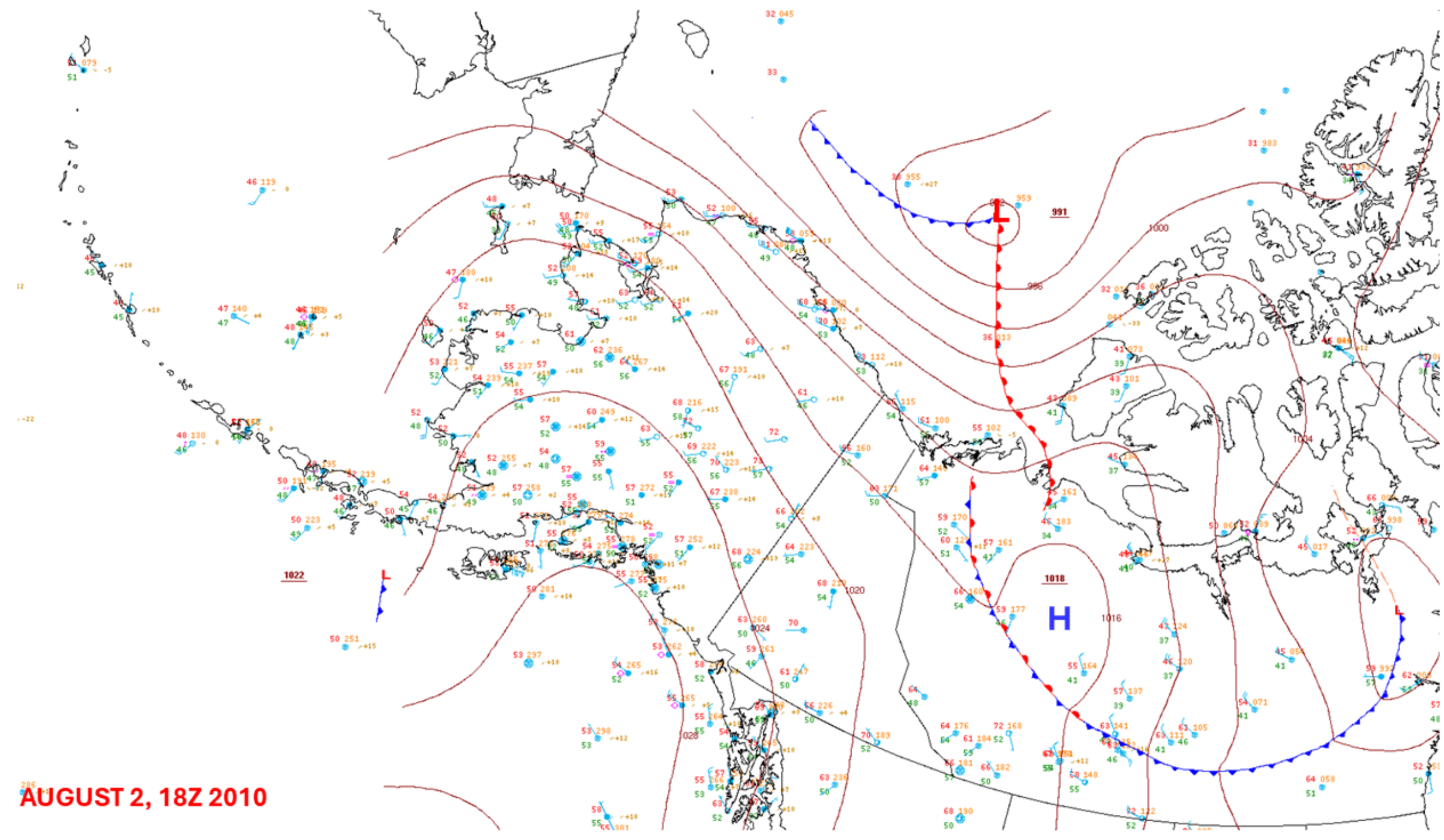

**Figure S37.** Surface weather map for 2 August 2010.

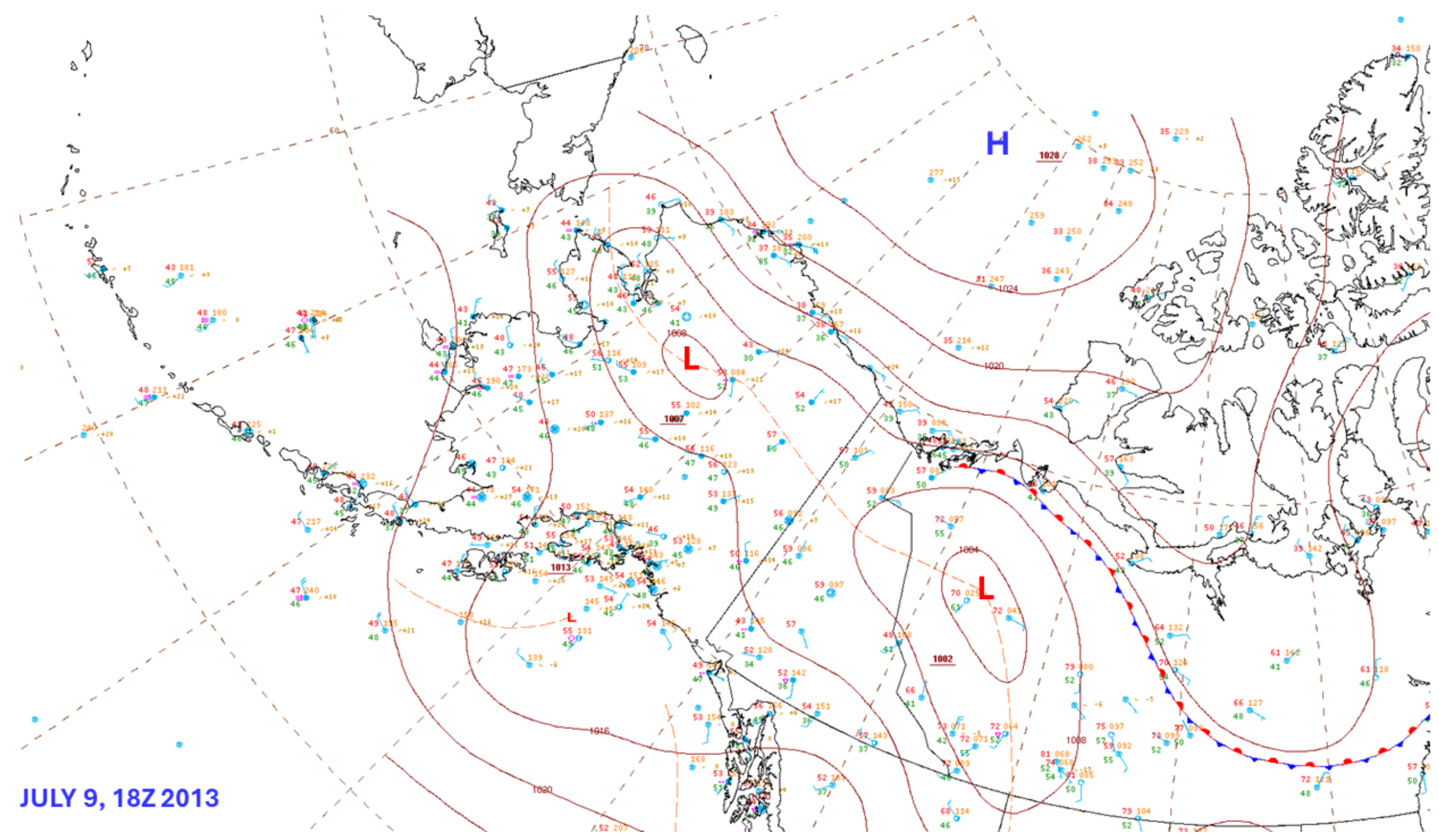

**Figure S38.** Surface weather map for 9 July 2013.

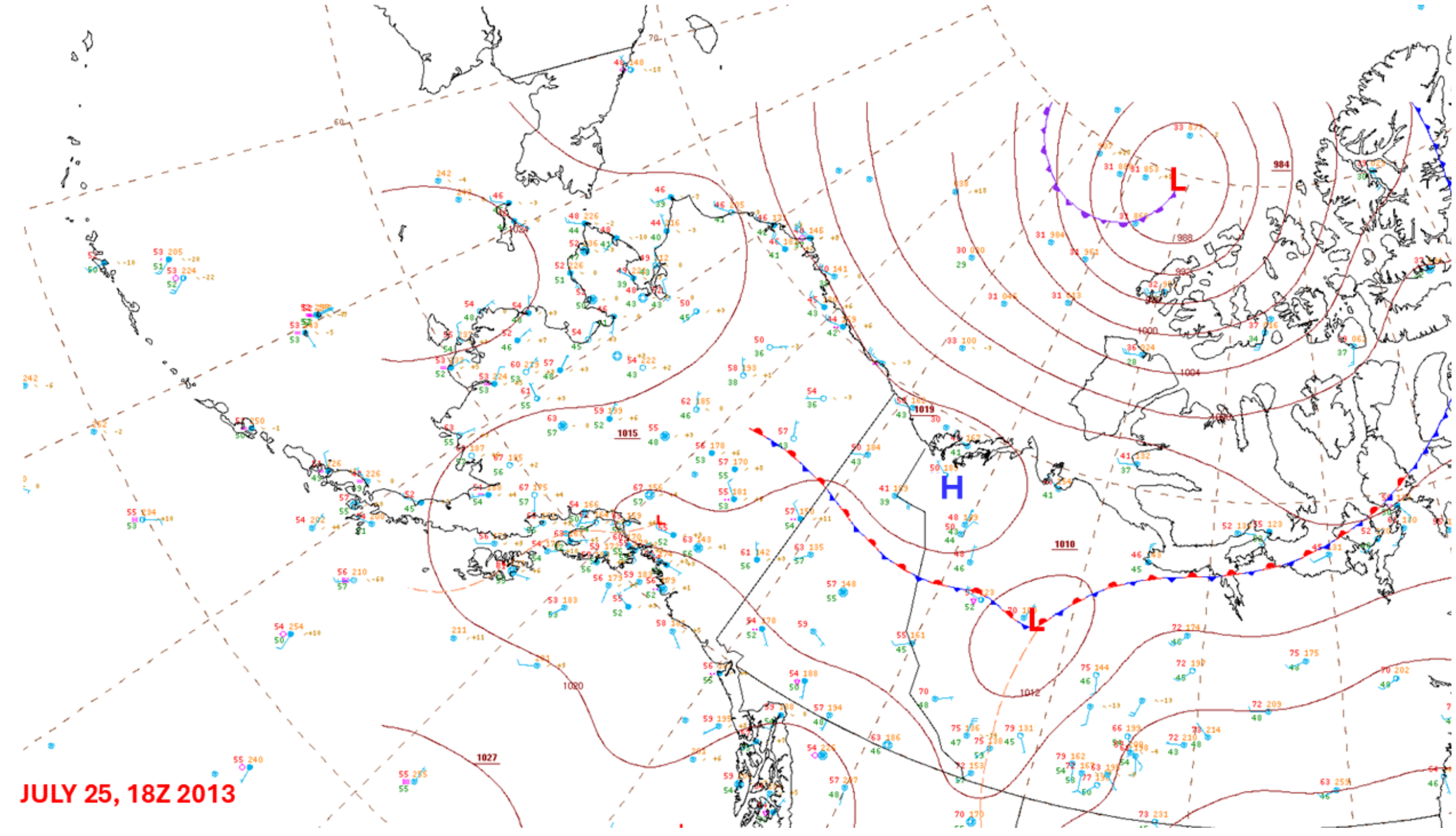

**Figure S39.** Surface weather map for 25 July 2013.

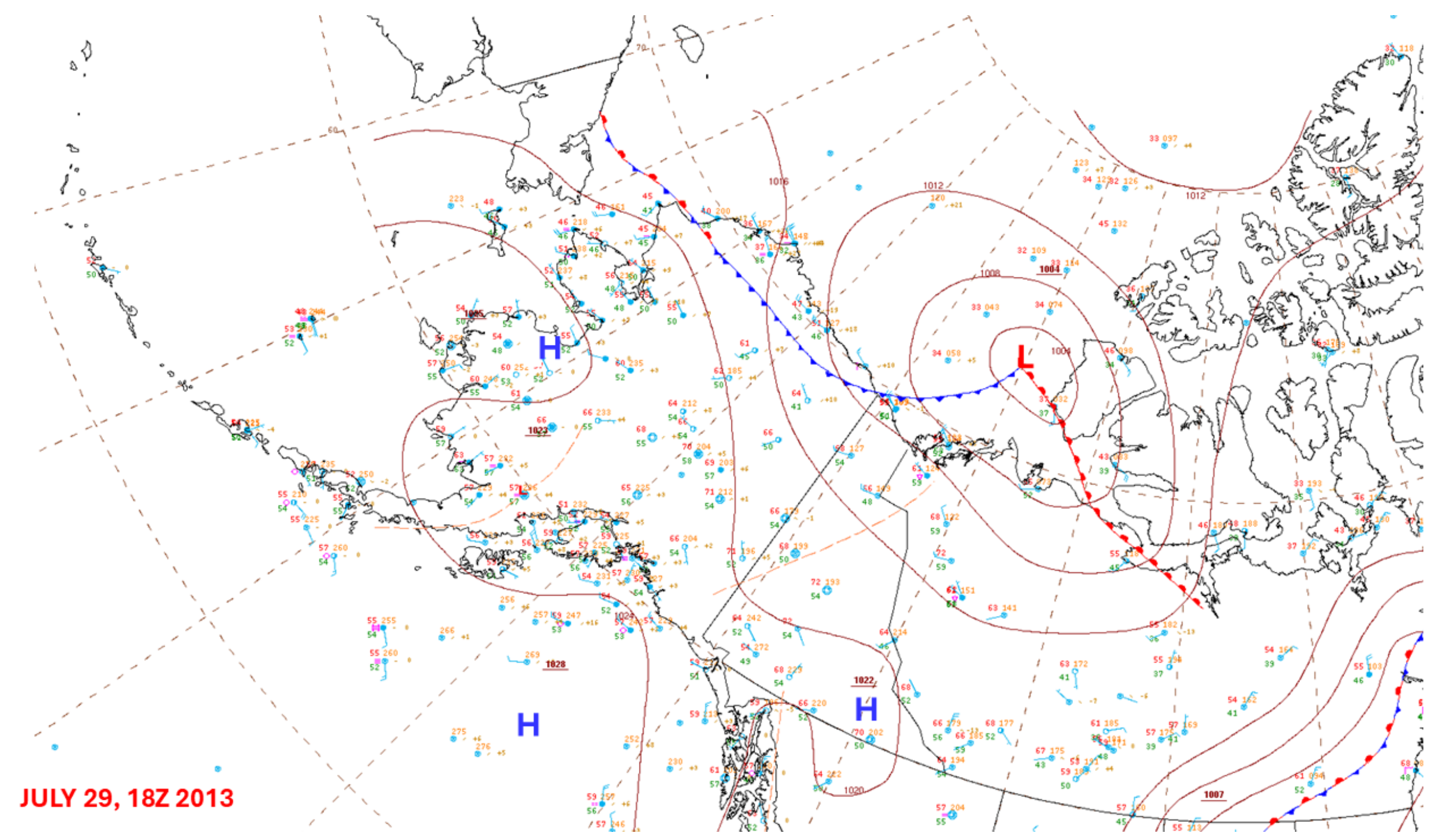

**Figure S40.** Surface weather map for 29 July 2013.

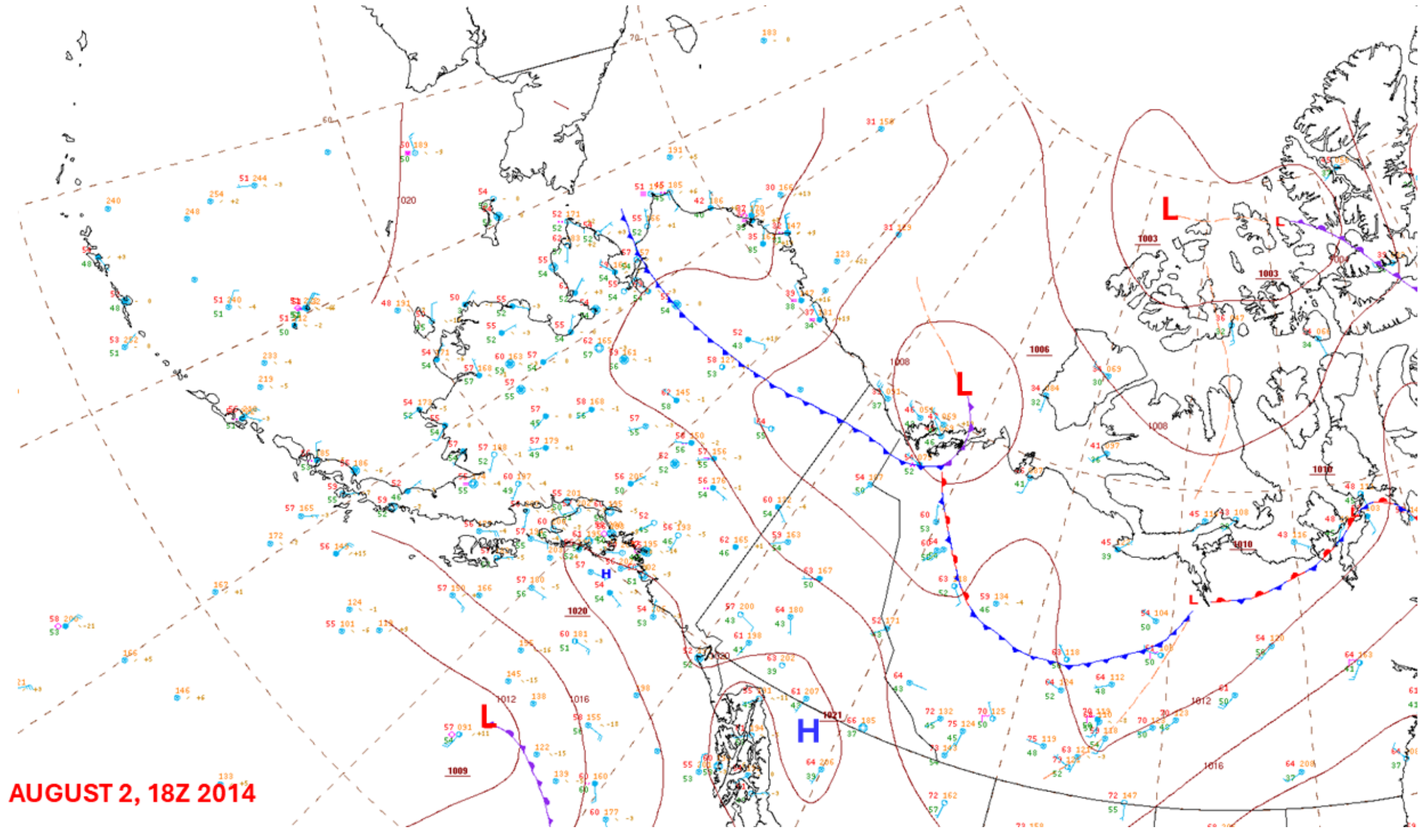

**Figure S41.** Surface weather map for 2 August 2014.

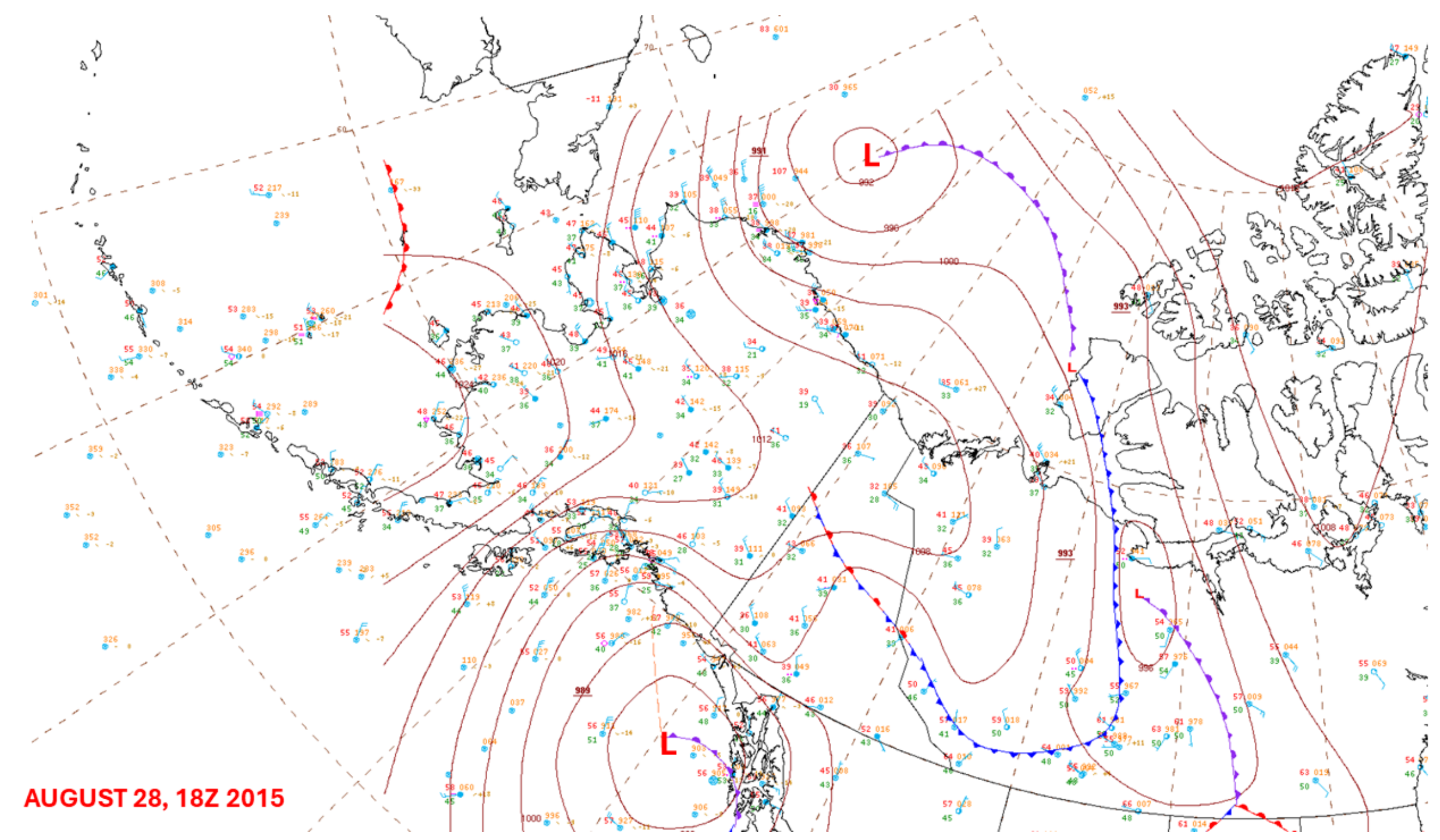

**Figure S42.** Surface weather map for 28 August 2015.

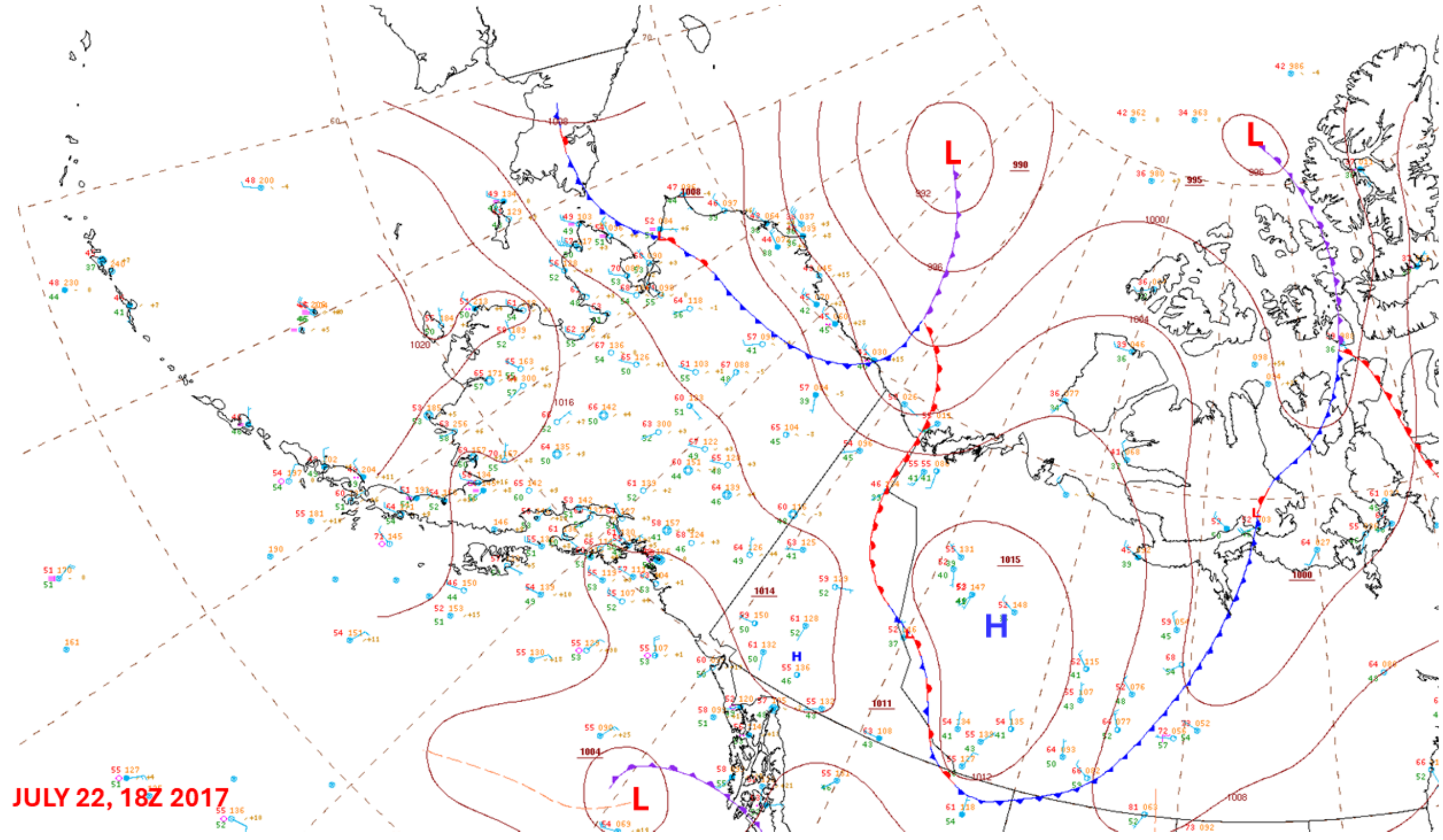

**Figure S43.** Surface weather map for 22 July 2017.

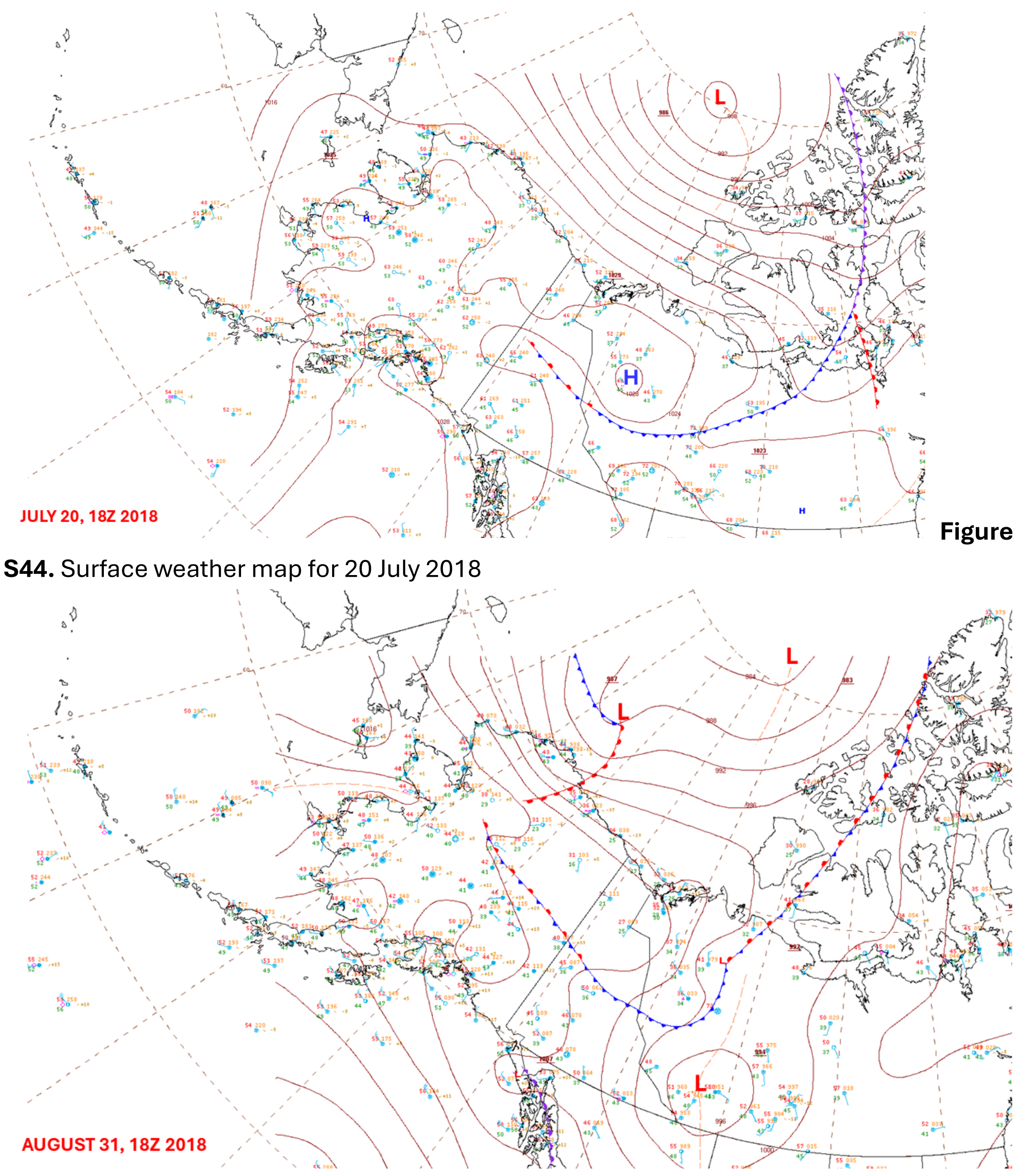

**Figure S44.** Surface weather map for 20 July 2018

**Figure S45.** Surface weather map for 31 August 2018

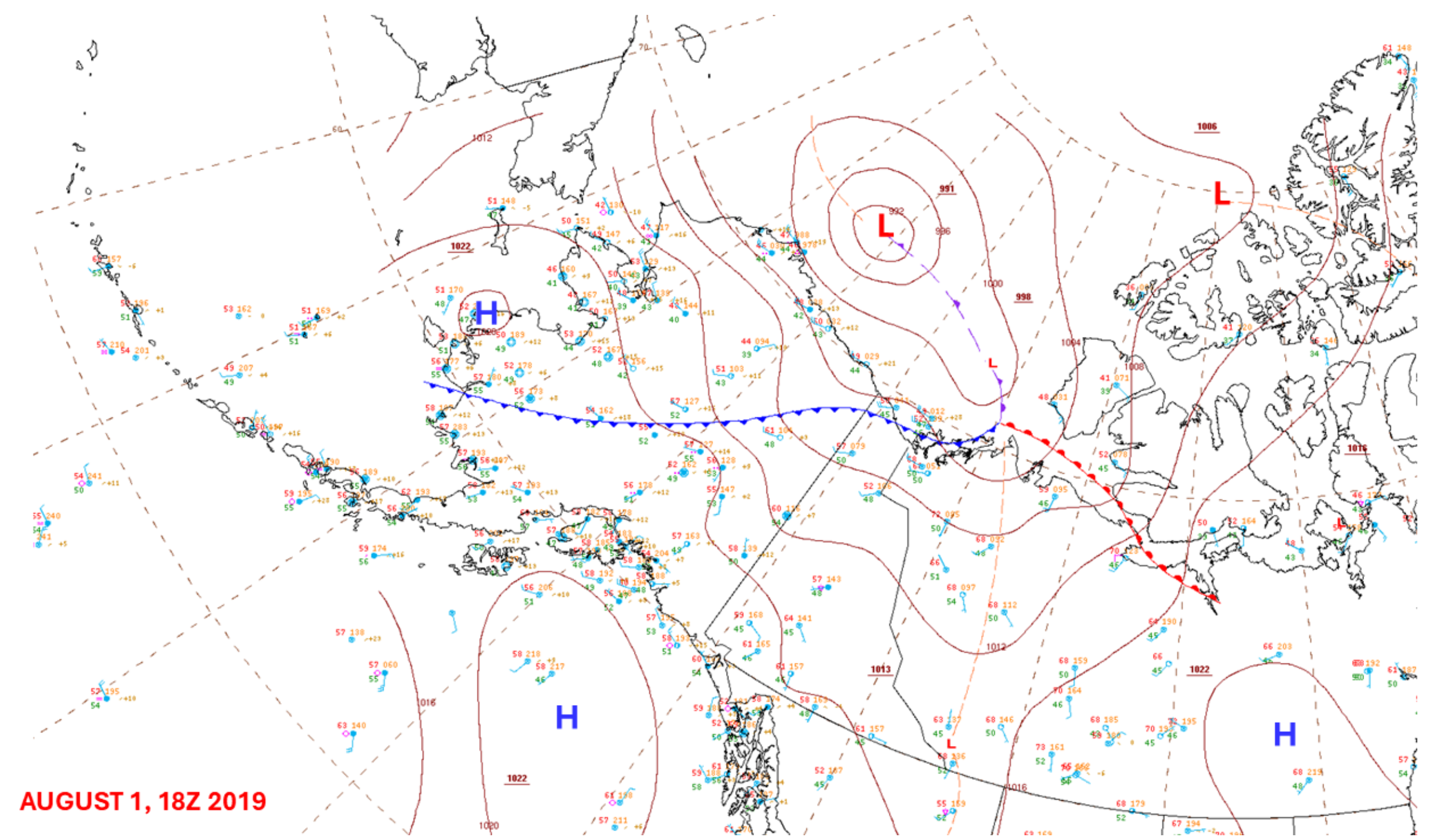

**Figure S46.** Surface analysis for 1 August 2019.

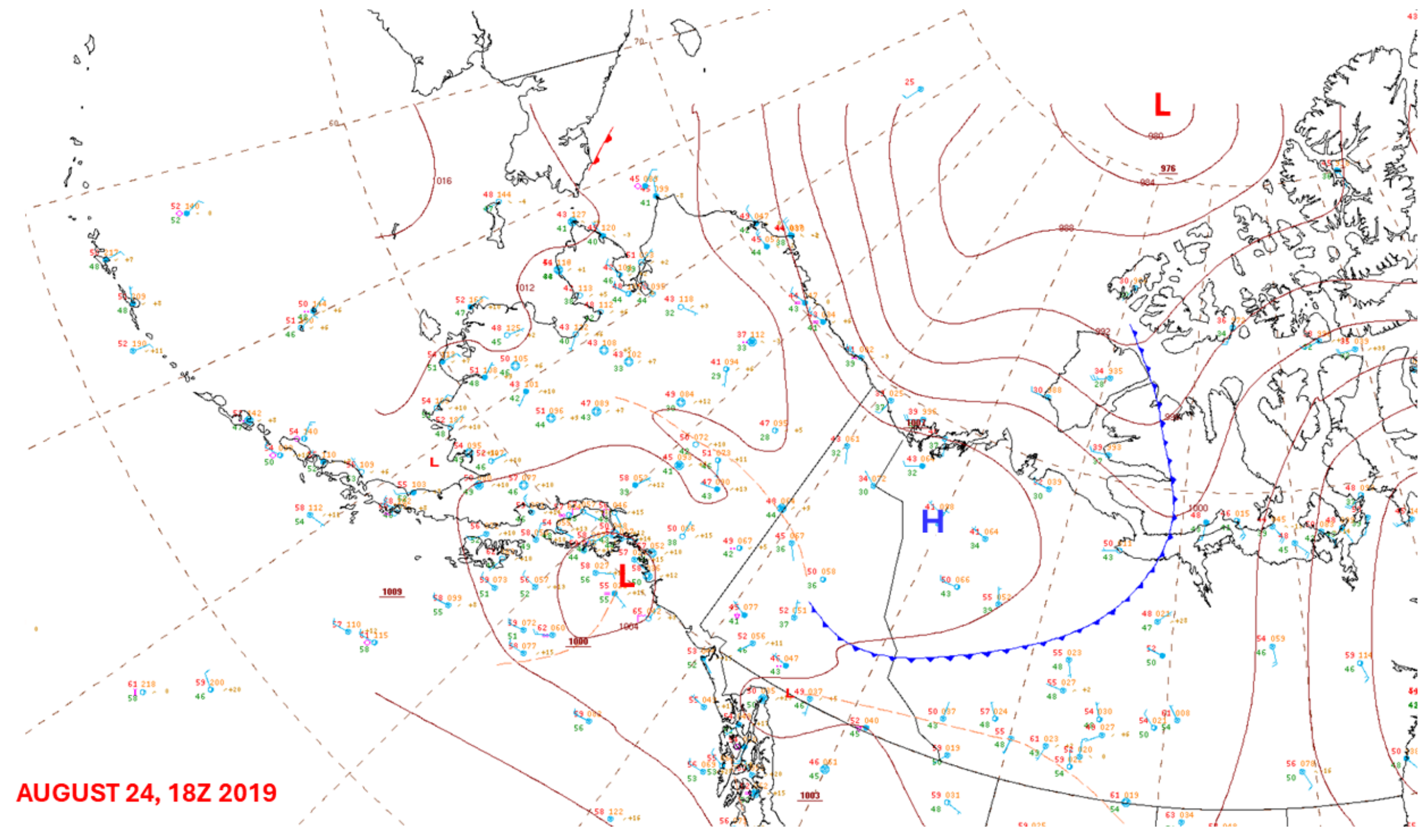

**Figure S47.** Surface analysis for 24 August 2019.

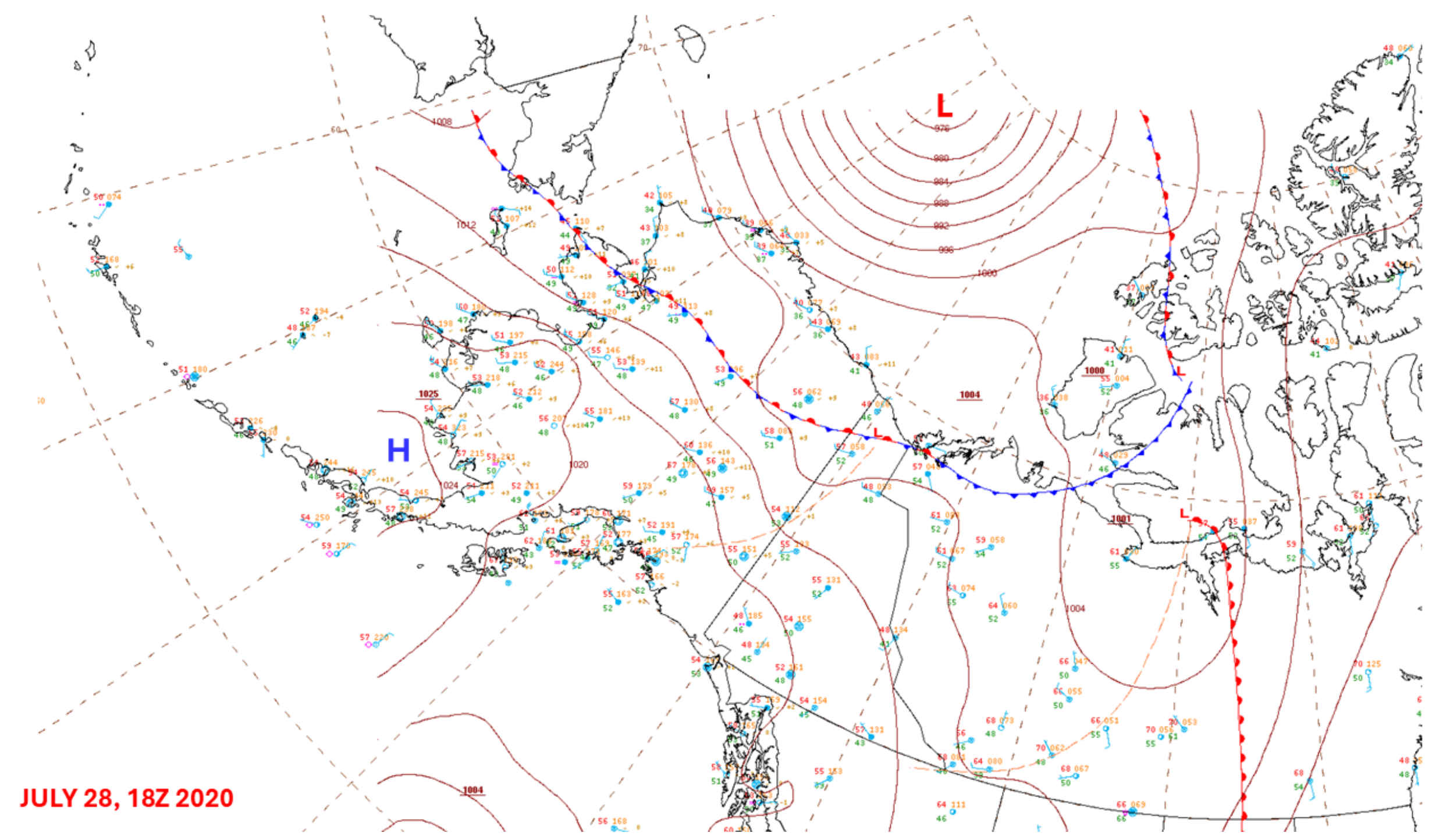

**Figure S48.** Surface analysis for 28 July 2020.

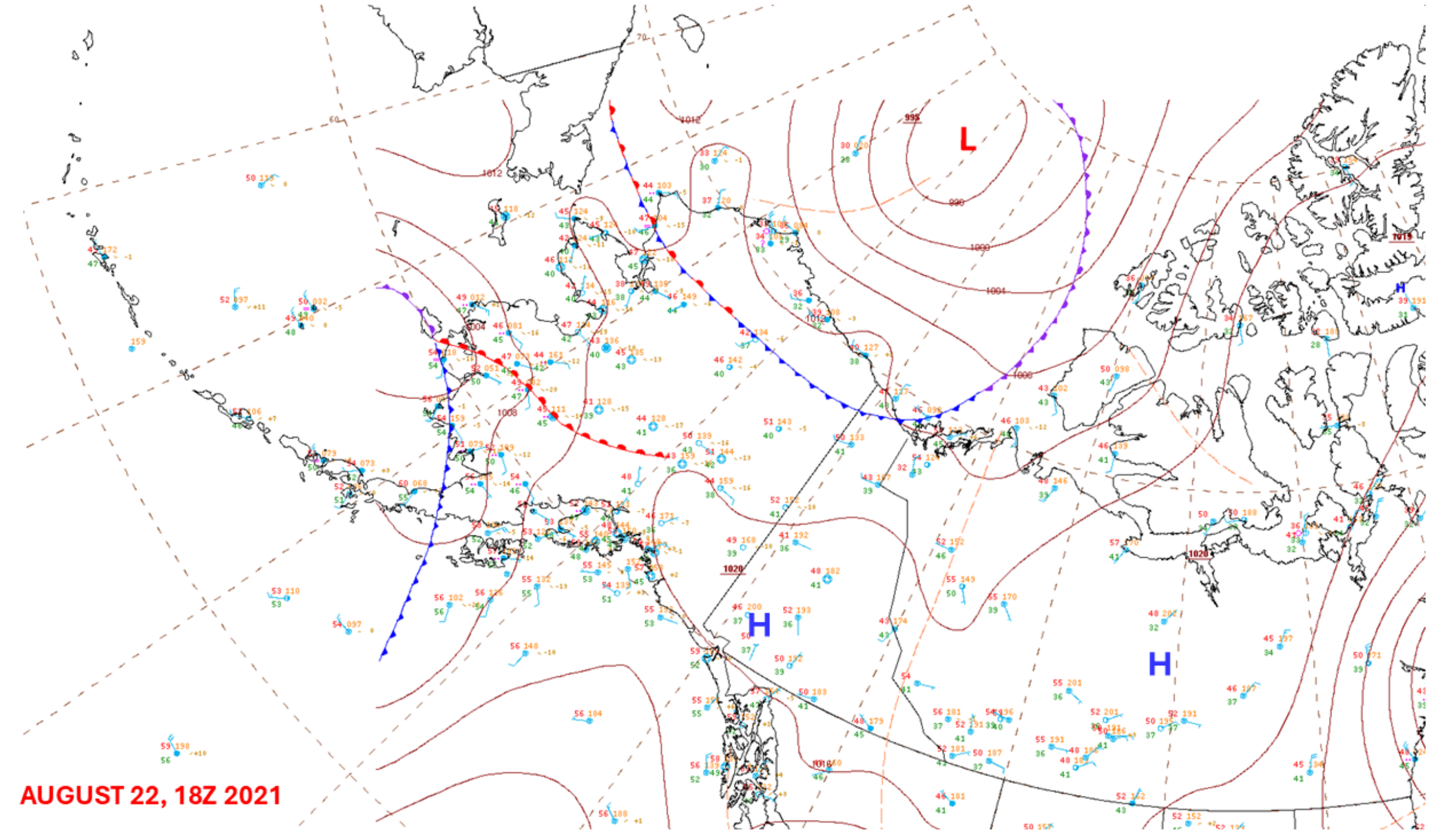

**Figure S49.** Surface analysis for 22 August 2021.

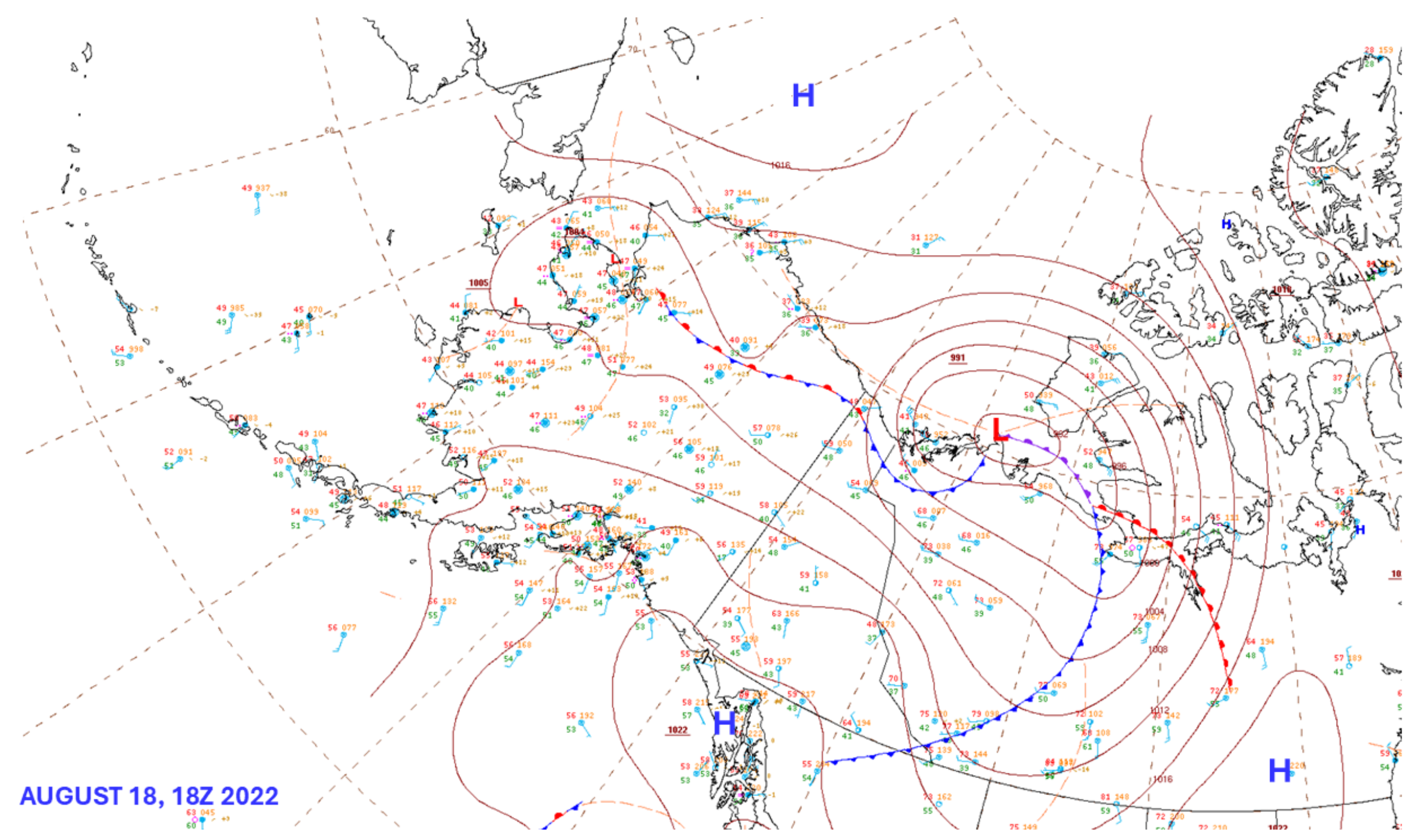

**Figure S50.** Surface analysis for 18 August 2022.

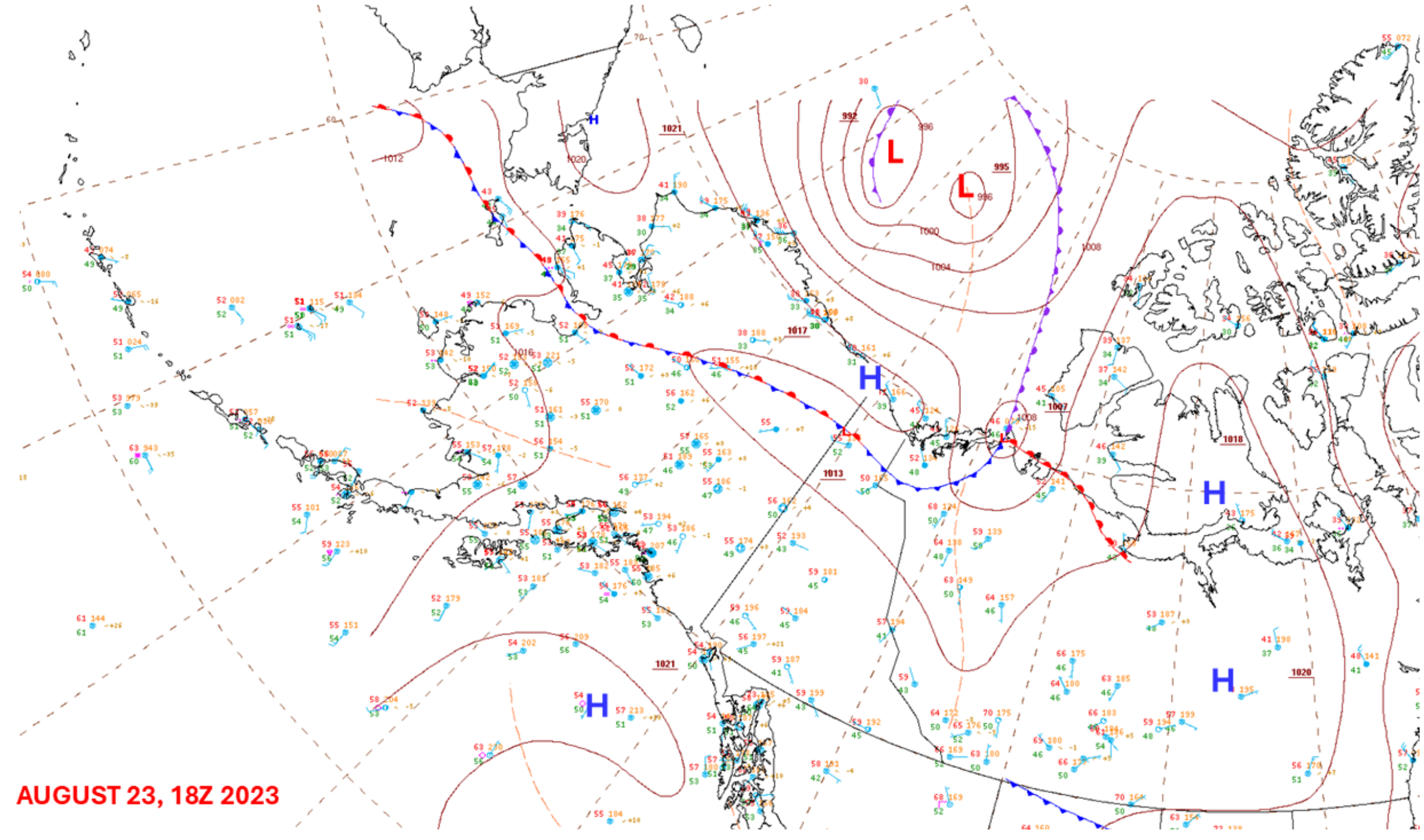

**Figure S51.** Surface analysis for 23 August 2023.

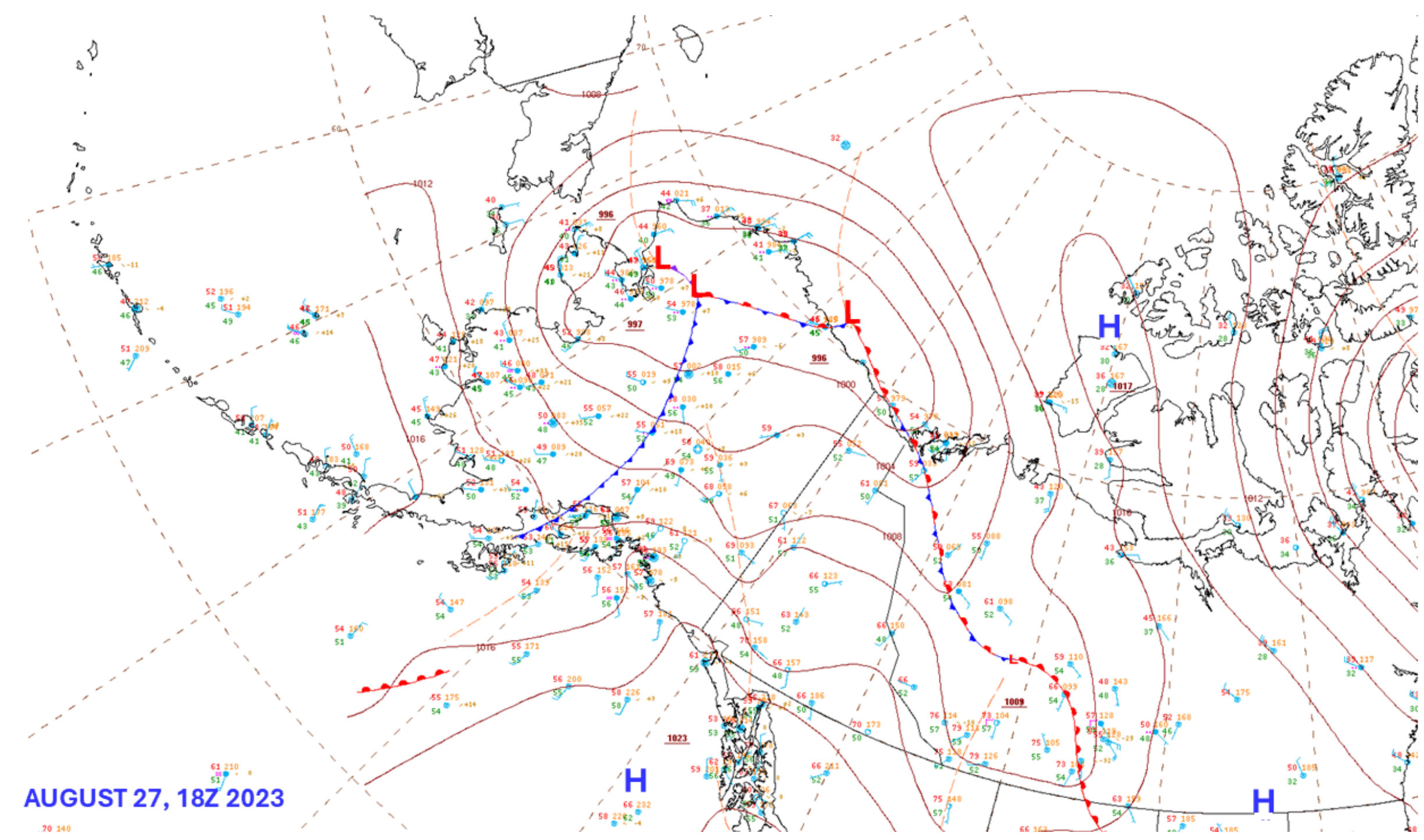

**Figure S52.** Surface analysis for 27 August 2023.

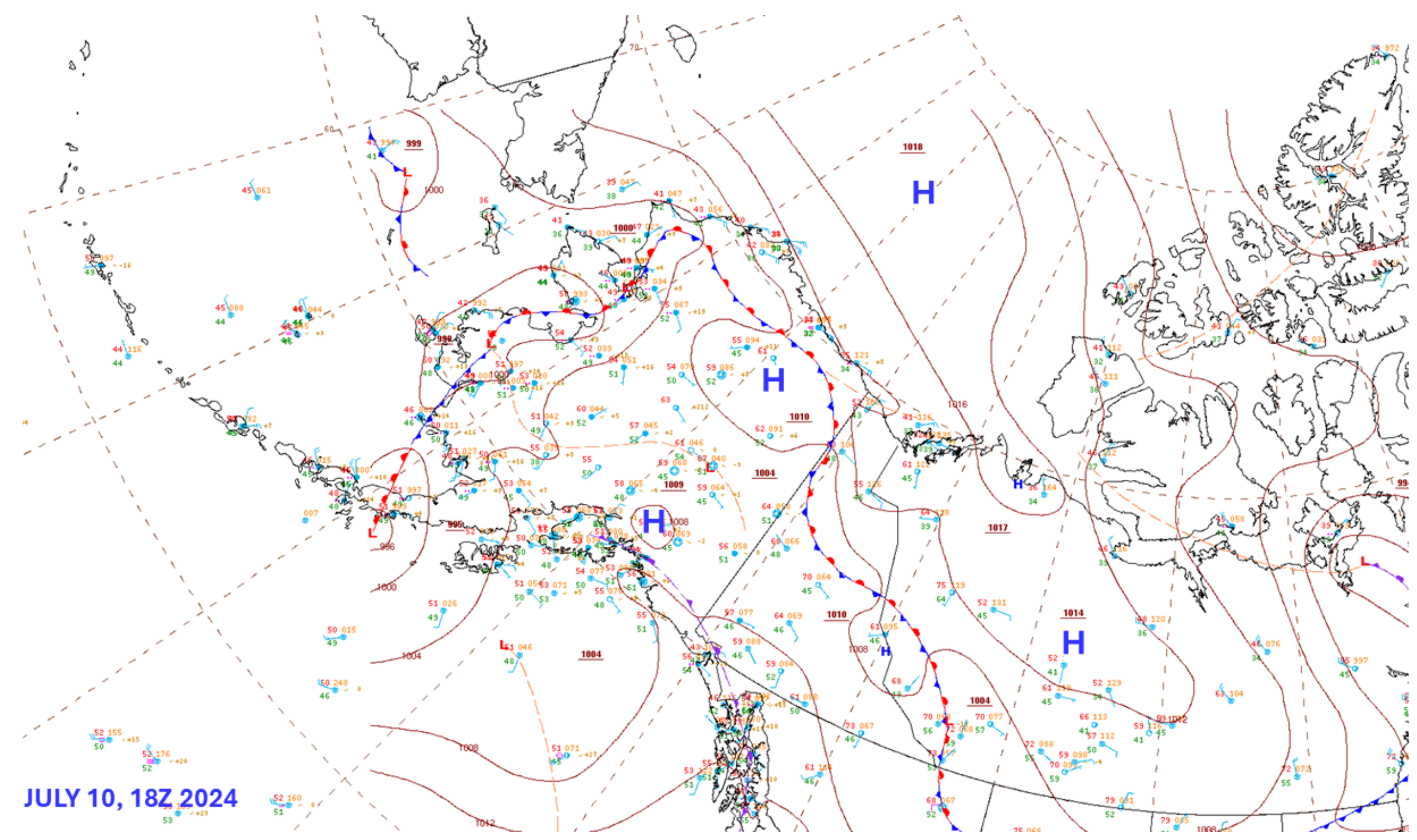

**Figure S53.** Surface analysis for 10 July 2024.

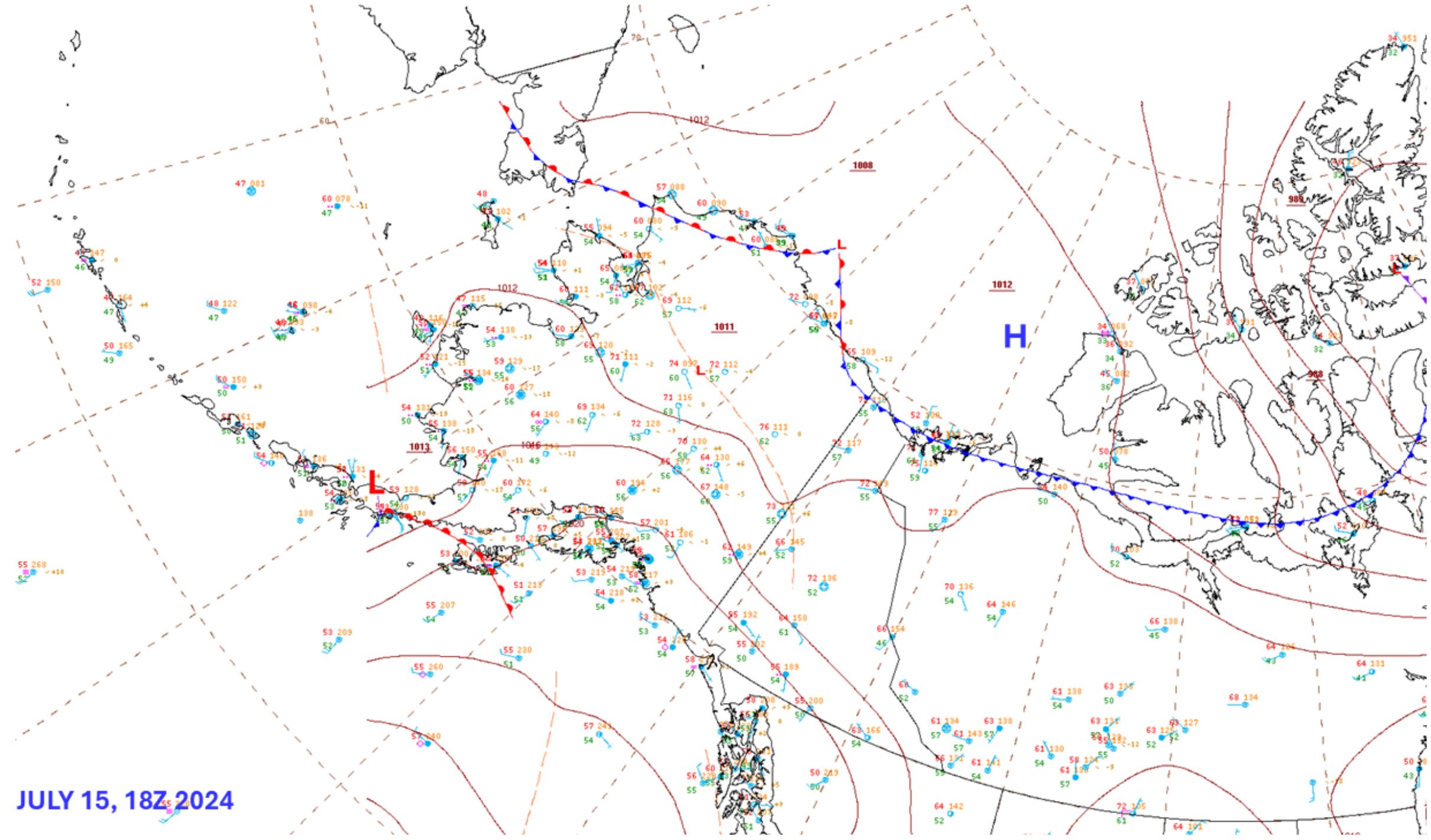

**Figure S54.** Surface analysis for 15 July 2024.

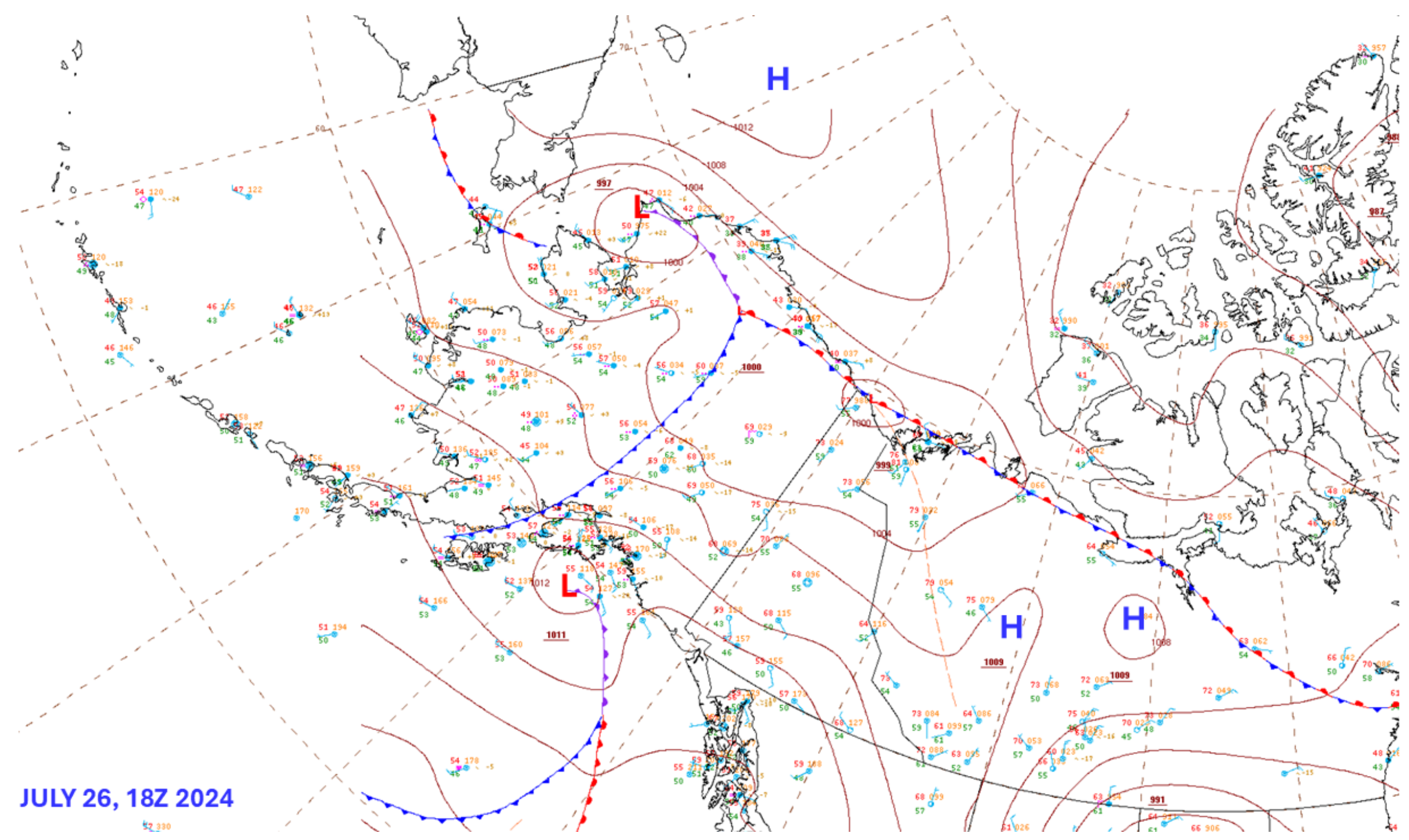

**Figure S55.** Surface analysis for 26 July 2024.

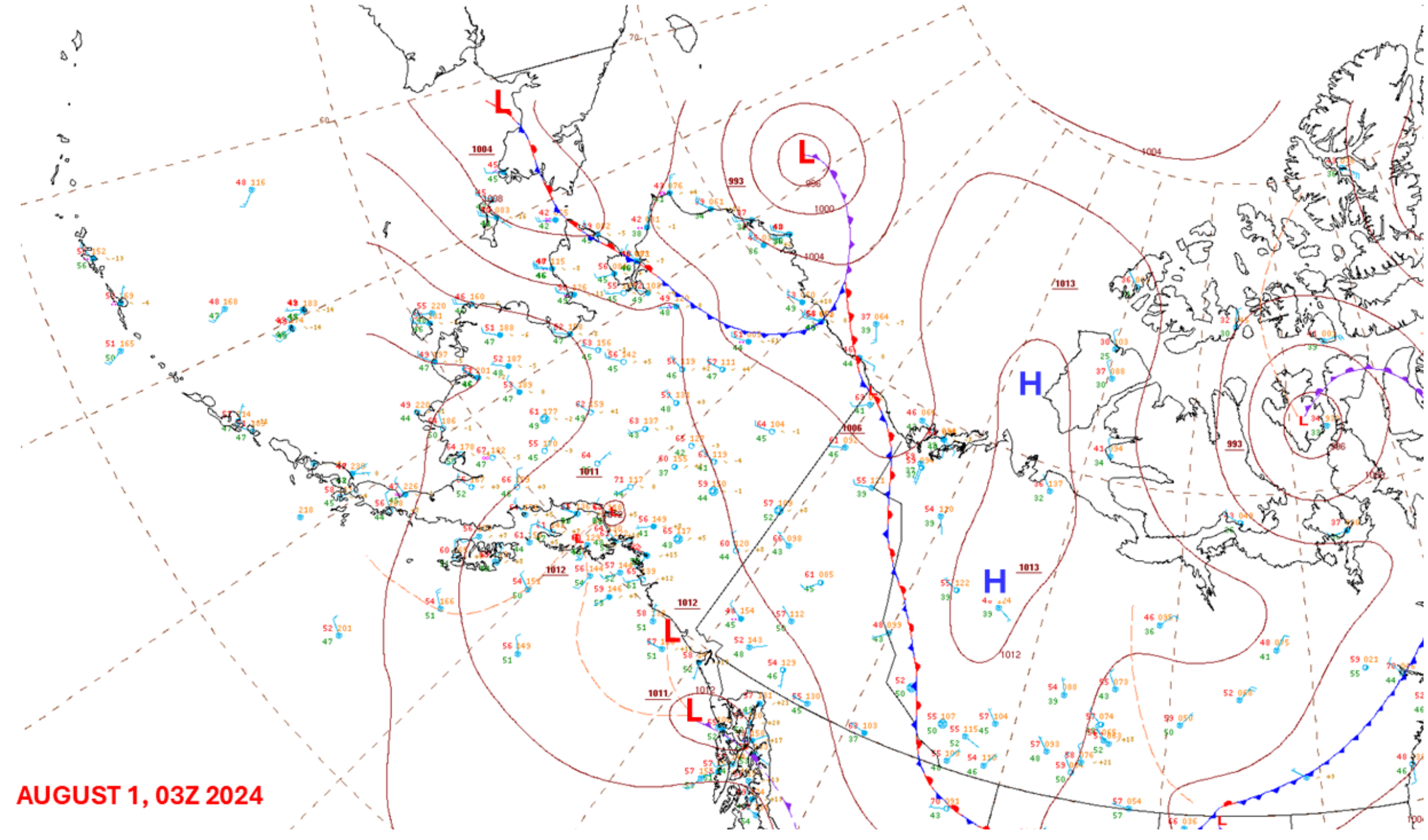

**Figure S56.** Surface analysis for 1 August 2024.

## S4. Regression results for all years (2000-2025)

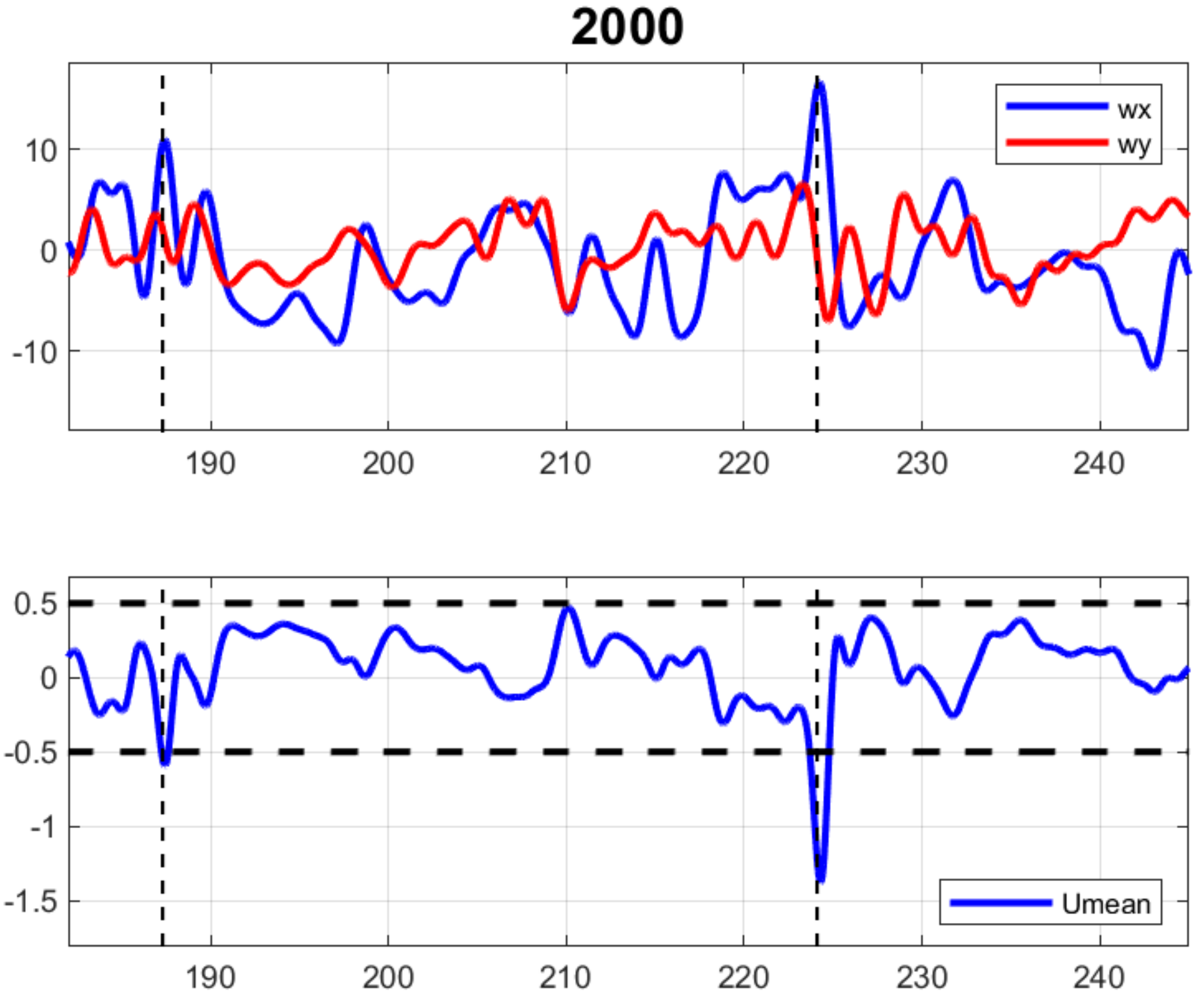

**Figure S57.** Regression results for summer of 2000 (July – August). Top panel: low-pass filtered wind velocity components (m/s). The east and north wind velocity components are blue and red lines, respectively. The axis is Julian days of the year. Lower panel: the regression produced low-pass filtered exchange velocity (m/s) at the Eluitkak Pass. The vertical dashed lines in both panels indicate the timing of the maximum velocity magnitude exceeding the threshold (0.5 m/s). If no vertical dashed line is present, there is no extreme event for that year.

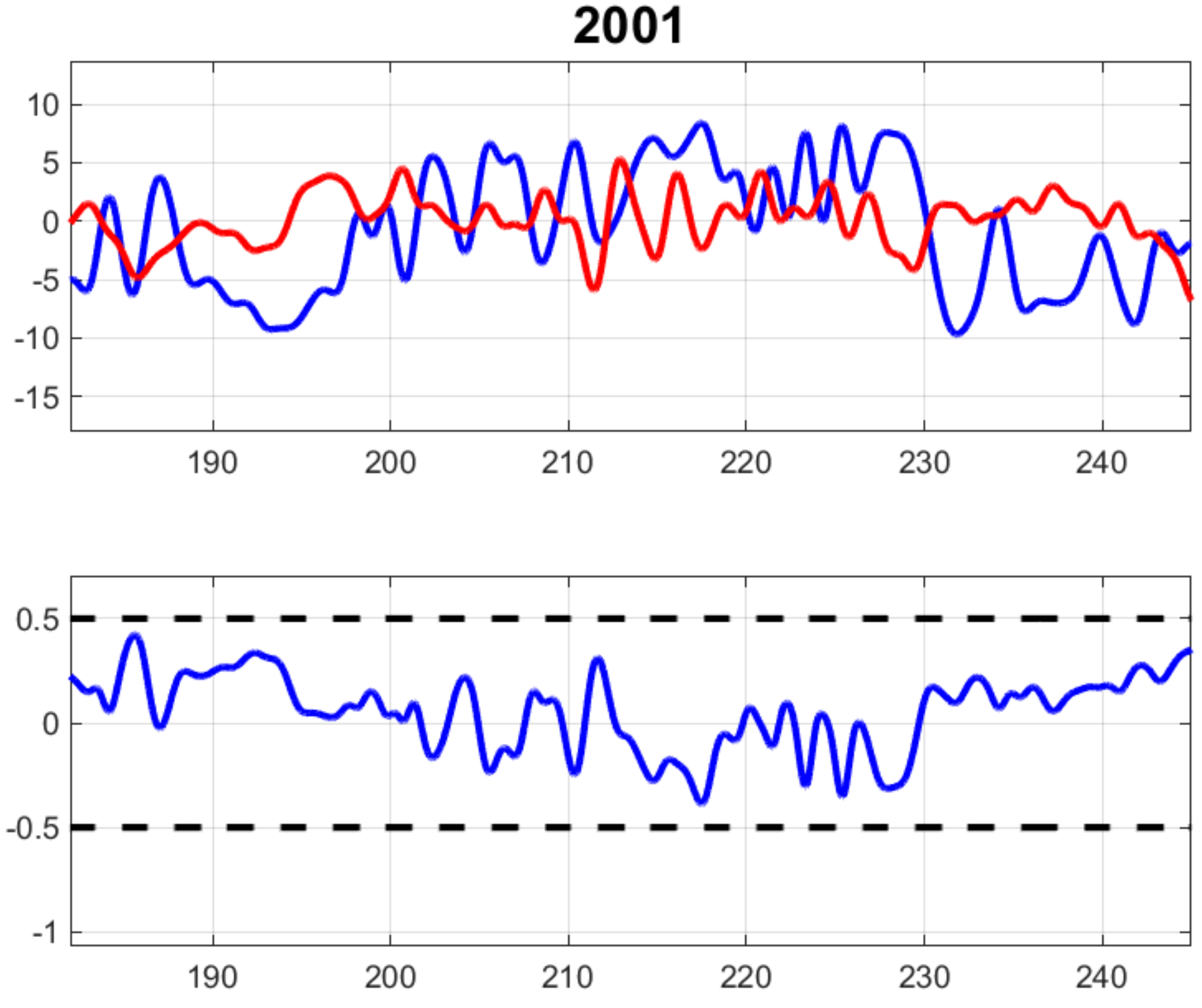

**Figure S58.** Regression results for summer of 2001 (July – August). Top panel: low-pass filtered wind velocity components (m/s). The east and north wind velocity components are blue and red lines, respectively. The axis is Julian days of the year. Lower panel: the regression produced low-pass filtered exchange velocity (m/s) at the Eluitkak Pass. The vertical dashed lines in both panels indicate the timing of the maximum velocity magnitude exceeding the threshold (0.5 m/s). If no vertical dashed line is present, there is no extreme event for that year.

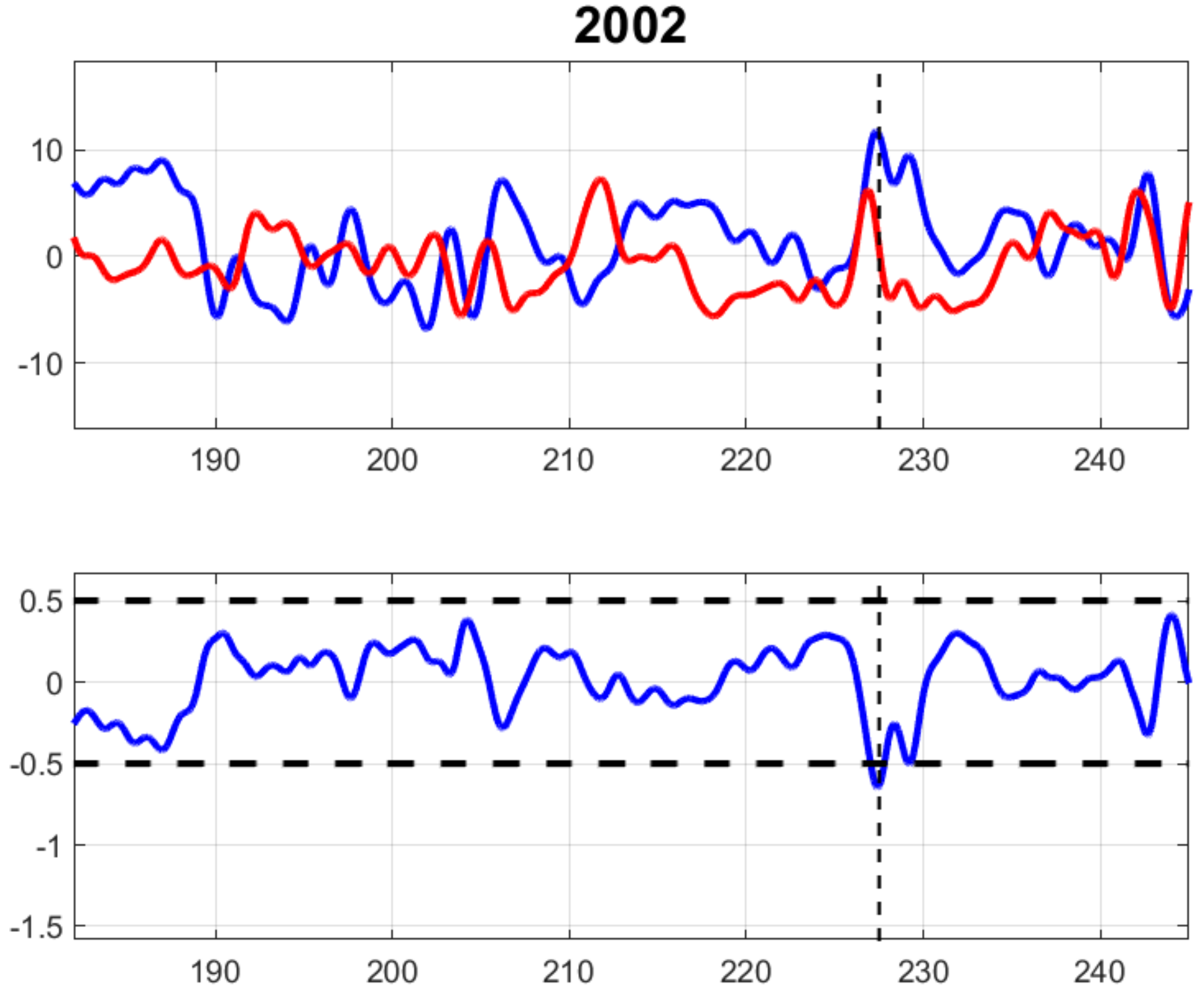

**Figure S59.** Regression results for summer of 2002 (July – August). Top panel: low-pass filtered wind velocity components (m/s). The east and north wind velocity components are blue and red lines, respectively. The axis is Julian days of the year. Lower panel: the regression produced low-pass filtered exchange velocity (m/s) at the Eluitkak Pass. The vertical dashed lines in both panels indicate the timing of the maximum velocity magnitude exceeding the threshold (0.5 m/s). If no vertical dashed line is present, there is no extreme event for that year.

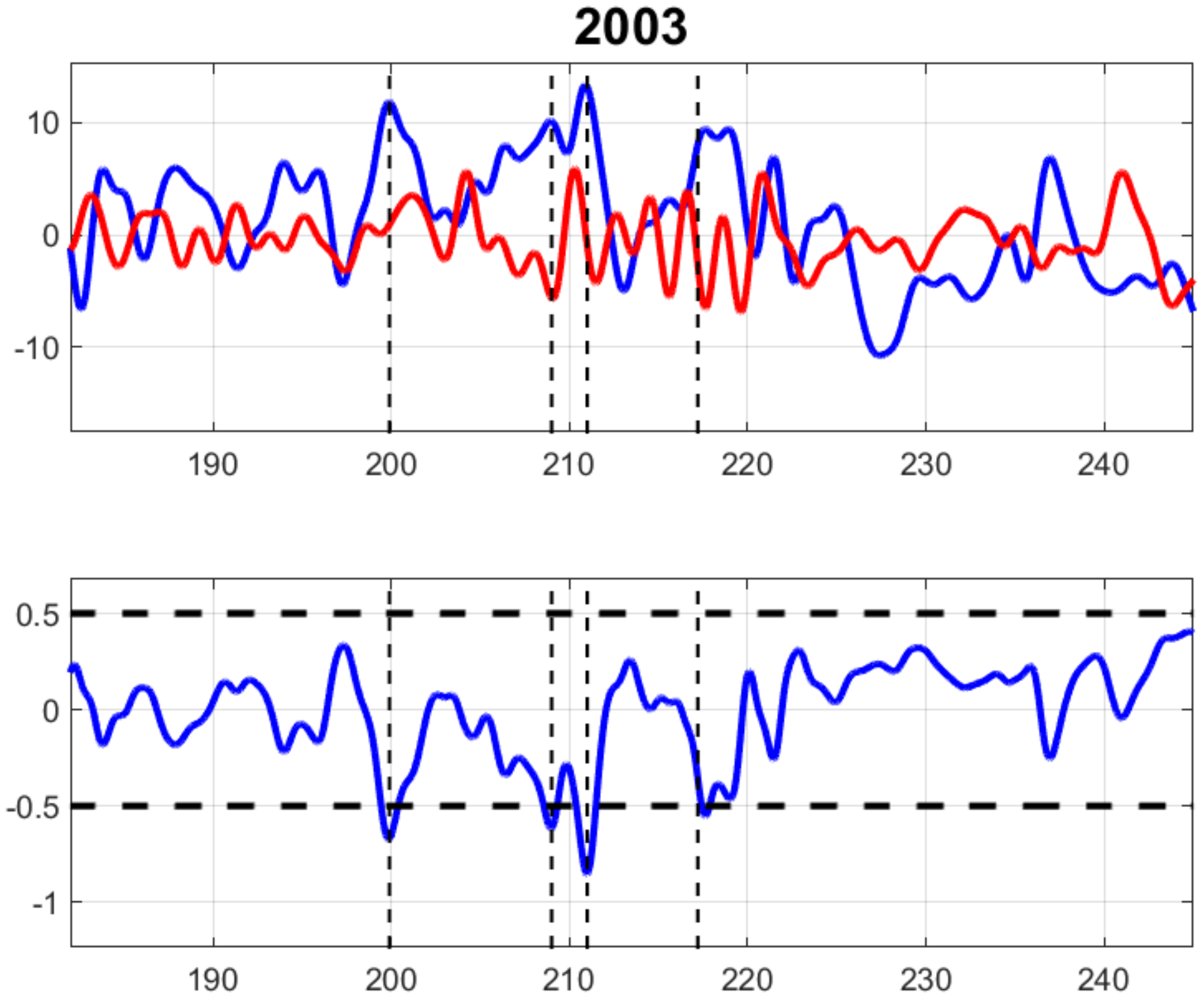

**Figure S60.** Regression results for summer of 2003 (July – August). Top panel: low-pass filtered wind velocity components (m/s). The east and north wind velocity components are blue and red lines, respectively. The axis is Julian days of the year. Lower panel: the regression produced low-pass filtered exchange velocity (m/s) at the Eluitkak Pass. The vertical dashed lines in both panels indicate the timing of the maximum velocity magnitude exceeding the threshold (0.5 m/s). If no vertical dashed line is present, there is no extreme event for that year.

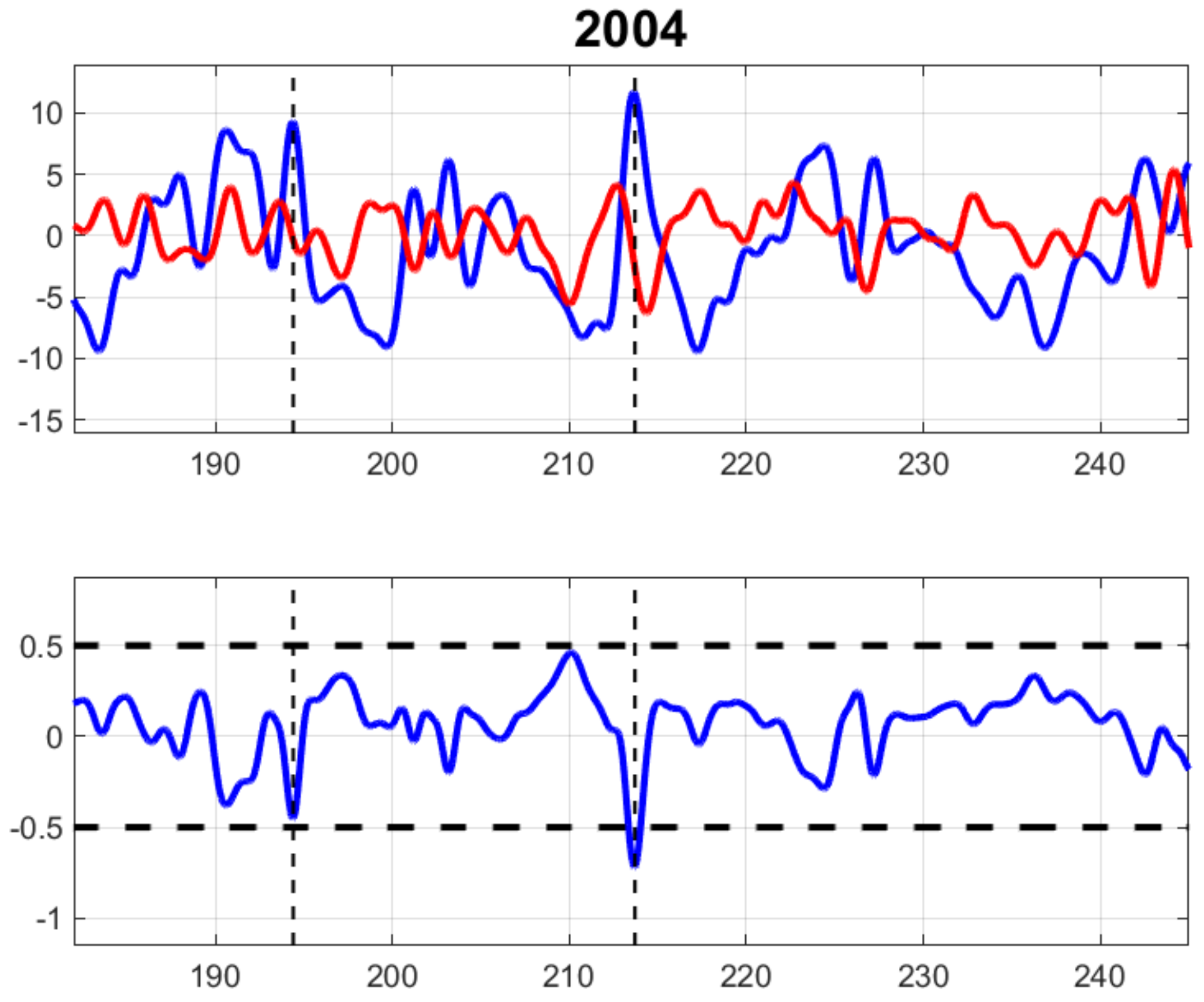

**Figure S61.** Regression results for summer of 2004 (July – August). Top panel: low-pass filtered wind velocity components (m/s). The east and north wind velocity components are blue and red lines, respectively. The axis is Julian days of the year. Lower panel: the regression produced low-pass filtered exchange velocity (m/s) at the Eluitkak Pass. The vertical dashed lines in both panels indicate the timing of the maximum velocity magnitude exceeding the threshold (0.5 m/s). If no vertical dashed line is present, there is no extreme event for that year.

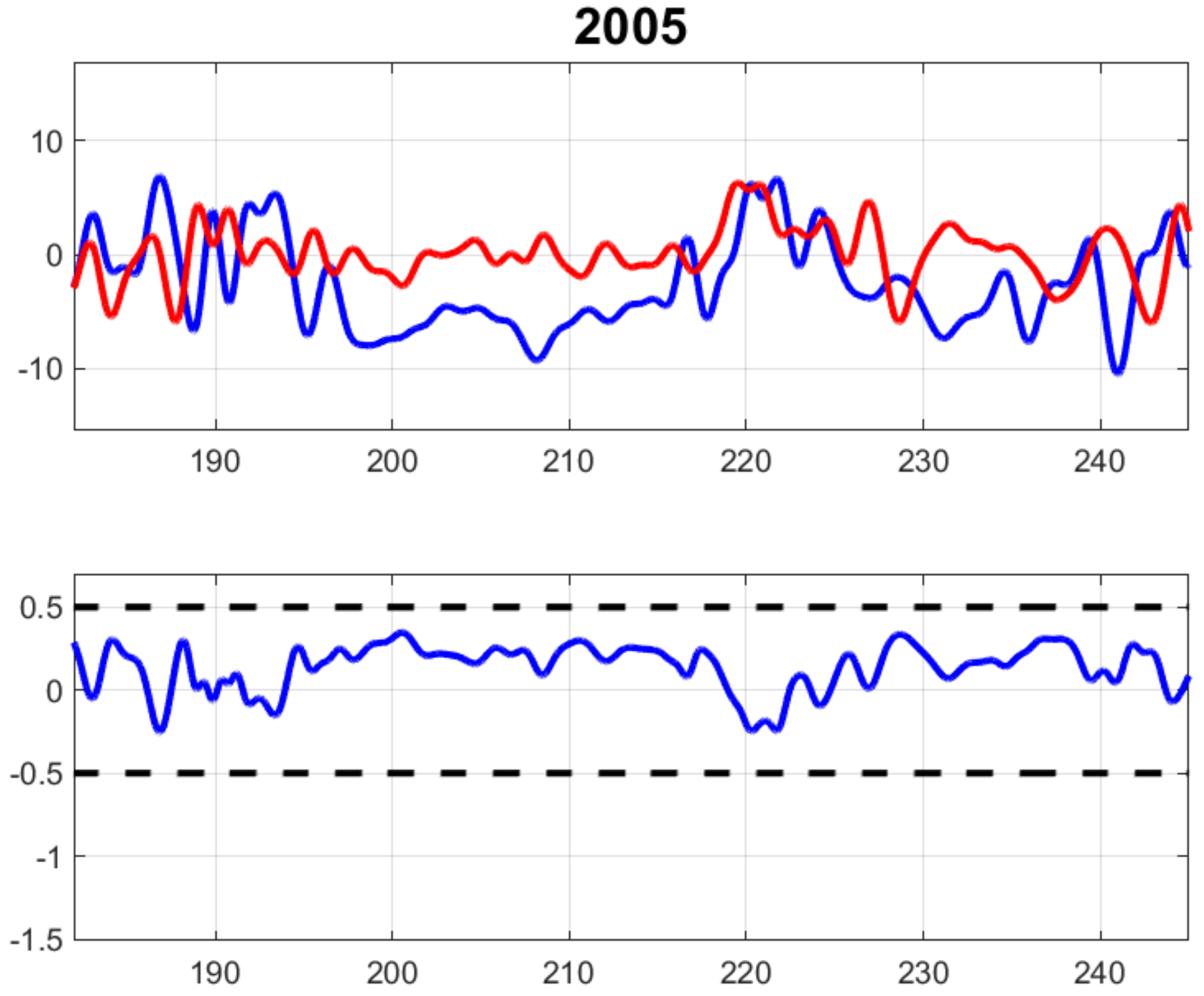

**Figure S62.** Regression results for summer of 2005 (July – August). Top panel: low-pass filtered wind velocity components (m/s). The east and north wind velocity components are blue and red lines, respectively. The axis is Julian days of the year. Lower panel: the regression produced low-pass filtered exchange velocity (m/s) at the Eluitkak Pass. The vertical dashed lines in both panels indicate the timing of the maximum velocity magnitude exceeding the threshold (0.5 m/s). If no vertical dashed line is present, there is no extreme event for that year.

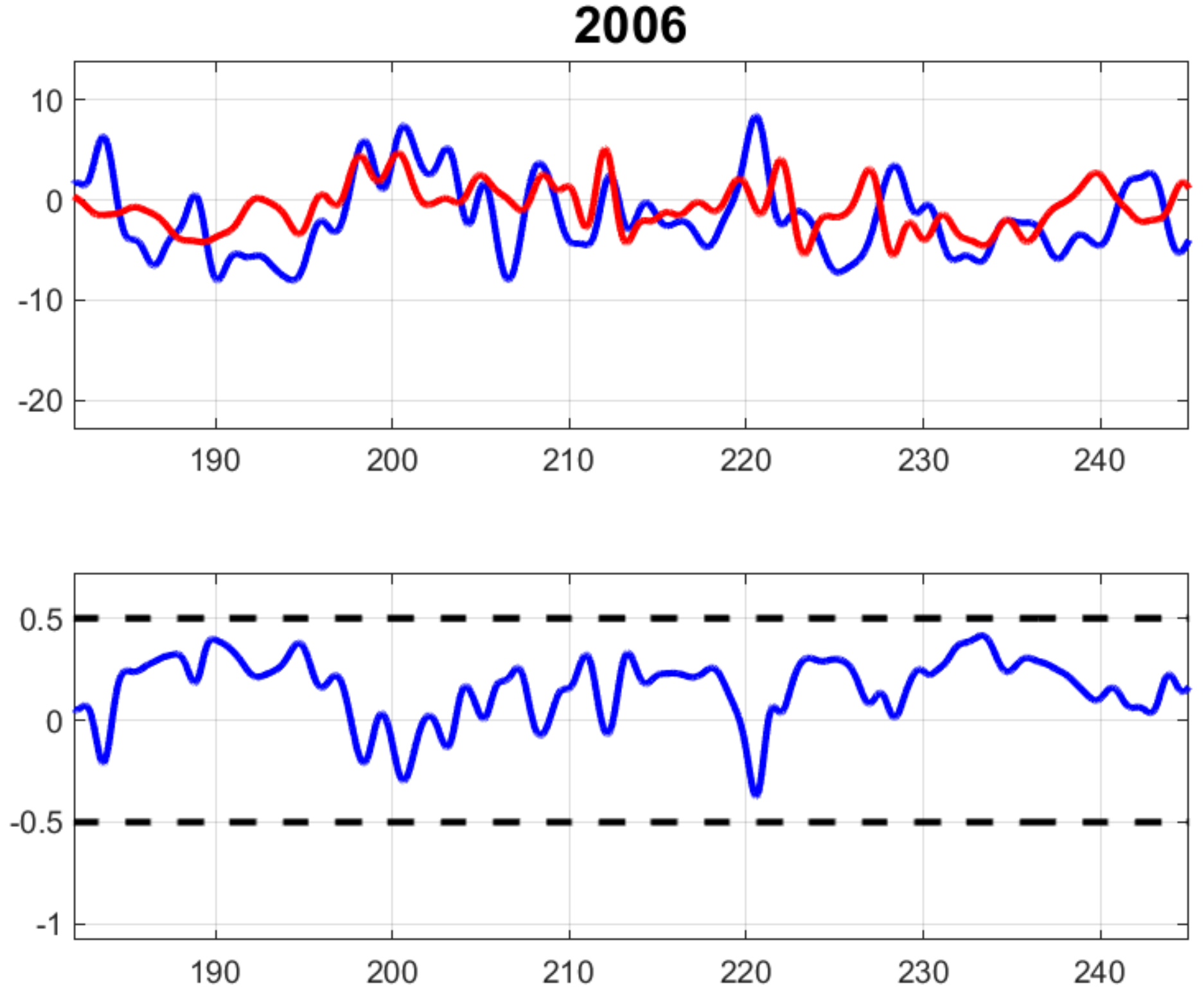

**Figure S63.** Regression results for summer of 2006 (July – August). Top panel: low-pass filtered wind velocity components (m/s). The east and north wind velocity components are blue and red lines, respectively. The axis is Julian days of the year. Lower panel: the regression produced low-pass filtered exchange velocity (m/s) at the Eluitkak Pass. The vertical dashed lines in both panels indicate the timing of the maximum velocity magnitude exceeding the threshold (0.5 m/s). If no vertical dashed line is present, there is no extreme event for that year.

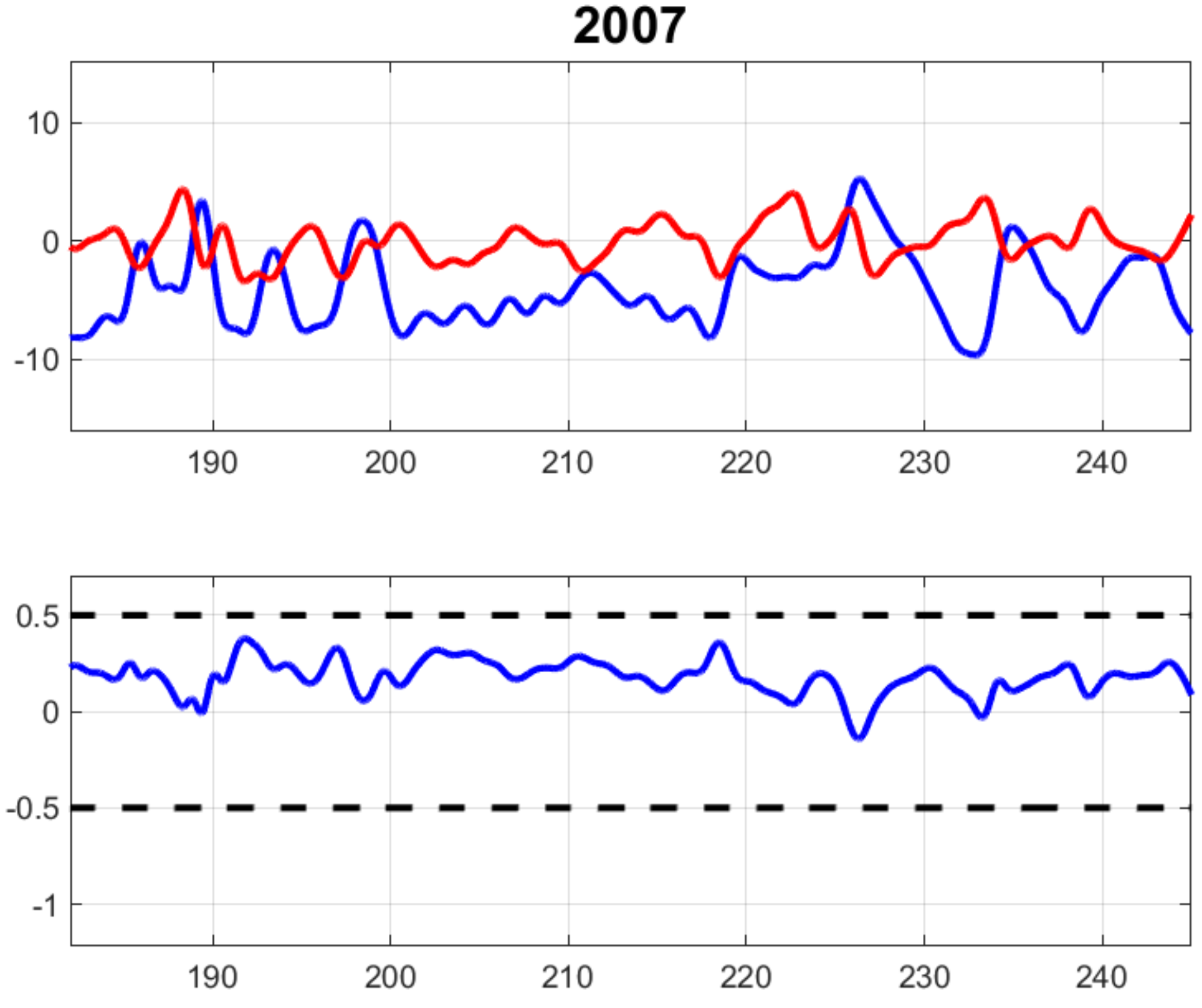

**Figure S64.** Regression results for summer of 2007 (July – August). Top panel: low-pass filtered wind velocity components (m/s). The east and north wind velocity components are blue and red lines, respectively. The axis is Julian days of the year. Lower panel: the regression produced low-pass filtered exchange velocity (m/s) at the Eluitkak Pass. The vertical dashed lines in both panels indicate the timing of the maximum velocity magnitude exceeding the threshold (0.5 m/s). If no vertical dashed line is present, there is no extreme event for that year.

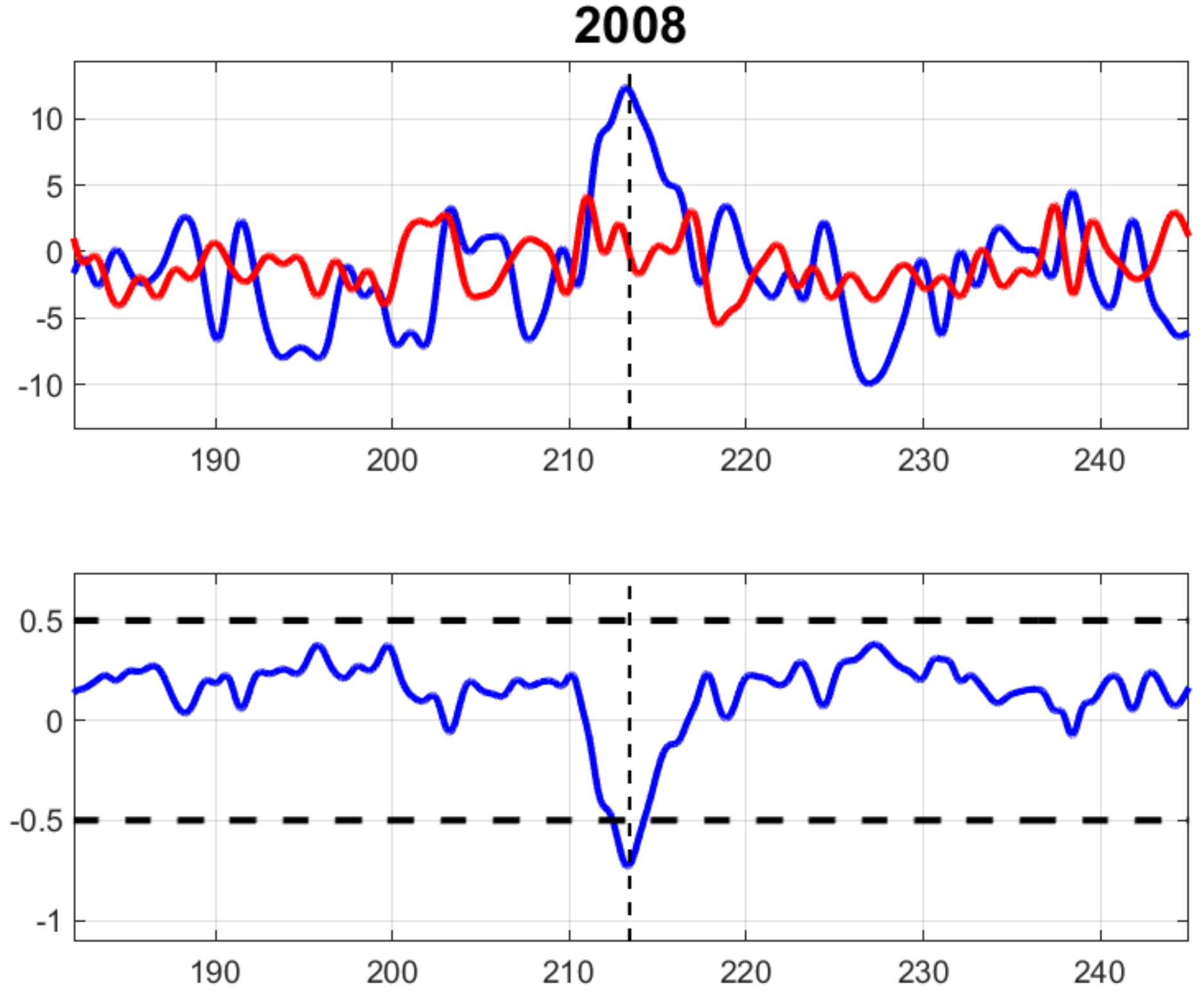

**Figure S65.** Regression results for summer of 2008 (July – August). Top panel: low-pass filtered wind velocity components (m/s). The east and north wind velocity components are blue and red lines, respectively. The axis is Julian days of the year. Lower panel: the regression produced low-pass filtered exchange velocity (m/s) at the Eluitkak Pass. The vertical dashed lines in both panels indicate the timing of the maximum velocity magnitude exceeding the threshold (0.5 m/s). If no vertical dashed line is present, there is no extreme event for that year.

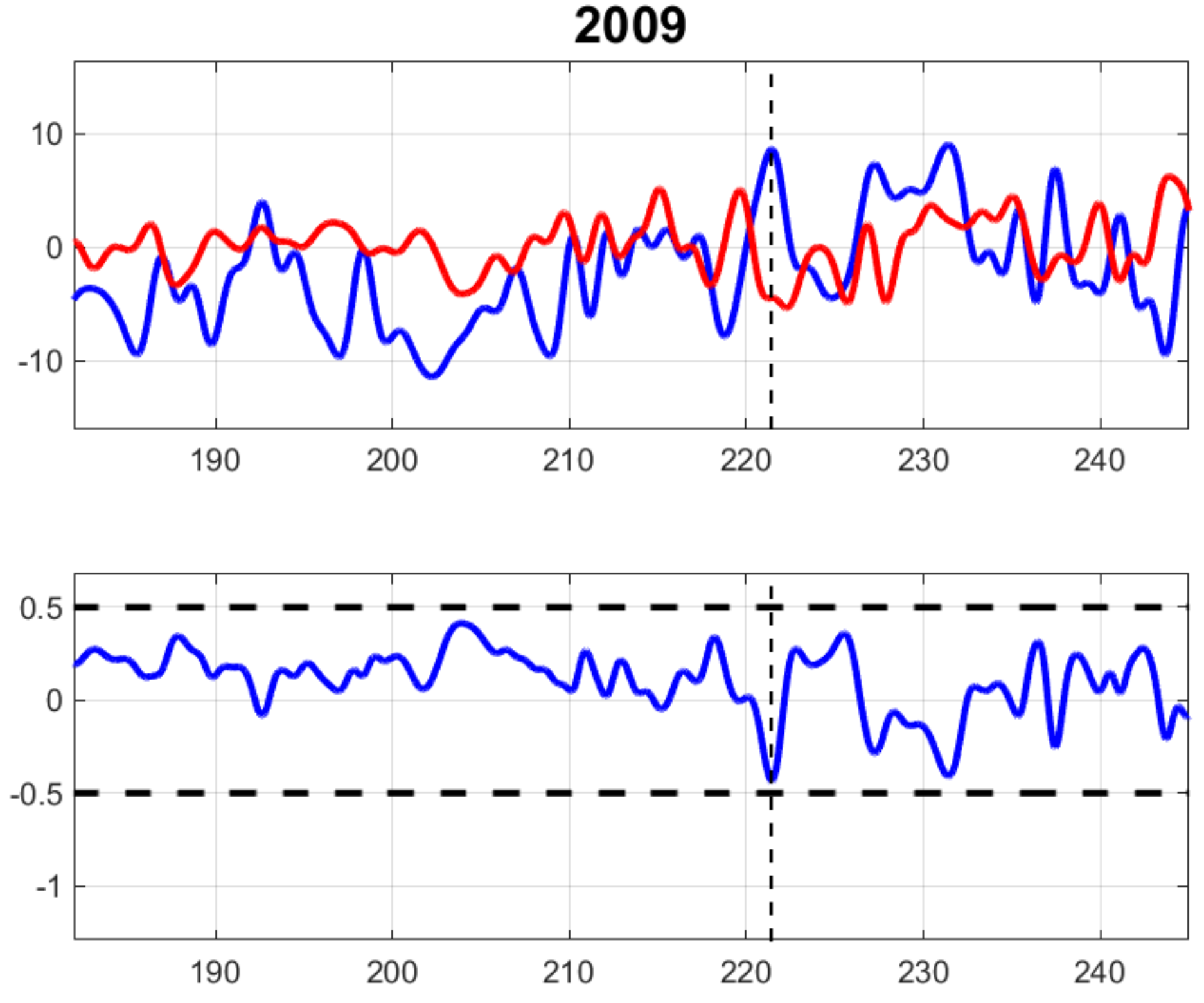

**Figure S66.** Regression results for summer of 2009 (July – August). Top panel: low-pass filtered wind velocity components (m/s). The east and north wind velocity components are blue and red lines, respectively. The axis is Julian days of the year. Lower panel: the regression produced low-pass filtered exchange velocity (m/s) at the Eluitkak Pass. The vertical dashed lines in both panels indicate the timing of the maximum velocity magnitude exceeding the threshold (0.5 m/s). If no vertical dashed line is present, there is no extreme event for that year.

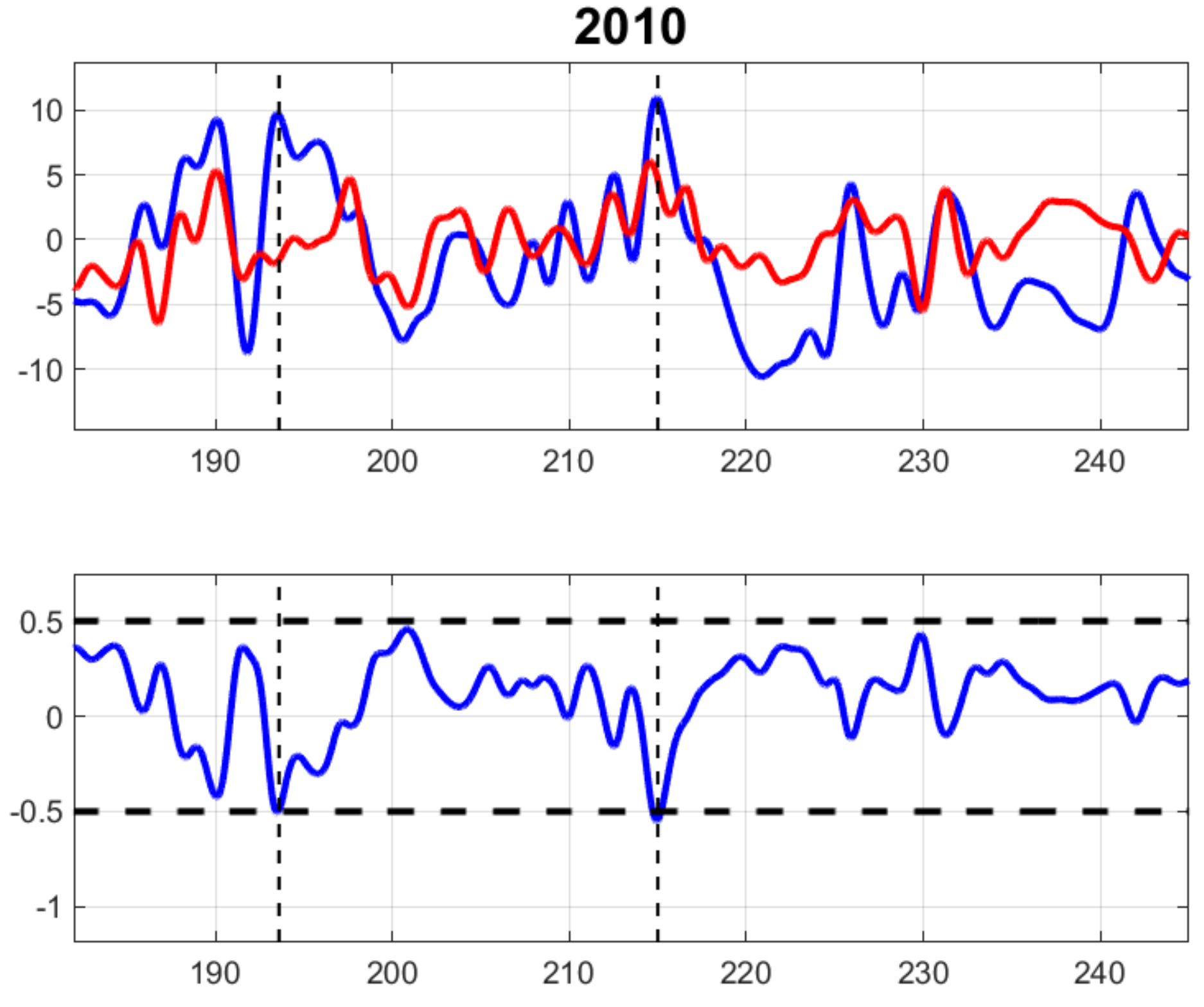

**Figure S67.** Regression results for summer of 2010 (July – August). Top panel: low-pass filtered wind velocity components (m/s). The east and north wind velocity components are blue and red lines, respectively. The axis is Julian days of the year. Lower panel: the regression produced low-pass filtered exchange velocity (m/s) at the Eluitkak Pass. The vertical dashed lines in both panels indicate the timing of the maximum velocity magnitude exceeding the threshold (0.5 m/s). If no vertical dashed line is present, there is no extreme event for that year.

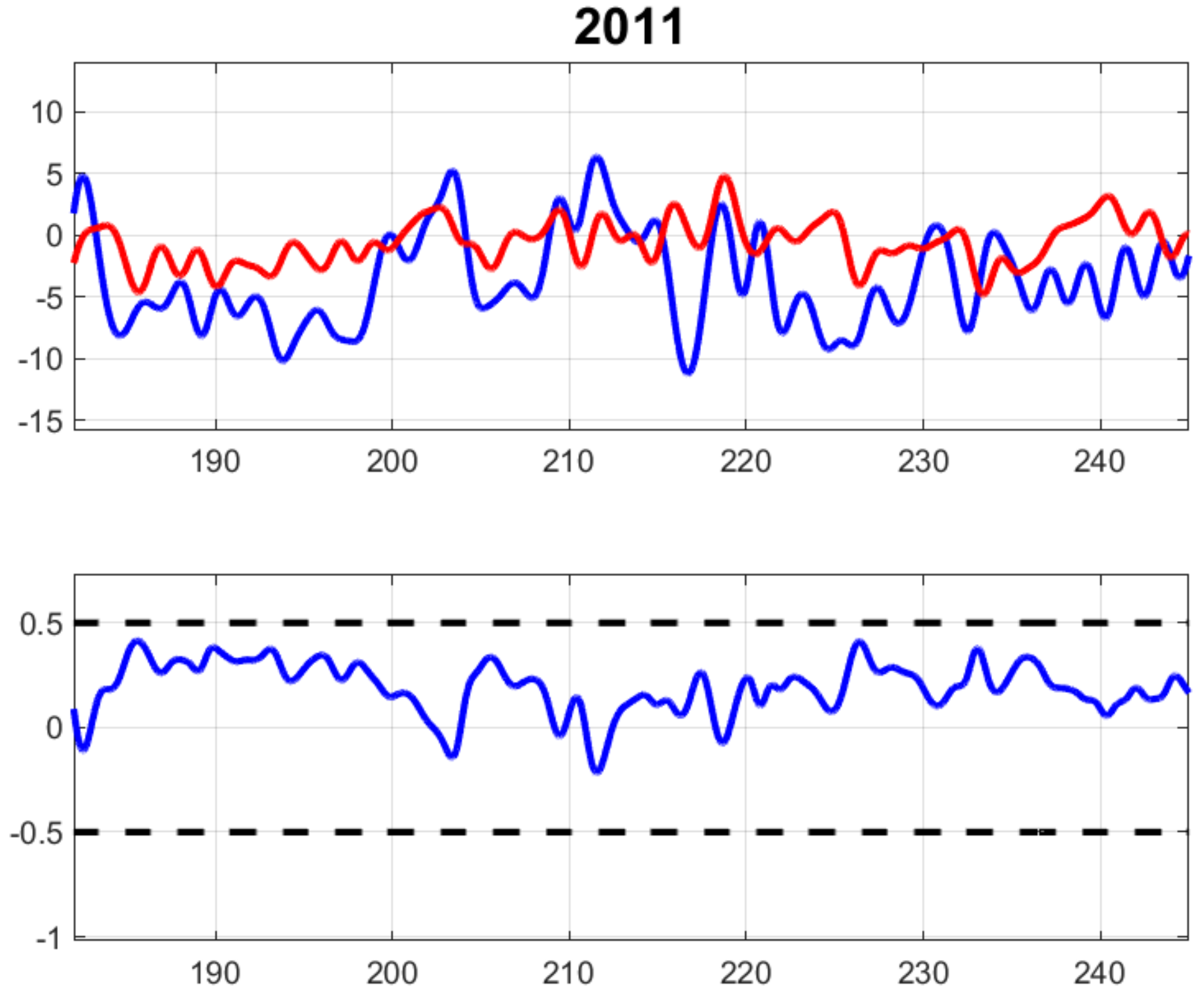

**Figure S68.** Regression results for summer of 2011 (July – August). Top panel: low-pass filtered wind velocity components (m/s). The east and north wind velocity components are blue and red lines, respectively. The axis is Julian days of the year. Lower panel: the regression produced low-pass filtered exchange velocity (m/s) at the Eluitkak Pass. The vertical dashed lines in both panels indicate the timing of the maximum velocity magnitude exceeding the threshold (0.5 m/s). If no vertical dashed line is present, there is no extreme event for that year.

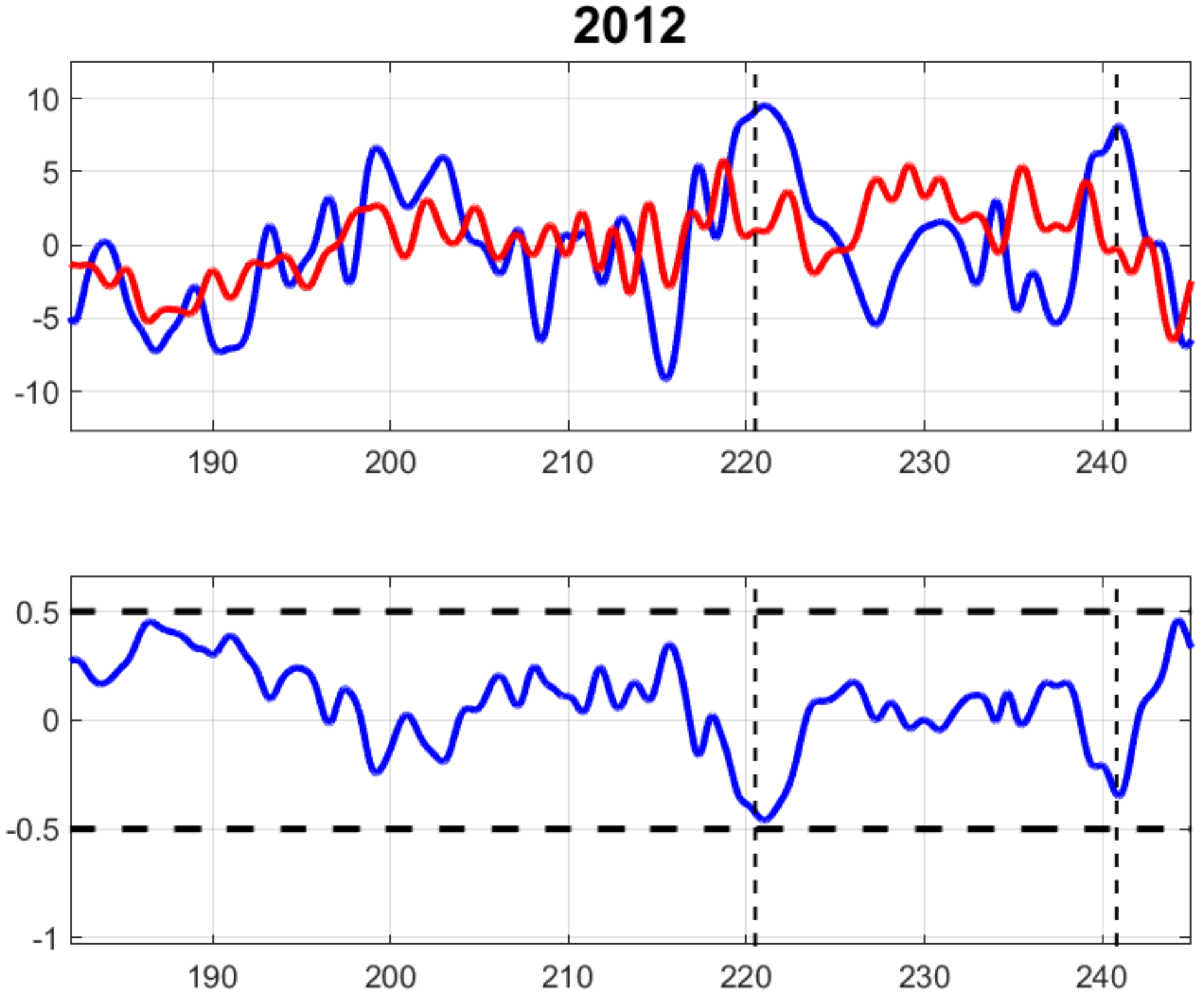

**Figure S69.** Regression results for summer of 2012 (July – August). Top panel: low-pass filtered wind velocity components (m/s). The east and north wind velocity components are blue and red lines, respectively. The axis is Julian days of the year. Lower panel: the regression produced low-pass filtered exchange velocity (m/s) at the Eluitkak Pass. The vertical dashed lines in both panels indicate the timing of the maximum velocity magnitude exceeding the threshold (0.5 m/s). If no vertical dashed line is present, there is no extreme event for that year.

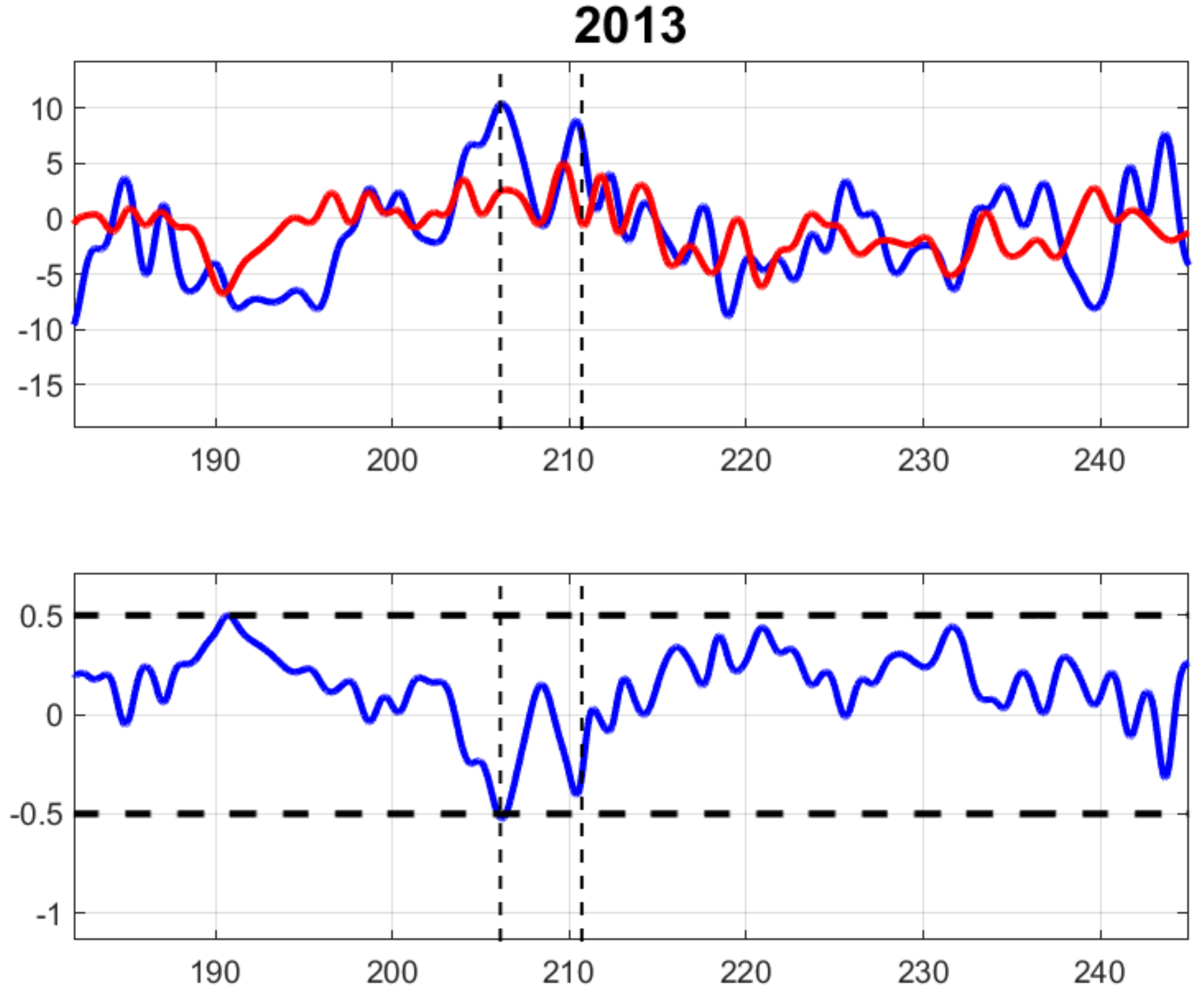

**Figure S70.** Regression results for summer of 2013 (July – August). Top panel: low-pass filtered wind velocity components (m/s). The east and north wind velocity components are blue and red lines, respectively. The axis is Julian days of the year. Lower panel: the regression produced low-pass filtered exchange velocity (m/s) at the Eluitkak Pass. The vertical dashed lines in both panels indicate the timing of the maximum velocity magnitude exceeding the threshold (0.5 m/s). If no vertical dashed line is present, there is no extreme event for that year.

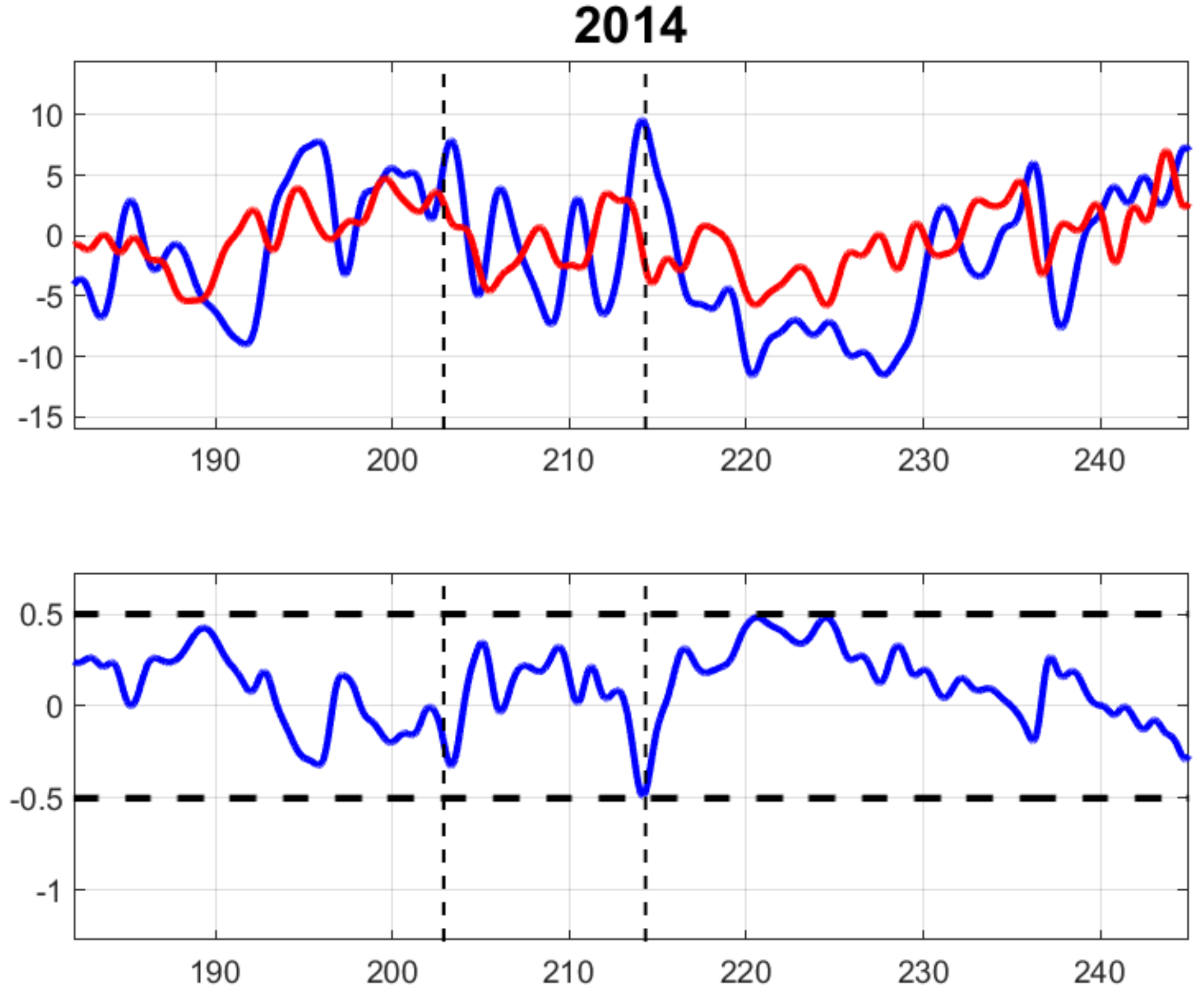

**Figure S71.** Regression results for summer of 2014 (July – August). Top panel: low-pass filtered wind velocity components (m/s). The east and north wind velocity components are blue and red lines, respectively. The axis is Julian days of the year. Lower panel: the regression produced low-pass filtered exchange velocity (m/s) at the Eluitkak Pass. The vertical dashed lines in both panels indicate the timing of the maximum velocity magnitude exceeding the threshold (0.5 m/s). If no vertical dashed line is present, there is no extreme event for that year.

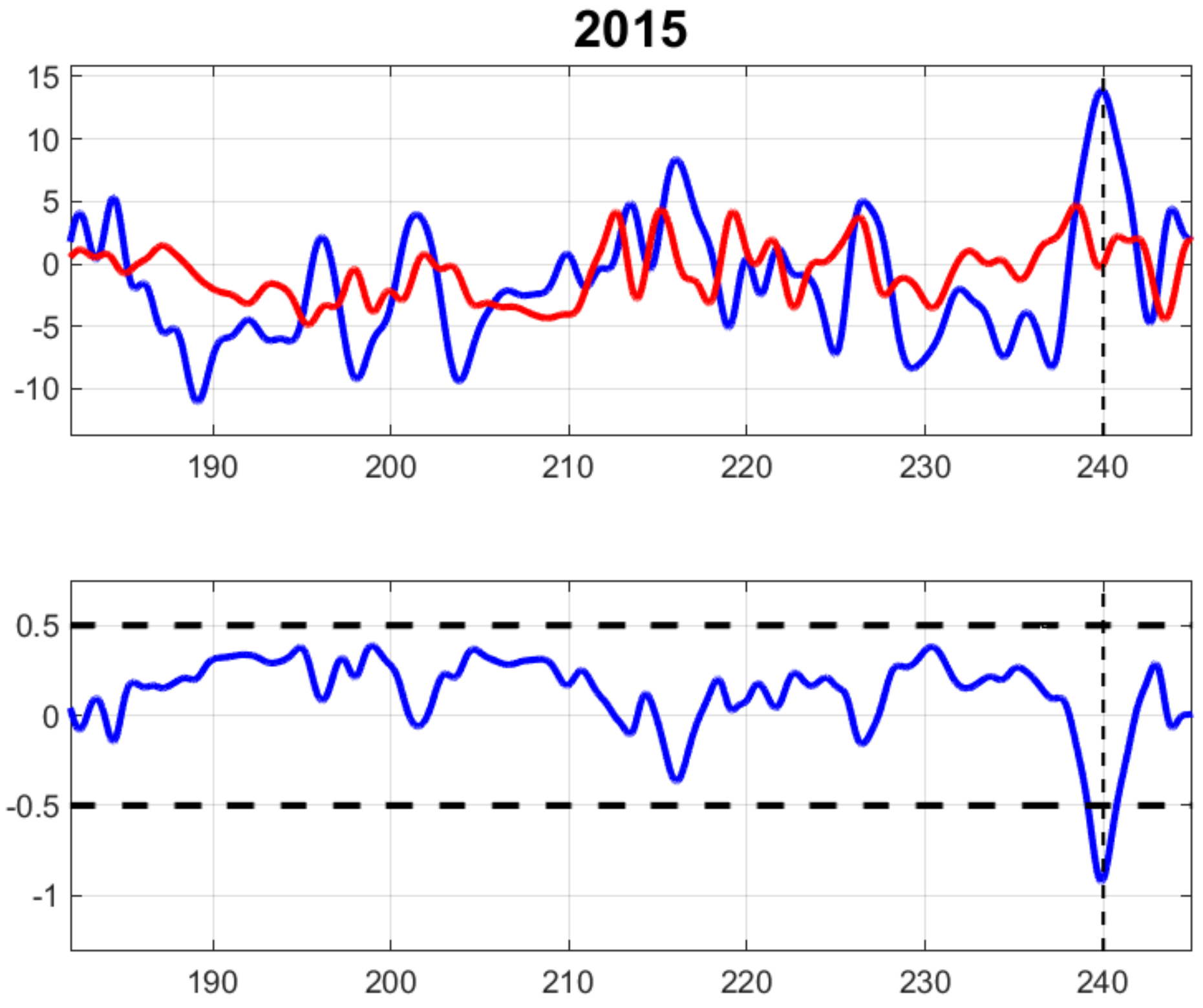

**Figure S72.** Regression results for summer of 2015 (July – August). Top panel: low-pass filtered wind velocity components (m/s). The east and north wind velocity components are blue and red lines, respectively. The axis is Julian days of the year. Lower panel: the regression produced low-pass filtered exchange velocity (m/s) at the Eluitkak Pass. The vertical dashed lines in both panels indicate the timing of the maximum velocity magnitude exceeding the threshold (0.5 m/s). If no vertical dashed line is present, there is no extreme event for that year.

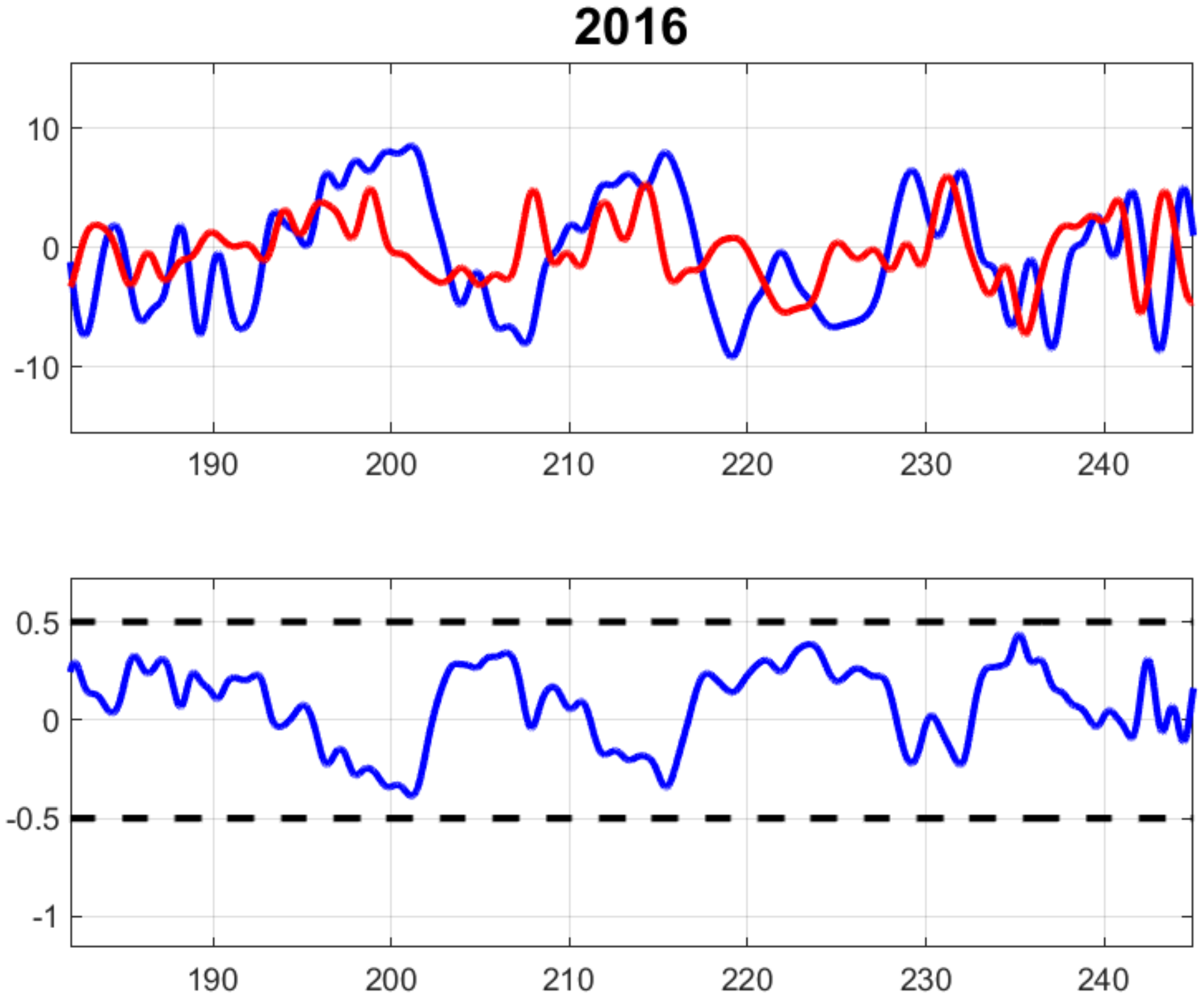

**Figure S73.** Regression results for summer of 2016 (July – August). Top panel: low-pass filtered wind velocity components (m/s). The east and north wind velocity components are blue and red lines, respectively. The axis is Julian days of the year. Lower panel: the regression produced low-pass filtered exchange velocity (m/s) at the Eluitkak Pass. The vertical dashed lines in both panels indicate the timing of the maximum velocity magnitude exceeding the threshold (0.5 m/s). If no vertical dashed line is present, there is no extreme event for that year.

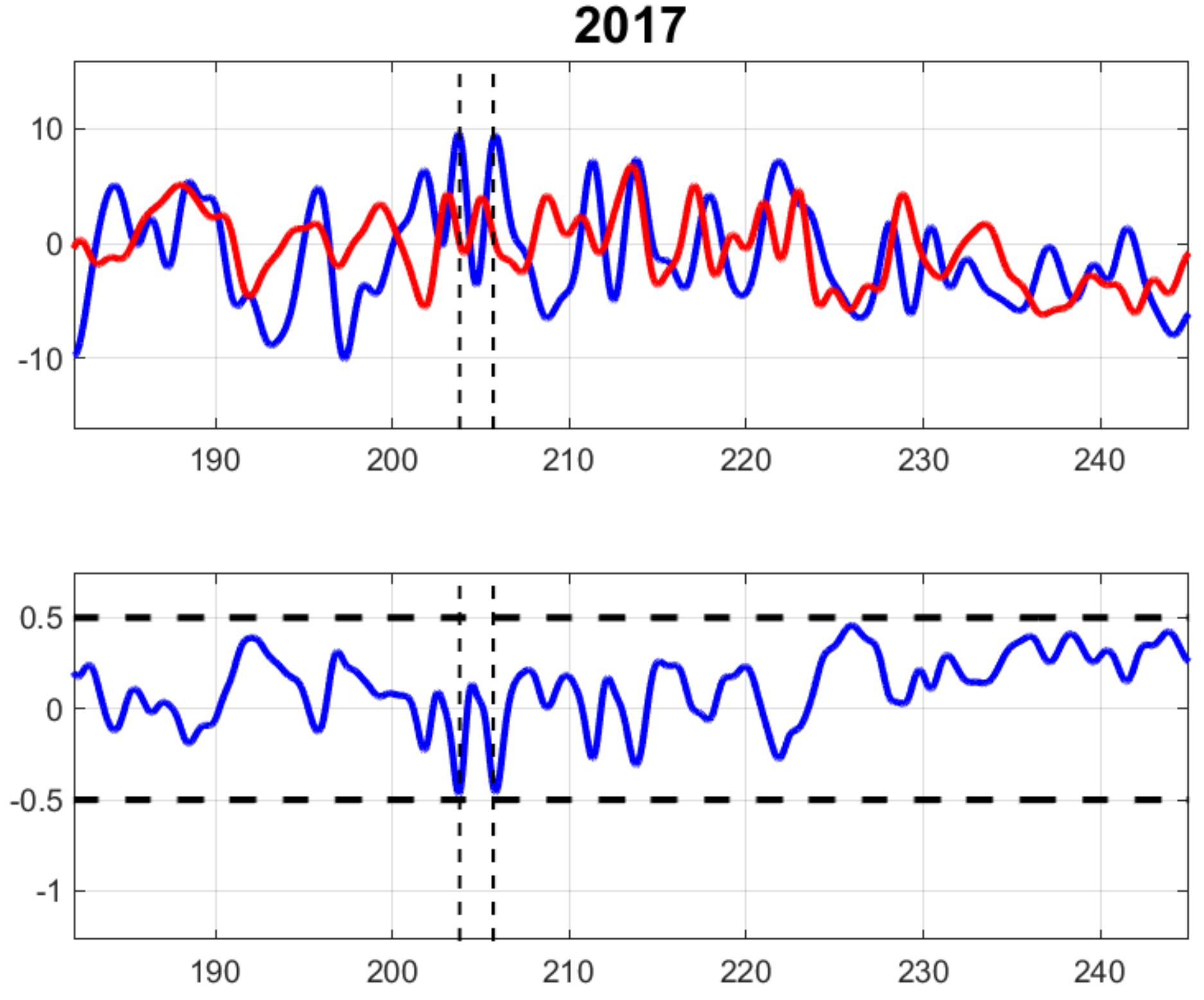

**Figure S74.** Regression results for summer of 2017 (July – August). Top panel: low-pass filtered wind velocity components (m/s). The east and north wind velocity components are blue and red lines, respectively. The axis is Julian days of the year. Lower panel: the regression produced low-pass filtered exchange velocity (m/s) at the Eluitkak Pass. The vertical dashed lines in both panels indicate the timing of the maximum velocity magnitude exceeding the threshold (0.5 m/s). If no vertical dashed line is present, there is no extreme event for that year.

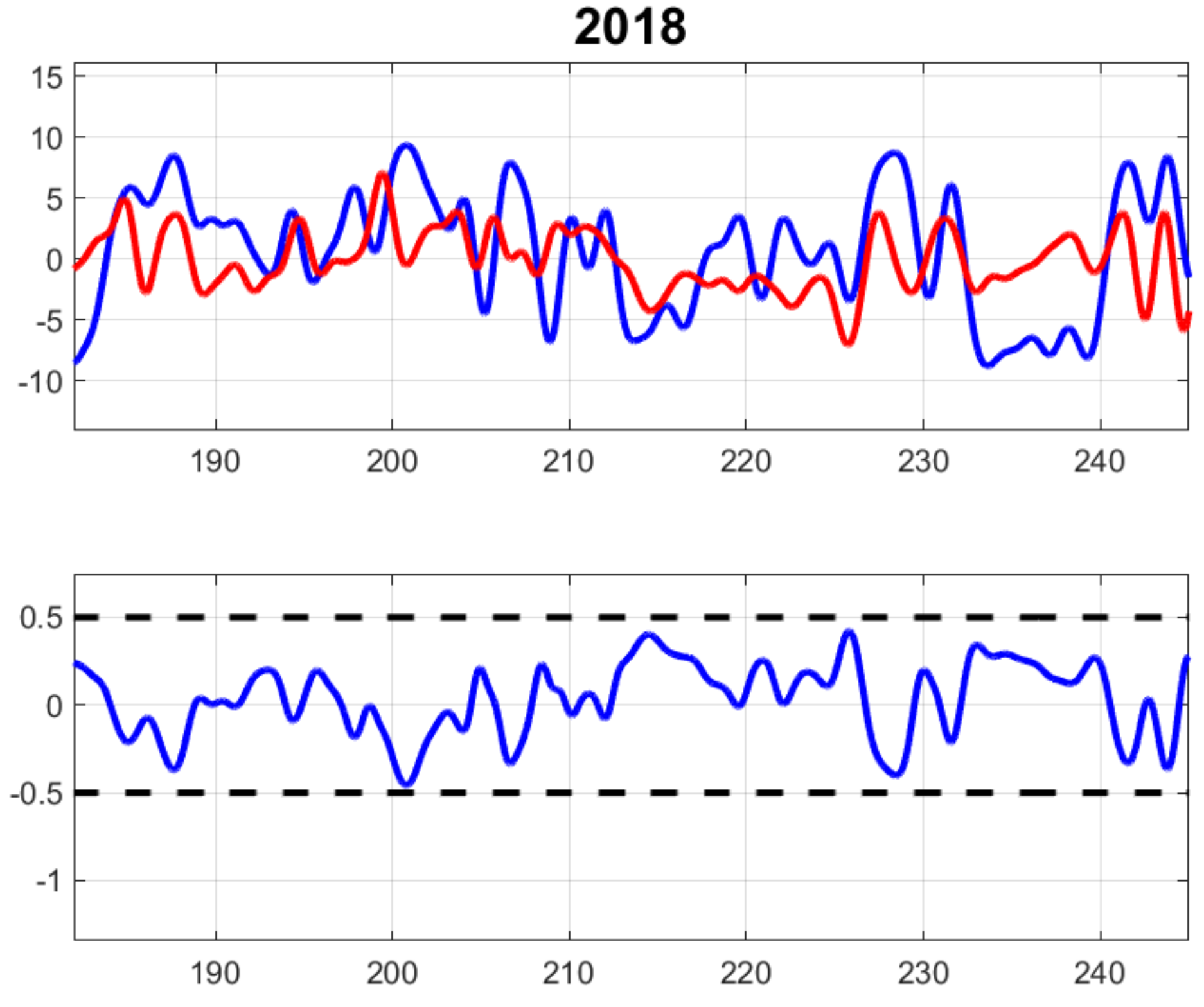

**Figure S75.** Regression results for summer of 2018 (July – August). Top panel: low-pass filtered wind velocity components (m/s). The east and north wind velocity components are blue and red lines, respectively. The axis is Julian days of the year. Lower panel: the regression produced low-pass filtered exchange velocity (m/s) at the Eluitkak Pass. The vertical dashed lines in both panels indicate the timing of the maximum velocity magnitude exceeding the threshold (0.5 m/s). If no vertical dashed line is present, there is no extreme event for that year.

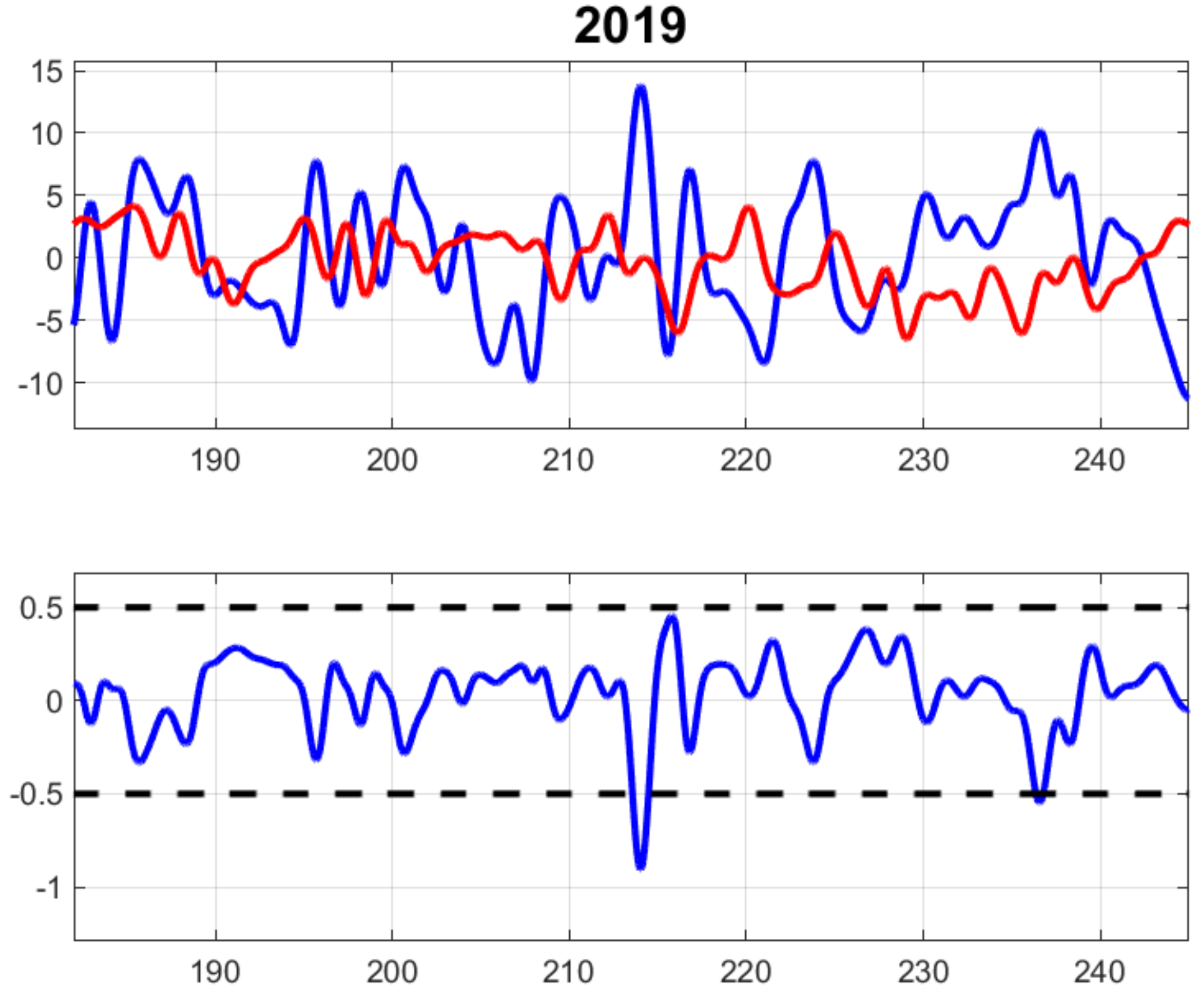

**Figure S76.** Regression results for summer of 2019 (July – August). Top panel: low-pass filtered wind velocity components (m/s). The east and north wind velocity components are blue and red lines, respectively. The axis is Julian days of the year. Lower panel: the regression produced low-pass filtered exchange velocity (m/s) at the Eluitkak Pass. The vertical dashed lines in both panels indicate the timing of the maximum velocity magnitude exceeding the threshold (0.5 m/s). If no vertical dashed line is present, there is no extreme event for that year.

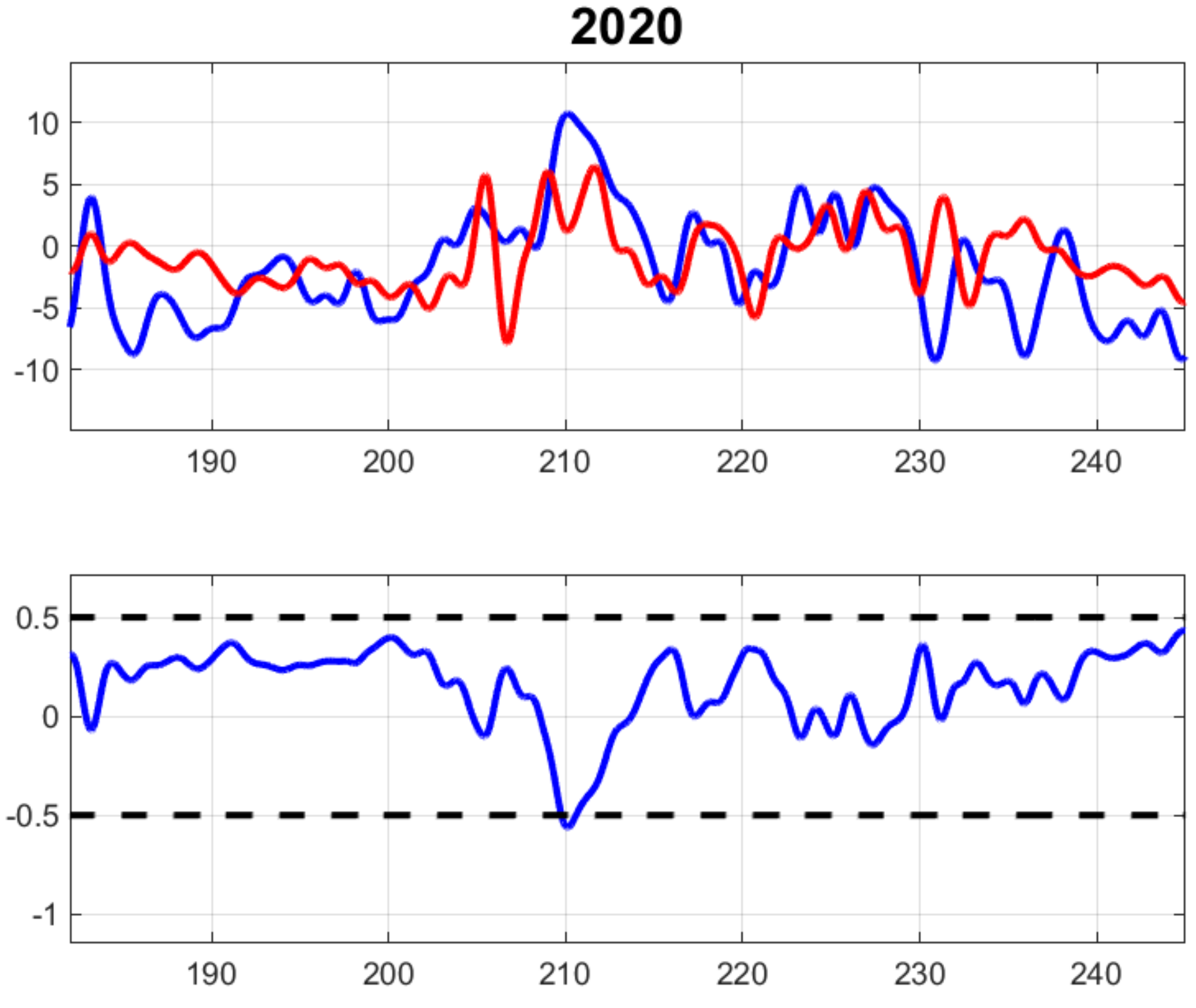

**Figure S77.** Regression results for summer of 2020 (July – August). Top panel: low-pass filtered wind velocity components (m/s). The east and north wind velocity components are blue and red lines, respectively. The axis is Julian days of the year. Lower panel: the regression produced low-pass filtered exchange velocity (m/s) at the Eluitkak Pass. The vertical dashed lines in both panels indicate the timing of the maximum velocity magnitude exceeding the threshold (0.5 m/s). If no vertical dashed line is present, there is no extreme event for that year.

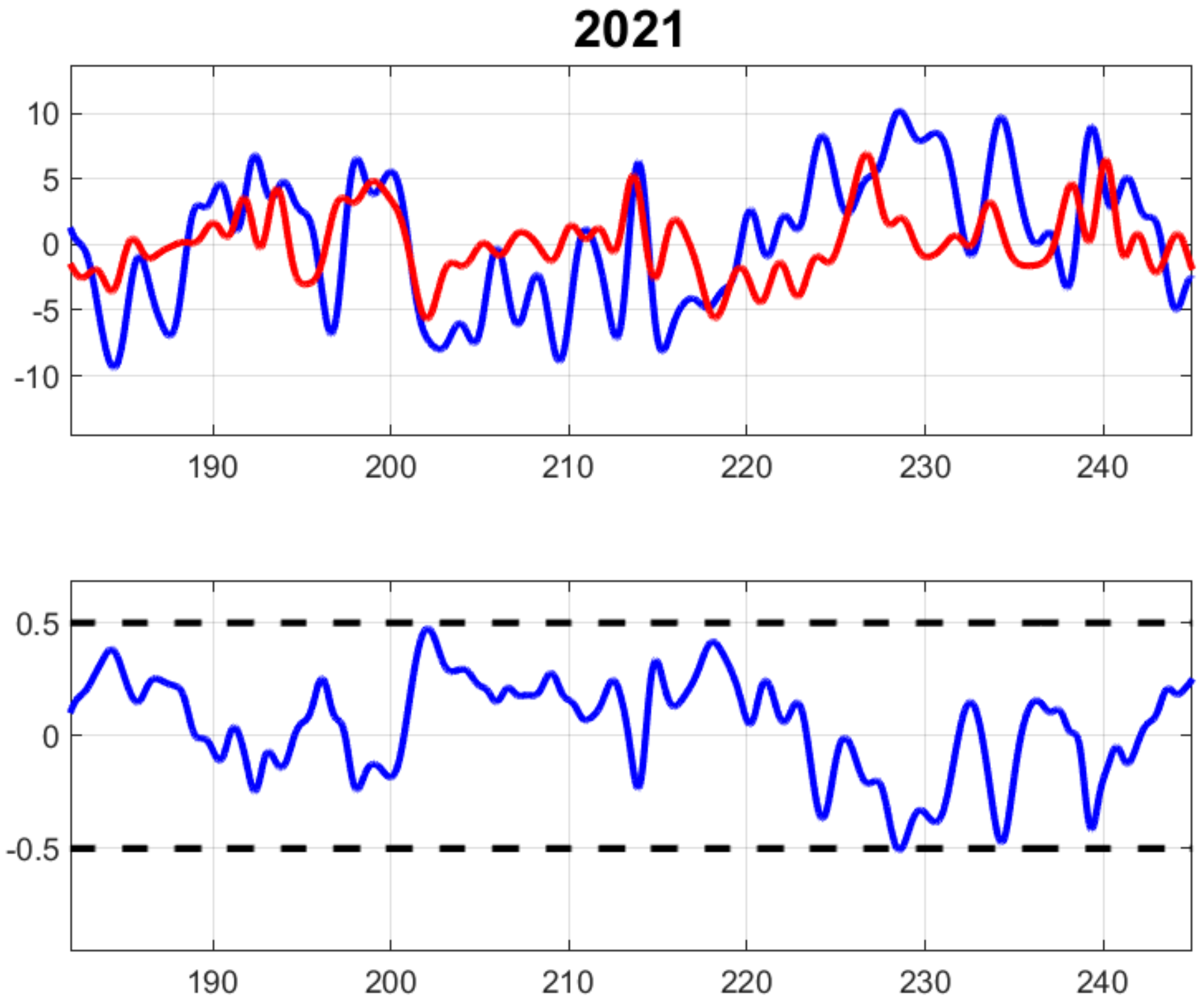

**Figure S78.** Regression results for summer of 2021 (July – August). Top panel: low-pass filtered wind velocity components (m/s). The east and north wind velocity components are blue and red lines, respectively. The axis is Julian days of the year. Lower panel: the regression produced low-pass filtered exchange velocity (m/s) at the Eluitkak Pass. The vertical dashed lines in both panels indicate the timing of the maximum velocity magnitude exceeding the threshold (0.5 m/s). If no vertical dashed line is present, there is no extreme event for that year.

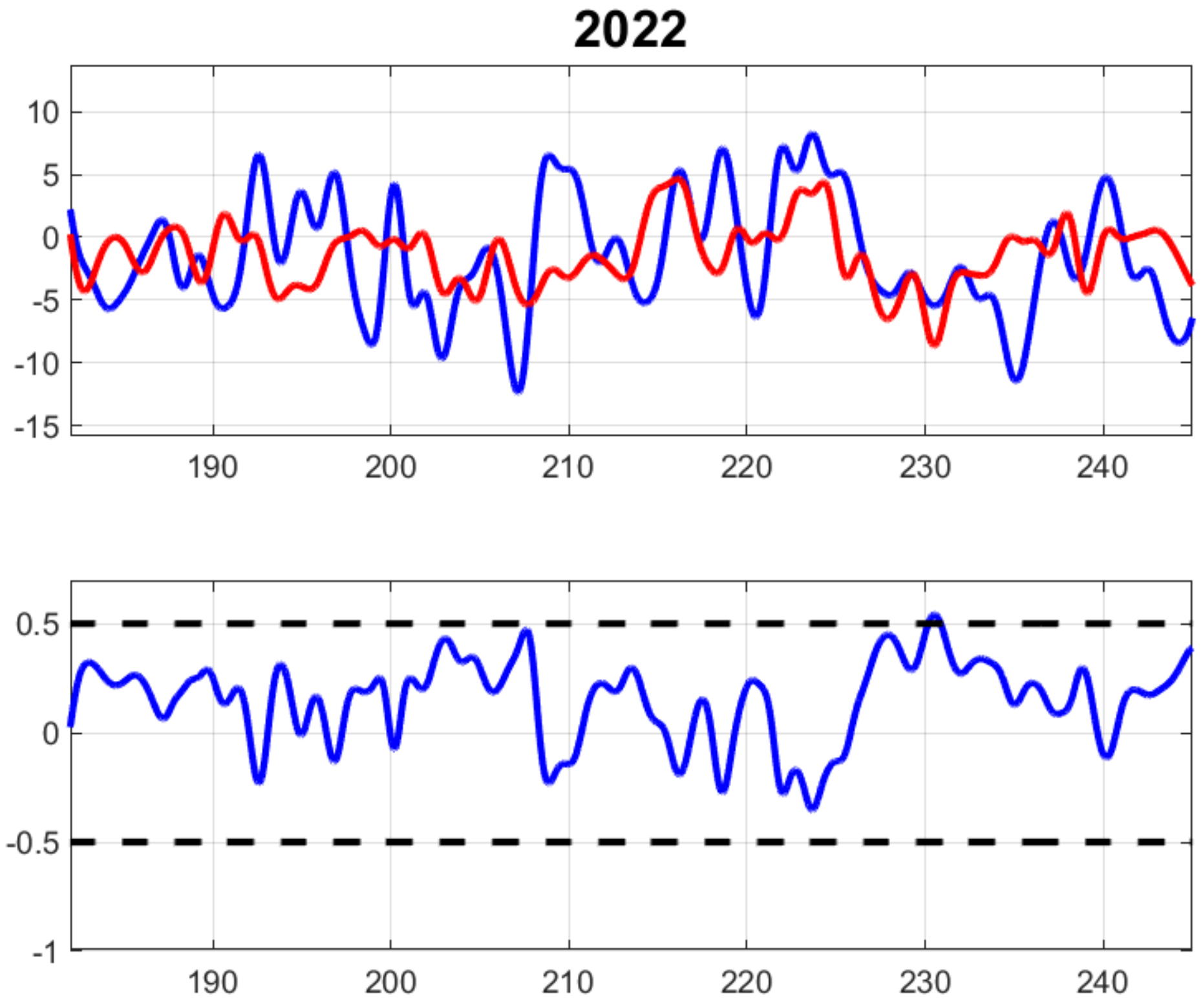

**Figure S79.** Regression results for summer of 2022 (July – August). Top panel: low-pass filtered wind velocity components (m/s). The east and north wind velocity components are blue and red lines, respectively. The axis is Julian days of the year. Lower panel: the regression produced low-pass filtered exchange velocity (m/s) at the Eluitkak Pass. The vertical dashed lines in both panels indicate the timing of the maximum velocity magnitude exceeding the threshold (0.5 m/s). If no vertical dashed line is present, there is no extreme event for that year.

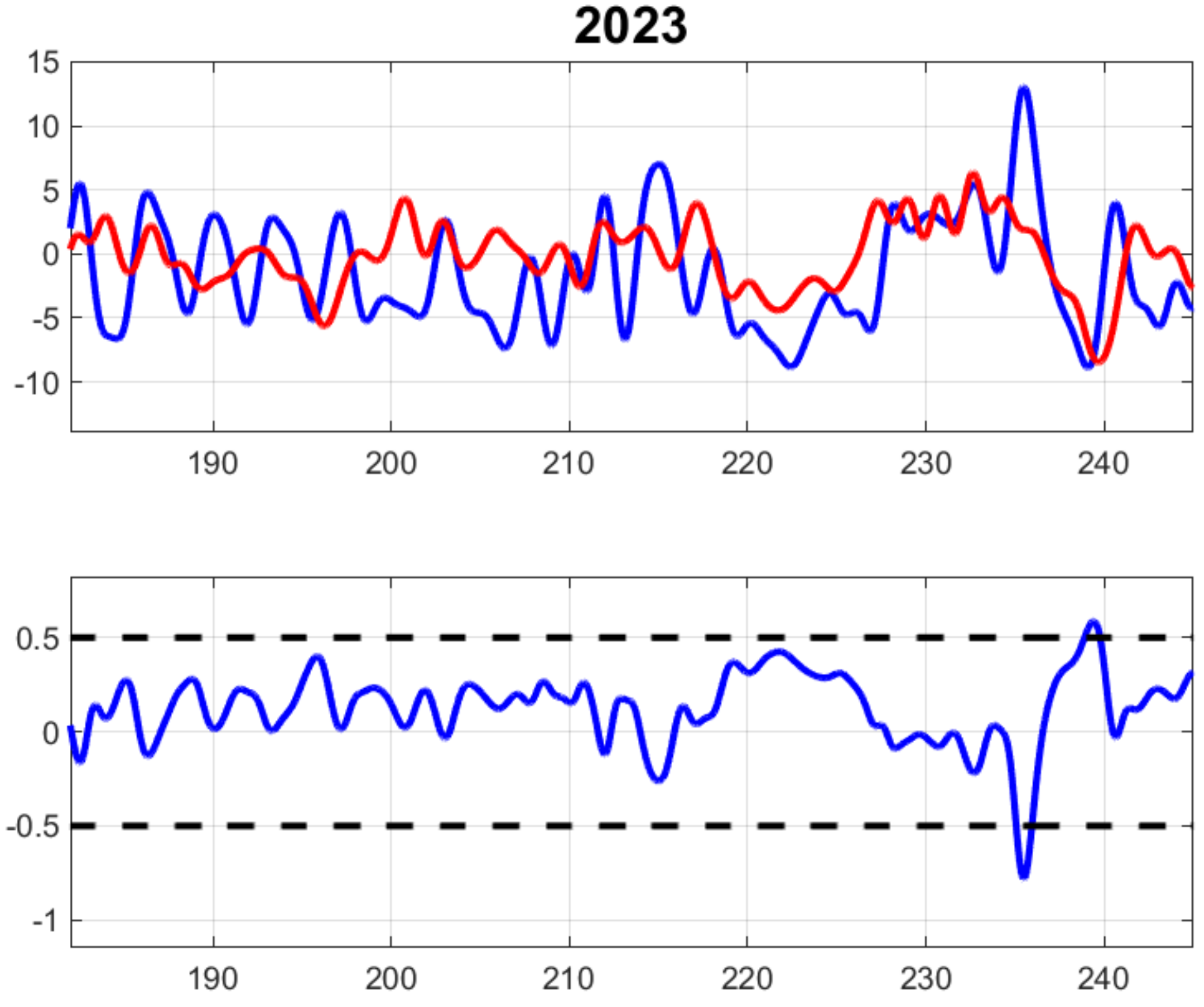

**Figure S80.** Regression results for summer of 2023 (July – August). Top panel: low-pass filtered wind velocity components (m/s). The east and north wind velocity components are blue and red lines, respectively. The axis is Julian days of the year. Lower panel: the regression produced low-pass filtered exchange velocity (m/s) at the Eluitkak Pass. The vertical dashed lines in both panels indicate the timing of the maximum velocity magnitude exceeding the threshold (0.5 m/s). If no vertical dashed line is present, there is no extreme event for that year.

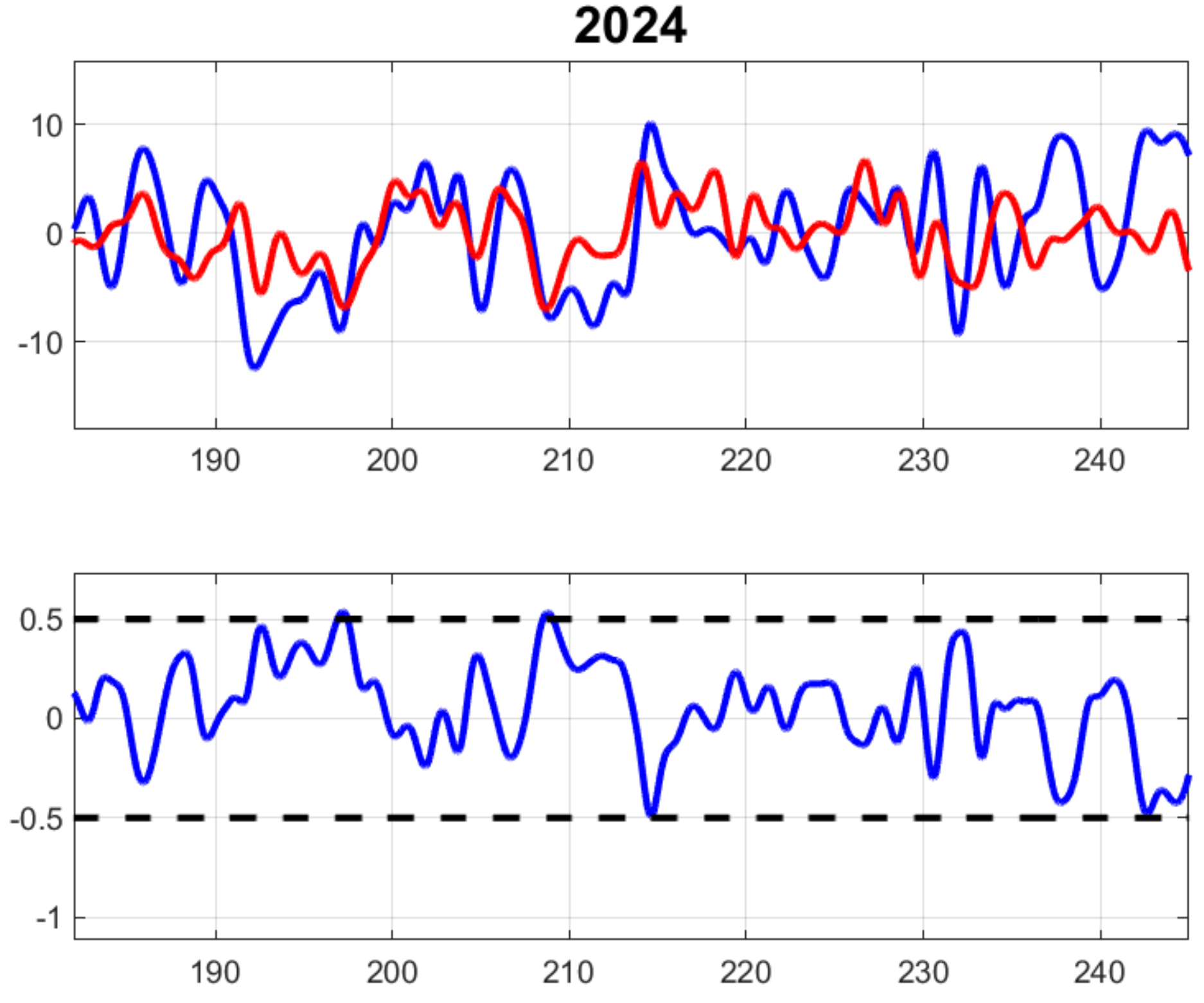

**Figure S81.** Regression results for summer of 2024 (July – August). Top panel: low-pass filtered wind velocity components (m/s). The east and north wind velocity components are blue and red lines, respectively. The axis is Julian days of the year. Lower panel: the regression produced low-pass filtered exchange velocity (m/s) at the Eluitkak Pass. The vertical dashed lines in both panels indicate the timing of the maximum velocity magnitude exceeding the threshold (0.5 m/s). If no vertical dashed line is present, there is no extreme event for that year.

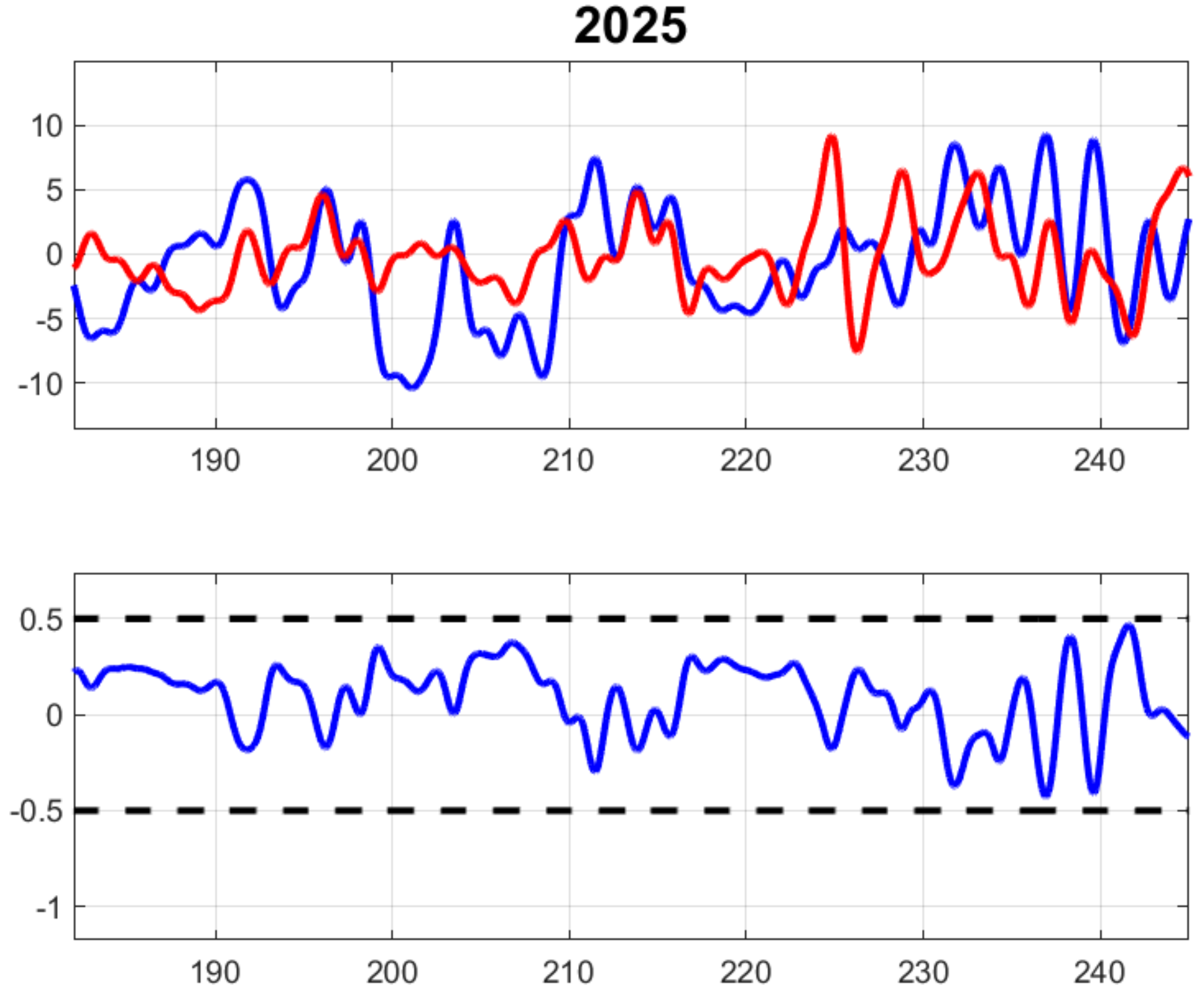

**Figure S82.** Regression results for summer of 2025 (July – August). Top panel: low-pass filtered wind velocity components (m/s). The east and north wind velocity components are blue and red lines, respectively. The axis is Julian days of the year. Lower panel: the regression produced low-pass filtered exchange velocity (m/s) at the Eluitkak Pass. The vertical dashed lines in both panels indicate the timing of the maximum velocity magnitude exceeding the threshold (0.5 m/s). If no vertical dashed line is present, there is no extreme event for that year.

## S5. Results of ERA5 data analysis for a larger region.

The paper presents an analysis of the ERA5 reanalysis data from 2000 to 2025, covering 26 years for July and August. The selected region covers more than 2.78 million $km^2$, spanning latitudes from 65°N to 80°N and longitudes from 170°W to 120°W. This represents an area greater than one-eighth of the Arctic domain. This region was specifically chosen to encompass the operational domain of the Beaufort High and Arctic cyclonic systems directly influencing northern coastal Alaska. Incorporating distant regions, such as the Russian sector or northern Europe, would introduce far-field atmospheric processes not directly related to the local momentum balance of northern Alaska.

To test the spatial sensitivity of the chosen domain, we conducted an experiment expanding the ERA5 analysis region by 15.5%, covering latitudes from 65°N to 82°N and longitudes from 172°W to 118°W.

To calculate the surface area bounded by latitude and longitude lines on the Earth, we integrate the spherical surface area element:

$$A = R^2 \int_{\phi_1}^{\phi_2} d\,\phi \int_{\lambda_1}^{\lambda_2} \cos\theta \; d\theta = R^2(\lambda_2 - \lambda_1)(\sin\phi_2 - \sin\phi_1) \qquad (S1)$$

Here, $R \approx$ 6,371 km is the mean radius of Earth, $\theta$ represents latitude in radians, and $\lambda$ represents longitude in radians.

**1. Original Region Calculation**

- Longitude range: $-170°$ to $-120°$ ($\Delta\lambda = 50° = 0.872665$ rad)
- Latitude range: 65°N to 80°N ($\phi_1 = 65°, \phi_2 = 80°$)

$$\sin(80°) - \sin(65°) \approx 0.984808 - 0.906308 = 0.078500 \qquad (S2)$$

$$A_{\text{original}} = (6371)^2 \times 0.872665 \times 0.078500 \approx 2{,}781{,}586 \text{ km}^2 \qquad (S3)$$

Using equation (S1), the original area is determined to be approximately $2.78 \times 10^6$ $km^2$ (or 2.78 million $km^2$).

**2. Expanded Region Calculation**

- Longitude range: $-172°$ to $-118°$ ($\Delta\lambda = 54° = 0.942478$ rad)
- Latitude range: 65°N to 82°N ($\phi_1 = 65°, \phi_2 = 82°$)

$$\sin(82°) - \sin(65°) \approx 0.990268 - 0.906308 = 0.083960 \quad (S4)$$

$$A_{\text{new}} = (6371)^2 \times 0.942478 \times 0.083960 \approx 3{,}212{,}058\ \text{km}^2 \qquad (S5)$$

The new area is approximately $3.21 \times 10^6$ km$^2$ (or 3.21 million km$^2$).

**3. Percentage Increase**

Using (S3) and (S5), we have

$$\text{Increase} = \frac{A_{\text{new}} - A_{\text{original}}}{A_{\text{original}}} \times 100 = \frac{3{,}212{,}058 - 2{,}781{,}586}{2{,}781{,}586} \times 100 \approx 15.5\% \qquad (S6)$$

Expanding the domain boundaries by $4°$ in longitude and $2°$ in latitude increases total spatial coverage by $15.5\%$.

The resulting EOF patterns are virtually identical: EOF1 changes slightly from $65.9\%$ to $64.4\%$ variance explained, EOF2 shifts from $16.7\%$ to $16.9\%$, and EOF3 remains at $8.3\%$. The corresponding time-series indices and standard deviation metrics remain entirely consistent across both domains, as illustrated in the following figures.

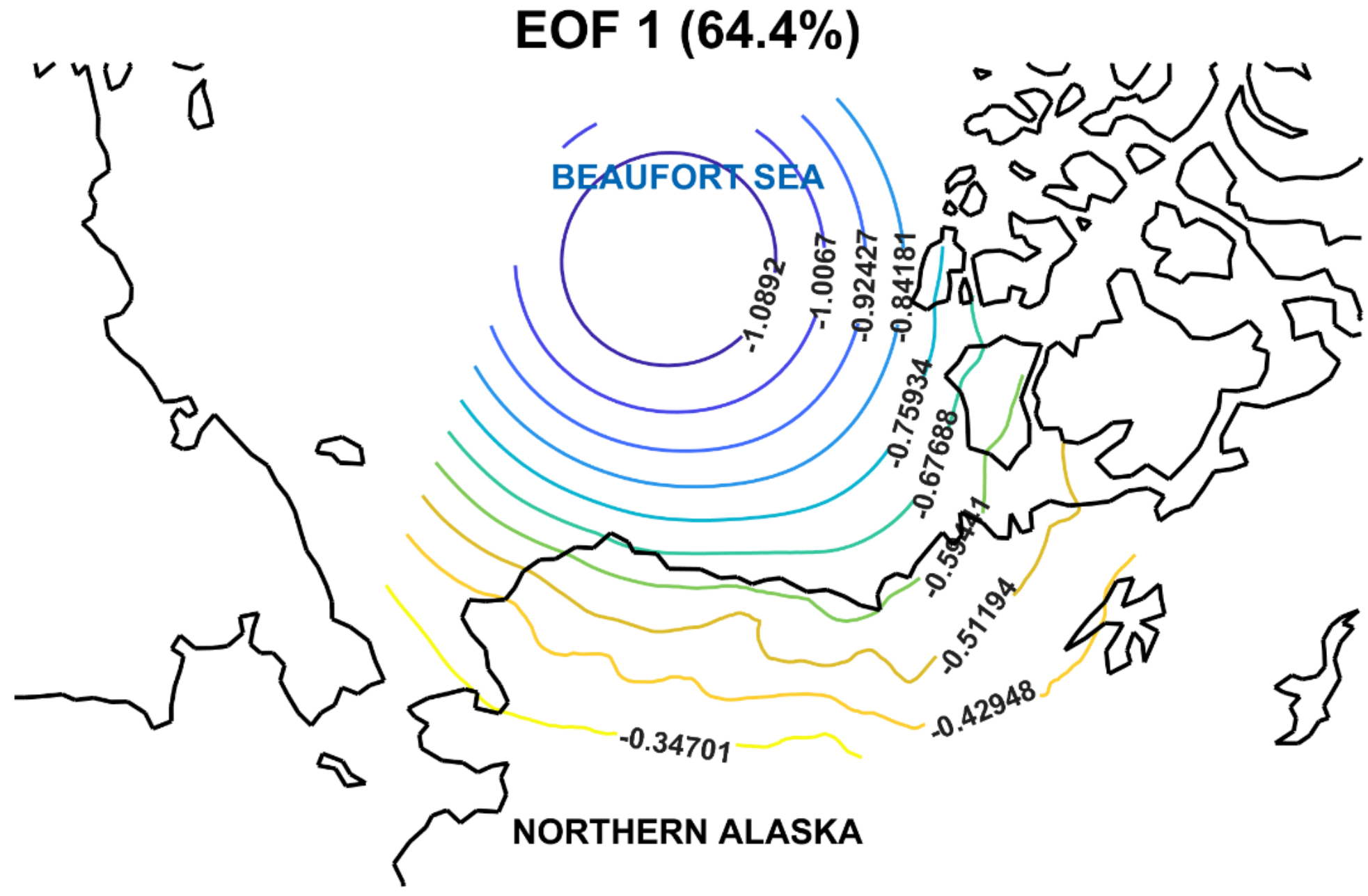

**Figure S83.** EOF1 for the larger data region of ERA5.

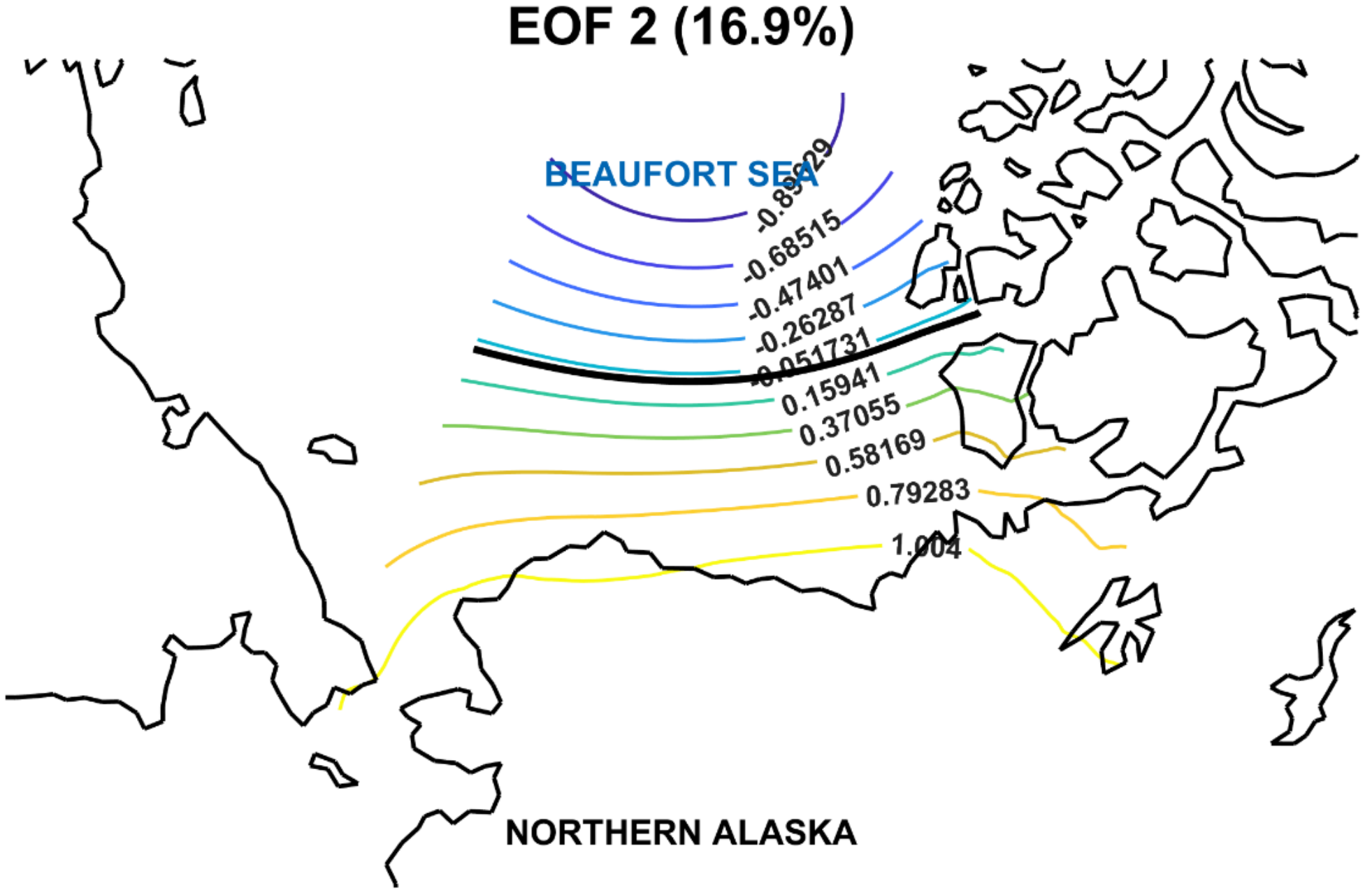

**Figure S84.** EOF2 for the larger data region of ERA5.

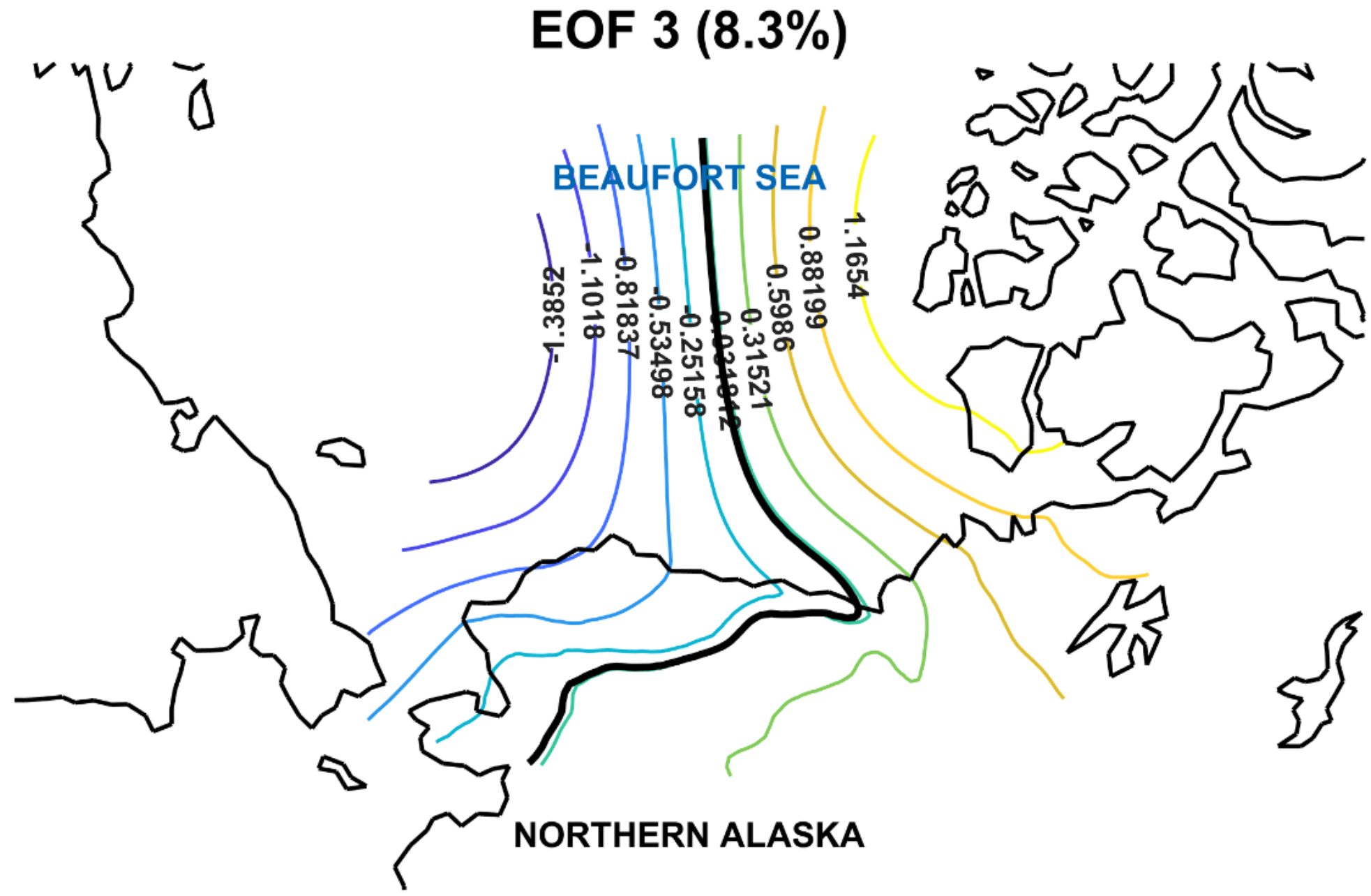

**Figure S85.** EOF3 for the larger data region of ERA5.

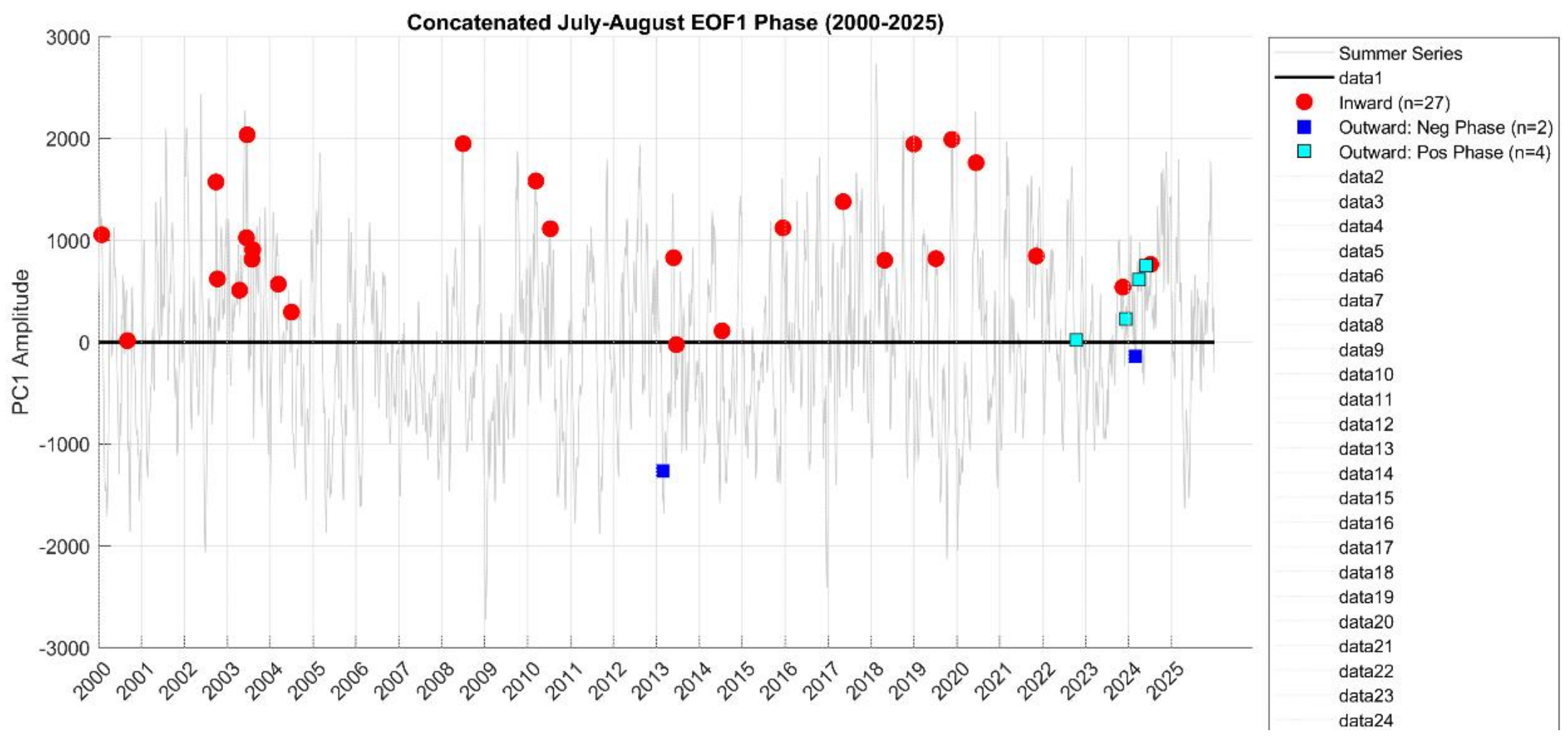

**Figure S86.** PC1 for the larger data region of ERA5.

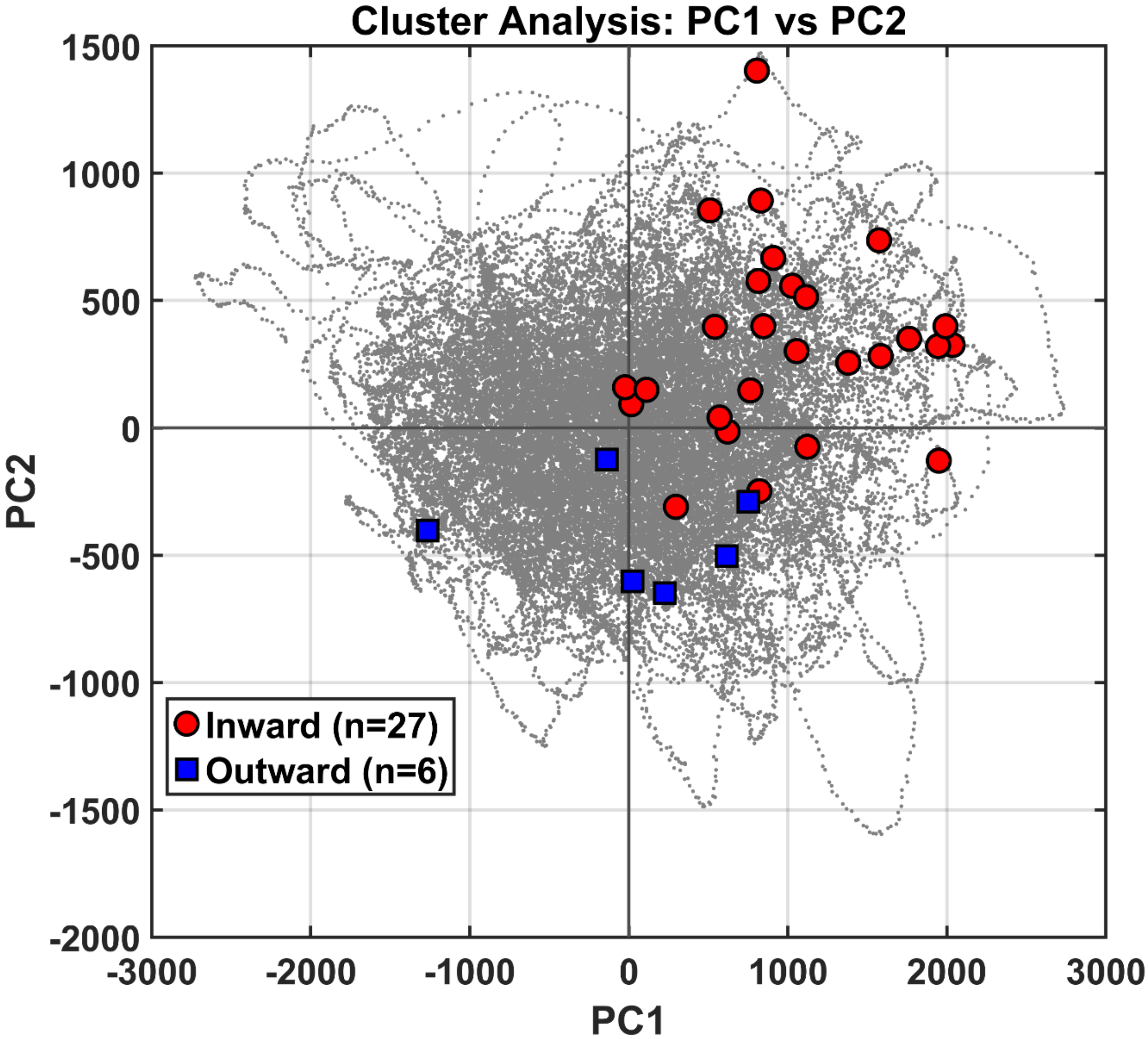

**Figure S87.** PC1 vs PC2 cluster analysis for the larger data region of ERA5.

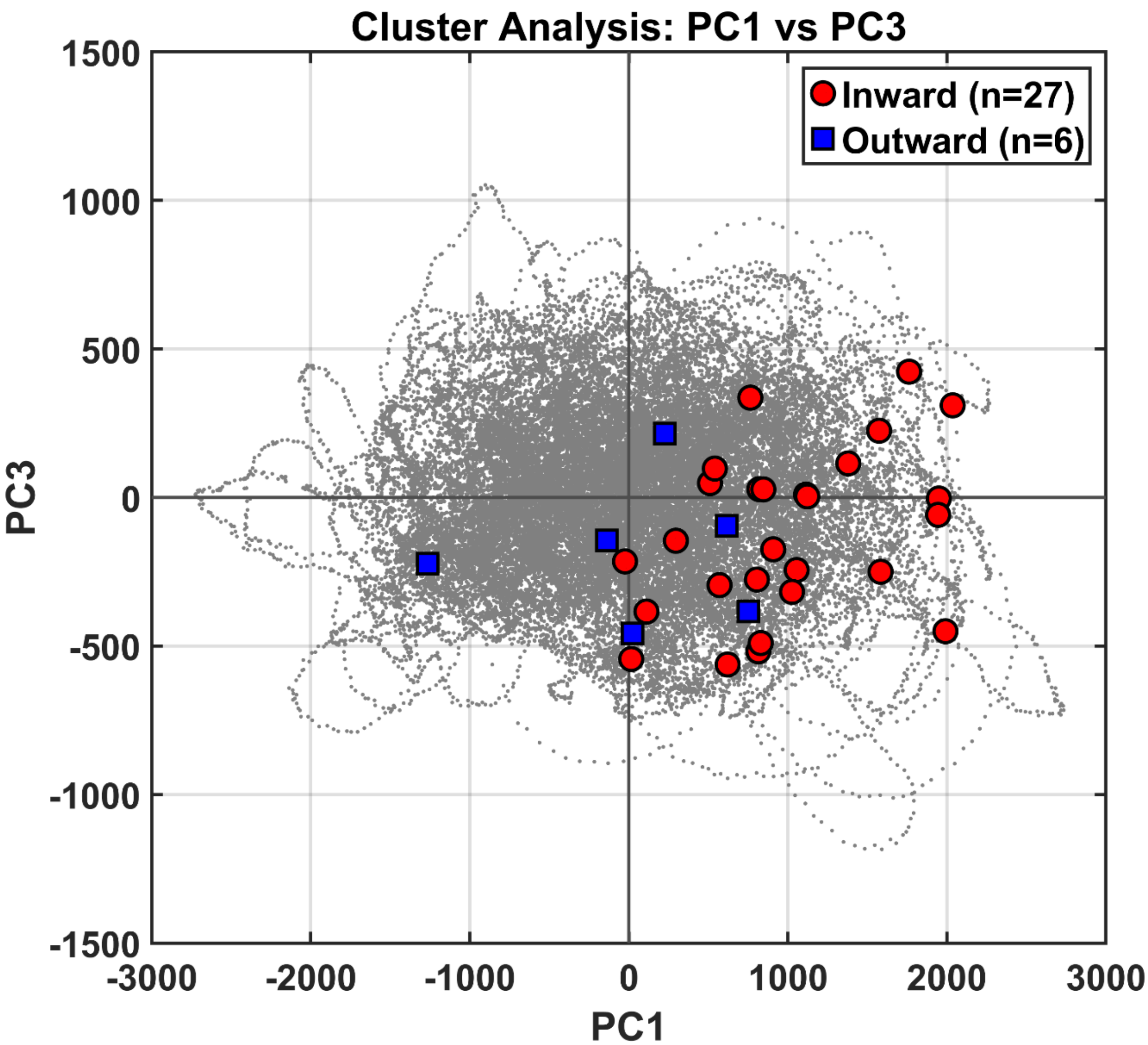

**Figure S88.** PC1 vs PC3 cluster analysis for the larger data region of ERA5.

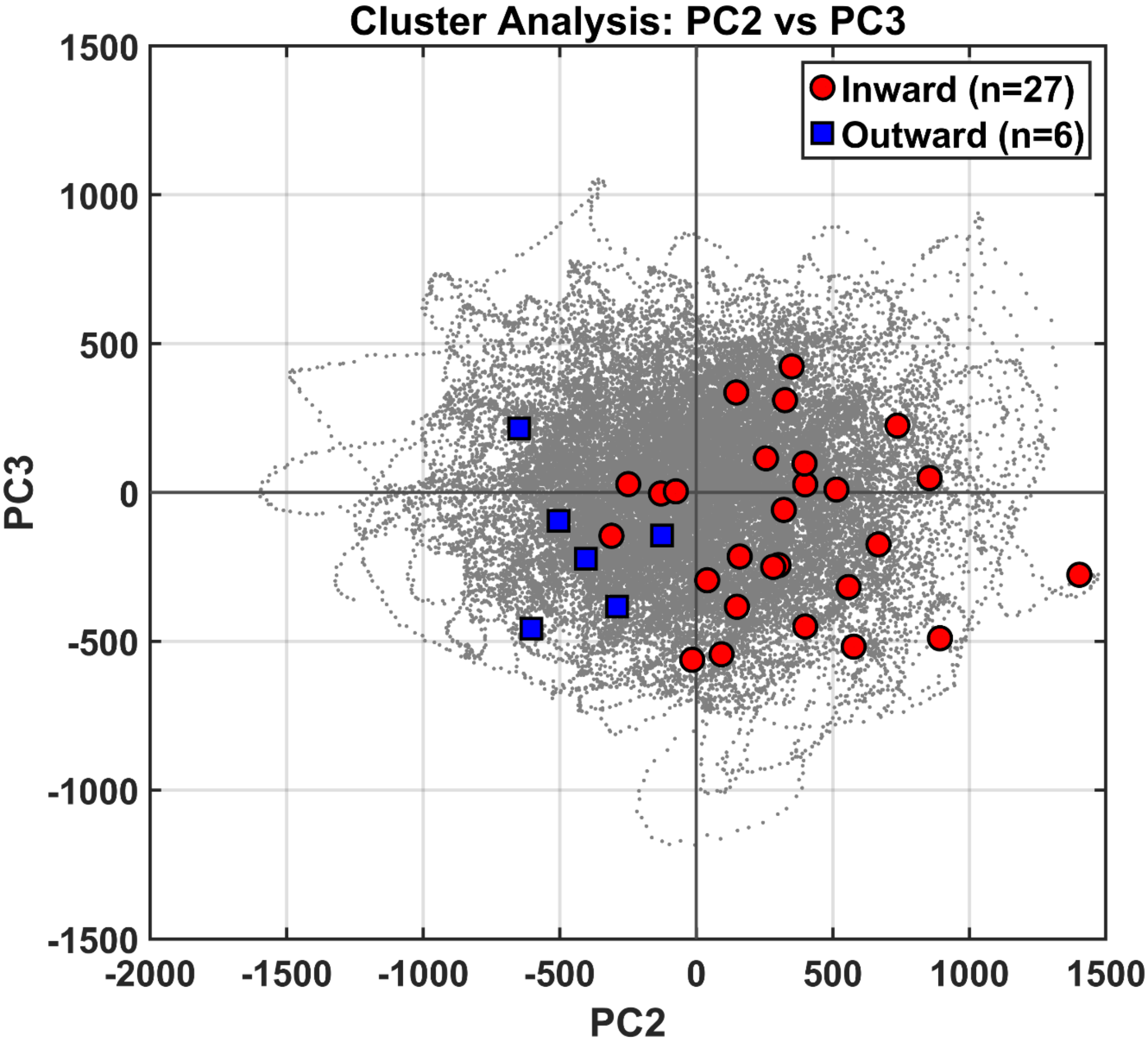

**Figure S89.** PC2 vs PC3 cluster analysis for the larger data region of ERA5.

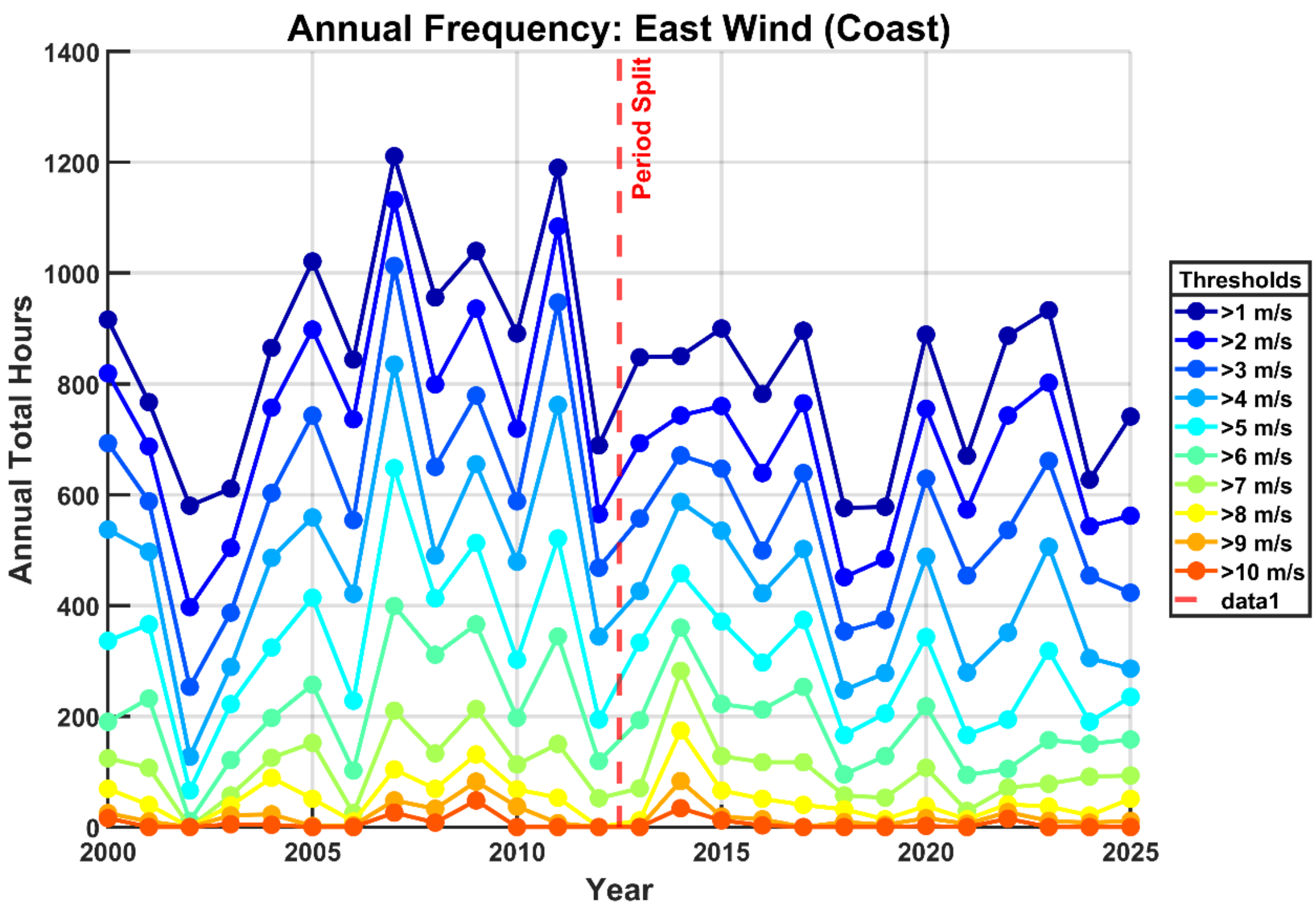

**Figure S90.** Trends in wind frequency between 2000-2025 for the larger data region of ERA5: East Wind (Coast).

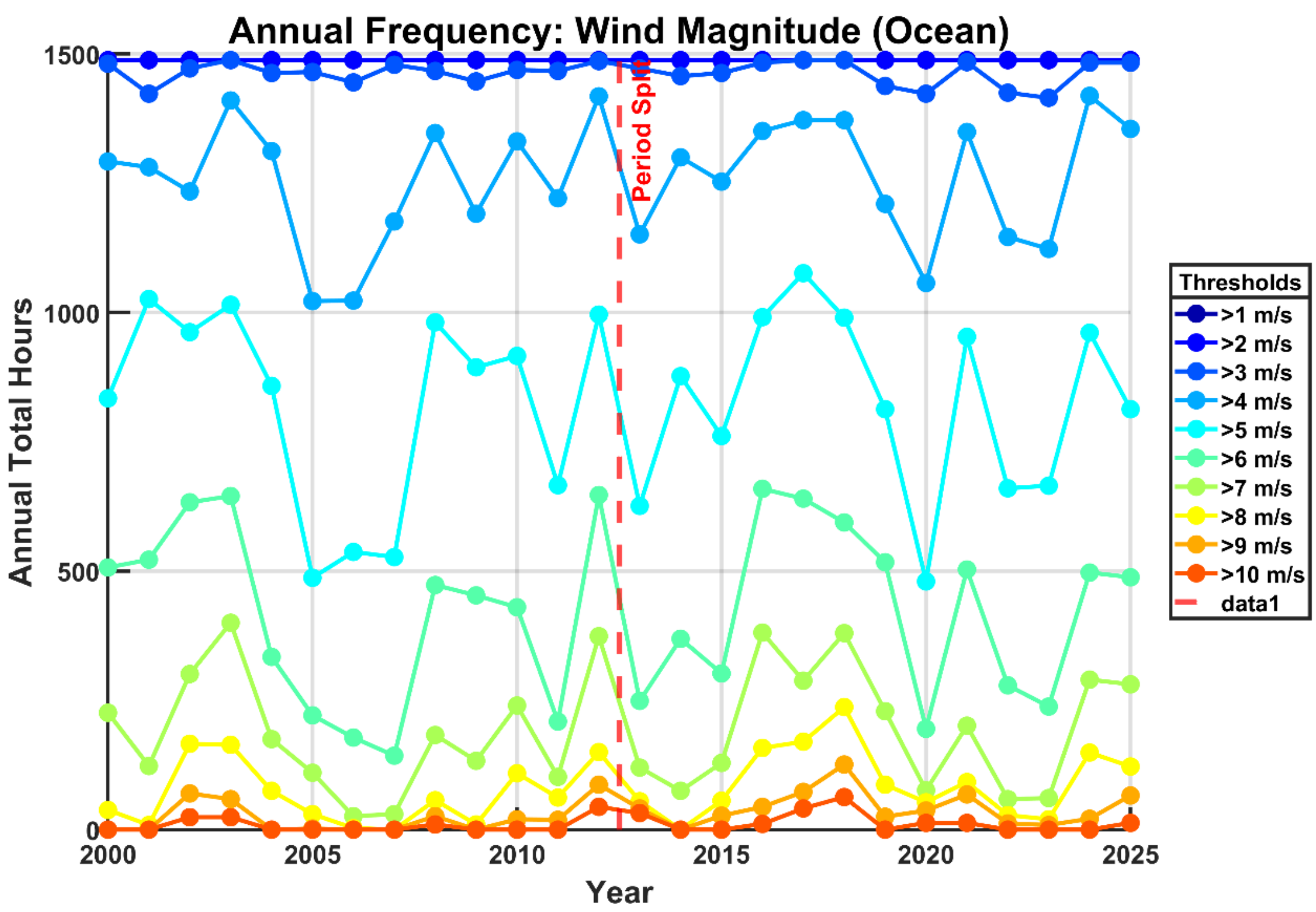

**Figure S91.** Trends in wind frequency between 2000-2025 for the larger data region of ERA5: Wind Magnitude (Ocean).

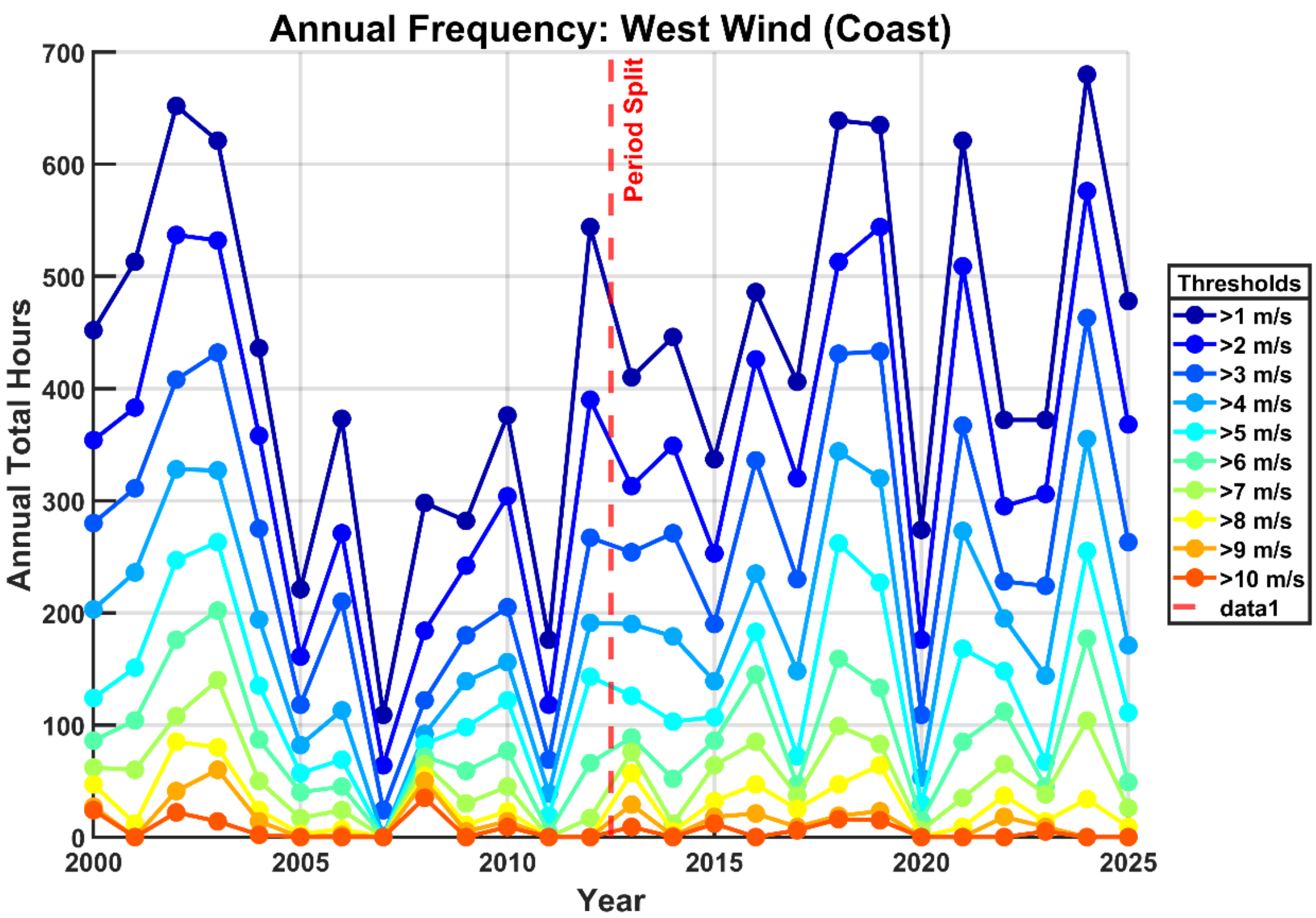

**Figure S92.** Trends in wind frequency between 2000-2025 for the larger data region of ERA5: West Wind (Coast).

# S6. The annual ice-free period averages of indices and their Five-year rolling standard deviations

**Annual means of indices during the ice-free period (July-August) for each year.**

**The indices are:**
1 -- EWWE (the annual ice-free period means of the East-West Wind Excursion) -- unit: km
2 -- IOFE (the annual ice-free period means of the In-Out Flow Excursion) -- unit: km
3 -- PRATIO (the annual ice-free period means of the positive flow ratio or fraction, e.g., 0.78 = 78% of the flow is positive or out of the lagoon)

```
>> [year EWWE/10000   IOFE/10000   PRATIO]

ans =

2000  -7.7155   0.4214   0.7800
2001  -7.9566   0.3980   0.7900
2002  -4.6786   0.1972   0.7000
2003  -5.5525   0.2503   0.7100
2004 -10.0238   0.5174   0.8500
2005  -9.5059   0.4793   0.8400
2006  -5.0040   0.3622   0.7700
2007 -11.8829   0.5302   0.9000
2008  -5.6162   0.4331   0.8300
2009  -7.2497   0.4063   0.8000
2010  -8.4919   0.4870   0.8200
2011  -8.3202   0.4901   0.8300
2012  -2.7152   0.3733   0.7500
2013  -6.1609   0.4485   0.7900
2014  -7.3152   0.4726   0.8100
2015  -7.0563   0.4845   0.7800
2016  -8.2872   0.4636   0.7200
2017  -5.7498   0.3491   0.7800
2018  -7.6736   0.4185   0.7600
2019  -6.3173   0.3616   0.7500
2020  -4.9771   0.3799   0.7700
2021  -4.7615   0.3968   0.7300
2022  -5.6891   0.3461   0.7600
2023  -5.2293   0.4334   0.7600
2024  -5.3915   0.4168   0.7700
2025  -5.4502   0.4475   0.7900
```

**The five-year rolling standard deviation computation variables of the annual ice-free period means of the indices are:**

1 -- stdEWWE (five-year rolling standard deviation of the annual ice-free period means of the East-West Wind Excursion) -- unit: km

2 -- stdIOFE (five-year rolling standard deviation of the annual ice-free period means of the In-Out Flow Excursion) -- unit: km

3 -- stdPRATIO (five-year rolling standard deviation of the annual ice-free period means of the positive flow ratio or fraction, e.g., 0.0619 = 6.19% variations)

```
>> [mid_point_year stdEWWE/1000 stdIOFE/1000 stdPRATIO]
ans =
2002  21.1437   1.3082   0.0619
2003  23.6352   1.4019   0.0705
2004  25.9227   1.3922   0.0702
2005  29.8460   1.1934   0.0744
2006  29.6952   0.6851   0.0466
2007  28.4852   0.6497   0.0487
2008  27.3515   0.6614   0.0483
2009  23.0107   0.4931   0.0378
2010  23.9512   0.5079   0.0336
2011  23.5819   0.5089   0.0311
2012  23.6297   0.4815   0.0316
2013  21.5366   0.4774   0.0303
2014  21.4603   0.4403   0.0354
2015   9.9868   0.5447   0.0336
2016   9.4101   0.5545   0.0332
2017  10.1753   0.5999   0.0249
2018  13.6343   0.4666   0.0230
2019  11.7153   0.2761   0.0192
2020  11.7373   0.2850   0.0152
2021   6.2058   0.3376   0.0152
2022   3.5994   0.3381   0.0164
2023   3.4540   0.3950   0.0217
```